\documentclass[a4paper,11pt]{article}
\pdfoutput=1

\usepackage{jcappub}

\usepackage[T1]{fontenc}
\usepackage[utf8]{inputenc} 
\usepackage[english]{babel}
\usepackage{float}

\title{The Morphology of the Anomalous Microwave Emission in the Planck 2015 data release}

\author[a,1]{S. von Hausegger}
\author[a,b]{H. Liu}

\affiliation[a]{The Niels Bohr Institute and Discovery Center,\\Blegdamsvej 17, 2100 Copenhagen Ø, Denmark}
\affiliation[b]{The Key laboratory of Particle and Astrophysics, Institute of High Energy Physics, CAS, China}

\emailAdd{s.vonhausegger@nbi.dk}
\emailAdd{liuhao@nbi.dk}

\abstract{We calculate weighted mosaic correlations between the recently published Planck 2015 foreground maps -  both anomalous microwave emission (AME) maps, free-free emission, synchrotron radiation and thermal dust emission. The weighting coefficients are constructed taking account of the signal-to-error ratio given by the data product.
Positive correlation is found for AME compared with thermal dust emission as well as synchrotron radiation.
We find AME and free-free emission tending to be anti-correlated, however, when investigating different scales, their relationship appears to be more complex. 
We argue that dust particles responsible for AME are pushed out of hot zones in the interstellar medium (ISM).}

\footnotetext[1]{Corresponding author.}

\keywords{CMBR experiments}

\begin{document}
\maketitle
\flushbottom


\section{Introduction}
\label{sec:intro}

Following the release of the PLANCK full mission data, cosmology and astro-particle physics are confronted with major challenges concerning the verification of the theories of inflation\cite{Guth1981,Mukhanov1981,Linde1982,Albrecht1982}. The rapid expansion at the earliest times of the Universe produces gravitational waves. This cosmological gravity wave background can be probed by precise measurements of the B-mode of polarization of the Cosmic Microwave Background (CMB), as done by \cite{Bicep2, Planck-BK}. 
Also future effort will be related to this mission by a variety of space- and ground-based, experiments (both ongoing and planned)
i.e. BICEP3/KECK, SPIDER, EBEX etc. which will be able to detect B-mode polarization with unprecedented pixel sensitivity.

All these experiments have to seriously take into account the recent Planck result \cite{PlanckIntXXX} that there are no zones in the sky, including the BICEP2/KECK region, which are free of contamination from galactic origin.
In addition, Anomalous Microwave Emission (AME) - a component not well understood and only recently added to foreground separation algorithms - further hampers one in obtaining a clean CMB map in both, intensity and polarization.  

In order to prepare for coming data, a number of fundamental and practical investigations have to be performed. The development of new strategies and methods for the removal of foregrounds - thermal dust emission at frequencies $\nu>100$ GHz as well as synchrotron radiation, free-free emission and AME at $\nu<100$ GHz - is urgently required. 

In this Paper we would like to analyze the morphology of the AME template in relation to other foreground components from the Planck 2015 data release \citep{Planck-foreground}.
In order to clarify the physics of AME, we cross-correlate its template with those of free-free emission, synchrotron radiation and thermal dust emission. 
For this we will use the mosaic correlation method (MCM), discussed in \cite{Verkhodanov2009,Verkhodanov2010,Hansen2012} with some modifications, described below. 
The specialty of the MCM is that it allows one to investigate correlations on different scales. In our analysis we restrict ourselves to regions outside the galactic plane in order to place focus on the non-trivial correlations. Furthermore, we weight the data determined by the signal-to-noise ratios of the respective maps. It is not surprising that we detect very strong correlations of AME with thermal dust and synchrotron emission, widely discussed in the literature \cite{Kogut1996, Oliveira1997, Leitch1997}.
However, we find significant anti-correlations between AME and free-free emission, especially on small scales ($\sim 1^{\circ}$). This can shed new light on the physics of AME. For larger scales ($\sim 7^{\circ}$) the results are strongly dependent on the weights chosen.

The structure of the paper is the following: Section 2 will briefly describe the production of the foreground templates 
by the Planck team. In Section 3 we will introduce the coefficient of correlation used 
in this analysis and as well as the weights used to account for errors. In section 4 we present the results of our analysis, with and without weighting the data. Lastly, section 5 discusses the outcome of our analysis.



\section{The Component Maps}
\label{sec:compmaps}

The Planck 2015 data release provided us with a selection of foreground maps in temperature and polarization of which we select five temperature maps for our analysis. In order to provide the reader with details about the physical processes behind respective components, we recap the essential assumptions for the calculations of the templates by the Planck group in \cite{Planck-foreground}.\footnote{Note, that not all formulas here are presented in their standard form but are rather adopted from \cite{Planck-foreground}.}\\

Synchrotron radiation is emitted by charged particles of relativistic velocities (mostly electrons from cosmic rays) accelerated in the interstellar, galactic, magnetic field. The spectrum arising from this radiation is assumed to have unbroken power law form for frequencies above 20 GHz. The Planck team used the GALPROP code \cite{Moskalenko1998, Strong2007, Orlando2013} to estimate the spectrum caused by cosmic ray electrons. A shift parameter $\nu_p$ acting as a free parameter for the subsequent optimization process was introduced to allow for the use of GALPROP at low frequencies. While this resulted in the determination of a template function $f_s(\nu)$, the 408 MHz Haslam map \cite{Haslam} was used to determine the amplitude $A_s$ of the brightness temperature for synchrotron radiation:
\begin{align}
T_{s}(\nu)=A_s\left(\frac{\nu_0}{\nu}\right)^2\frac{f_s(\frac{\nu}{\alpha})}{f_s(\frac{\nu_0}{\alpha})}
\label{synchroeq}
\end{align} 
where $\nu$, $\nu_0$ and $\alpha$ are the frequency, the reference frequency for the Haslam
408 MHz map, and the shift parameter ($\alpha=0.26$) respectively.\\

AME, a phenomenon detected by \cite{Oliveira1997,Kogut1997,Leitch1997}, currently is assumed to be sufficiently well described by spinning dust - rotating dust grains with an electric dipole moment
 which emit radiation at microwave frequencies, as proposed by \cite{Erickson1957,DL1998a,DL1998b}. 
This model \cite{Haimoud2009, Silsbee2011, SPDUST2} was used by the Planck team to fit the data. Effectively altering the width of the peak in the AME spectrum,  
the template was chosen to contain two spinning dust components, rather than only one, in order to fit the data properly: One had the freedom to change peak frequency  over the entire sky only constrained by a Gaussian prior centered around 19~GHz with a width of 3~GHz, the other kept its peak frequency constant over the sky but could choose it freely fitting the data the best which resulted in a peak frequency of 33.35~GHz. The two components were evaluated at 22.8~GHz and 41.0~GHz respectively, providing the two AME maps (from now on referred to as AME1 and AME2). 
It was further noted that large systematic uncertainties correlated with synchrotron and free-free emission are associated with the spinning dust component fit.\\

Free-free emission arises from hot electrons being accelerated in intergalactic gas 
and regions of ionized hydrogen. The corresponding antenna temperature $s_{ff}$ (in $\mu K_{RJ}$) for free-free emission is given by  
\begin{align}
s_{ff}\left(\nu\right)=10^6\,T_e\left(1-e^{\tau_{ff}}\right),\qquad\tau_{ff}=5.468\cdot10^{-2}\,T^{-3/2}_e\,\nu_9^{-2}\cdot EM\cdot g_{ff}\left(T_e\right),
\label{free}
\end{align}
\vspace{-0.5cm}
\begin{align}
g_{ff}\left(T_e\right)=\log\left\{exp\left[5.96-\frac{\sqrt{3}}{\pi}\log\left(\nu_9T^{-3/2}_4\right)\right]+e\right\}, \qquad T_4\equiv T/10^4 K,\qquad\nu_9\equiv\nu/1 GHz\nonumber
\end{align}
where $\tau_{ff}$ is the optical depth for free-free emission, $g_{ff}$ is the corresponding Gaunt factor and $EM$ is the emissivity measure.
The model chosen for the template used by Planck includes 
two free parameters, the emissivity measure $EM$ given by the line-of-sight integration $\int_{LOS} dl \,n_e^2$ and the electron temperature $T_e$. In Planck's component separation algorithm Commander priors were chosen as $log(EM)\sim \text{Uniform}(-\infty,+\infty)$ and $T_e\sim N(7000\,K\pm 500 \,K)$.\\

The thermal dust model used is a one component gray body model with a varying spectral index:
\begin{eqnarray}
 s_d(\nu)= A_d\left(\frac{\nu}{\nu_d}\right)^{1+\beta_d}\frac{\exp(\gamma\nu_d)-1}{\exp(\gamma\nu)-1},\hspace{0.3cm}\gamma=\frac{h}{kT_d}
\label{dust}
\end{eqnarray}
where $\nu_d=545$ GHz, $\beta_d$ is  the spectral index, $T_d$ is the dust temperature, $A_d$ is an amplitude (including terms of i.e. the black body spectrum as well as the optical depth), $h$ is the Planck constant and $k$ is the Boltzman constant. The priors were chosen as $\beta_d\sim N(1.55\pm 0.1)$ and $T_d\sim N(23\pm 3$K). It is noted that the cosmic infrared background will be absorbed by this model as well.\\

\begin{figure}[floatfix,h!]
\centering
\begin{minipage}{0.24\textwidth}
\centering
\includegraphics[trim=0.3cm 2cm 0.3cm 2cm,clip=true,width=5cm]{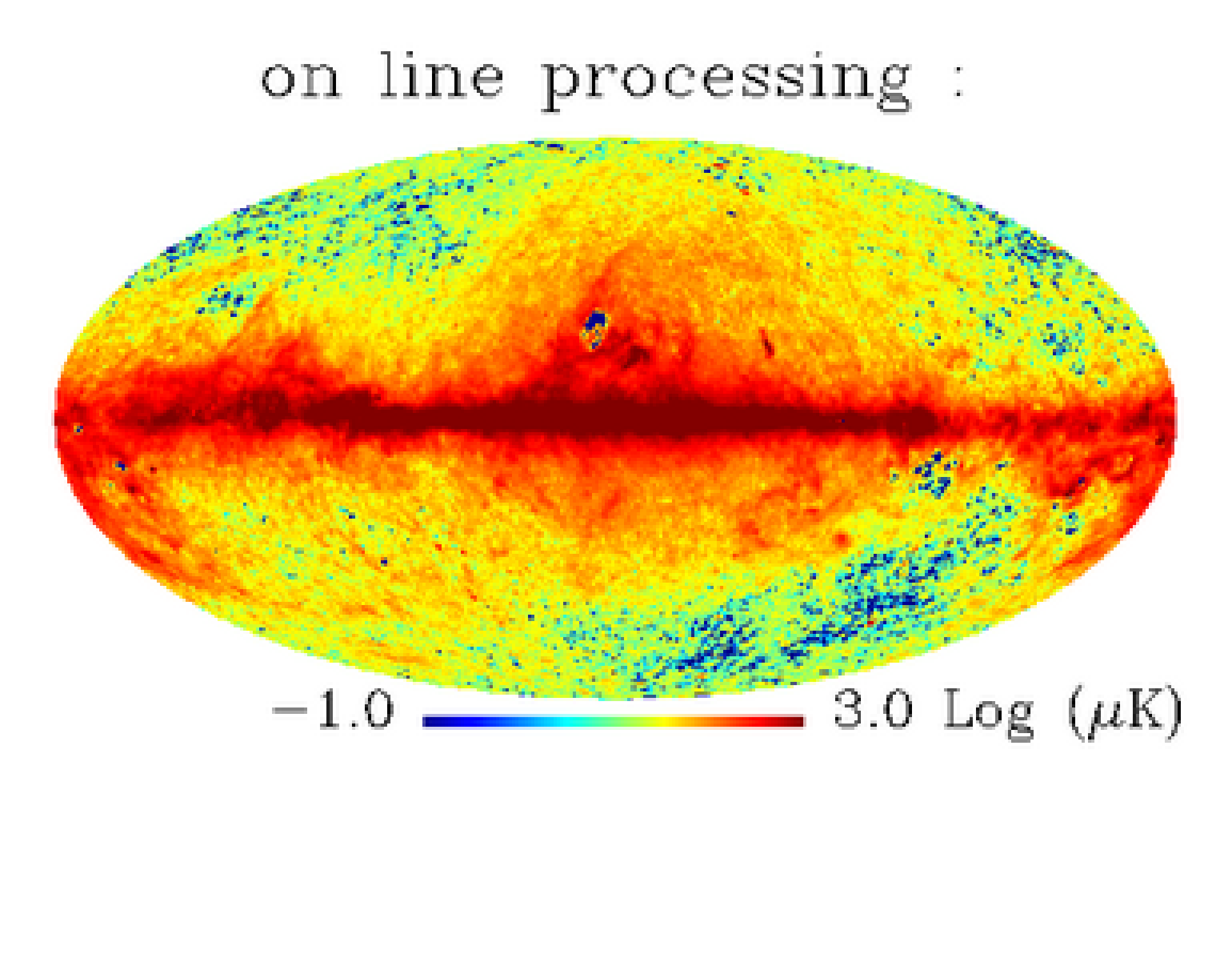}
\end{minipage}
\hspace{1cm}
\begin{minipage}{0.24\textwidth}
\centering
\includegraphics[trim=0.3cm 2cm 0.3cm 2cm,clip=true,width=5cm]{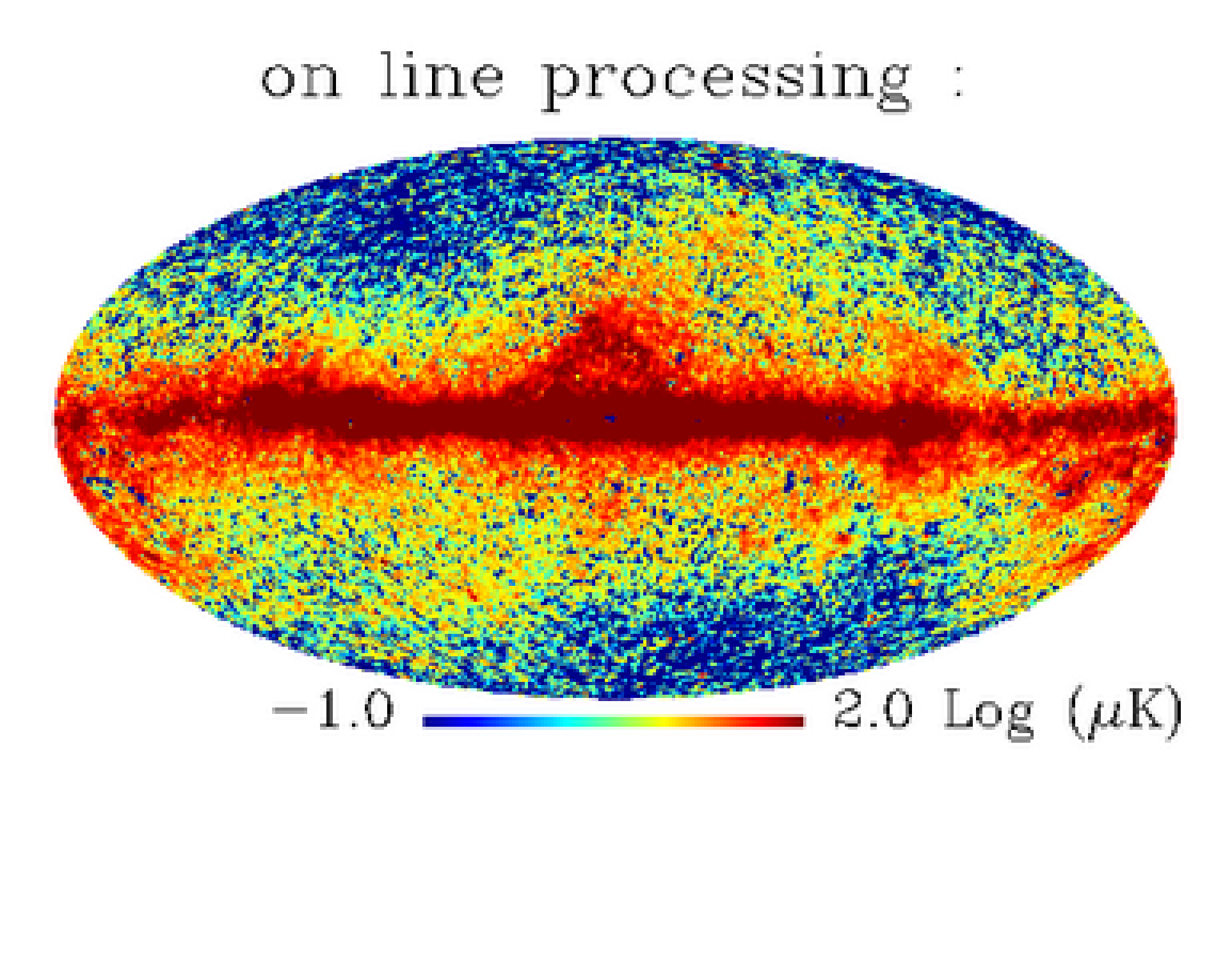}
\end{minipage}
\hspace{1cm}
\centering
\begin{minipage}{0.24\textwidth}
\centering
\includegraphics[trim=0.3cm 2cm 0.3cm 2cm,clip=true,width=5cm]{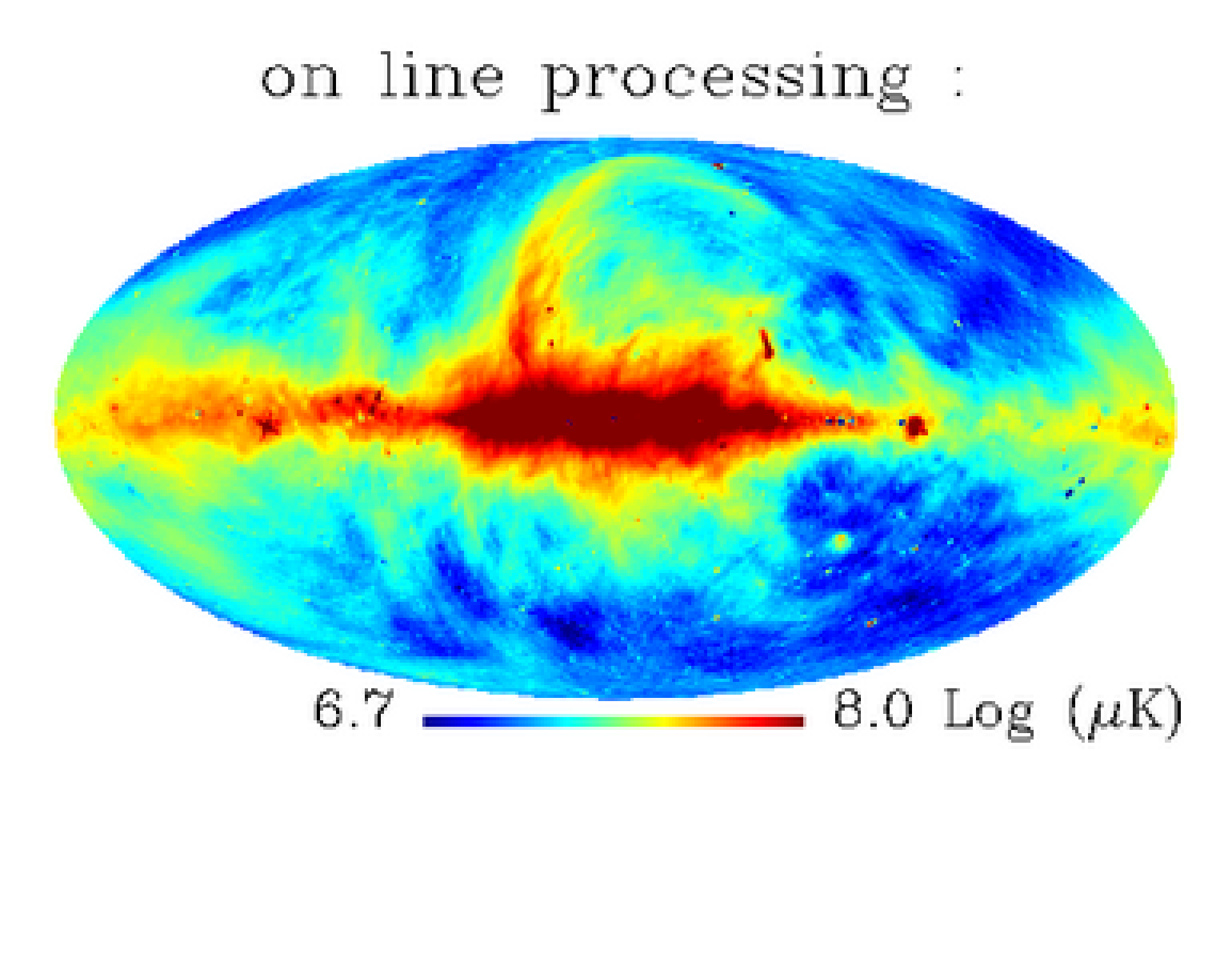}
\end{minipage}
\begin{minipage}{0.24\textwidth}
\centering
\includegraphics[trim=0.3cm 2cm 0.3cm 2cm,clip=true,width=5cm]{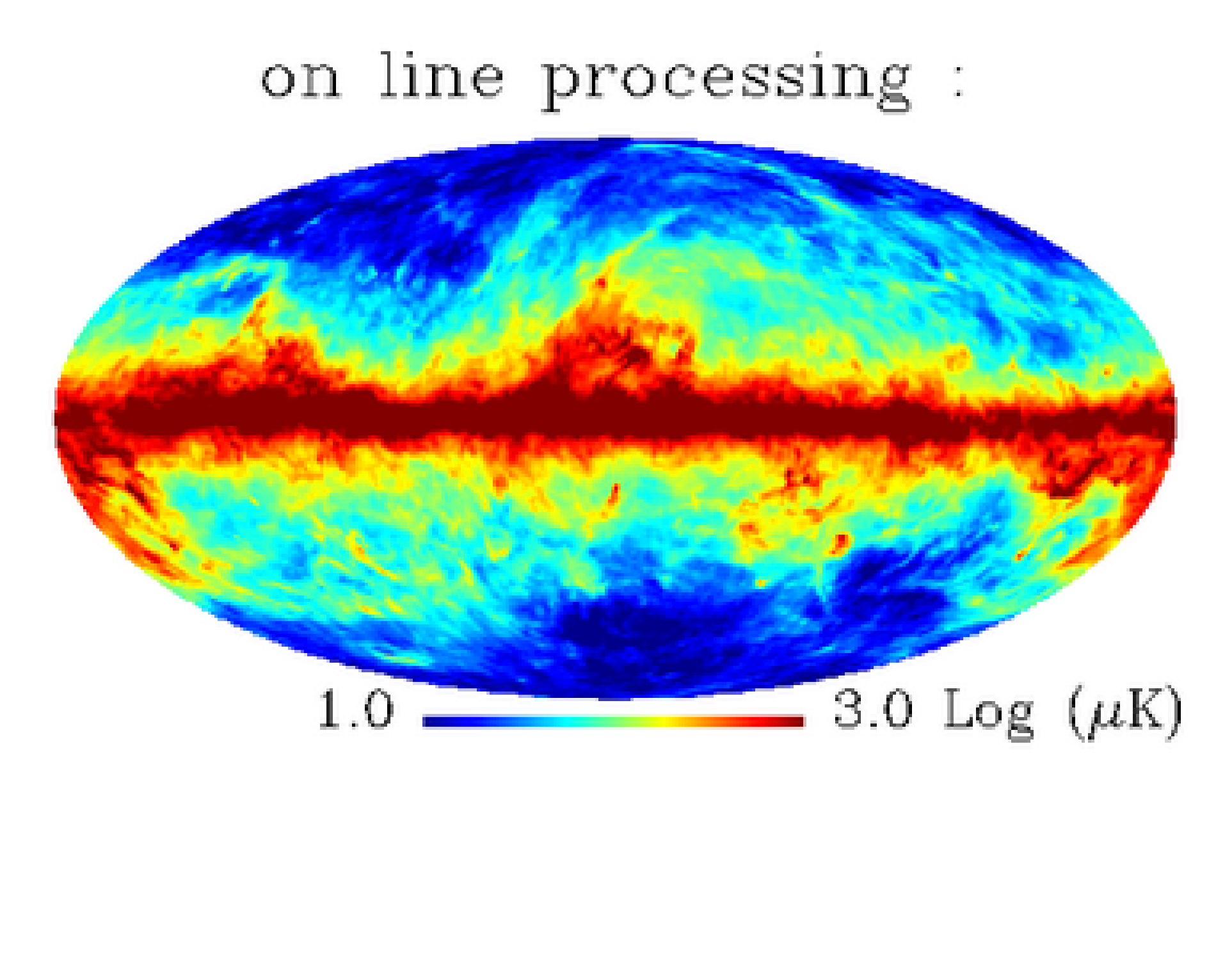}
\end{minipage}
\hspace{1cm}
\begin{minipage}{0.24\textwidth}
\centering
\includegraphics[trim=0.3cm 2cm 0.3cm 2cm,clip=true,width=5cm]{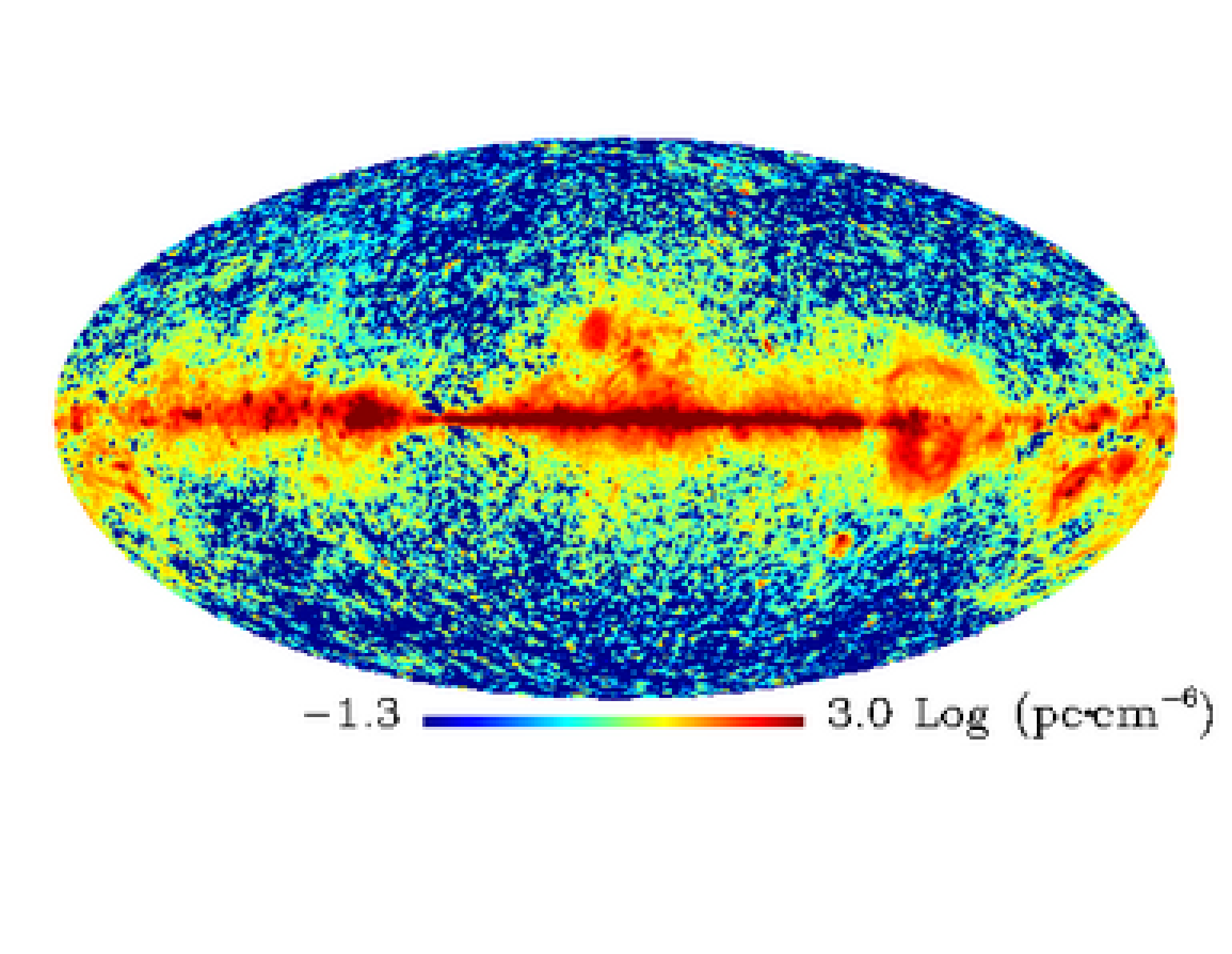}
\end{minipage}
\caption{The Planck 2015 foreground templates. From left to right and top to bottom: AME1, AME2, synchrotron radiation, thermal dust emission and free-free emission. Here, the maximum likelihood maps are shown.}
\label{foregroundmaps}
\end{figure}

All discussed maps can be seen in figure \ref{foregroundmaps}. Unlike the others, the free-free emission map comes in units of $pc\, cm^{-6}$ for the emissivity measure than in $K$ for the brightness temperature. However, as can be seen from eq. \ref{free}, since $\tau_{ff}$ is very small, $s_{ff}(\nu)$ is approximately proportional to $\tau_{ff}$, and thereby proportional to $EM$. Therefore, the morphology of the temperature map is hardly any different from that of the EM map which justifies the use of the EM map for our correlations.

All foreground maps come with various extensions. Amongst them we find the map obtained by a maximum likelihood method within the Commander algorithm, a mean value map and a RMS map including the standard deviations for the mean value map. In our analysis, we will use the maximum likelihood map for calculations not including the errors. When regarding the error map, however, for obvious reasons, we will use the mean value map.


\section{Mosaic Correlation}

In general, a correlation of two maps in pixel space measures common spatial properties. However, these properties might only show up on certain scales. Recently, in a work by Hensley and Draine \cite{Hensley2015} correlations between
Planck's AME and parameters of thermal dust emission were analyzed for the masked
sky, based on a maximum likelihood approach. The conclusion given in this paper, that
AME most likely originates from magnetic dust rather than spinning dust, cannot be ignored in future work regarding foreground estimation.
The MCM, as mentioned in sec. \ref{sec:intro}, provides a means of mapping out regions of correlation on different scales. Since we also attempt to weight the data with appropriate coefficients, related to the errors given by the data product, we introduce the weighted correlation coefficient. Further, we also describe a method of assessing the significance of found correlations suiting our purposes.

\subsection{Weighted Mosaic Correlation}

The Mosaic Correlation method computes the correlation between two maps in defined regions, which collectively cover the entire sky without overlapping. The Healpix \cite{Healpix} pixelization offers a convenient way of doing so: pixels at lower resolution than the original resolution of the maps mark those regions in which the correlation should be computed. Only one value of the correlation coefficient is obtained per "mother pixel". Thereby a full map of correlation values is created. Choosing a different resolution of course changes the size of the mother pixels and thus the scale at which the correlation maps are investigated. In the following we denote a mother pixel by $\Omega$. Eq. \ref{cov} defines the weighted and unbiased sample covariance between two signals $S_1$ and $S_2$ in $\Omega$. $w_i(p)$ stands for the weight of signal $i$ in pixel $p$.
\begin{align}
\label{cov}
cov(\Omega,S_1,S_2)&=f(\Omega)\cdot\sum_{p\in\Omega}\sqrt{w_1(p)w_2(p)}\left(S_1(p)-\overline{S_{1,w}(\Omega)}\right)\left(S_2(p)-\overline{S_{2,w}(\Omega)}\right)\\
f(\Omega)&=\frac{\sum_{p\in\Omega}\sqrt{w_1(p)w_2(p)}}{\left(\sum_{p\in\Omega}\sqrt{w_1(p)w_2(p)}\right)^2-\sum_{p\in\Omega}w_1(p)w_2(p)}\nonumber
\end{align}
$f(\Omega)$  reduces to its well known form of $f(\Omega)=1/\left(N_{\Omega}-1\right)$ if we set \mbox{$w_i(p)=1, \forall p\in\Omega, \forall i$,} where $N_{\Omega}$ is the number of pixels within $\Omega$. With $\overline{S_{i,w}(\Omega)}$ we denote the weighted average of $S_i(p)$ in $\Omega$, defined as:
\begin{align}
\overline{S_{i,w}(\Omega)}=\frac{\sum_{p\in\Omega}w_iS_i(p)}{\sum_{p\in\Omega}w_i}
\end{align}
Finally, the weighted and unbiased correlation coefficient $K$ in $\Omega$ can be calculated as:
\begin{align}
K(\Omega)=\frac{cov(\Omega,S_1,S_2)}{\sqrt{cov(\Omega,S_1,S_1)\cdot cov(\Omega,S_2,S_2)}}
\label{coeff}
\end{align}

Note, that due to the normalization, the monopole component of the respective maps has no impact on K - only the structure in the sky is evaluated.

\subsection{Weights}
\label{sec:weights}

Comparing the provided RMS maps with the respective signals reveals that the errors are of considerable order. Especially the free-free component's relative uncertainty is very high at high galactic latitudes. We adopt for our analysis a weighting coefficient whose rigor can be tuned by choice of the parameter $n$:
\begin{align}
w_i(p)=\frac{1}{1+\left(\frac{\sigma_i(p)}{S_i(p)}\right)^n}
\label{weights}
\end{align}
$\sigma_i(p)$ denotes the RMS value for signal $S_i(p)$. Very high values of $\sigma_i(p)/S_i(p)$ will result in a weight close to 0 whereas low ratios approach 1. Furthermore, larger values of $n$ correspond to weights chosen with greater rigor. A very high $n$ will effectively produce a mask only regarding regions of a signal-to-error ratio larger than 1.

\begin{figure}[floatfix,h!]
\centering
\begin{minipage}{0.24\textwidth}
\centering
\includegraphics[trim=0cm 3cm 0cm 2cm,clip=true,width=5cm]{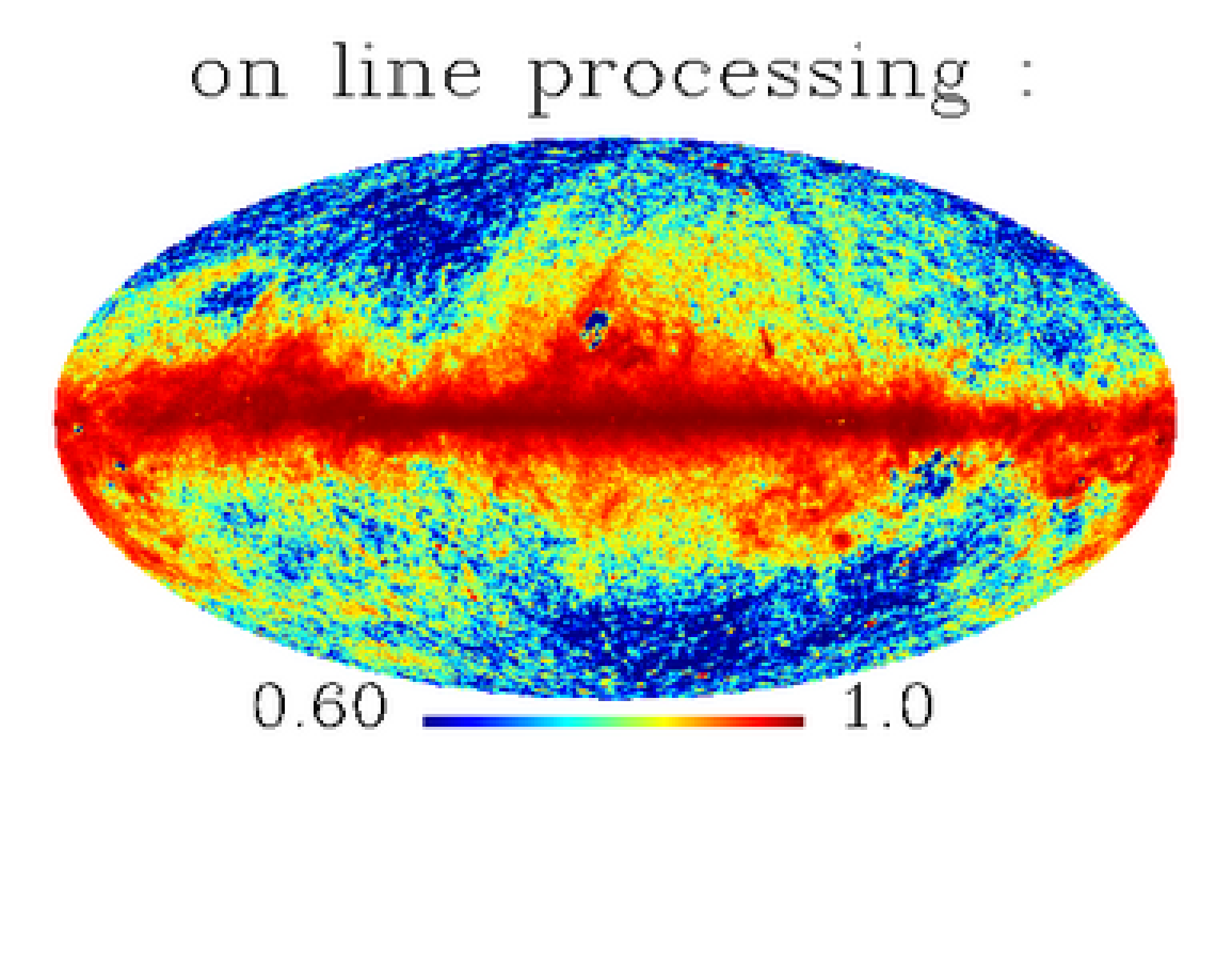}
\end{minipage}
\hspace{1cm}
\begin{minipage}{0.24\textwidth}
\centering
\includegraphics[trim=0cm 3cm 0cm 2cm,clip=true,width=5cm]{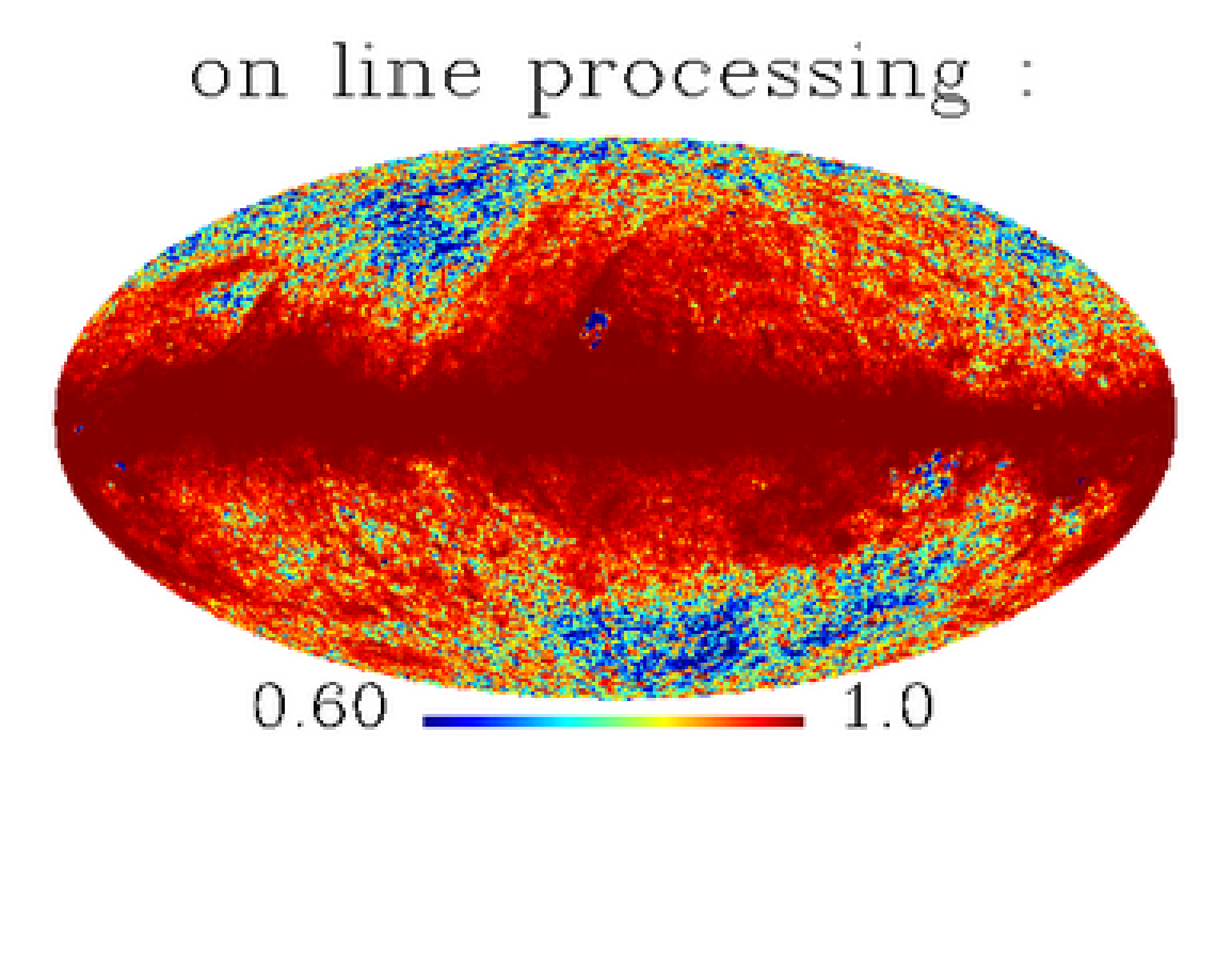}
\end{minipage}
\hspace{1cm}
\centering
\begin{minipage}{0.24\textwidth}
\centering
\includegraphics[trim=0cm 3cm 0cm 2cm,clip=true,width=5cm]{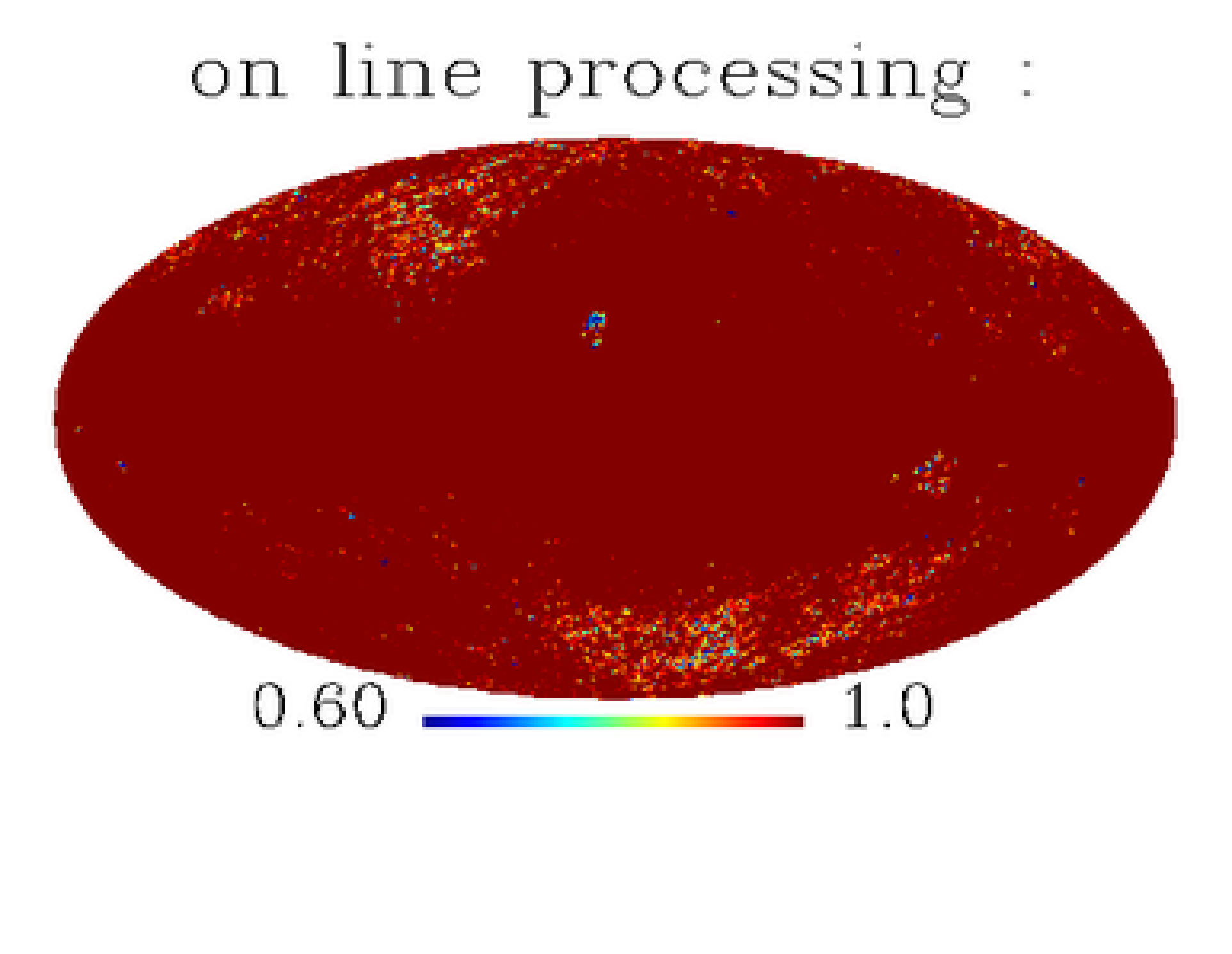}
\end{minipage}
\begin{minipage}{0.24\textwidth}
\centering
\includegraphics[trim=0cm 3cm 0cm 2cm,clip=true,width=5cm]{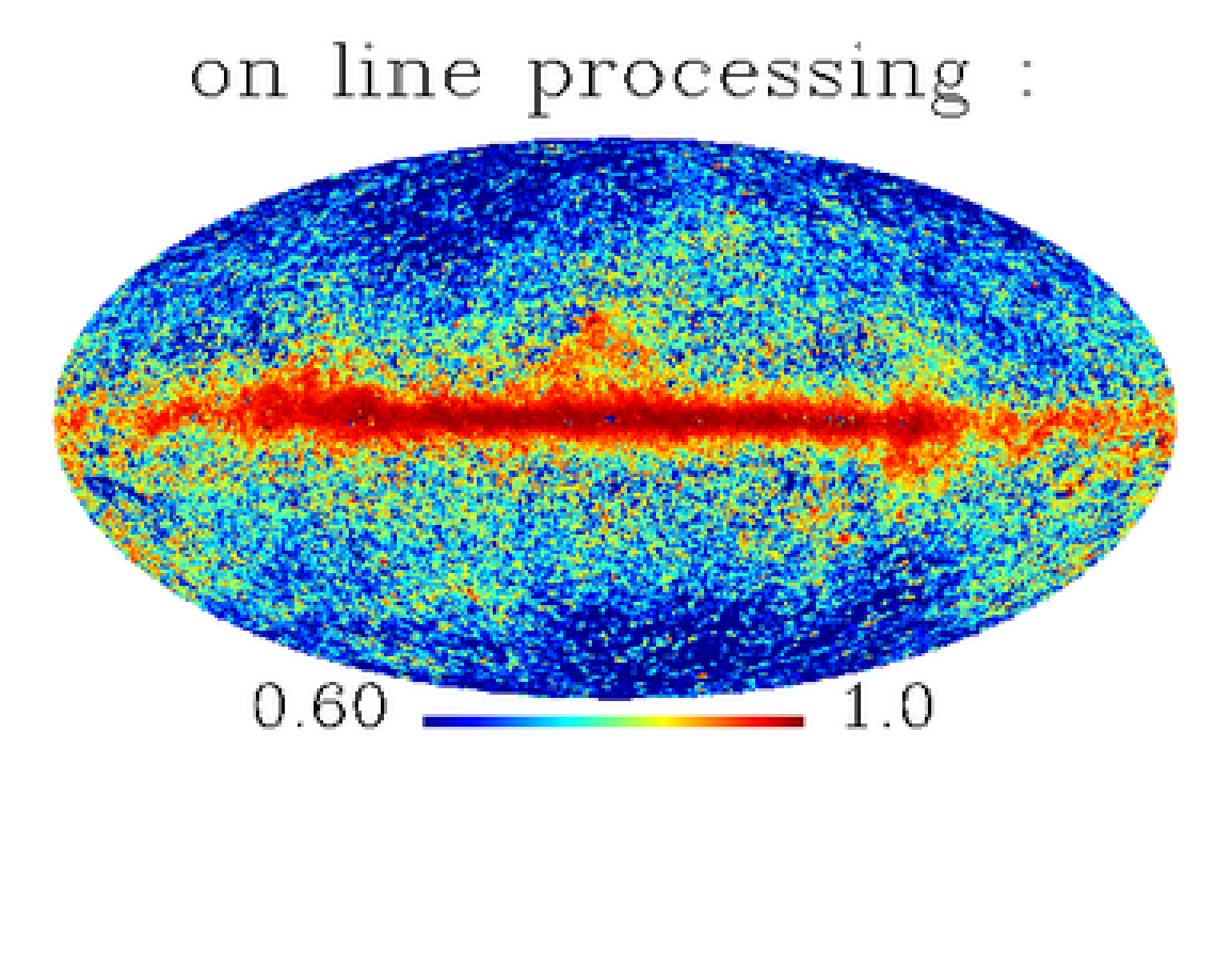}
\end{minipage}
\hspace{1cm}
\begin{minipage}{0.24\textwidth}
\centering
\includegraphics[trim=0cm 3cm 0cm 2cm,clip=true,width=5cm]{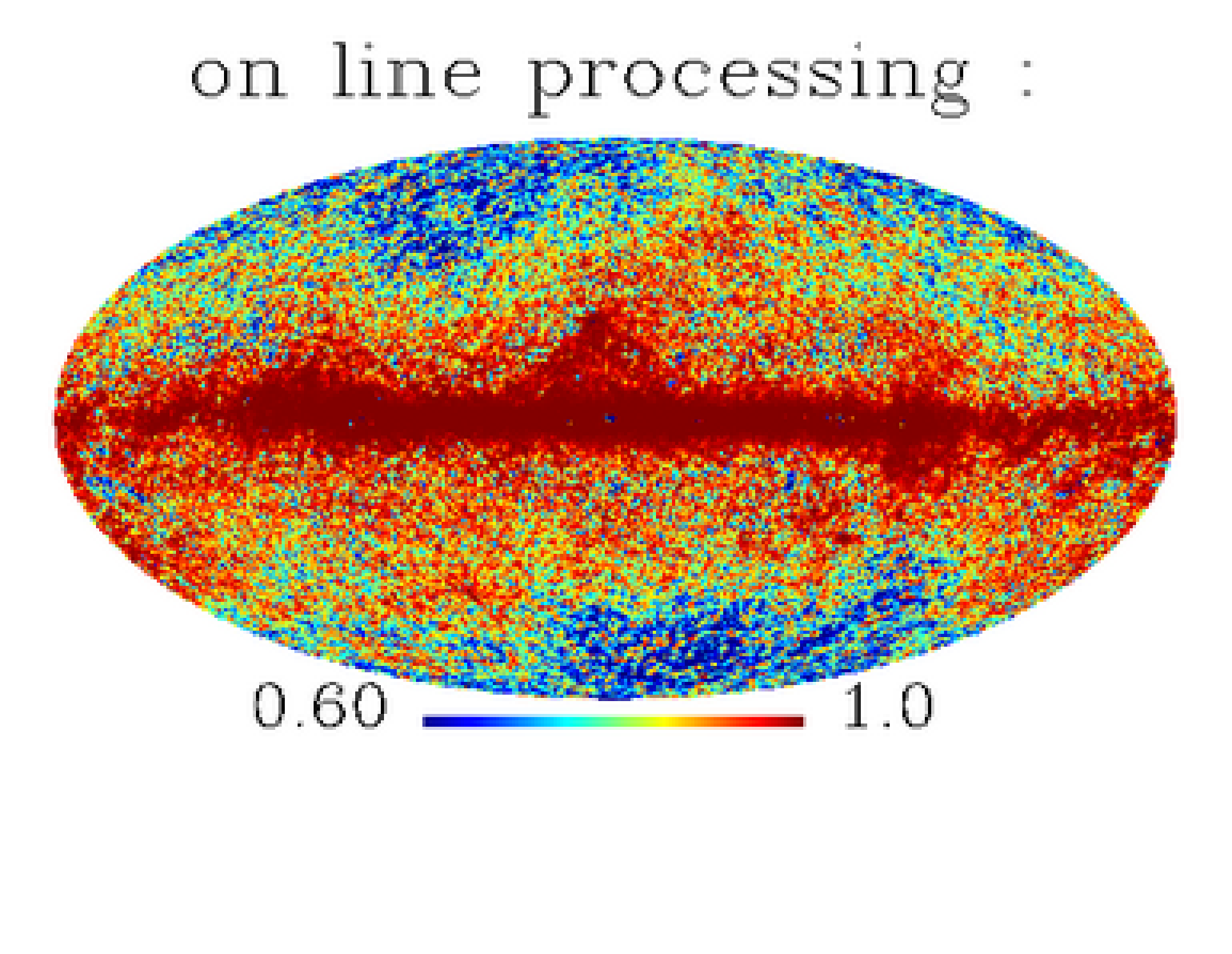}
\end{minipage}
\hspace{1cm}
\centering
\begin{minipage}{0.24\textwidth}
\centering
\includegraphics[trim=0cm 3cm 0cm 2cm,clip=true,width=5cm]{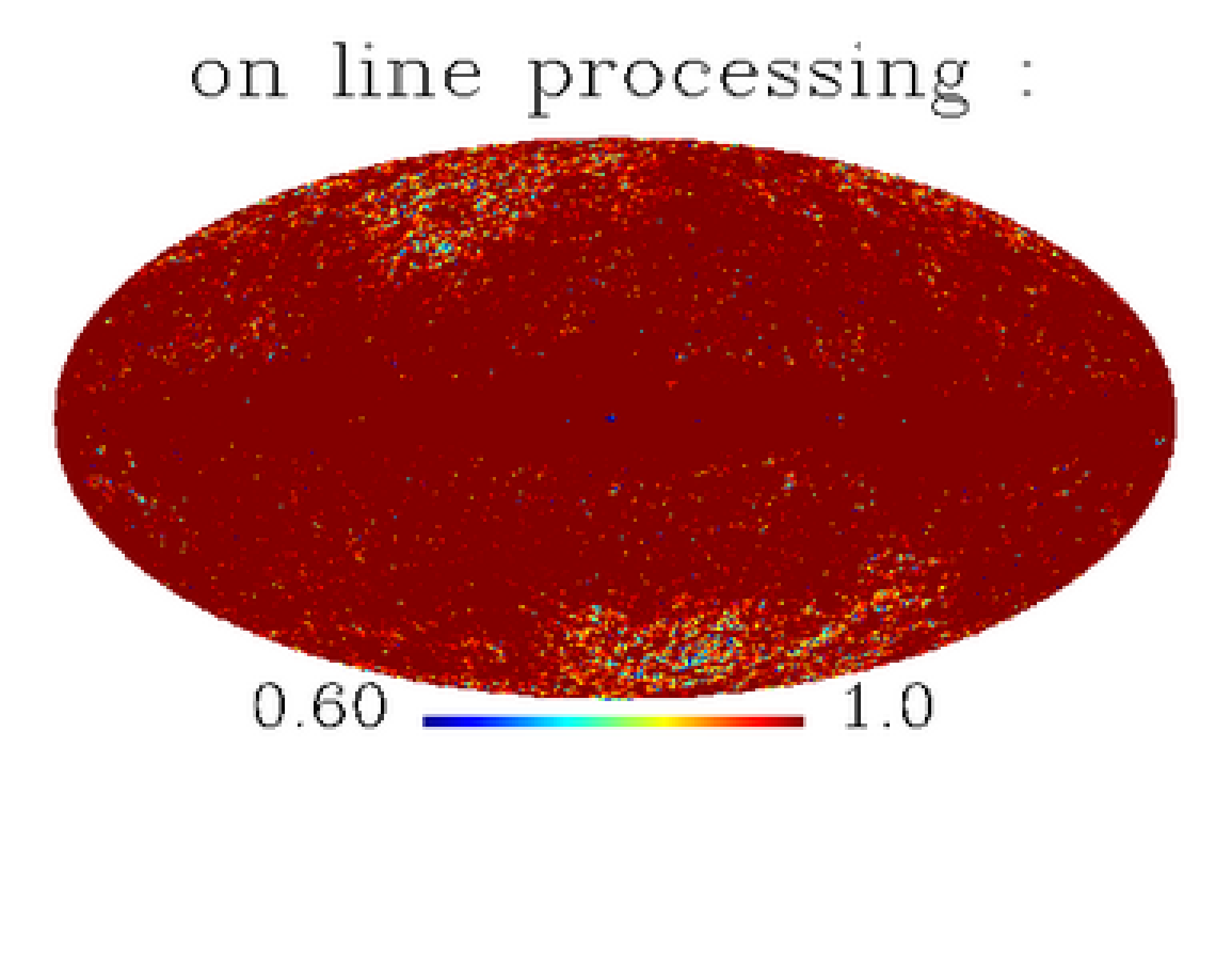}
\end{minipage}
\begin{minipage}{0.24\textwidth}
\centering
\includegraphics[trim=0cm 3cm 0cm 2cm,clip=true,width=5cm]{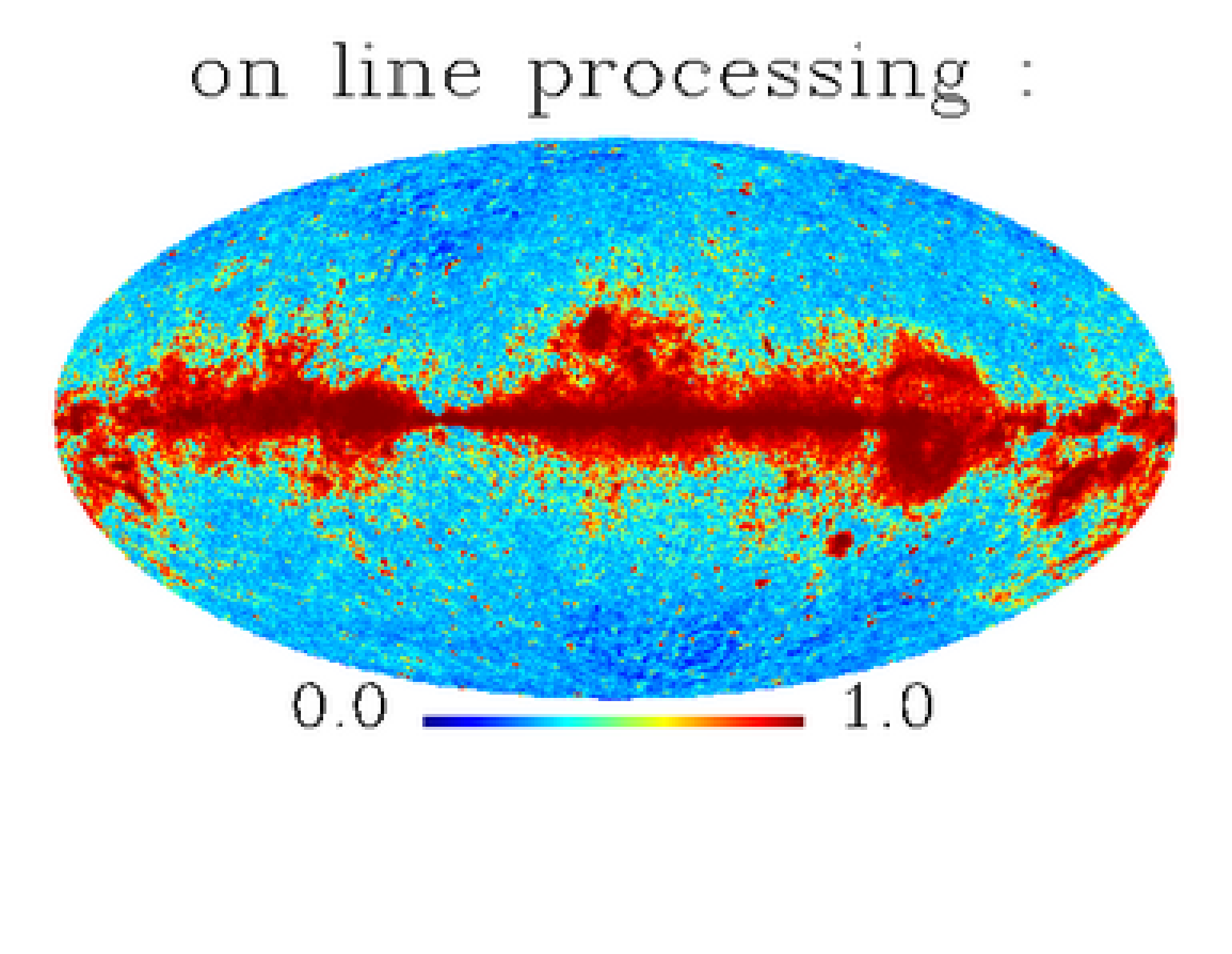}
\end{minipage}
\hspace{1cm}
\begin{minipage}{0.24\textwidth}
\centering
\includegraphics[trim=0cm 3cm 0cm 2cm,clip=true,width=5cm]{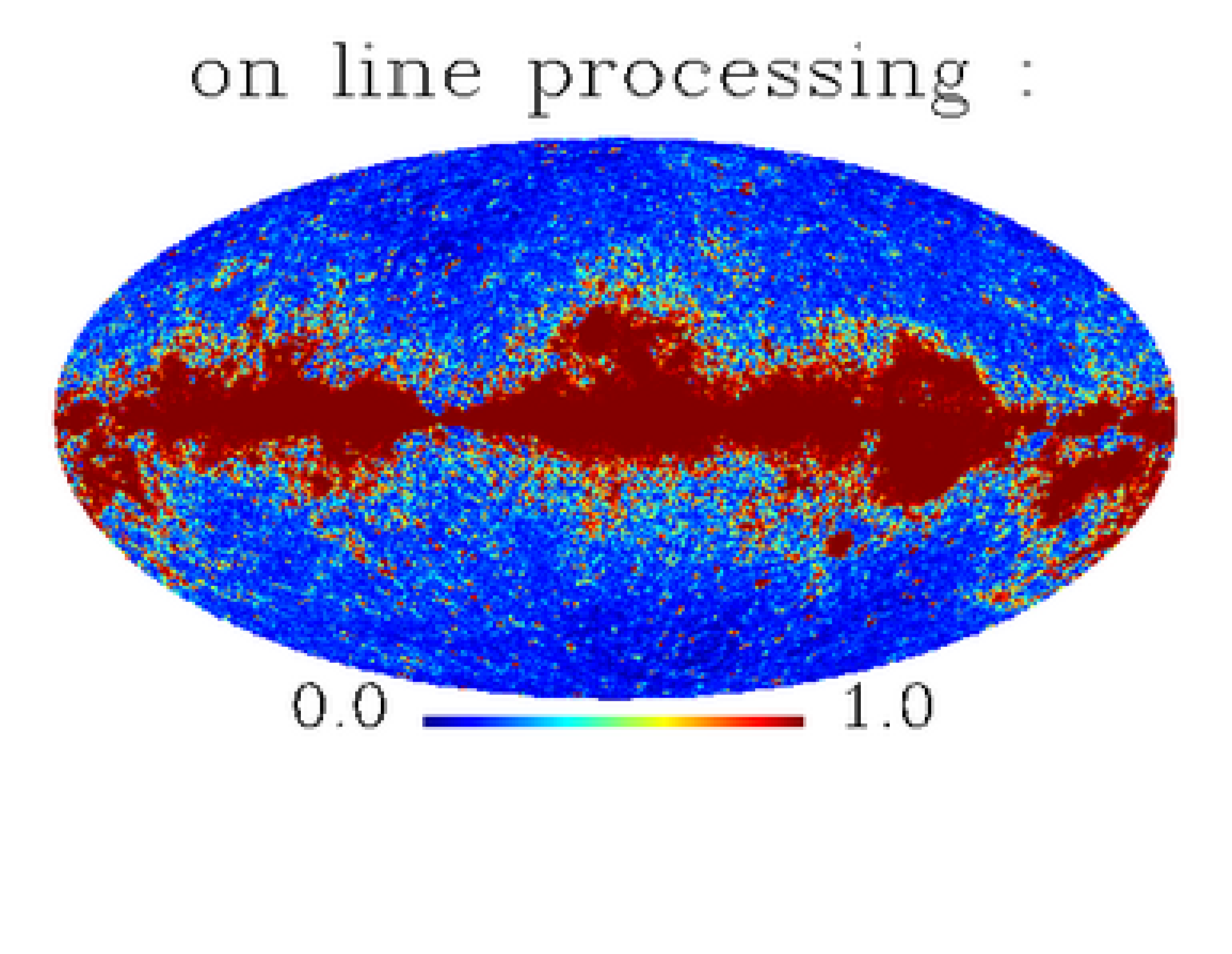}
\end{minipage}
\hspace{1cm}
\centering
\begin{minipage}{0.24\textwidth}
\centering
\includegraphics[trim=0cm 3cm 0cm 2cm,clip=true,width=5cm]{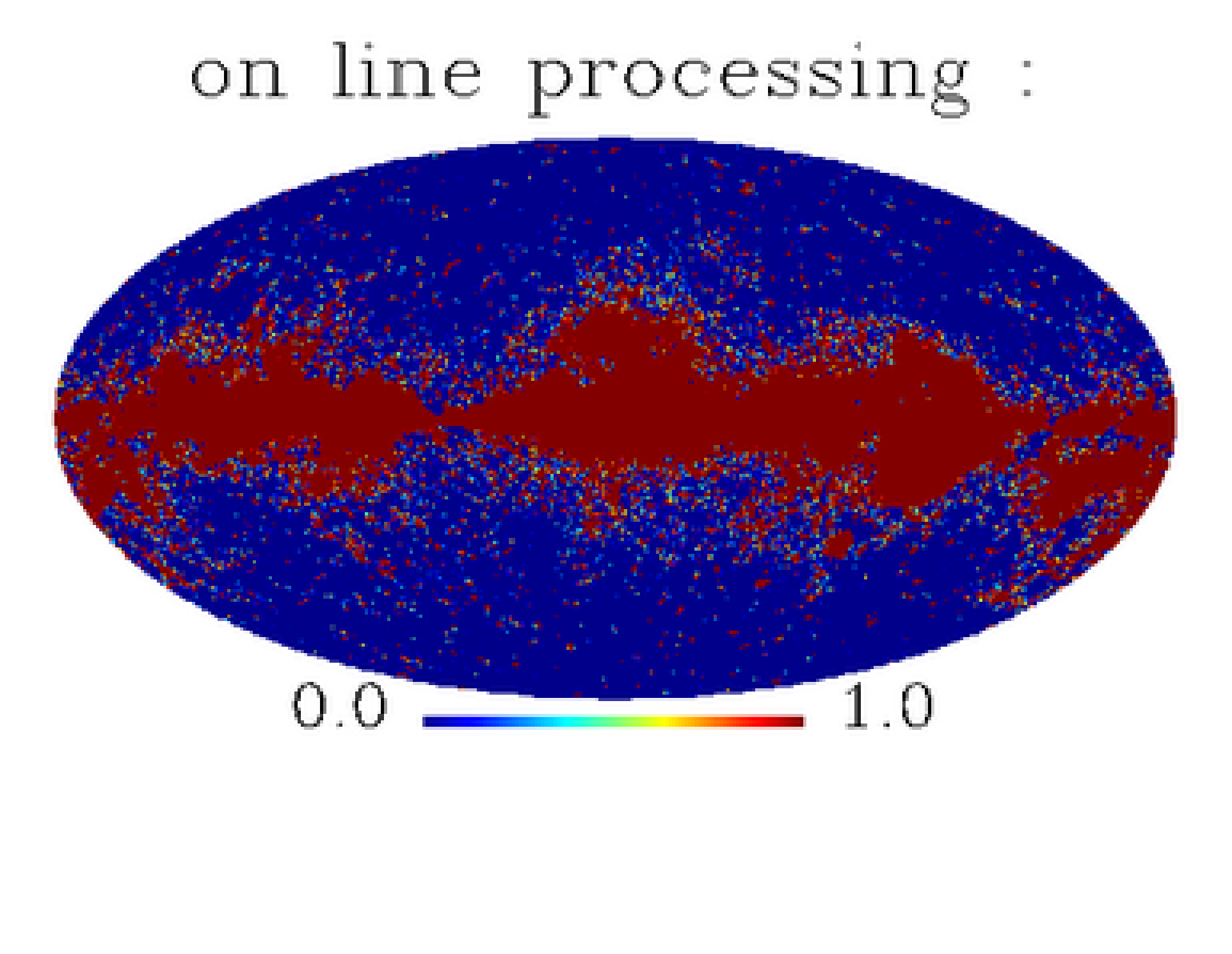}
\end{minipage}
\begin{minipage}{0.24\textwidth}
\centering
\includegraphics[trim=0cm 3cm 0cm 2cm,clip=true,width=5cm]{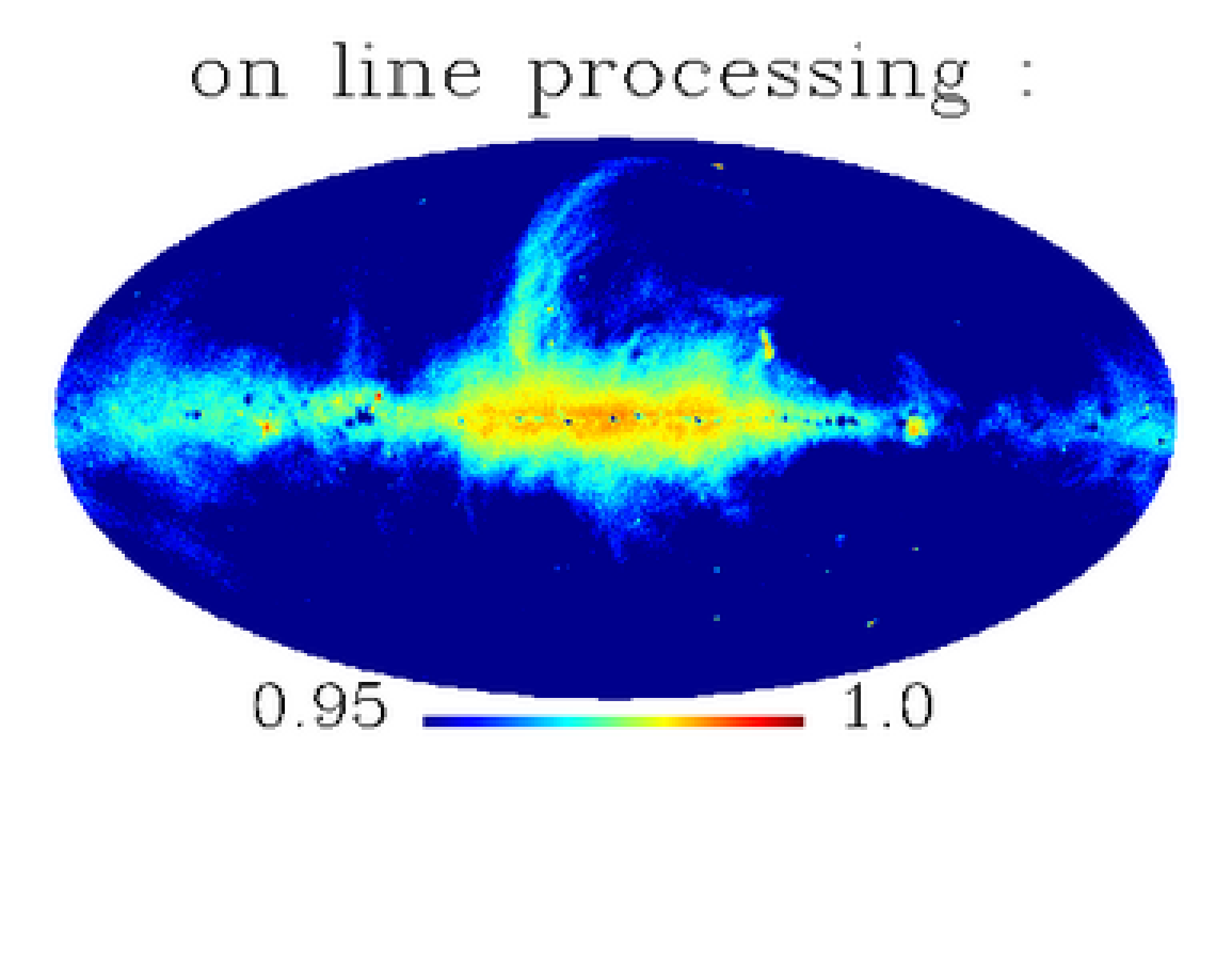}
\end{minipage}
\hspace{1cm}
\begin{minipage}{0.24\textwidth}
\centering
\includegraphics[trim=0cm 3cm 0cm 2cm,clip=true,width=5cm]{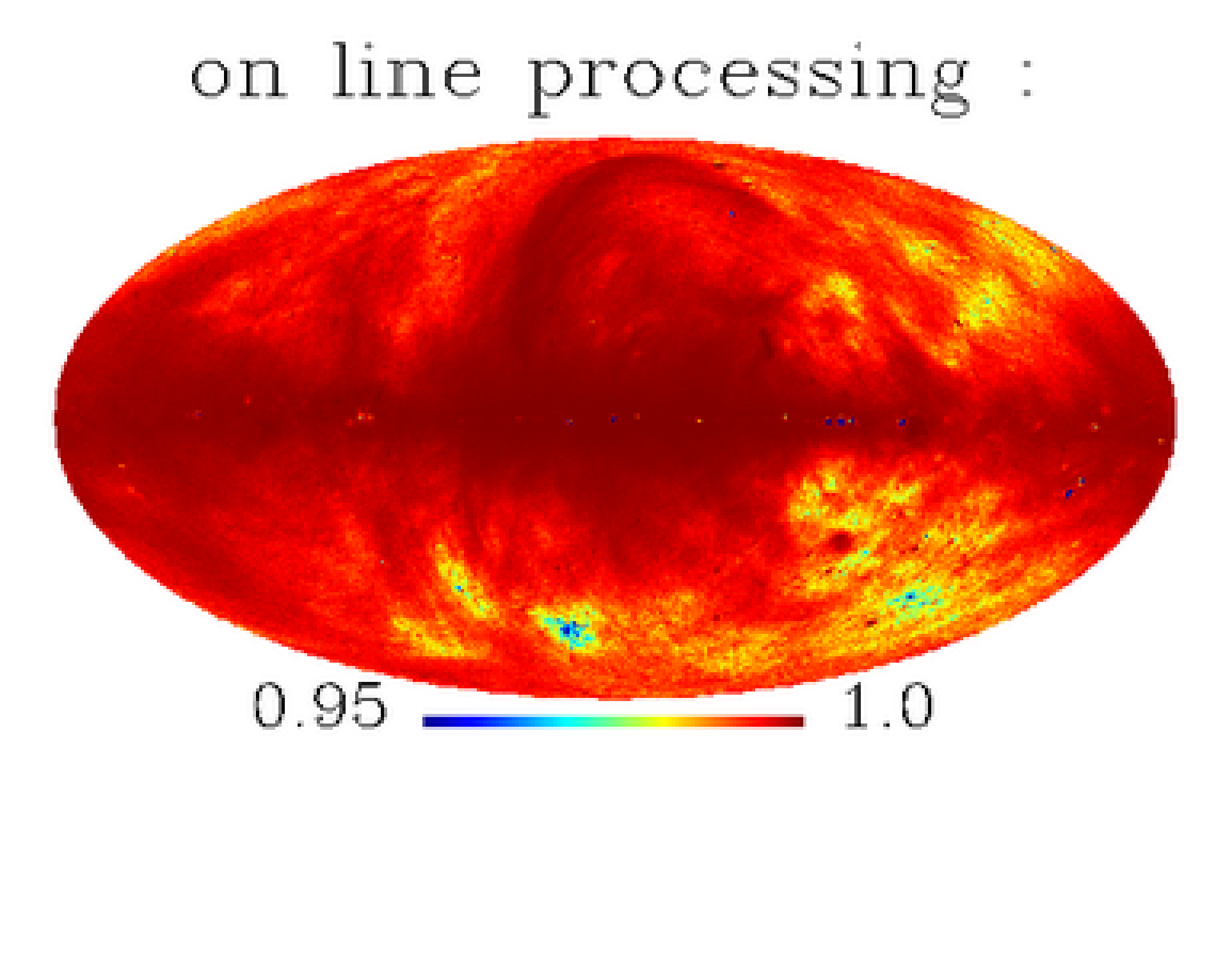}
\end{minipage}
\hspace{1cm}
\centering
\begin{minipage}{0.24\textwidth}
\centering
\includegraphics[trim=0cm 3cm 0cm 2cm,clip=true,width=5cm]{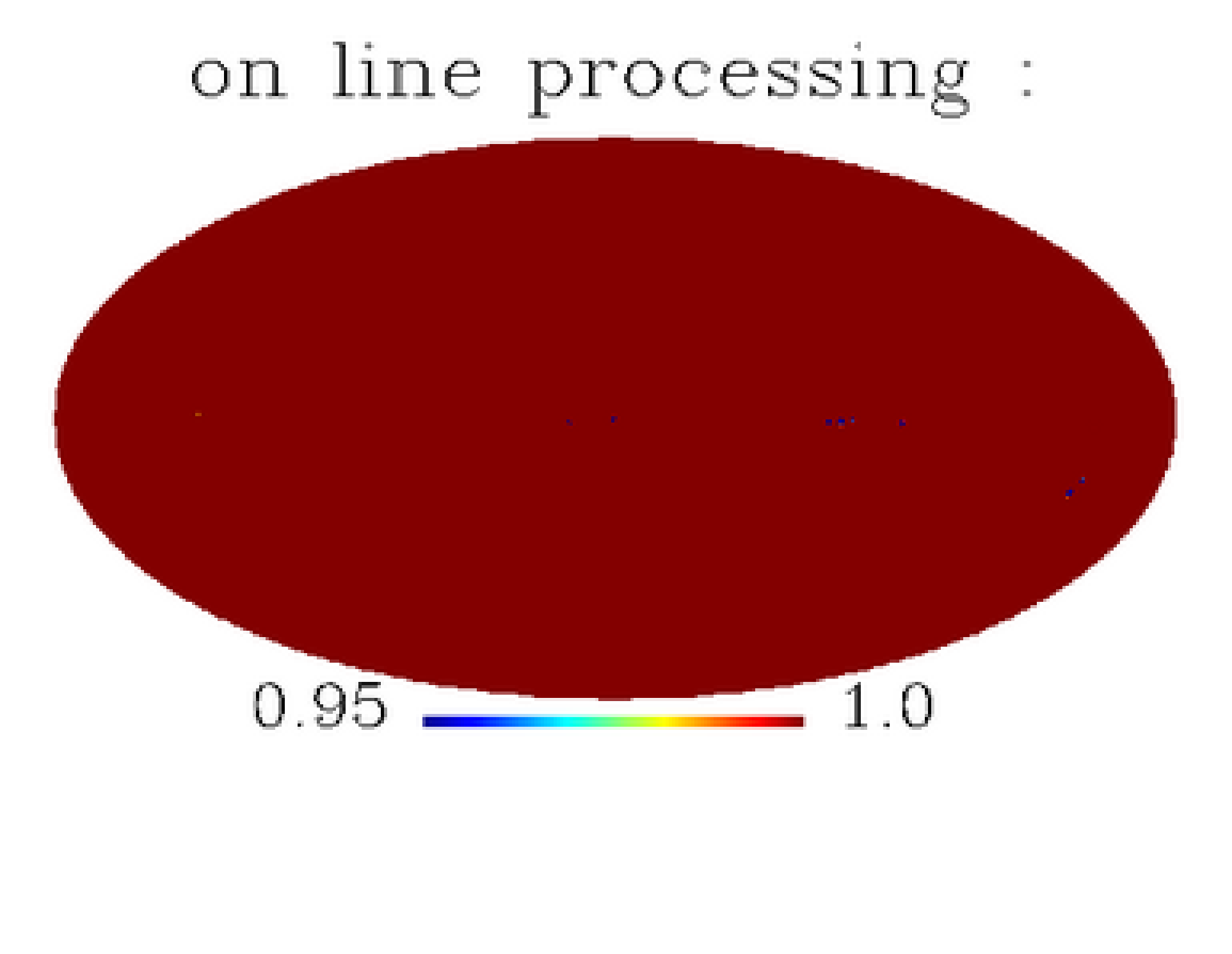}
\end{minipage}
\begin{minipage}{0.24\textwidth}
\centering
\includegraphics[trim=0cm 3cm 0cm 2cm,clip=true,width=5cm]{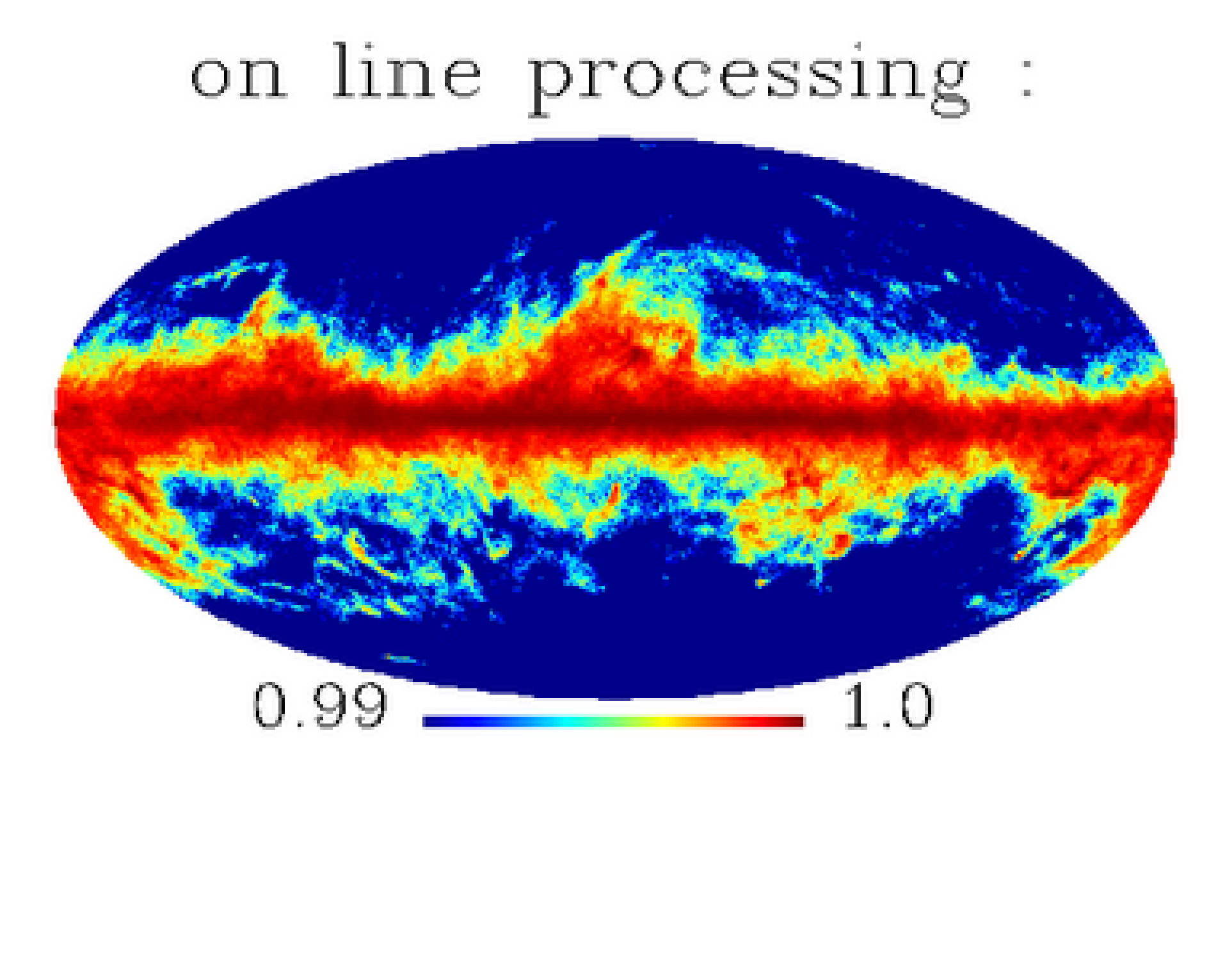}
\end{minipage}
\hspace{1cm}
\begin{minipage}{0.24\textwidth}
\centering
\includegraphics[trim=0cm 3cm 0cm 2cm,clip=true,width=5cm]{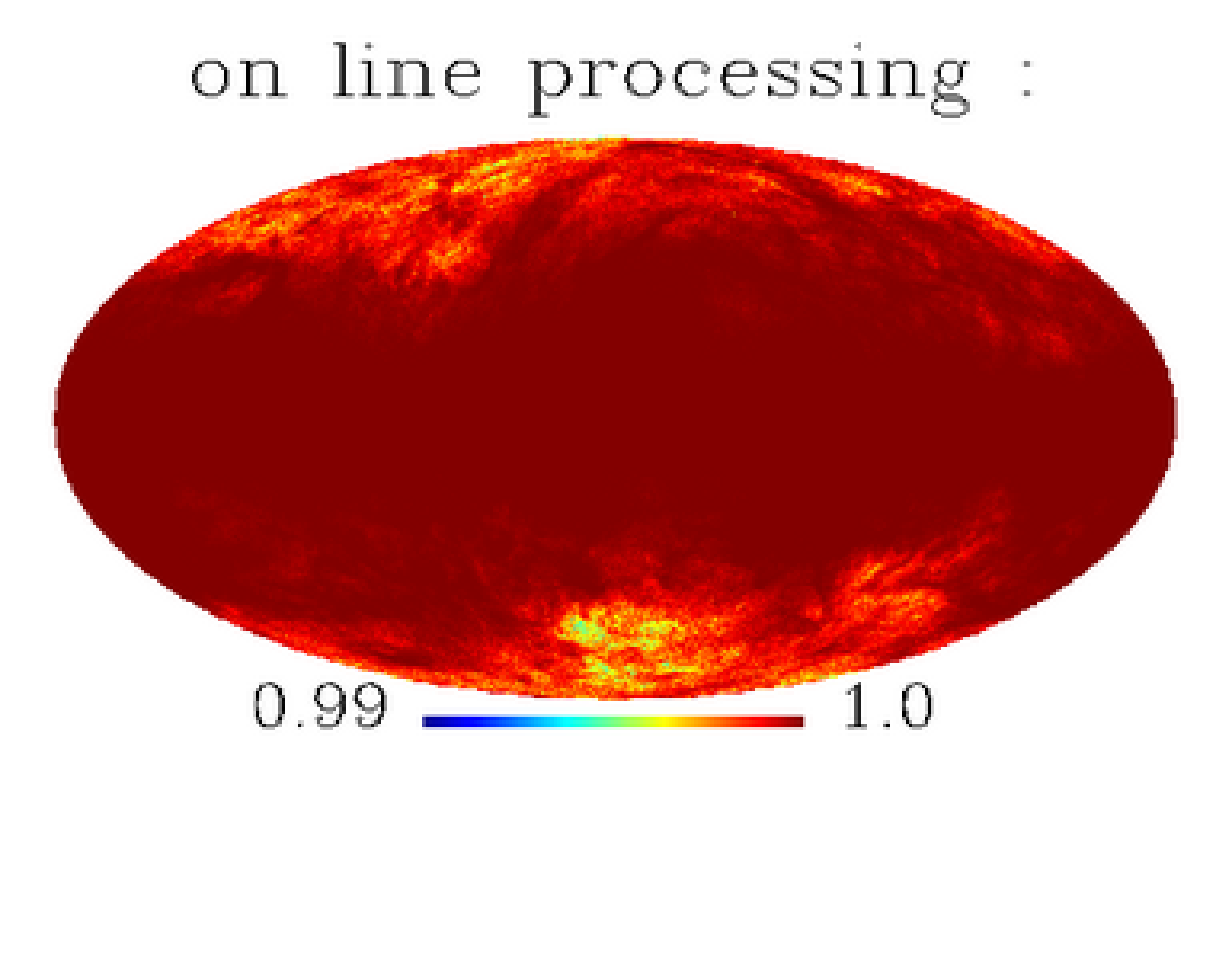}
\end{minipage}
\hspace{1cm}
\centering
\begin{minipage}{0.24\textwidth}
\centering
\includegraphics[trim=0cm 3cm 0cm 2cm,clip=true,width=5cm]{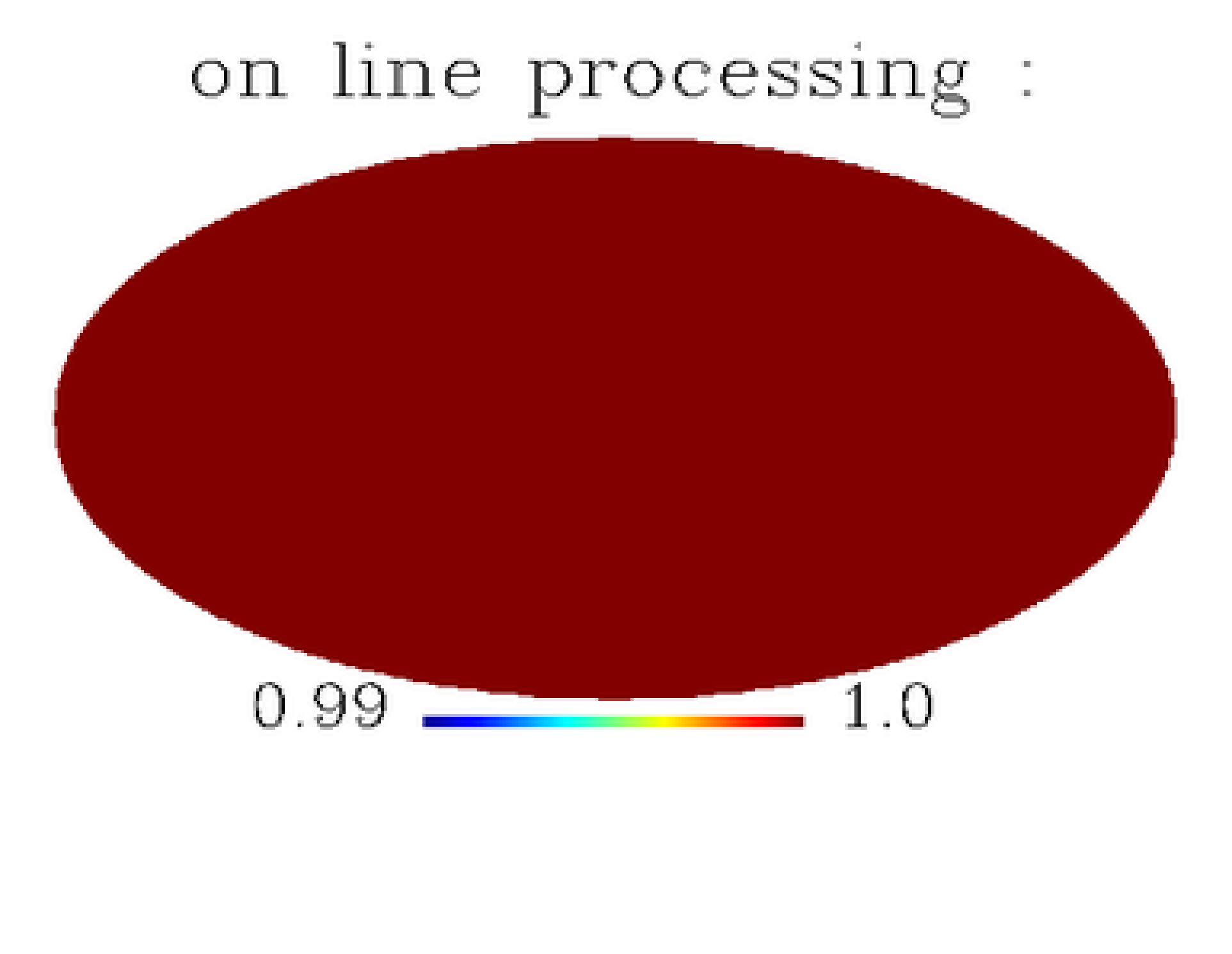}
\end{minipage}
\caption{Maps of the weights calculated for the respective foregrounds. From top to bottom: AME, AME2, free-free emission, synchrotron emission and thermal dust emission. From left to right: $n=1$, $n=2$ and $n=10$}
\vspace{0.5cm}
\label{n}
\end{figure}

\begin{figure}
\centering
\begin{minipage}{0.31\textwidth}
\includegraphics[trim=0cm 0cm 0cm 0cm,clip=true, width=5cm]{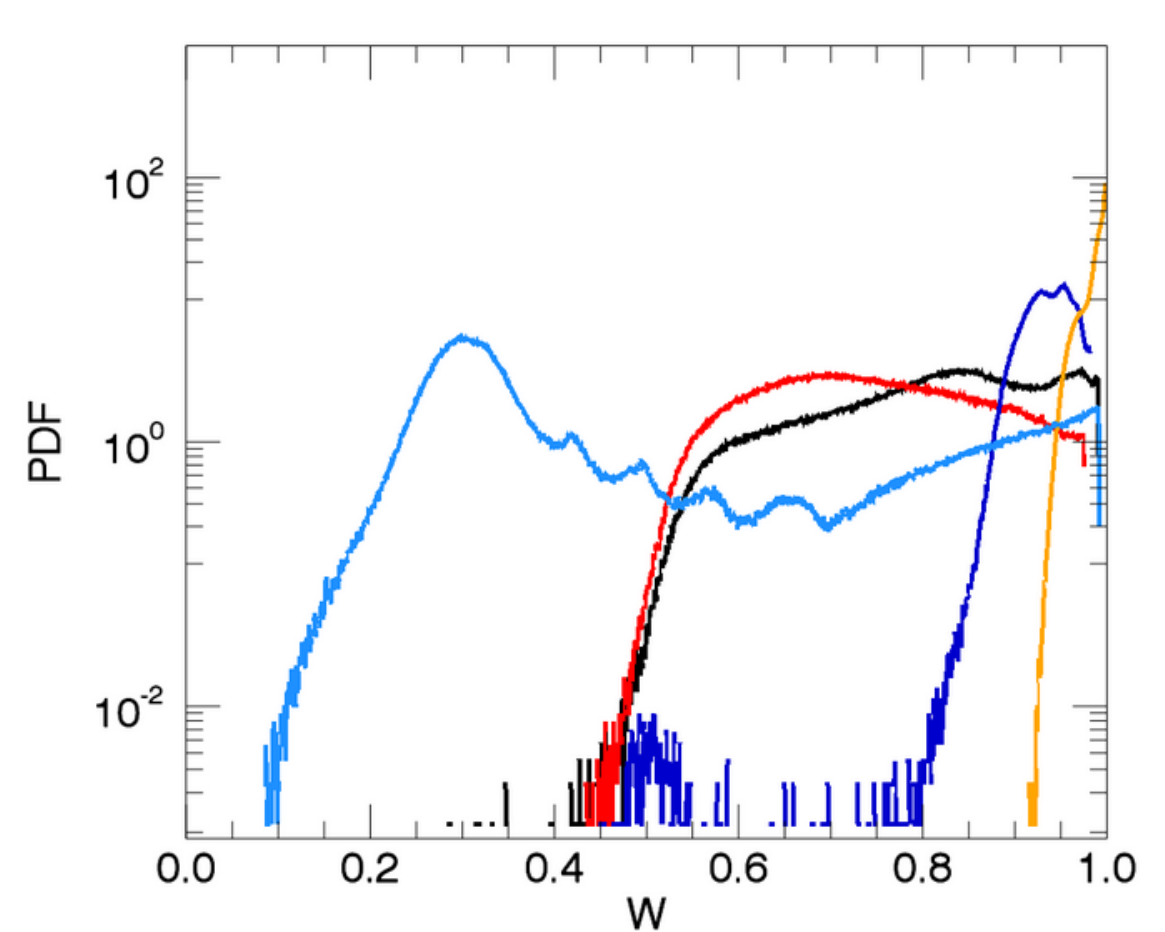}
\end{minipage}
\hfill
\begin{minipage}{0.31\textwidth}
\includegraphics[trim=0cm 0cm 0cm 0cm,clip=true, width=5cm]{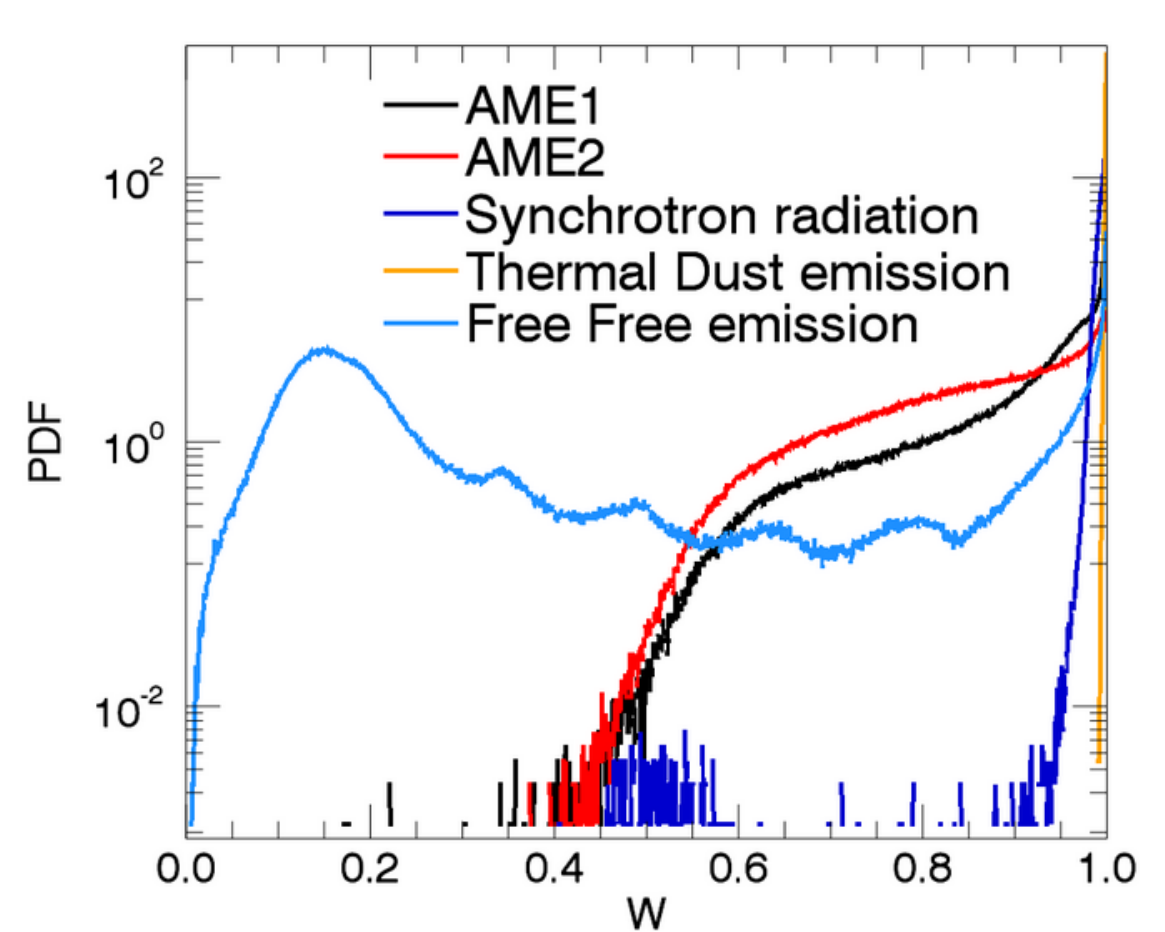}
\end{minipage}
\hfill
\begin{minipage}{0.31\textwidth}
\includegraphics[trim=0cm 0cm 0cm 0cm,clip=true, width=5cm]{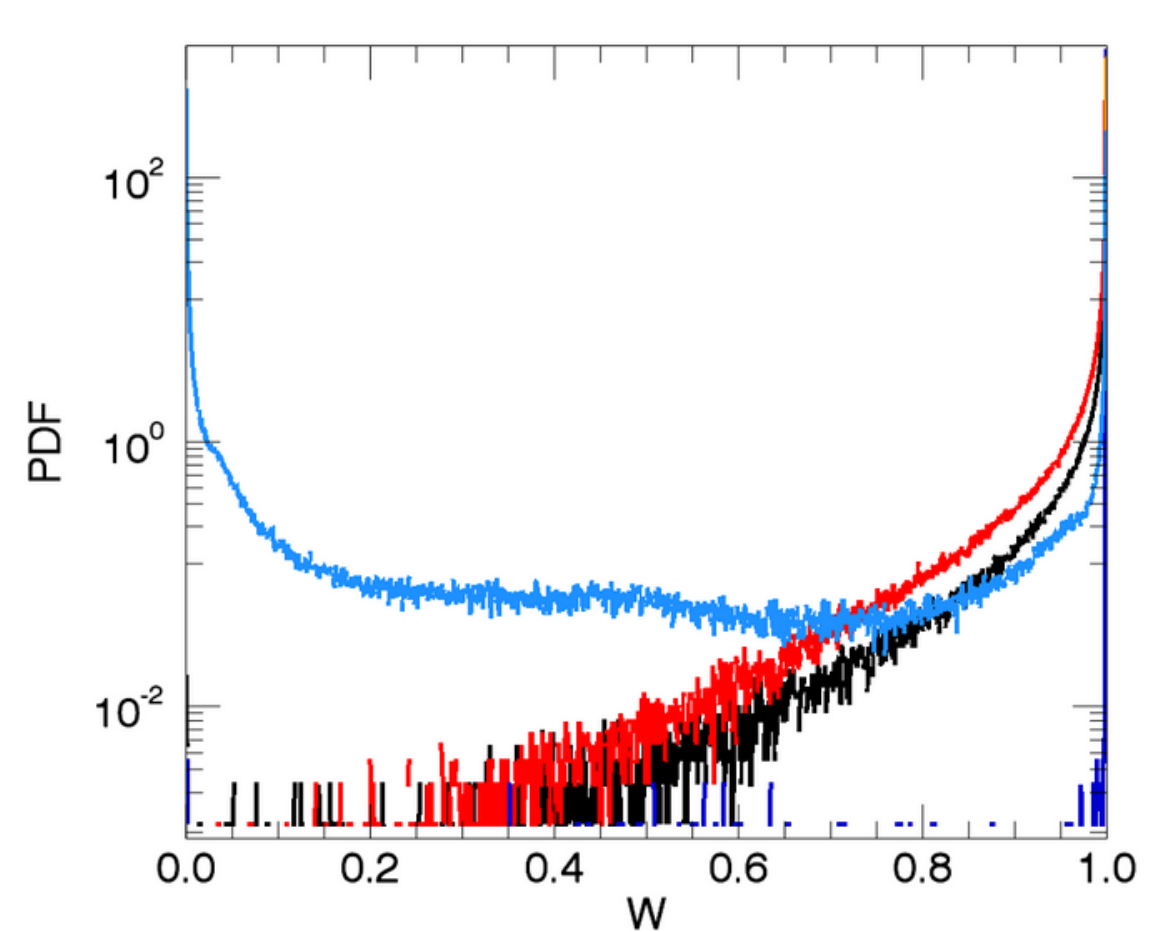}
\end{minipage}
\caption{PDFs of the values of the weights for the respective foregrounds. From left to right: $n=1$, $n=2$ and $n=10$. The legend for all plots can be found in the middle panel.}
\vspace{1cm}
\label{weightshist}
\end{figure}

In our following analysis of weighted correlations we will restrict ourselves to $n=1, 2$ and as an exemplary choice of a high $n$, $n=10$. Figure \ref{n} shows maps of the resulting weights for each of the choices of $n$ for every foreground. We notice that free-free emission has by far the lowest signal-to-error ratio (disregarding the galactic plane). The weights of the two AME signals show slight domination of the errors at the ecliptic poles, whereas the thermal dust and the synchrotron map appear to be the cleanest (Note the ranges of the color scale of the respective maps). For another view on the distribution of weights for the different choices of $n$, see figure \ref{weightshist}.

\subsection{Assessment of correlations.}
\label{sec:assess}

\begin{figure}[floatfix,h!]
\begin{minipage}{0.48\textwidth}
\includegraphics[width=7cm]{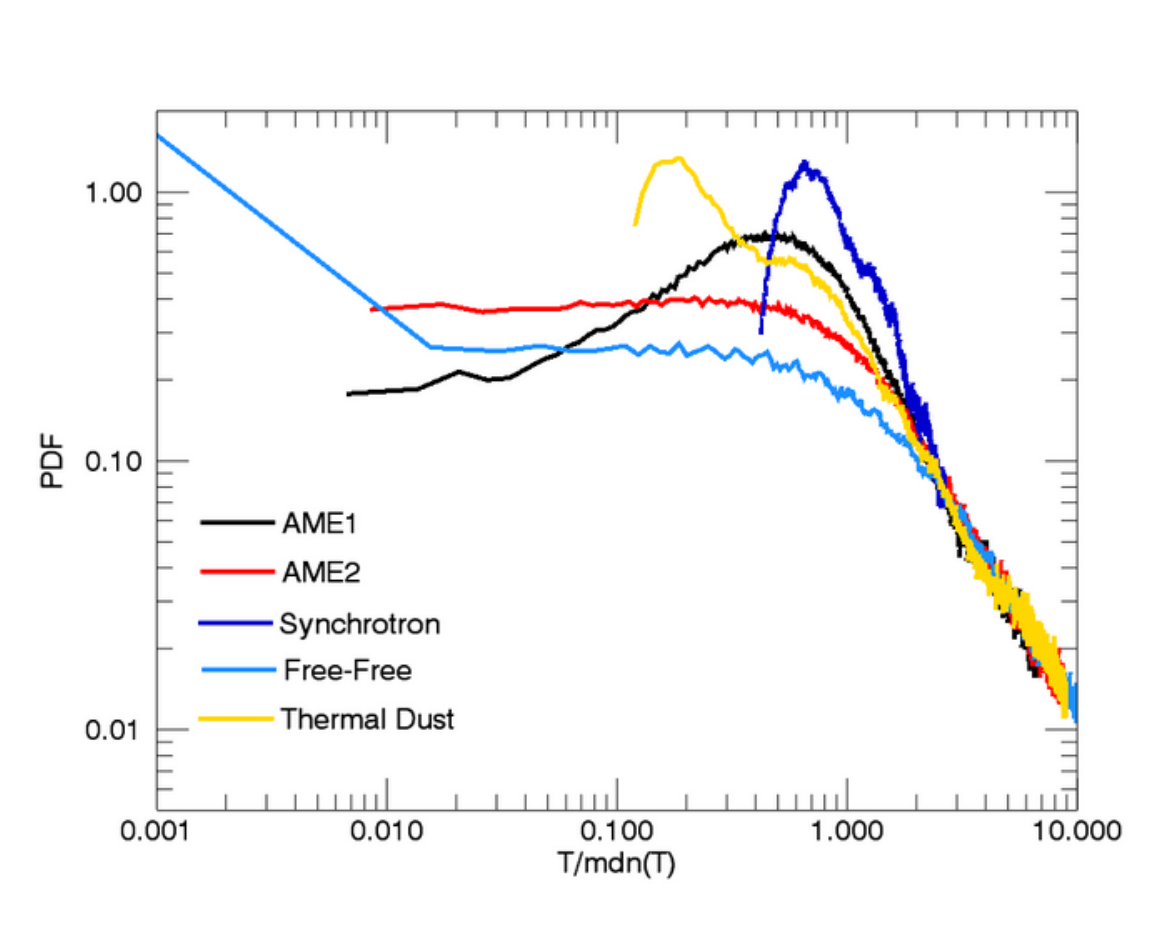}
\end{minipage}
\hfill
\begin{minipage}{0.48\textwidth}
\includegraphics[width=7cm]{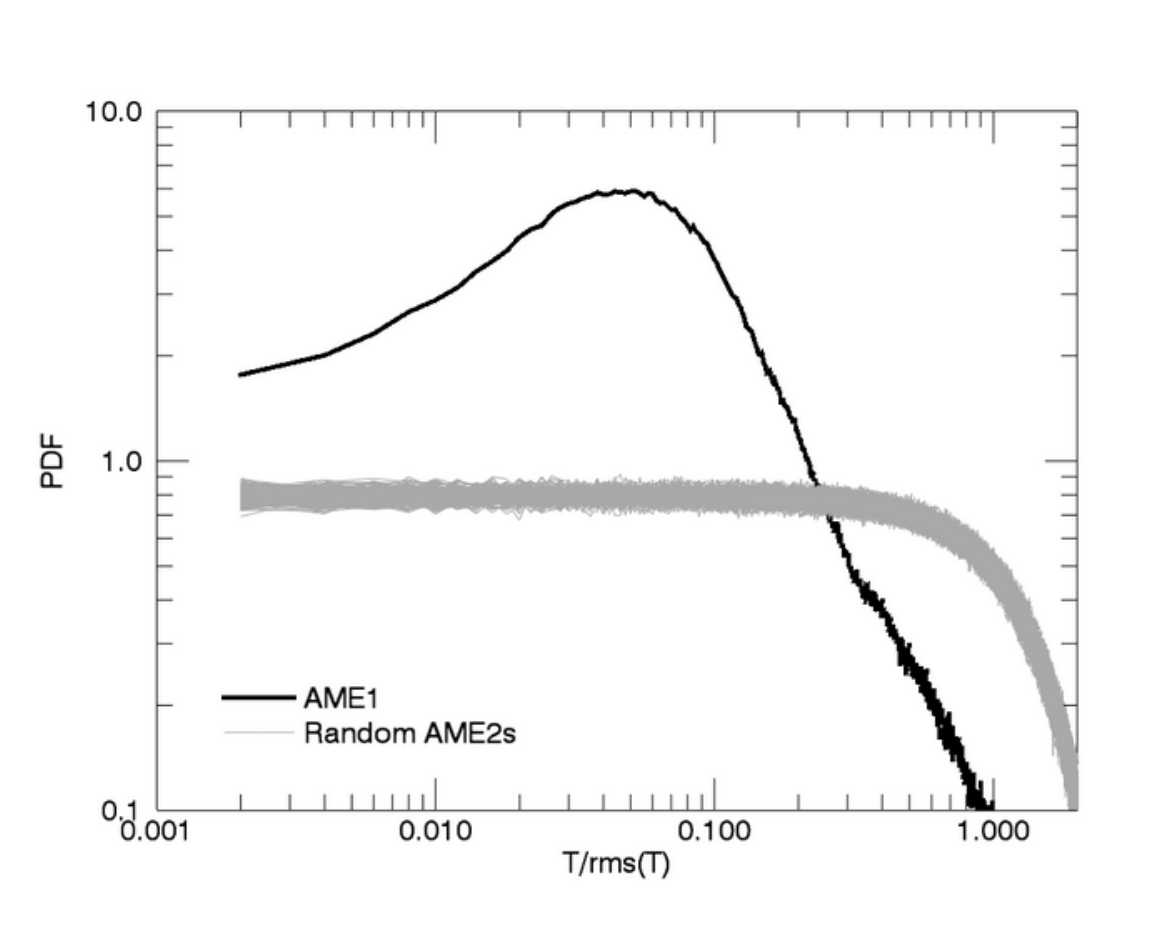}
\end{minipage}
\caption{Left: Probability density functions for used foreground templates. Right: Probability density function of AME1 compared with that of a randomly generated AME1 sky as described in the text.} 
\label{prob}
\end{figure}

To assess the significance of found correlations we define a statistical ensemble of realizations, based on general properties of the maps $S_1$ and $S_2$. We would like to
emphasize that both these maps represent statistically inhomogeneous and non-Gaussian signals. Hereto, all foreground's PDFs can be seen in the left panel of figure \ref{prob}. \footnote{In order to make the PDFs of all foregrounds well visible in one plot, we actually show the PDFs of the foreground's temperature relative to its median value. Additionally, we exclude the 10\% pixels with the highest values as well as the 1\% pixels with the lowest values.}

In this paper, we focus on the properties of the AME maps in comparison to the synchrotron, free-free and thermal dust emission maps.
We therefore create artificial signals based on the respective AME maps, as described below, which we then correlate with the remaining foreground maps for comparison with the real correlation. 
Firstly, we decompose the AME signal into spherical harmonics:
\begin{align}
 S^{AME}(\theta,\phi)=\sum_{l=0}^{l_{max}}\sum_{m=-l}^l |S_{lm}|e^{i\Phi_{lm}}Y_{lm}(\theta,\phi)
\label{prob1}
\end{align}
where $|S_{lm}|$ are the amplitudes and $\Phi_{lm}$ are the phases of each $(l,m)$ mode, respectively. The power spectrum of $S^{AME}$-map is then given by
$C^{AME}_l=\frac{1}{2l+1}\sum_{m=-l}^l|S_{l,m}|^2$, independent of the phases $\Phi_{lm}$. For a statistically inhomogeneous and non-Gaussian signal the statistical
distribution of phases is far from a uniform distribution, with different correlations between $(l,m)$ and $(l+k, m+n)$ modes for various $k,n=0,1,2...$ . 
At the same time, detecting a common morphology in two maps, say AME and another foreground map, means that the phases $\Phi_{lm}$ and the phases for the foreground signal, $\Psi_{lm}$, are strongly
correlated, $\Phi_{l,m}\simeq \Psi_{lm}$. The last statement gives us an idea of how to generate a statistical ensemble of realizations of artificial AME maps, which
could only have chance correlations with synchrotron, free-free or thermal dust maps. For that, we extract the amplitudes of the AME maps $|S_{lm}|$ 
from Eq(\ref{prob1}) and select the phases $\psi_{lm}$ randomly from a uniform distribution within $[0,2\pi]$. Combining $|S_{lm}|$ and $\psi_{lm}$ to $S^{rand}_{lm}=|S_{lm}|\exp(i\psi_{lm})$ for each realization. The right panel in figure \ref{prob} shows the PDFs of all 1000 random AME1 maps, generated as just described, compared to the actual signal.\footnote{Again for better visibility, we show PDFs of the temperature T relative to the RMS of T. Since the random realizations have both, negative and positive values, for this plot, we consider the absolute value of the maps.} The same process has been applied to the AME2 map (not shown here).

\section{K-maps for AME and synchrotron, free-free and dust emission maps}
In this section we summarize the results of mosaic cross-correlations for the two AME skies with synchrotron, free-free and thermal dust maps. $N_{side}=256$ for all input maps which defines the area of each pixel to $q_p=4\pi/N_{p}$, where $N_{p}$ is the number of pixels for $N_{side}$. The sizes of the mother pixels, $\Omega$, are chosen according to the lower resolutions \mbox{$N'_{side}=64,\,32,\,16,\,8$} which correspond to the sizes $Q_{\Omega}=N_{\Omega}q_p$, with $N_{\Omega}=16,\,64,\,256,\,1024$. $N_{\Omega}$ is the number of pixels within one mother pixel for a given $N'_{side}$. The angular extent of each pixel in the original resolution is $\theta_p\approx14$ arcmin and for the mother pixels $\theta'_p\approx$ 55 arcmin, 1.8$^{\circ}$, 3.7$^{\circ}$ and 7.4$^{\circ}$.

Before showing the results of the weighted correlations, we choose to neglect the errors of both maps to obtain a regular, unweighted correlation. For this, we use the maximum likelihood map rather than the mean value map, as already mentioned in sec. \ref{sec:compmaps}. In the subsequent section we then include the errors and weight the mean value data with the three different coefficients introduced in section \ref{sec:weights}, corresponding to different levels of rigor towards the data.

\subsection{Unweighted correlations}

\begin{figure}[H]
\centering
\begin{minipage}{0.24\textwidth}
\centering
\includegraphics[trim=0cm 0.9cm 0cm 0.5cm,clip=true,width=4cm]{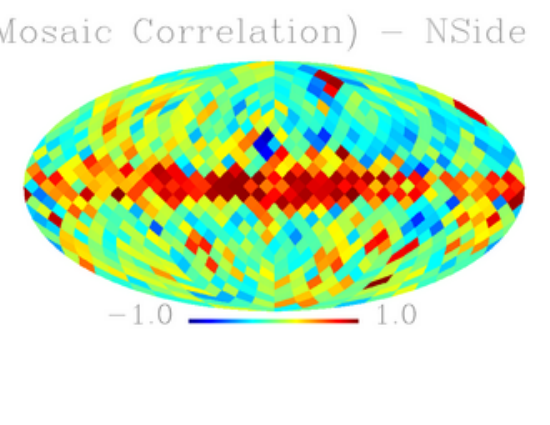}
\end{minipage}
\hfill
\begin{minipage}{0.24\textwidth}
\centering
\includegraphics[trim=0cm 0.9cm 0cm 0.5cm,clip=true,width=4cm]{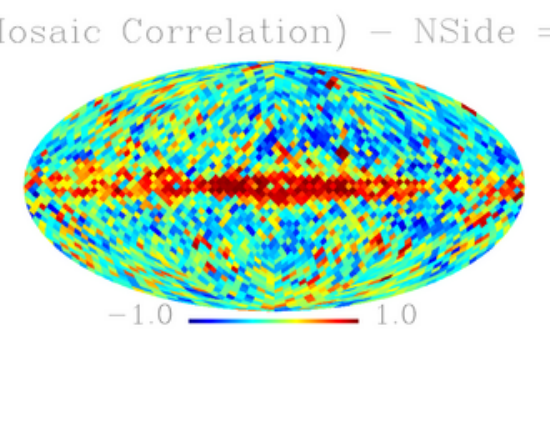}
\end{minipage}
\hfill
\centering
\begin{minipage}{0.24\textwidth}
\centering
\includegraphics[trim=0cm 0.9cm 0cm 0.5cm,clip=true,width=4cm]{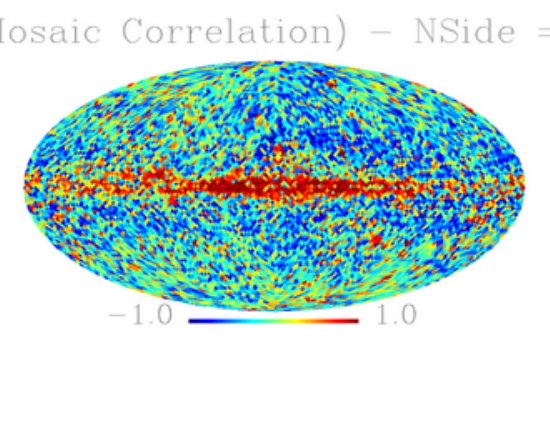}
\end{minipage}
\hfill
\begin{minipage}{0.24\textwidth}
\centering
\includegraphics[trim=0cm 0.9cm 0cm 0.5cm,clip=true,width=4cm]{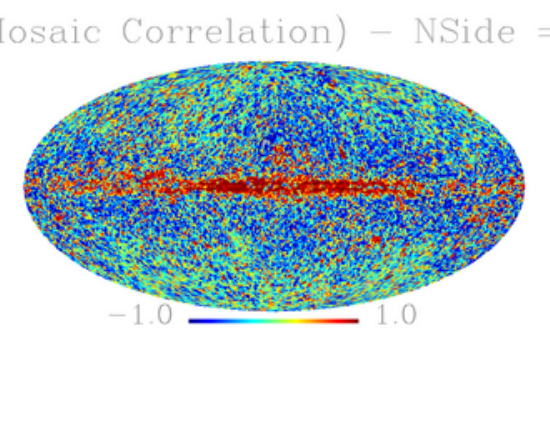}
\end{minipage}
\centering
\begin{minipage}{0.24\textwidth}
\centering
\includegraphics[trim=0cm 0.9cm 0cm 0.5cm,clip=true,width=4cm]{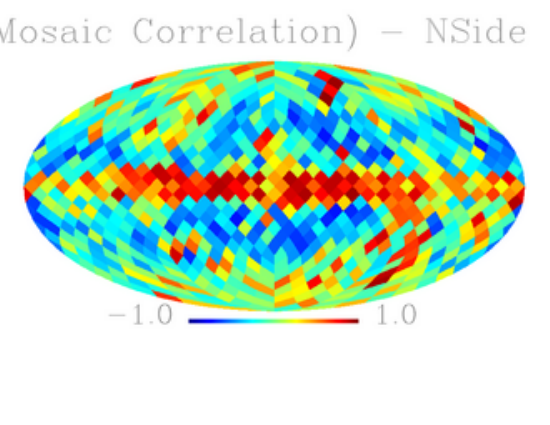}
\end{minipage}
\hfill
\begin{minipage}{0.24\textwidth}
\centering
\includegraphics[trim=0cm 0.9cm 0cm 0.5cm,clip=true,width=4cm]{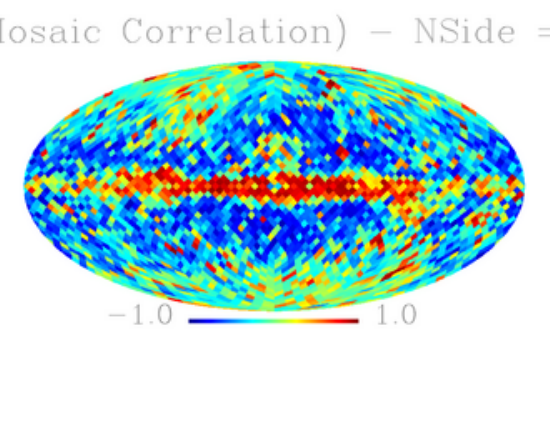}
\end{minipage}
\hfill
\centering
\begin{minipage}{0.24\textwidth}
\centering
\includegraphics[trim=0cm 0.9cm 0cm 0.5cm,clip=true,width=4cm]{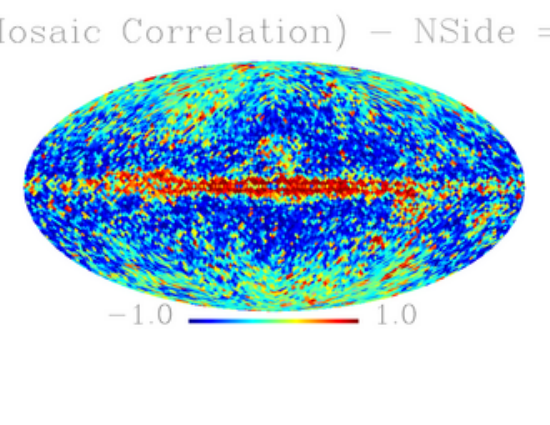}
\end{minipage}
\hfill
\begin{minipage}{0.24\textwidth}
\centering
\includegraphics[trim=0cm 0.9cm 0cm 0.5cm,clip=true,width=4cm]{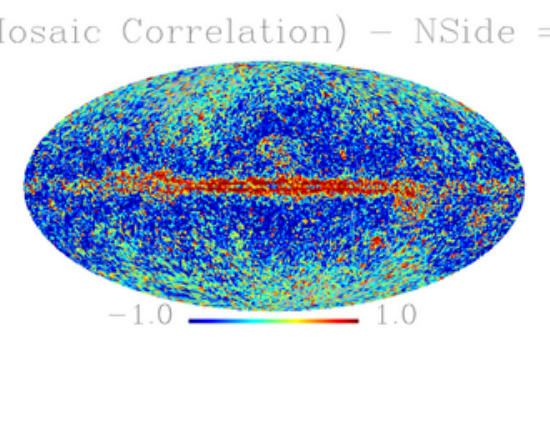}
\end{minipage}
\centering
\begin{minipage}{0.24\textwidth}
\centering
\includegraphics[trim=0cm 0.9cm 0cm 0.5cm,clip=true,width=4cm]{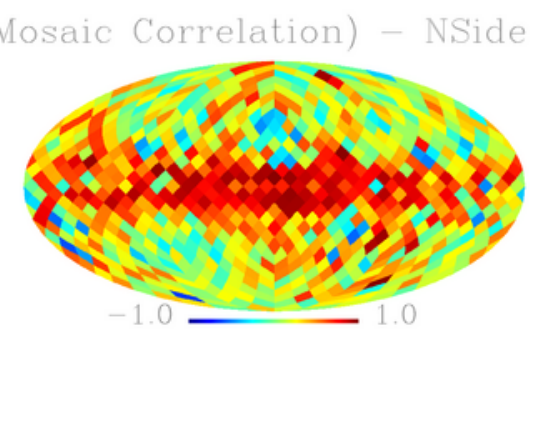}
\end{minipage}
\hfill
\begin{minipage}{0.24\textwidth}
\centering
\includegraphics[trim=0cm 0.9cm 0cm 0.5cm,clip=true,width=4cm]{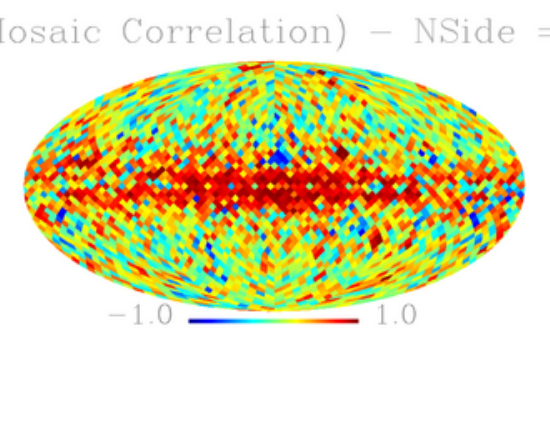}
\end{minipage}
\hfill
\centering
\begin{minipage}{0.24\textwidth}
\centering
\includegraphics[trim=0cm 0.9cm 0cm 0.5cm,clip=true,width=4cm]{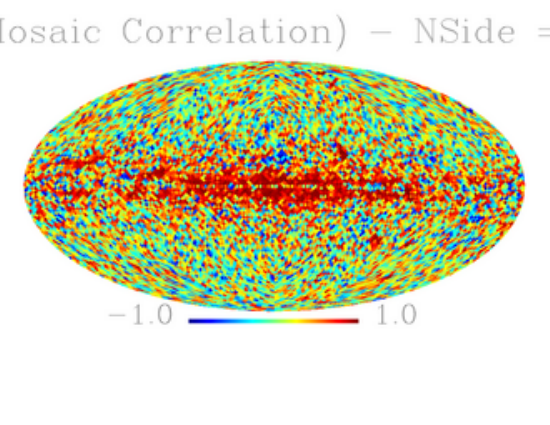}
\end{minipage}
\hfill
\begin{minipage}{0.24\textwidth}
\centering
\includegraphics[trim=0cm 0.9cm 0cm 0.5cm,clip=true,width=4cm]{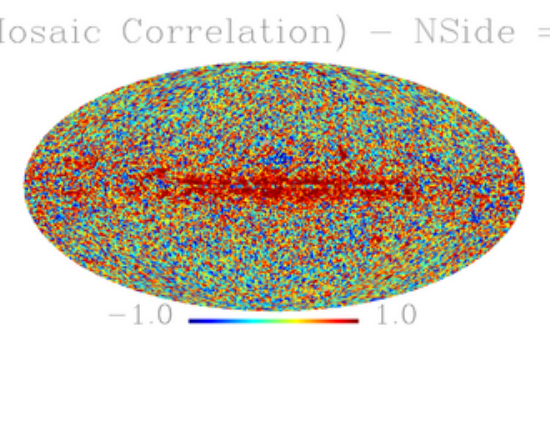}
\end{minipage}
\centering
\begin{minipage}{0.24\textwidth}
\centering
\includegraphics[trim=0cm 0.9cm 0cm 0.5cm,clip=true,width=4cm]{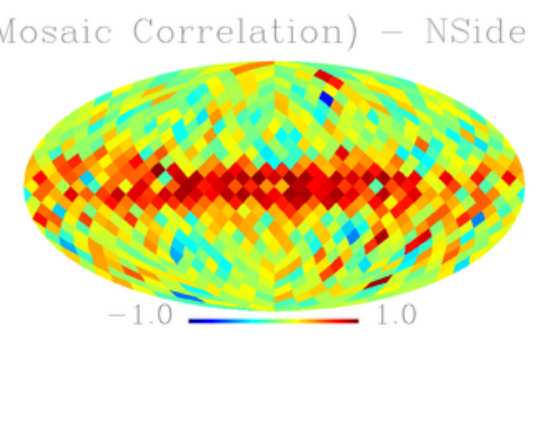}
\end{minipage}
\hfill
\begin{minipage}{0.24\textwidth}
\centering
\includegraphics[trim=0cm 0.9cm 0cm 0.5cm,clip=true,width=4cm]{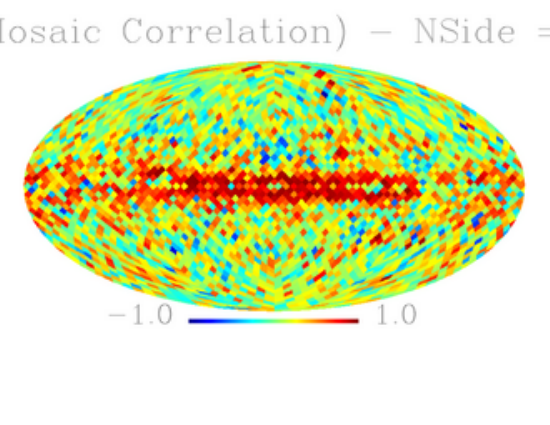}
\end{minipage}
\hfill
\centering
\begin{minipage}{0.24\textwidth}
\centering
\includegraphics[trim=0cm 0.9cm 0cm 0.5cm,clip=true,width=4cm]{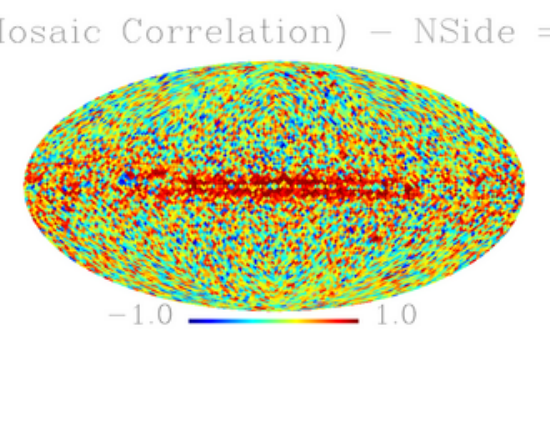}
\end{minipage}
\hfill
\begin{minipage}{0.24\textwidth}
\centering
\includegraphics[trim=0cm 0.9cm 0cm 0.5cm,clip=true,width=4cm]{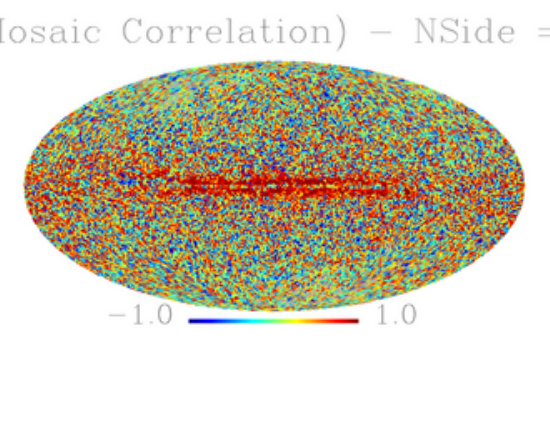}
\end{minipage}
\centering
\begin{minipage}{0.24\textwidth}
\centering
\includegraphics[trim=0cm 0.9cm 0cm 0.5cm,clip=true,width=4cm]{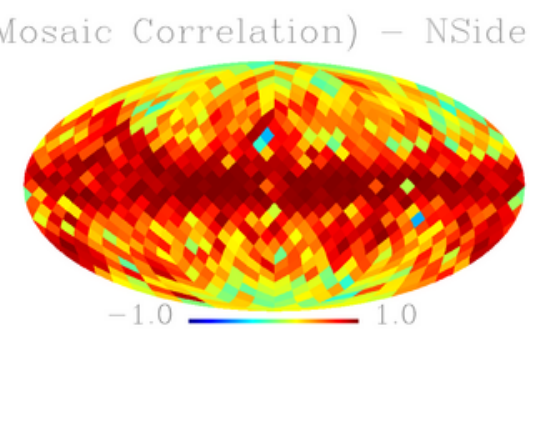}
\end{minipage}
\hfill
\begin{minipage}{0.24\textwidth}
\centering
\includegraphics[trim=0cm 0.9cm 0cm 0.5cm,clip=true,width=4cm]{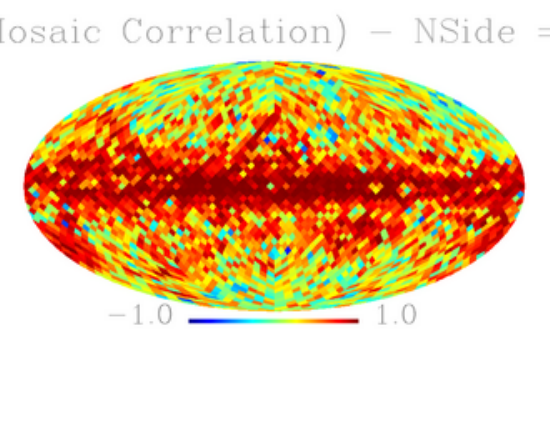}
\end{minipage}
\hfill
\centering
\begin{minipage}{0.24\textwidth}
\centering
\includegraphics[trim=0cm 0.9cm 0cm 0.5cm,clip=true,width=4cm]{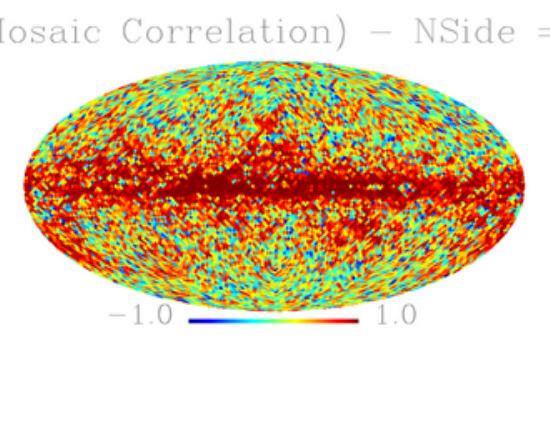}
\end{minipage}
\hfill
\begin{minipage}{0.24\textwidth}
\centering
\includegraphics[trim=0cm 0.9cm 0cm 0.5cm,clip=true,width=4cm]{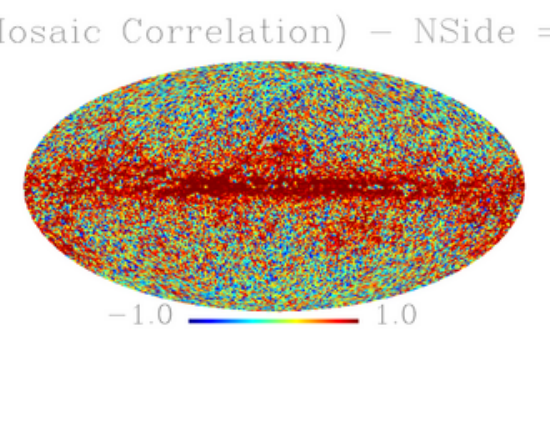}
\end{minipage}
\centering
\begin{minipage}{0.24\textwidth}
\centering
\includegraphics[trim=0cm 0.9cm 0cm 0.5cm,clip=true,width=4cm]{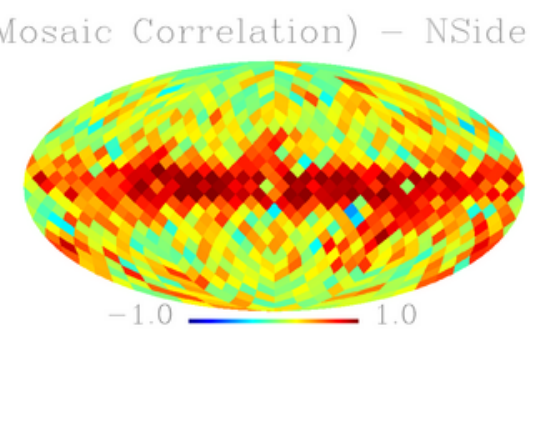}
\end{minipage}
\hfill
\begin{minipage}{0.24\textwidth}
\centering
\includegraphics[trim=0cm 0.9cm 0cm 0.5cm,clip=true,width=4cm]{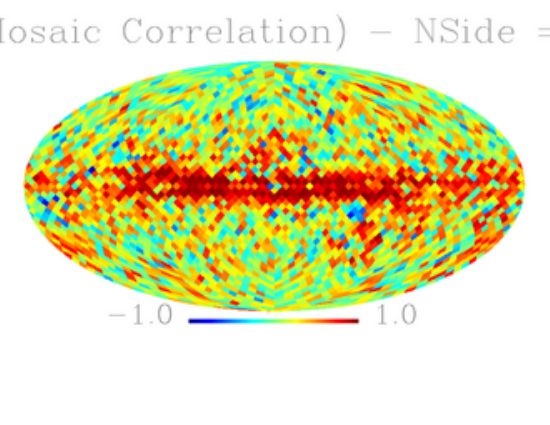}
\end{minipage}
\hfill
\centering
\begin{minipage}{0.24\textwidth}
\centering
\includegraphics[trim=0cm 0.9cm 0cm 0.5cm,clip=true,width=4cm]{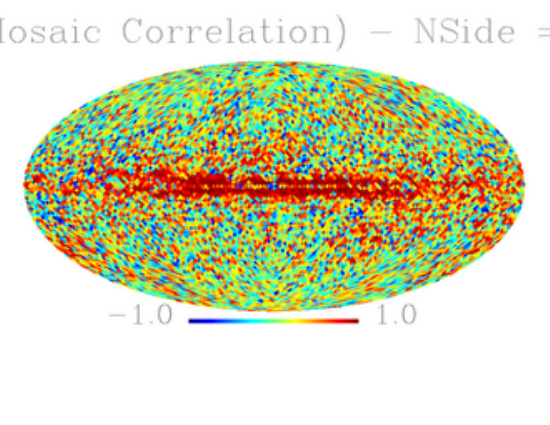}
\end{minipage}
\hfill
\begin{minipage}{0.24\textwidth}
\centering
\includegraphics[trim=0cm 0.9cm 0cm 0.5cm,clip=true,width=4cm]{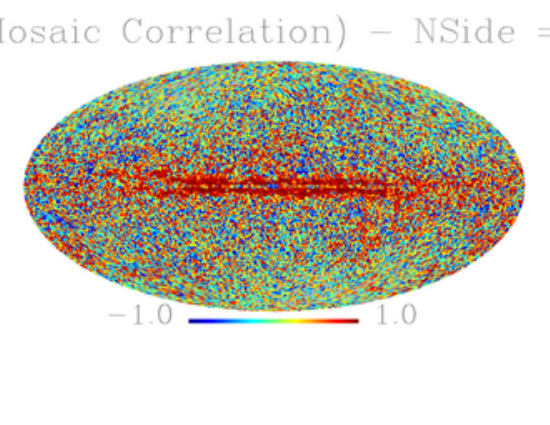}
\end{minipage}
\caption{From top to bottom: Unweighted correlation maps between AME1 and free-free emission, AME2 and free-free emission, AME1 and synchrotron radiation, AME2 and synchrotron radiation, AME1 and thermal dust emission, AME2 and thermal dust emission. From left to right: $\Omega$ contains 1024, 256, 64 and 16 pixel. Here, the maximum likelihood maps are used. The $\sigma_i$s were all assumed \mbox{to be zero.}}
\label{corrmapsno}
\end{figure}

\begin{figure}[H]
\begin{center}
\begin{minipage}{0.2\textwidth}
\centering
\includegraphics[trim=0cm 0cm 0cm 0cm,clip=true,width=4.1cm]{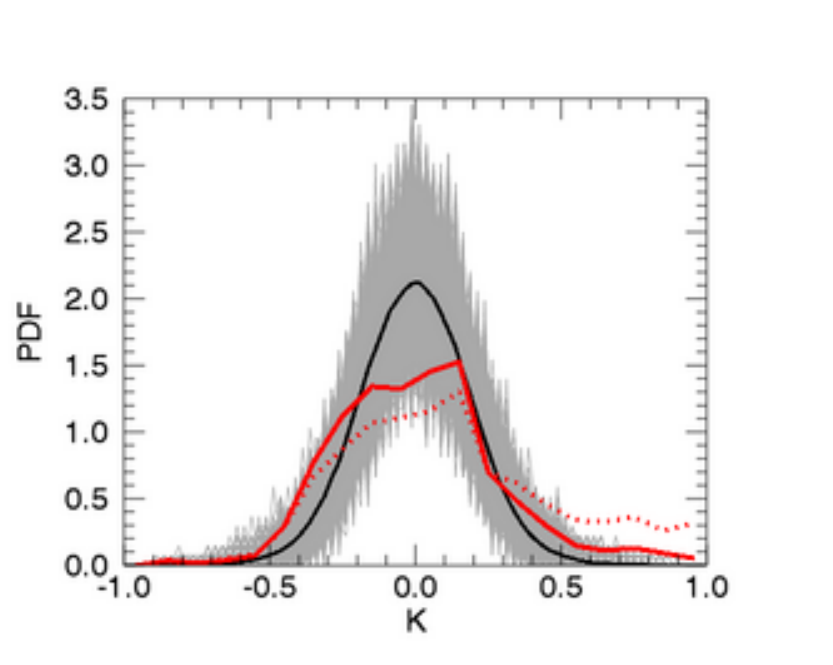}
\end{minipage}
\hspace{0.5cm}
\begin{minipage}{0.2\textwidth}
\centering
\includegraphics[trim=0cm 0cm 0cm 0cm,clip=true,width=4.1cm]{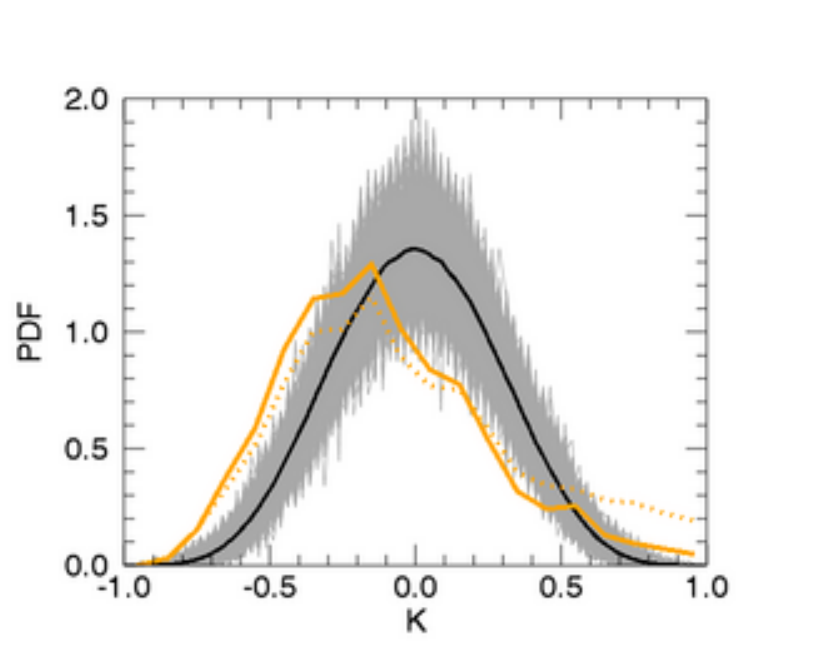}
\end{minipage}
\hspace{0.5cm}
\centering
\begin{minipage}{0.2\textwidth}
\centering
\includegraphics[trim=0cm 0cm 0cm 0cm,clip=true,width=4.1cm]{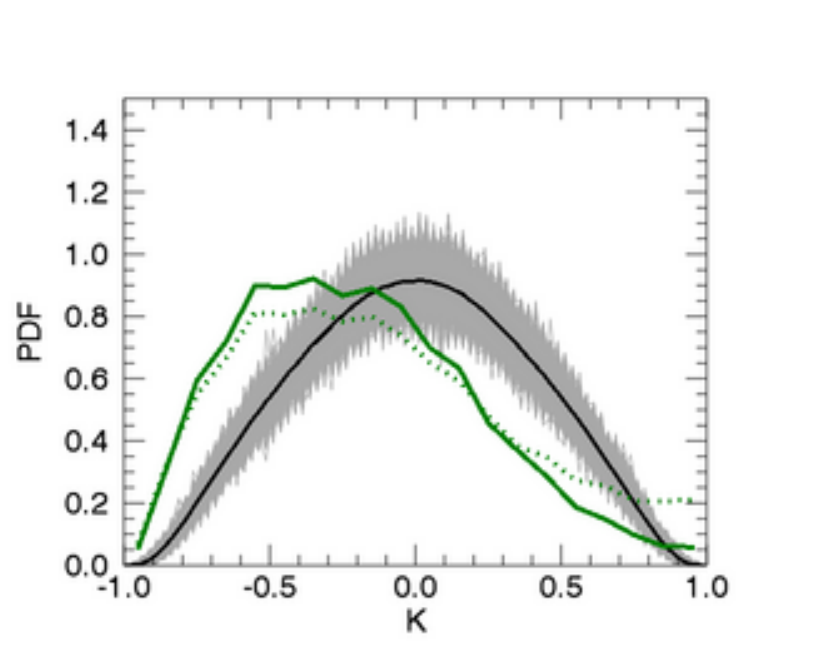}
\end{minipage}
\hspace{0.5cm}
\begin{minipage}{0.2\textwidth}
\centering
\includegraphics[trim=0cm 0cm 0cm 0cm,clip=true,width=4.1cm]{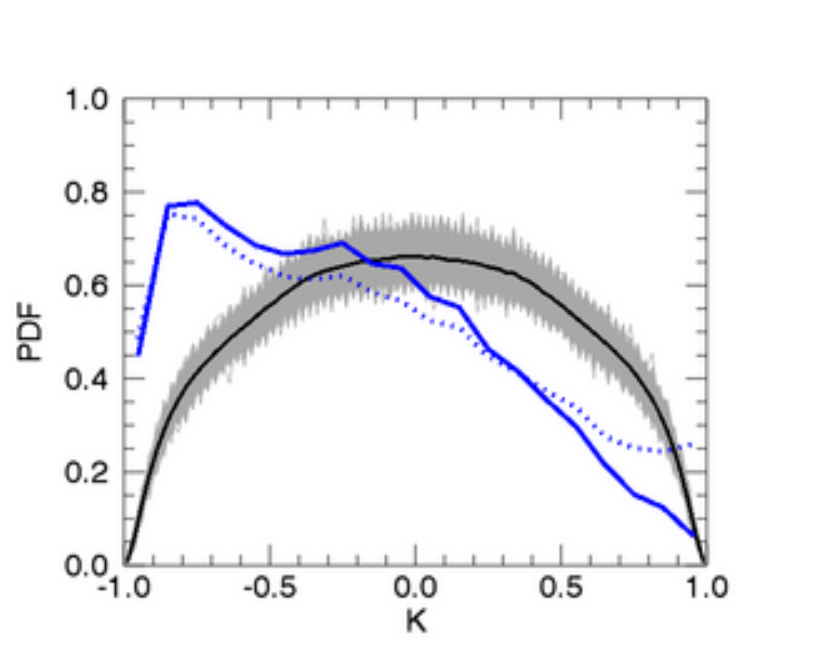}
\end{minipage}
\vspace{-0.2cm}
\begin{minipage}{0.2\textwidth}
\centering
\includegraphics[trim=0cm 0cm 0cm 0cm,clip=true,width=4.1cm]{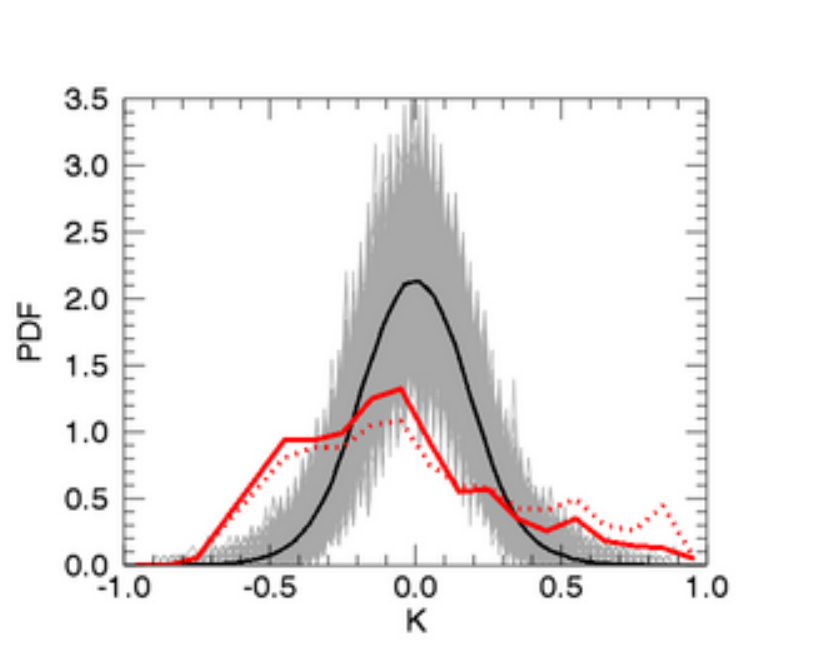}
\end{minipage}
\hspace{0.5cm}
\begin{minipage}{0.2\textwidth}
\centering
\includegraphics[trim=0cm 0cm 0cm 0cm,clip=true,width=4.1cm]{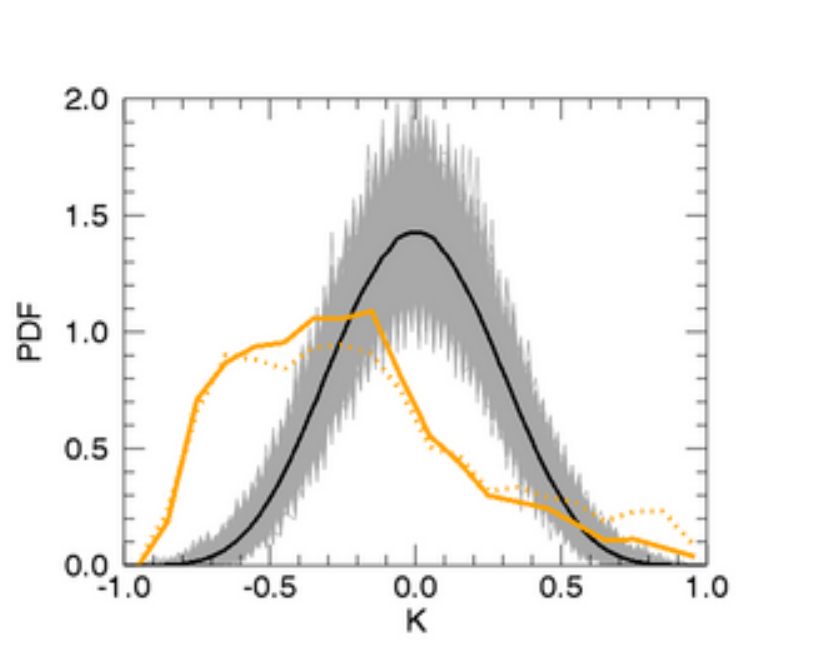}
\end{minipage}
\hspace{0.5cm}
\begin{minipage}{0.2\textwidth}
\centering
\includegraphics[trim=0cm 0cm 0cm 0cm,clip=true,width=4.1cm]{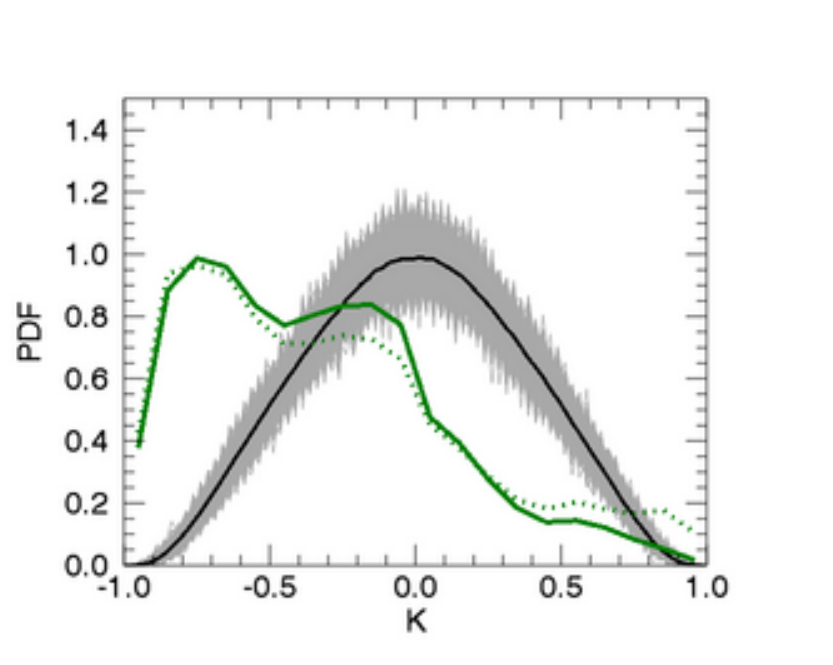}
\end{minipage}
\hspace{0.5cm}
\begin{minipage}{0.2\textwidth}
\centering
\includegraphics[trim=0cm 0cm 0cm 0cm,clip=true,width=4.1cm]{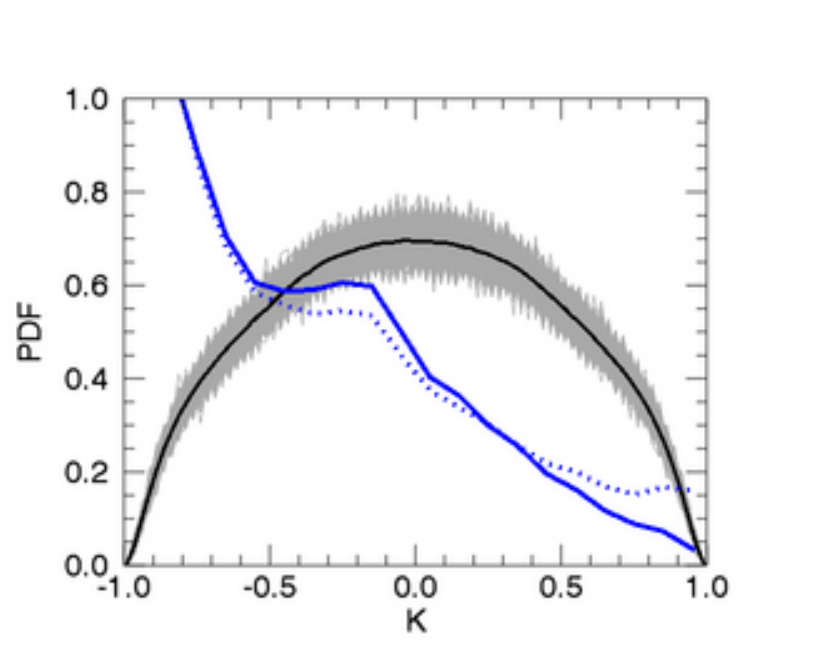}
\end{minipage}
\vspace{-0.2cm}
\begin{minipage}{0.2\textwidth}
\centering
\includegraphics[trim=0cm 0cm 0cm 0cm,clip=true,width=4.1cm]{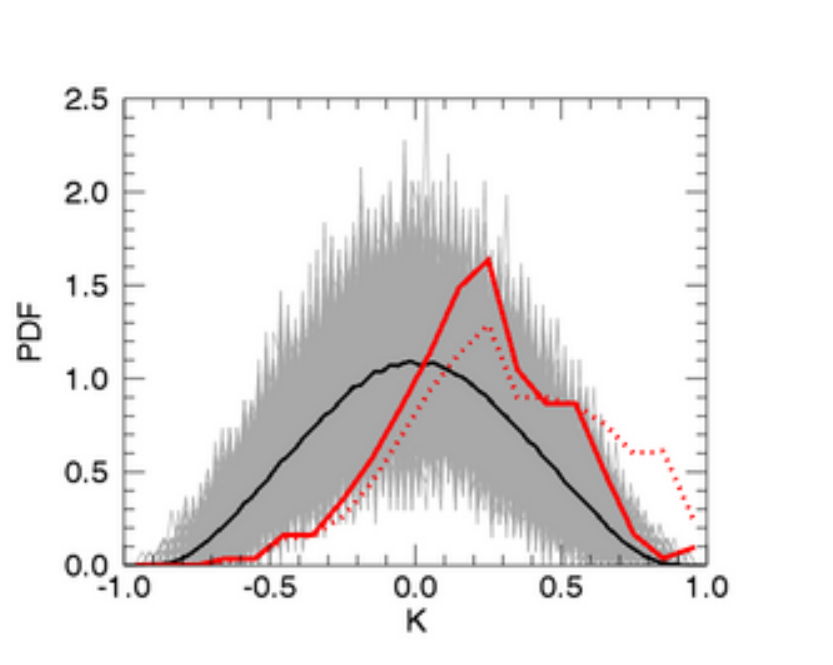}
\end{minipage}
\hspace{0.5cm}
\begin{minipage}{0.2\textwidth}
\centering
\includegraphics[trim=0cm 0cm 0cm 0cm,clip=true,width=4.1cm]{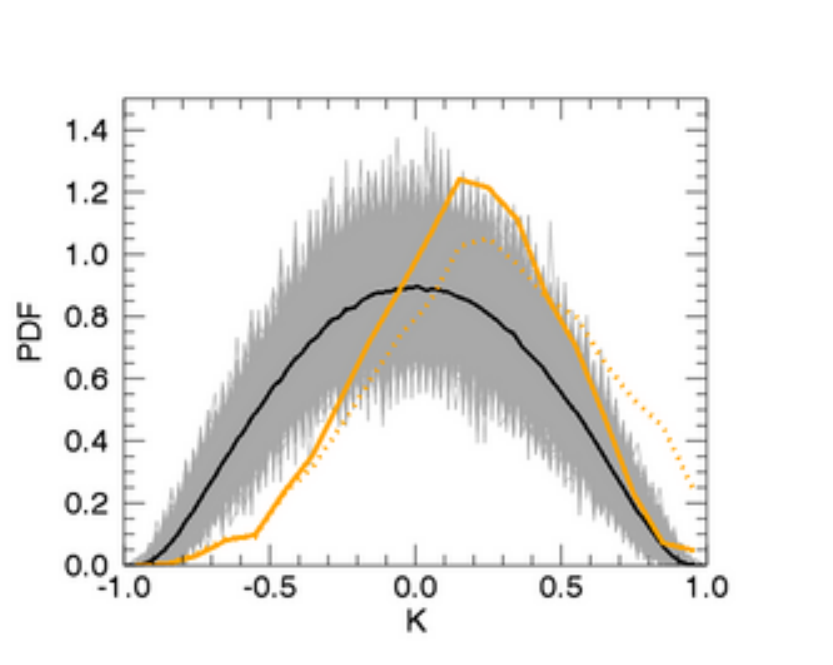}
\end{minipage}
\hspace{0.5cm}
\begin{minipage}{0.2\textwidth}
\centering
\includegraphics[trim=0cm 0cm 0cm 0cm,clip=true,width=4.1cm]{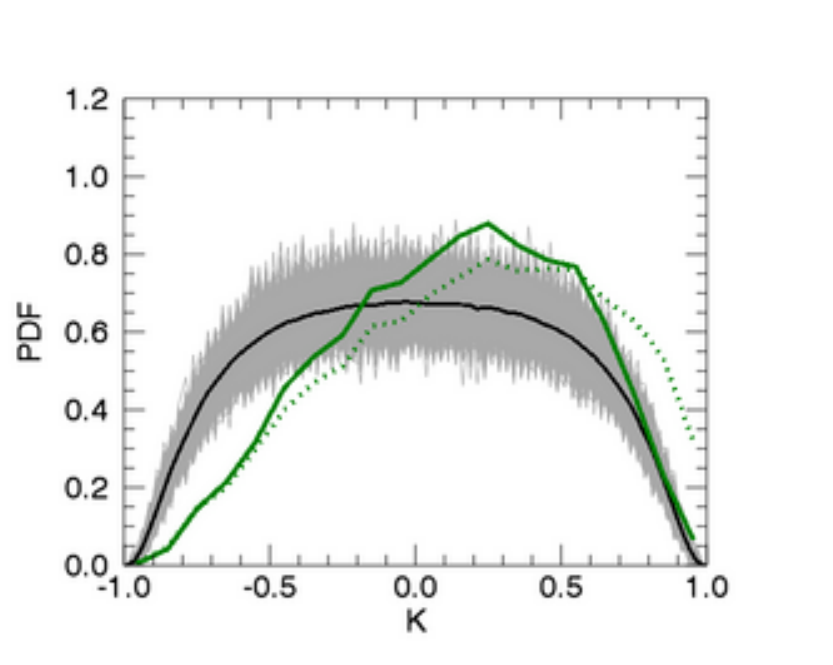}
\end{minipage}
\hspace{0.5cm}
\begin{minipage}{0.2\textwidth}
\centering
\includegraphics[trim=0cm 0cm 0cm 0cm,clip=true,width=4.1cm]{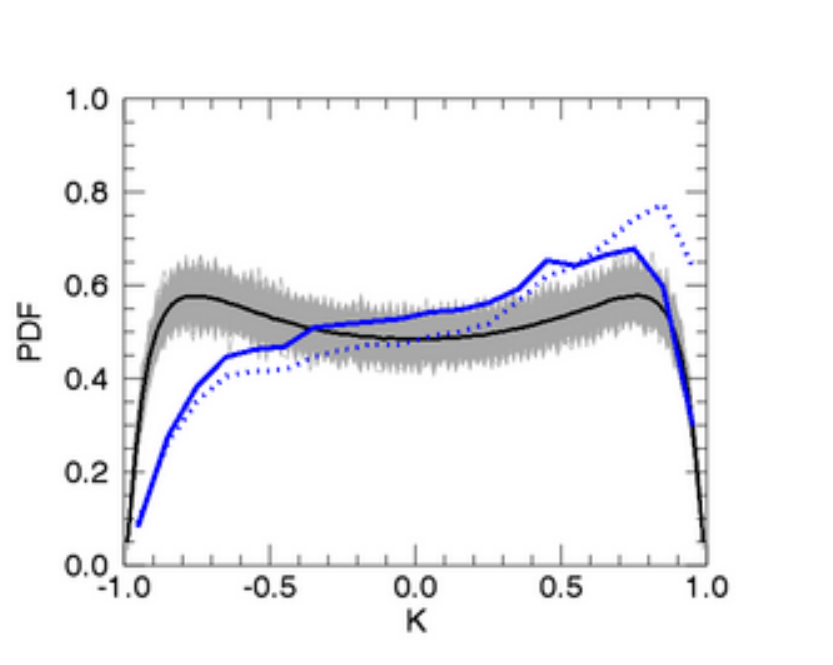}
\end{minipage}
\vspace{-0.2cm}
\begin{minipage}{0.2\textwidth}
\centering
\includegraphics[trim=0cm 0cm 0cm 0cm,clip=true,width=4.1cm]{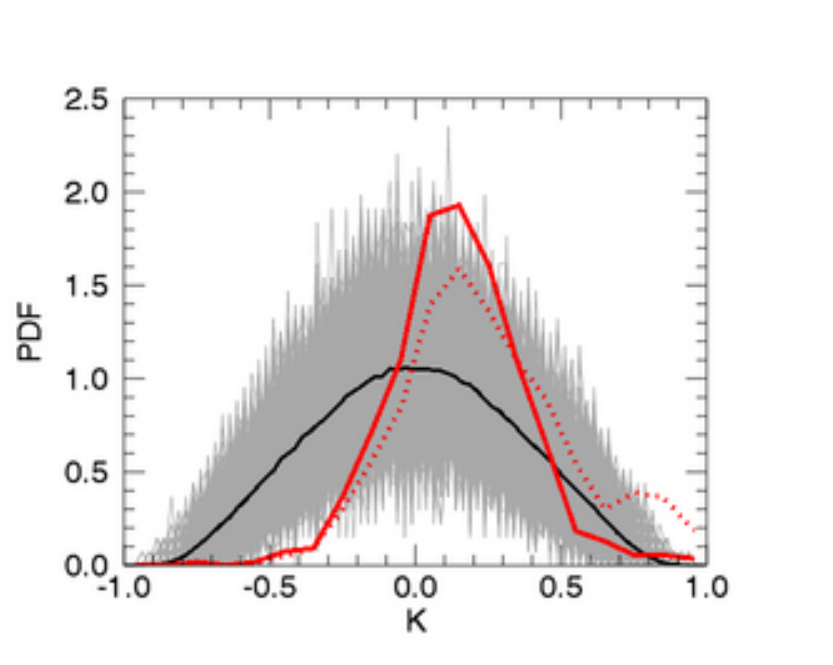}
\end{minipage}
\hspace{0.5cm}
\begin{minipage}{0.2\textwidth}
\centering
\includegraphics[trim=0cm 0cm 0cm 0cm,clip=true,width=4.1cm]{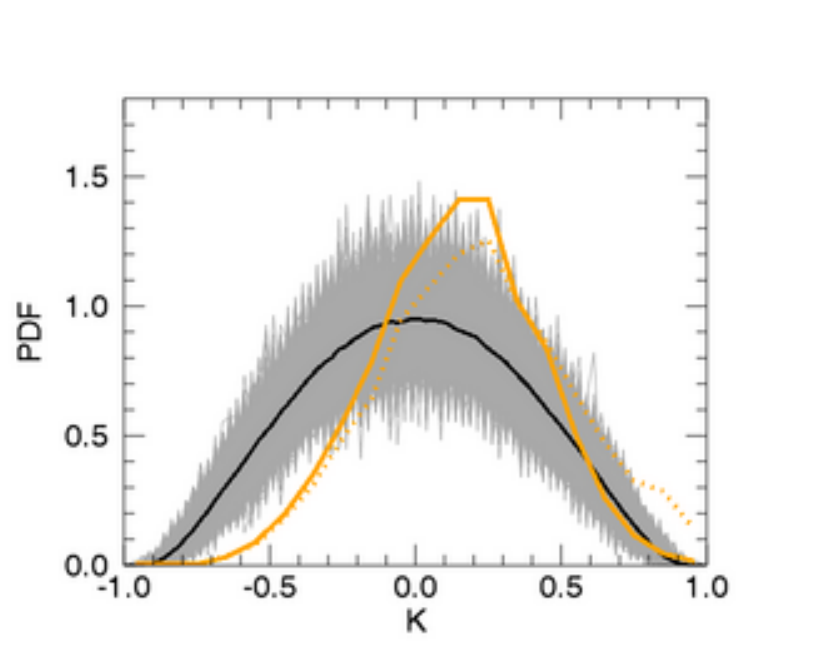}
\end{minipage}
\hspace{0.5cm}
\begin{minipage}{0.2\textwidth}
\centering
\includegraphics[trim=0cm 0cm 0cm 0cm,clip=true,width=4.1cm]{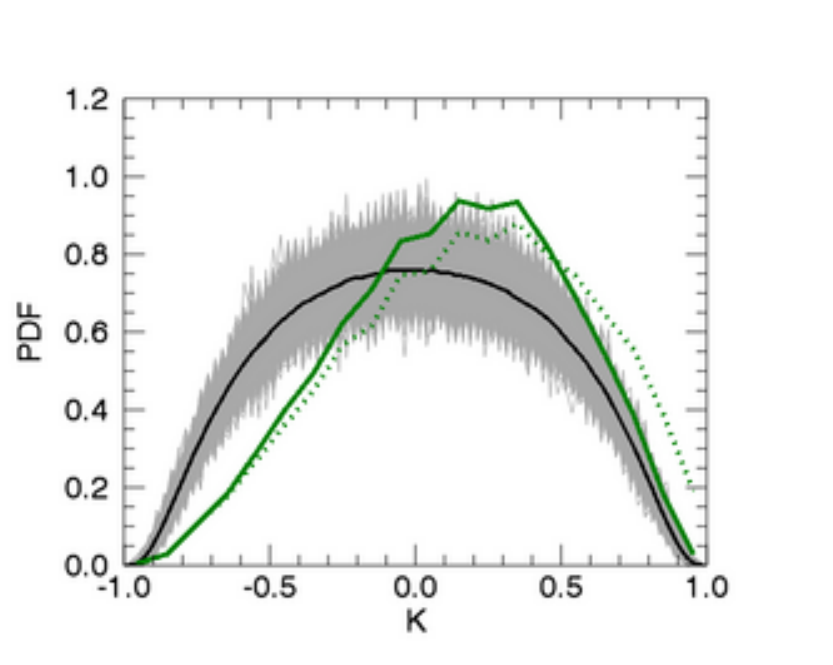}
\end{minipage}
\hspace{0.5cm}
\begin{minipage}{0.2\textwidth}
\centering
\includegraphics[trim=0cm 0cm 0cm 0cm,clip=true,width=4.1cm]{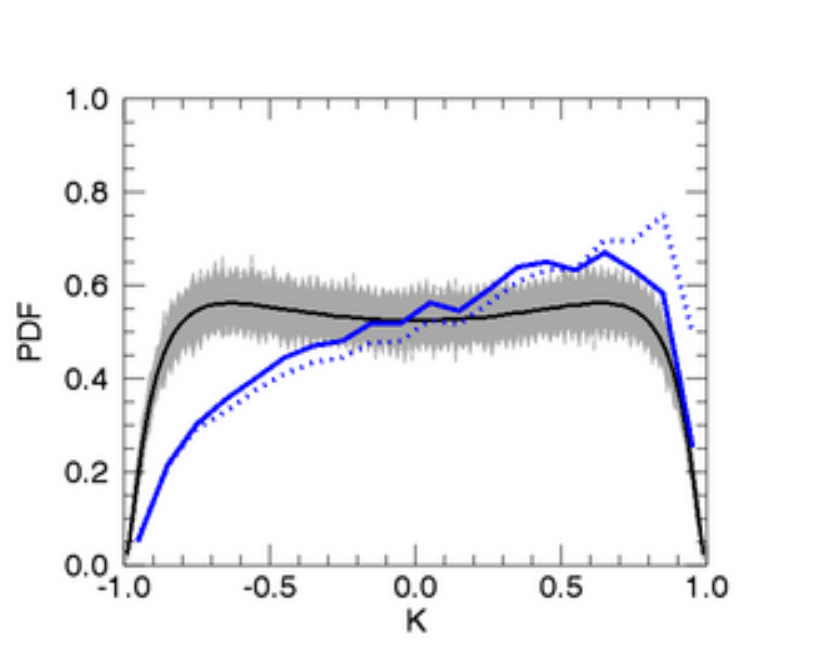}
\end{minipage}
\vspace{-0.2cm}
\begin{minipage}{0.2\textwidth}
\centering
\includegraphics[trim=0cm 0cm 0cm 0cm,clip=true,width=4.1cm]{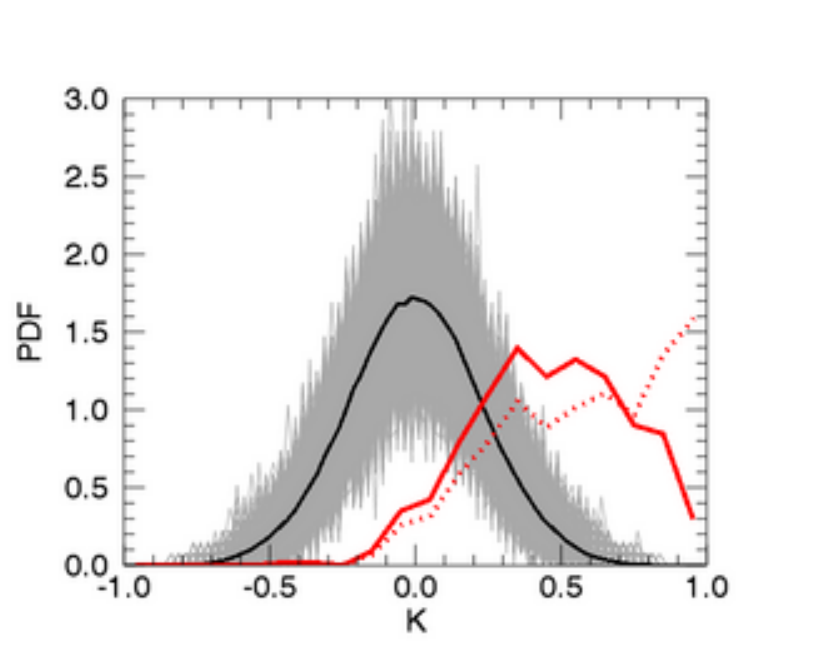}
\end{minipage}
\hspace{0.5cm}
\begin{minipage}{0.2\textwidth}
\centering
\includegraphics[trim=0cm 0cm 0cm 0cm,clip=true,width=4.1cm]{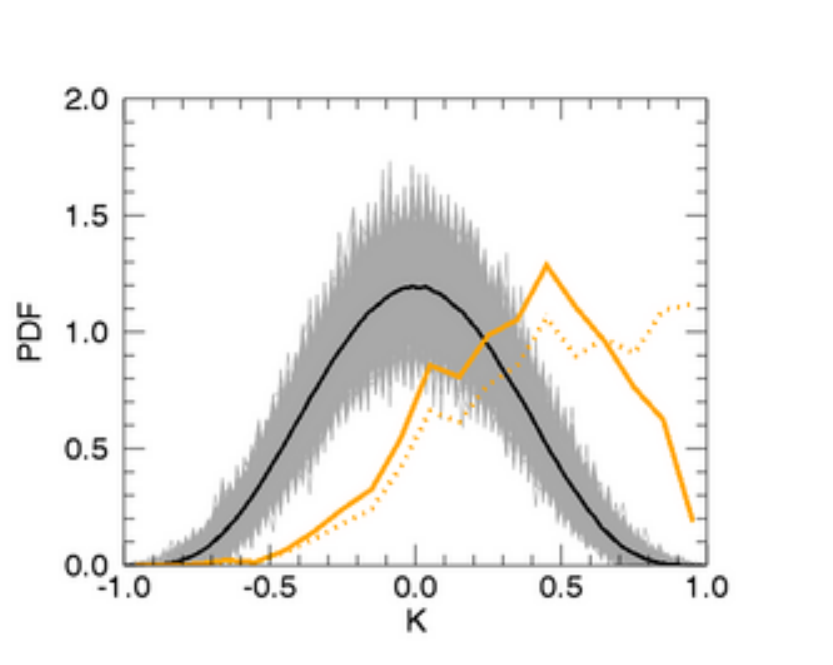}
\end{minipage}
\hspace{0.5cm}
\begin{minipage}{0.2\textwidth}
\centering
\includegraphics[trim=0cm 0cm 0cm 0cm,clip=true,width=4.1cm]{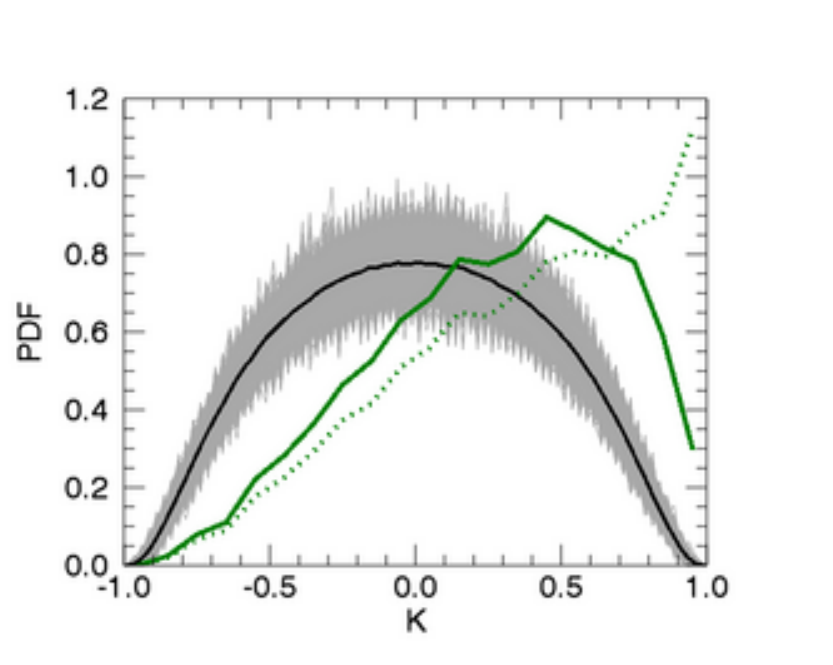}
\end{minipage}
\hspace{0.5cm}
\begin{minipage}{0.2\textwidth}
\centering
\includegraphics[trim=0cm 0cm 0cm 0cm,clip=true,width=4.1cm]{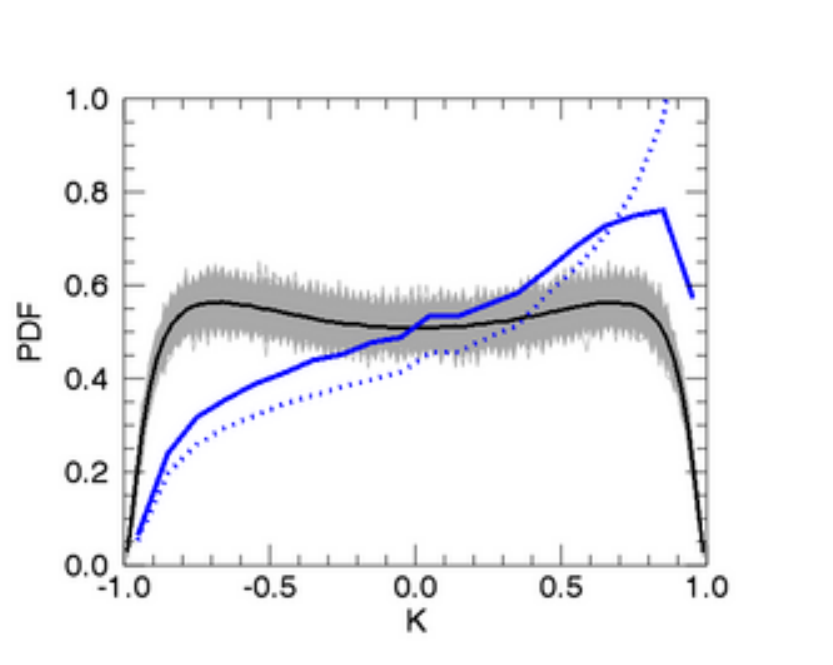}
\end{minipage}
\vspace{-0.2cm}
\begin{minipage}{0.2\textwidth}
\centering
\includegraphics[trim=0cm 0cm 0cm 0cm,clip=true,width=4.1cm]{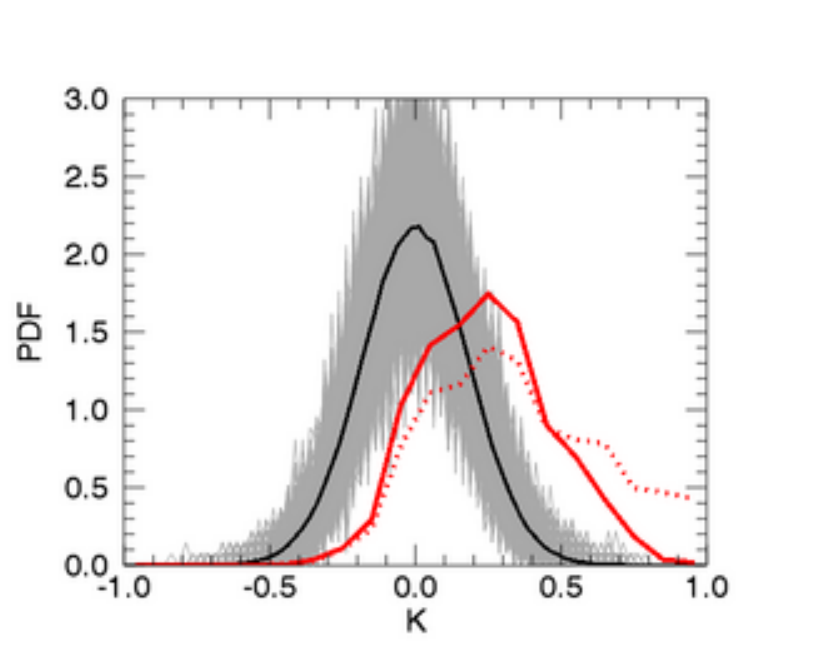}
\end{minipage}
\hspace{0.5cm}
\begin{minipage}{0.2\textwidth}
\centering
\includegraphics[trim=0cm 0cm 0cm 0cm,clip=true,width=4.1cm]{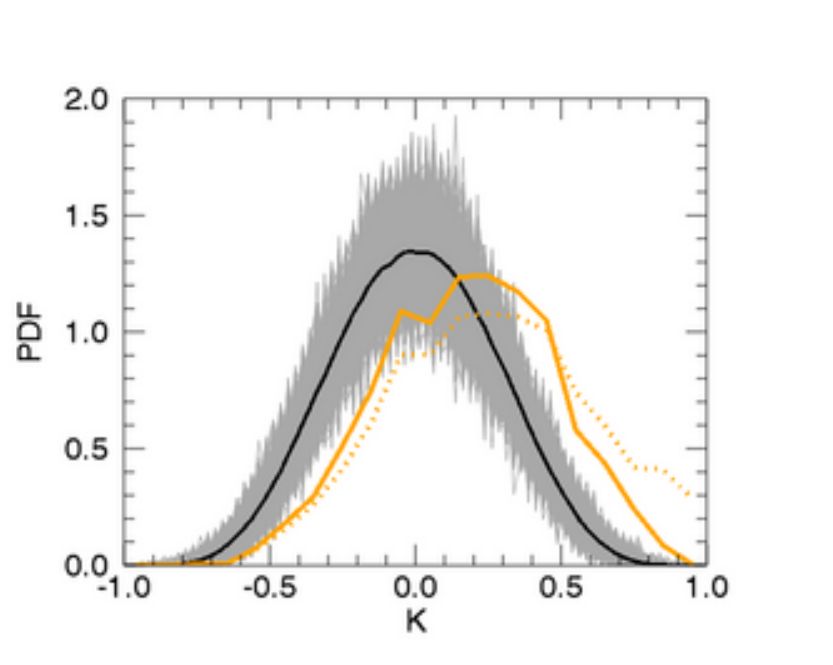}
\end{minipage}
\hspace{0.5cm}
\begin{minipage}{0.2\textwidth}
\centering
\includegraphics[trim=0cm 0cm 0cm 0cm,clip=true,width=4.1cm]{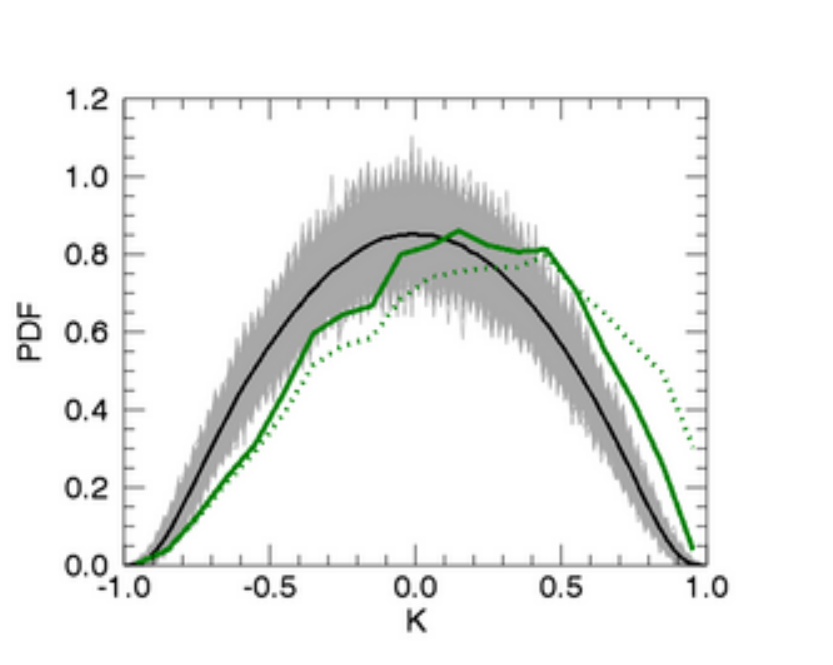}
\end{minipage}
\hspace{0.5cm}
\begin{minipage}{0.2\textwidth}
\centering
\includegraphics[trim=0cm 0cm 0cm 0cm,clip=true,width=4.1cm]{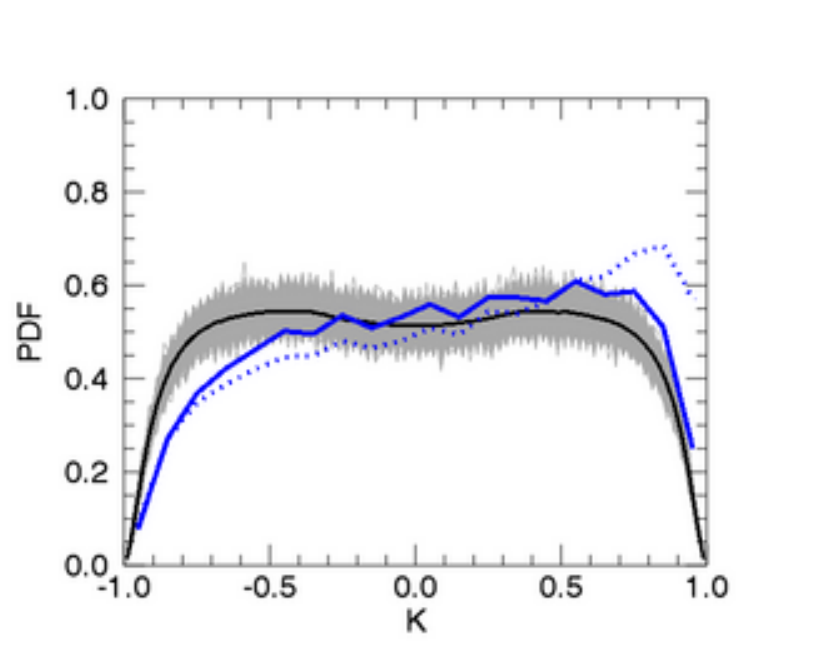}
\end{minipage}
\end{center}
\caption{PDFs of the unweighted correlation coefficients K for different combinations
of foregrounds and each of the $\Omega$ sizes.
From top to bottom: Correlation between AME1 and free-free emission, AME2 and free-free emission, 
AME1 and synchrotron radiation, AME2 and synchrotron radiation, AME1 and thermal dust emission, AME2 and thermal dust emission. 
Colored lines give distributions of K's from the actual foregrounds, solid for masked sky, dotted for unmasked sky. The gray lines
show all distributions of K's resulting from random simulations of AME1 or AME2 correlated with the respective foregrounds. 
Black lines are averages over all 1000 random realizations. Simulations are shown for the masked sky. 
From left to right: $\Omega$ contains 1024, 256, 64 and 16 pixels.}
\label{distno}
\end{figure}

We start our analysis with unweighted correlations, corresponding to the assumption that $\sigma_i=0$ and thus $w_i=1$. Figure \ref{corrmapsno} shows the correlation maps for the combinations of AME1 and AME2 with 
free-free emission, synchrotron radiation and thermal dust emission for all discussed pixelizations. Strong correlations are found for all combinations along the galactic plane. When showing the PDFs of each of the maps in figure \ref{distno}, we therefore plot both - a PDF of all pixels of the map as a dotted line and another as a solid line after excluding the galactic plane, simply taken to be a uniform band of 30$^{\circ}$ opening angle centered around the galactic plane. We notice that, as expected, high correlation values are removed. In addition, the average PDF over 1000 simulations, computed as described in section \ref{sec:assess}, are shown as a black line with those of all single simulations in gray indicating the spread.

\subsection{Weighted correlations}
\label{weightedcorr}

\begin{figure}[H]
\centering
\begin{minipage}{0.24\textwidth}
\centering
\includegraphics[trim=0cm 1.8cm 0cm 1.2cm,clip=true,width=4cm]{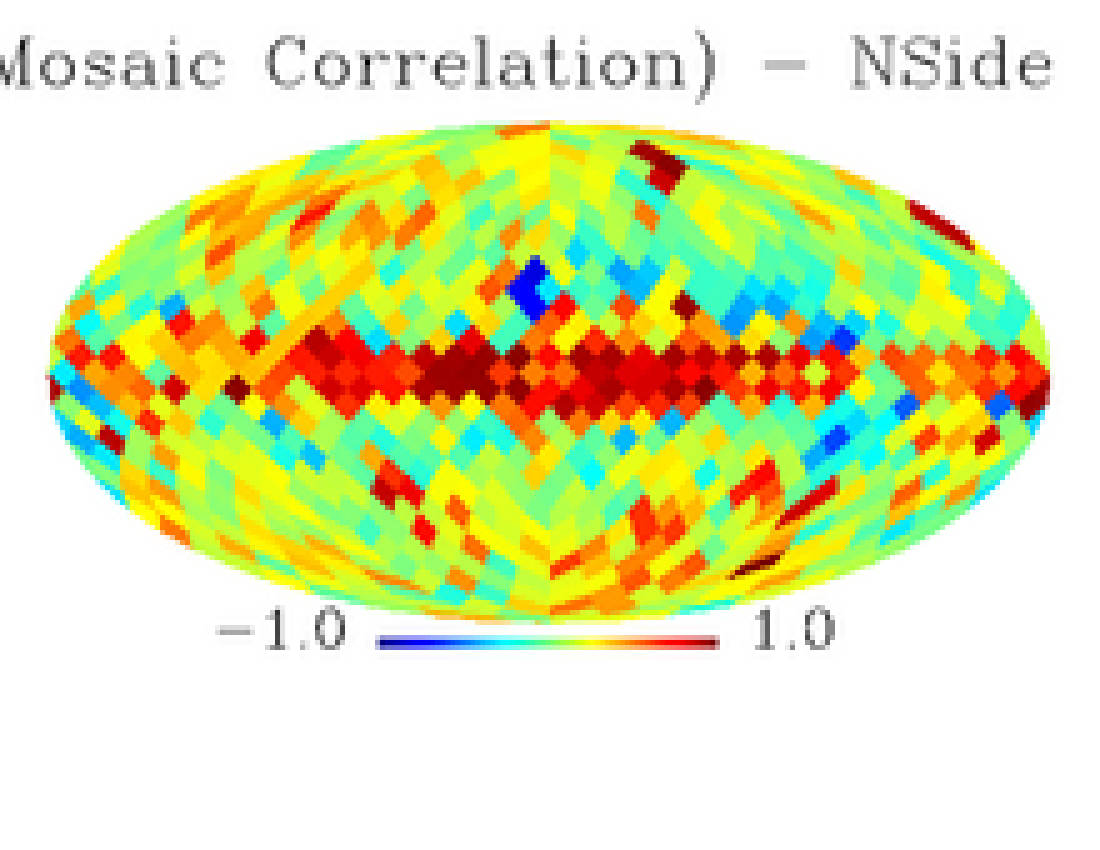}
\end{minipage}
\hfill
\begin{minipage}{0.24\textwidth}
\centering
\includegraphics[trim=0cm 1.8cm 0cm 1.2cm,clip=true,width=4cm]{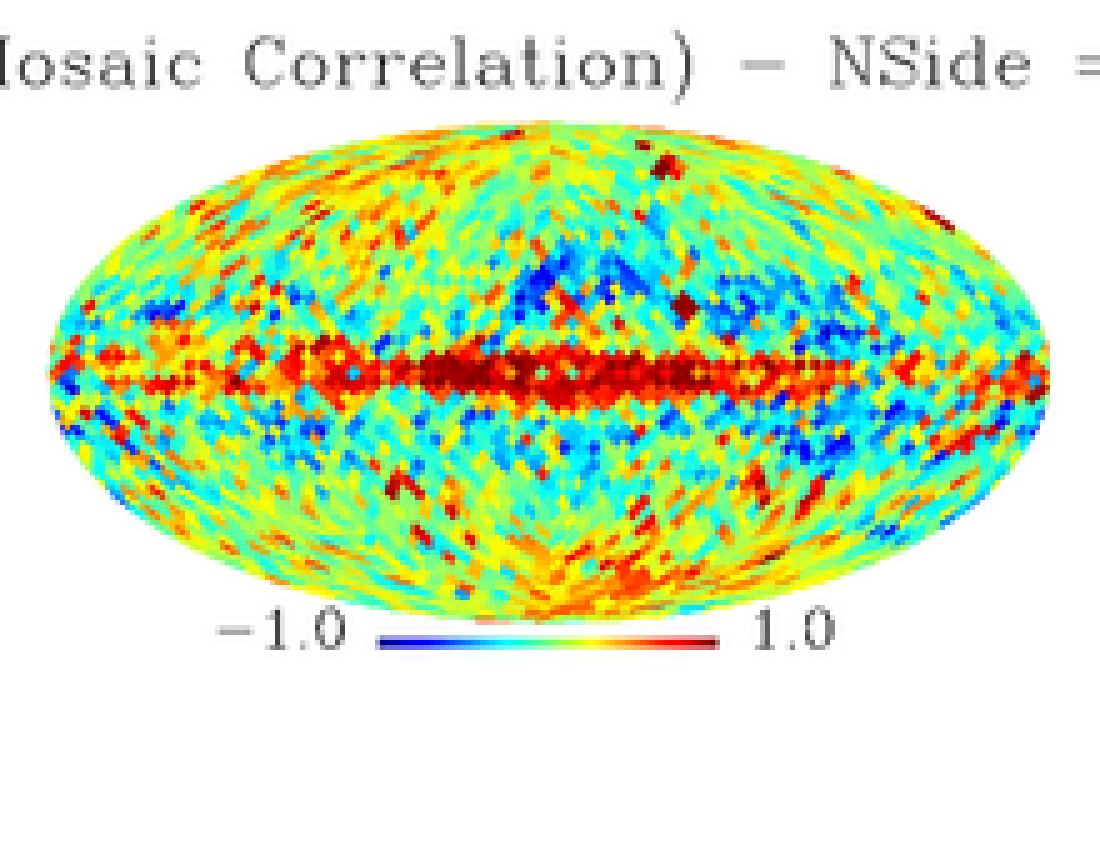}
\end{minipage}
\hfill
\centering
\begin{minipage}{0.24\textwidth}
\centering
\includegraphics[trim=0cm 1.8cm 0cm 1.2cm,clip=true,width=4cm]{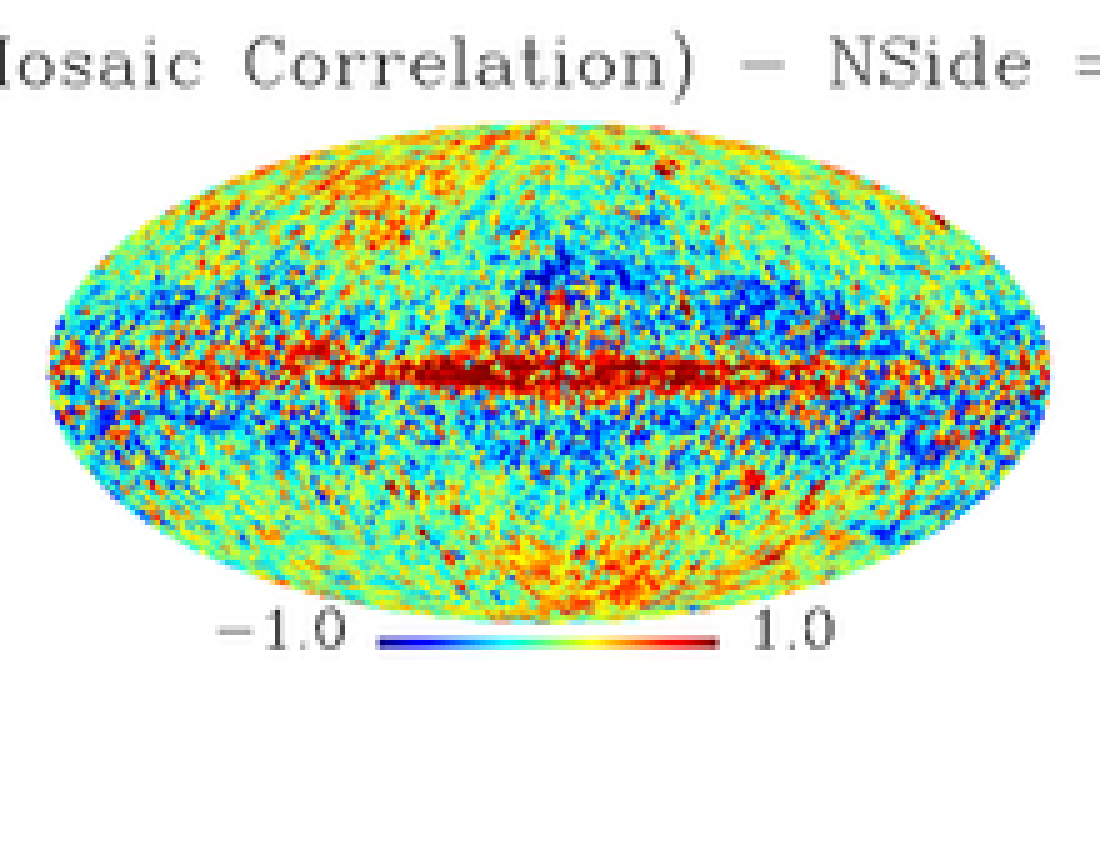}
\end{minipage}
\hfill
\begin{minipage}{0.24\textwidth}
\centering
\includegraphics[trim=0cm 1.8cm 0cm 1.2cm,clip=true,width=4cm]{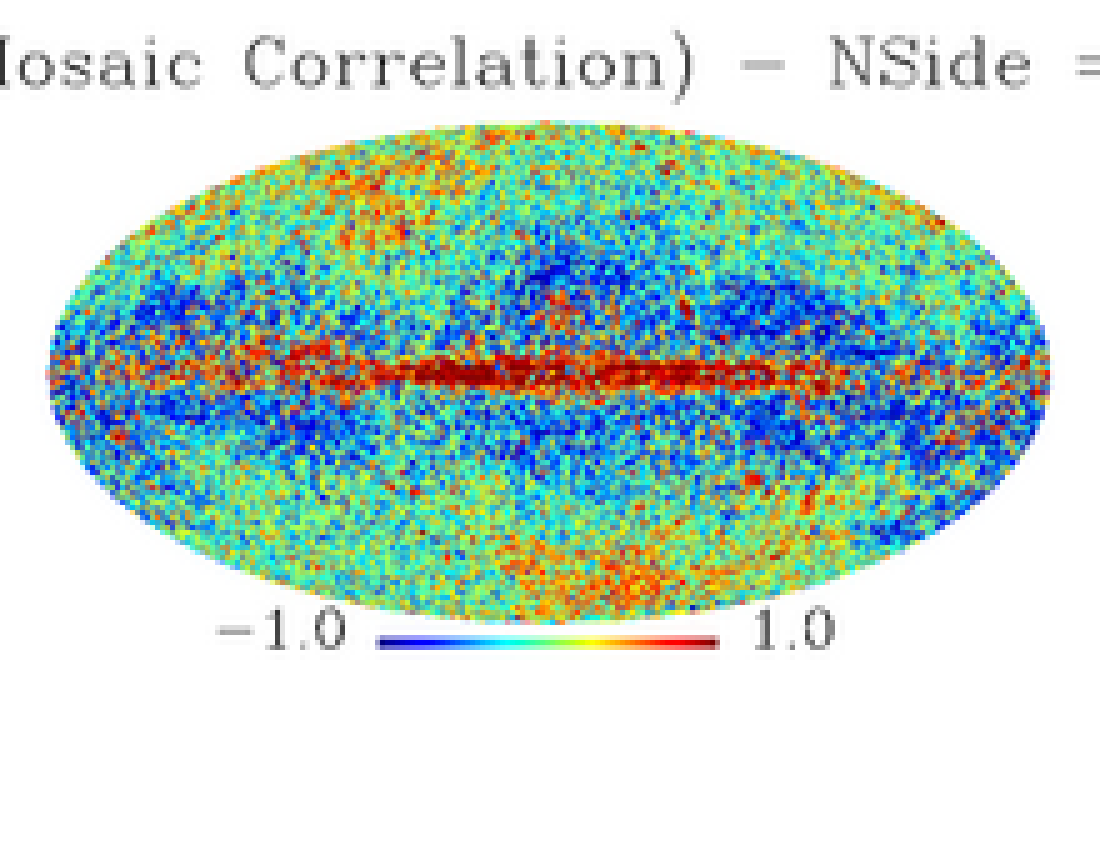}
\end{minipage}
\centering
\begin{minipage}{0.24\textwidth}
\centering
\includegraphics[trim=0cm 1.8cm 0cm 1.2cm,clip=true,width=4cm]{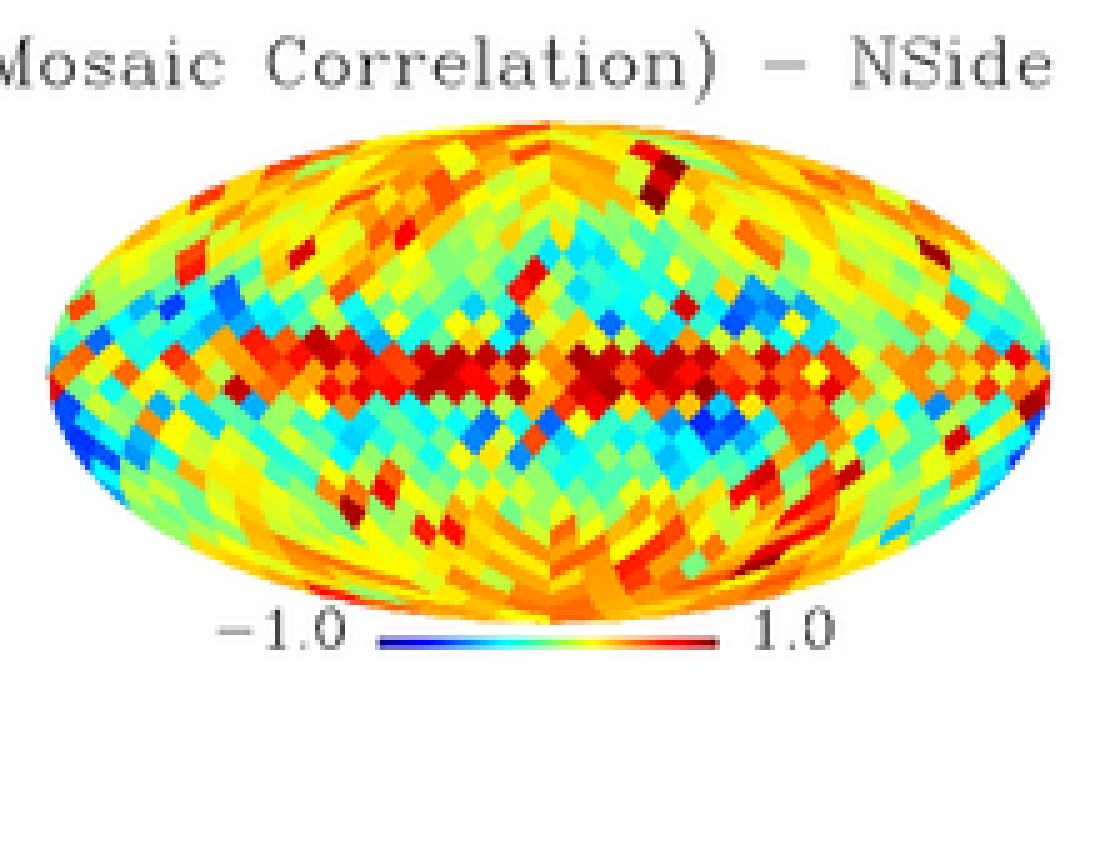}
\end{minipage}
\hfill
\begin{minipage}{0.24\textwidth}
\centering
\includegraphics[trim=0cm 1.8cm 0cm 1.2cm,clip=true,width=4cm]{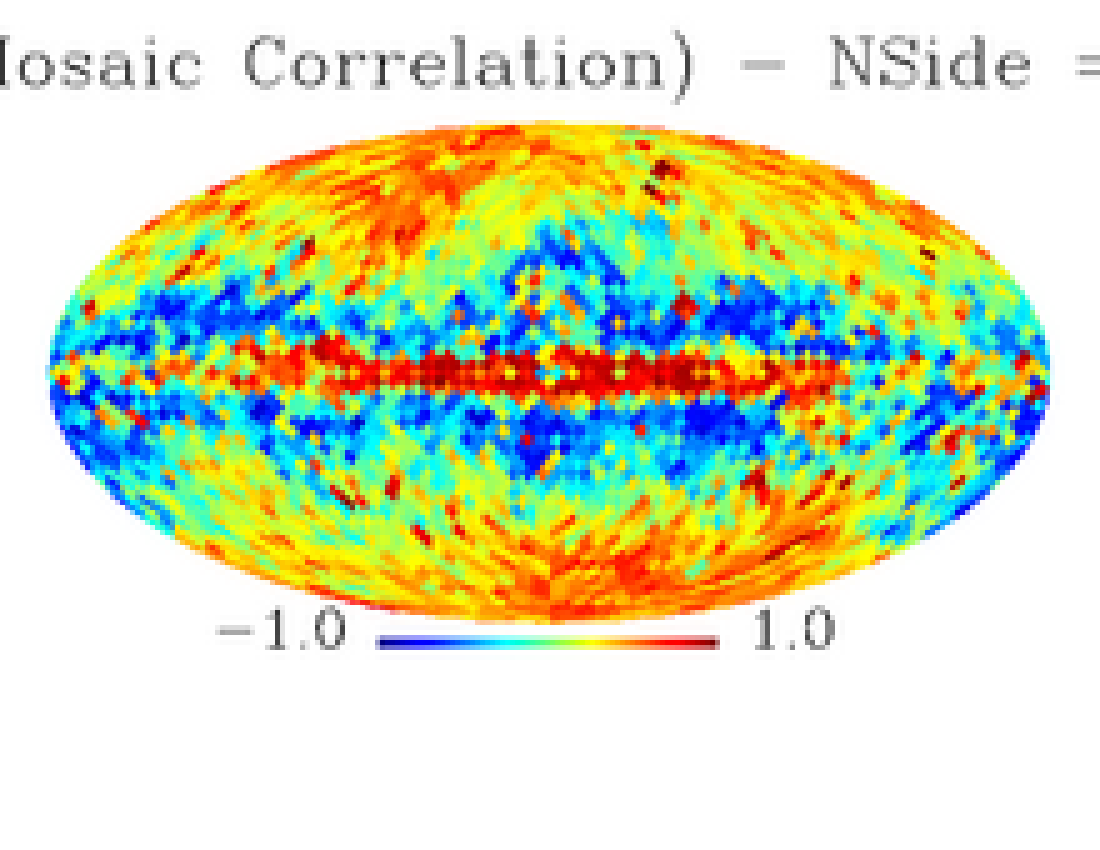}
\end{minipage}
\hfill
\centering
\begin{minipage}{0.24\textwidth}
\centering
\includegraphics[trim=0cm 1.8cm 0cm 1.2cm,clip=true,width=4cm]{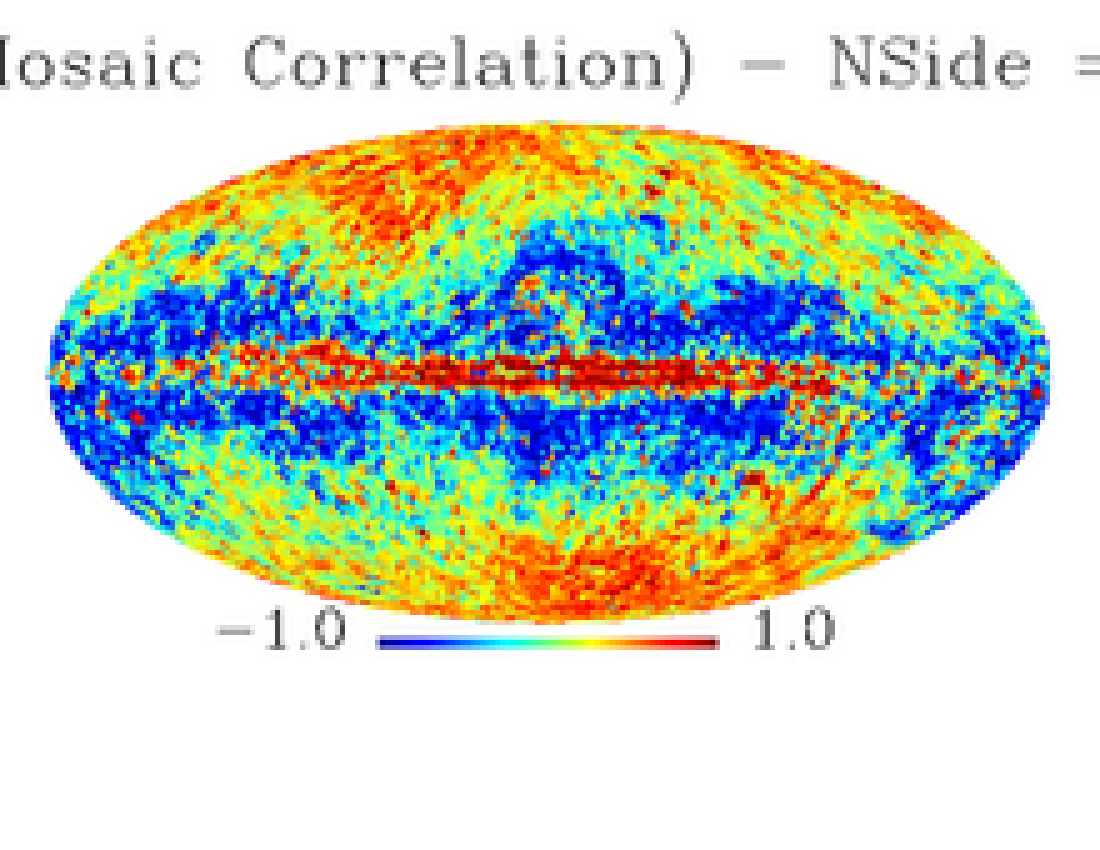}
\end{minipage}
\hfill
\begin{minipage}{0.24\textwidth}
\centering
\includegraphics[trim=0cm 1.8cm 0cm 1.2cm,clip=true,width=4cm]{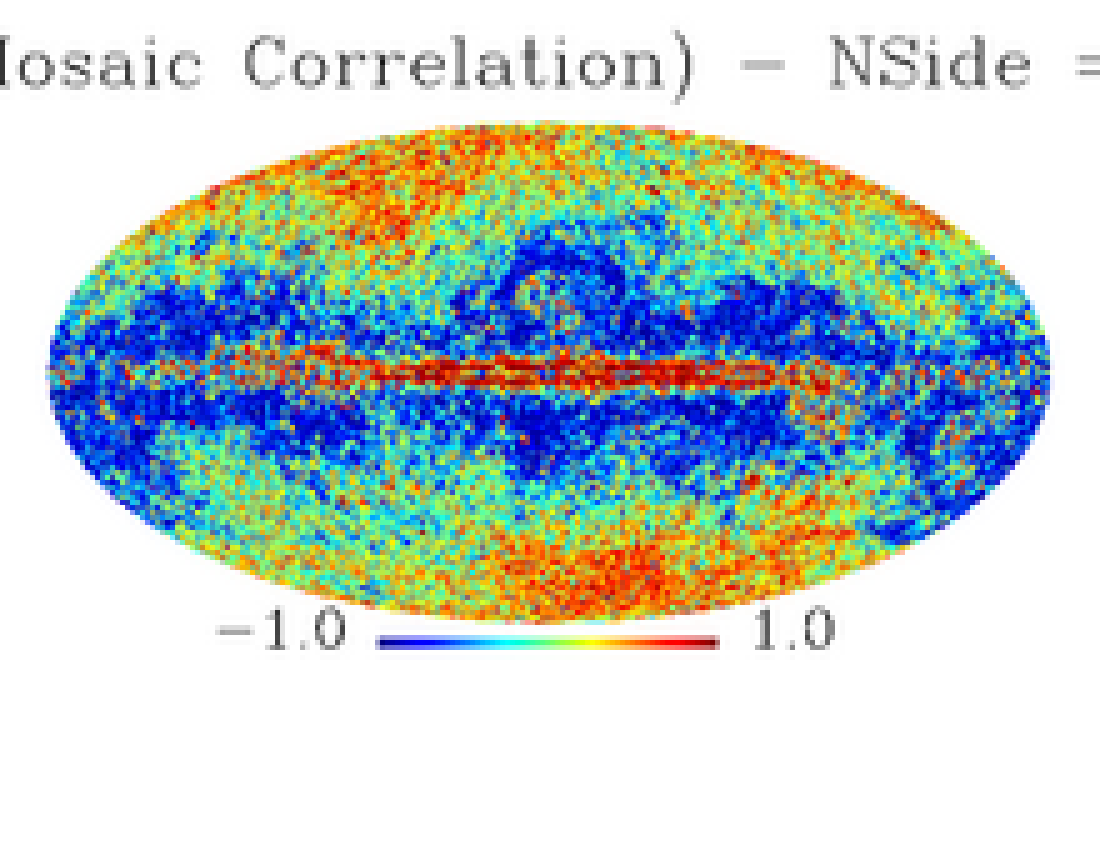}
\end{minipage}
\caption{From top to bottom: Weighted-Correlation maps between AME1 and free-free emission, AME2 and free-free emission. From left to right: $\Omega$ contains 1024, 256, 64 and 16 pixel. Here, we weighted with $w_i(n=1)$.}
\label{corrmaps1}
\end{figure}

\begin{figure}[H]
\centering
\begin{minipage}{0.24\textwidth}
\centering
\includegraphics[trim=0cm 1.8cm 0cm 1.2cm,clip=true,width=4cm]{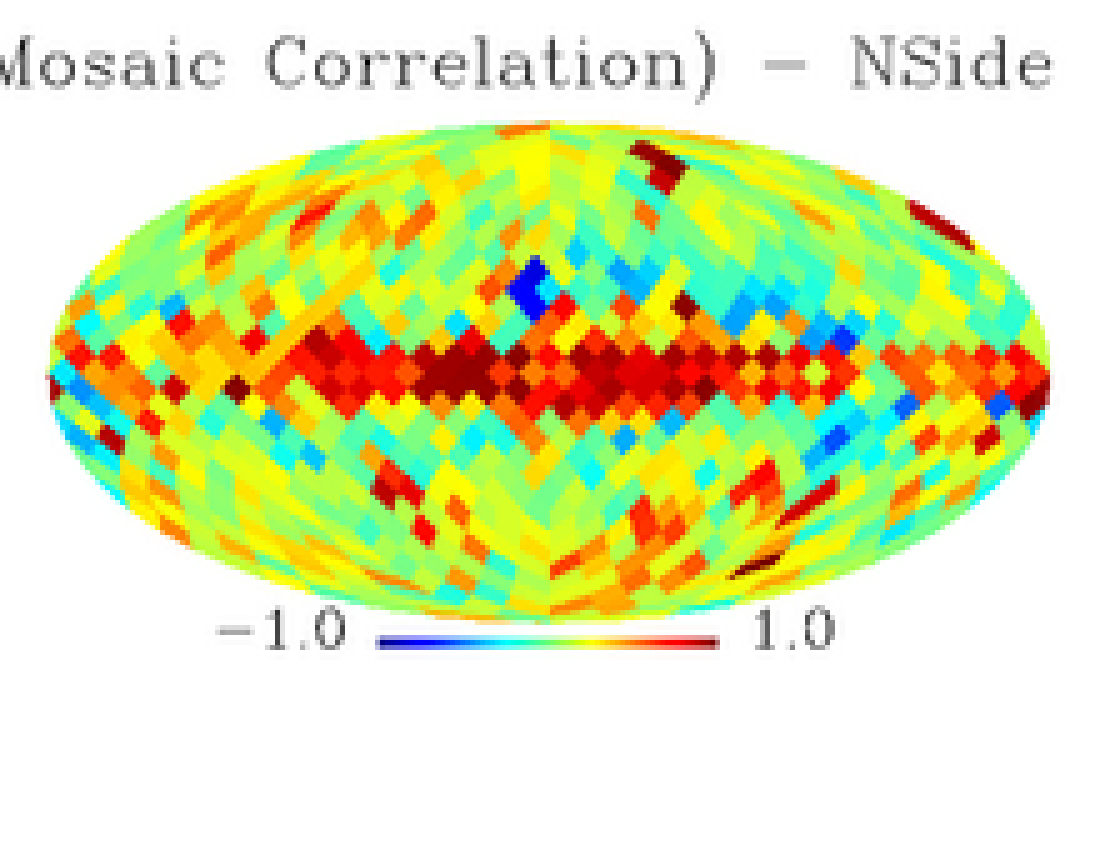}
\end{minipage}
\hfill
\begin{minipage}{0.24\textwidth}
\centering
\includegraphics[trim=0cm 1.8cm 0cm 1.2cm,clip=true,width=4cm]{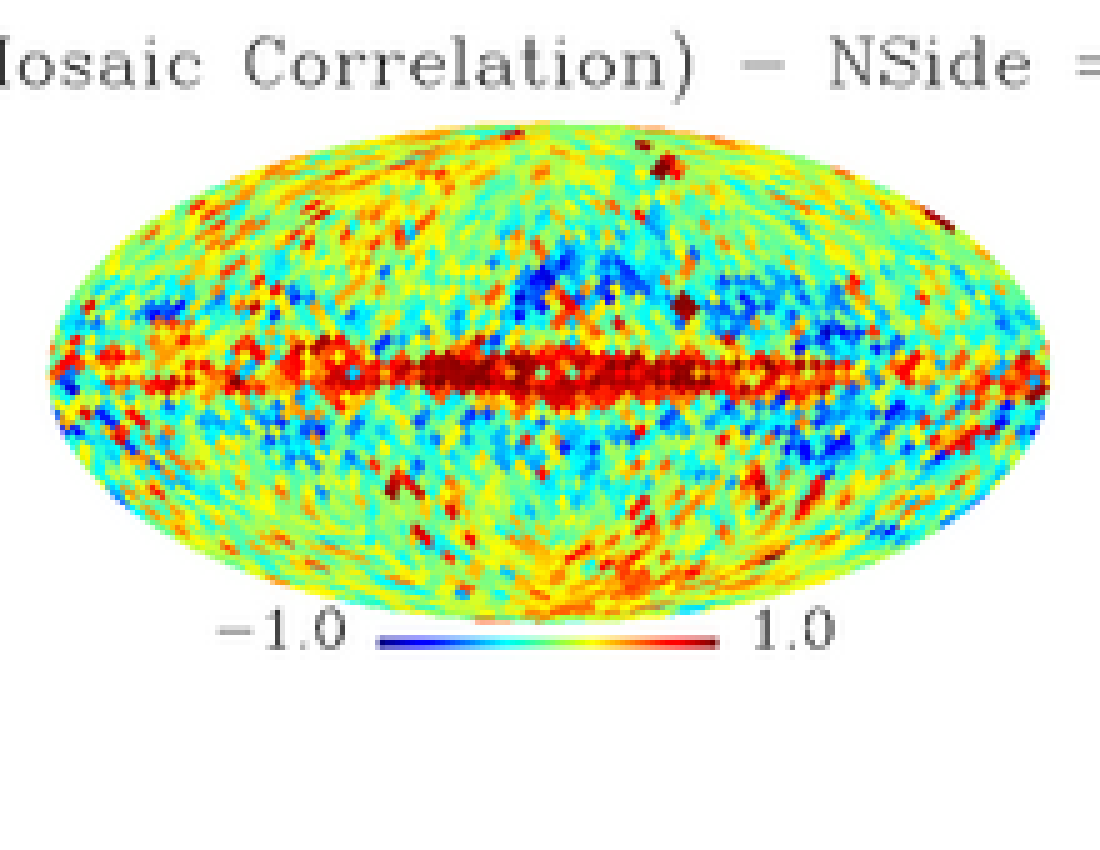}
\end{minipage}
\hfill
\centering
\begin{minipage}{0.24\textwidth}
\centering
\includegraphics[trim=0cm 1.8cm 0cm 1.2cm,clip=true,width=4cm]{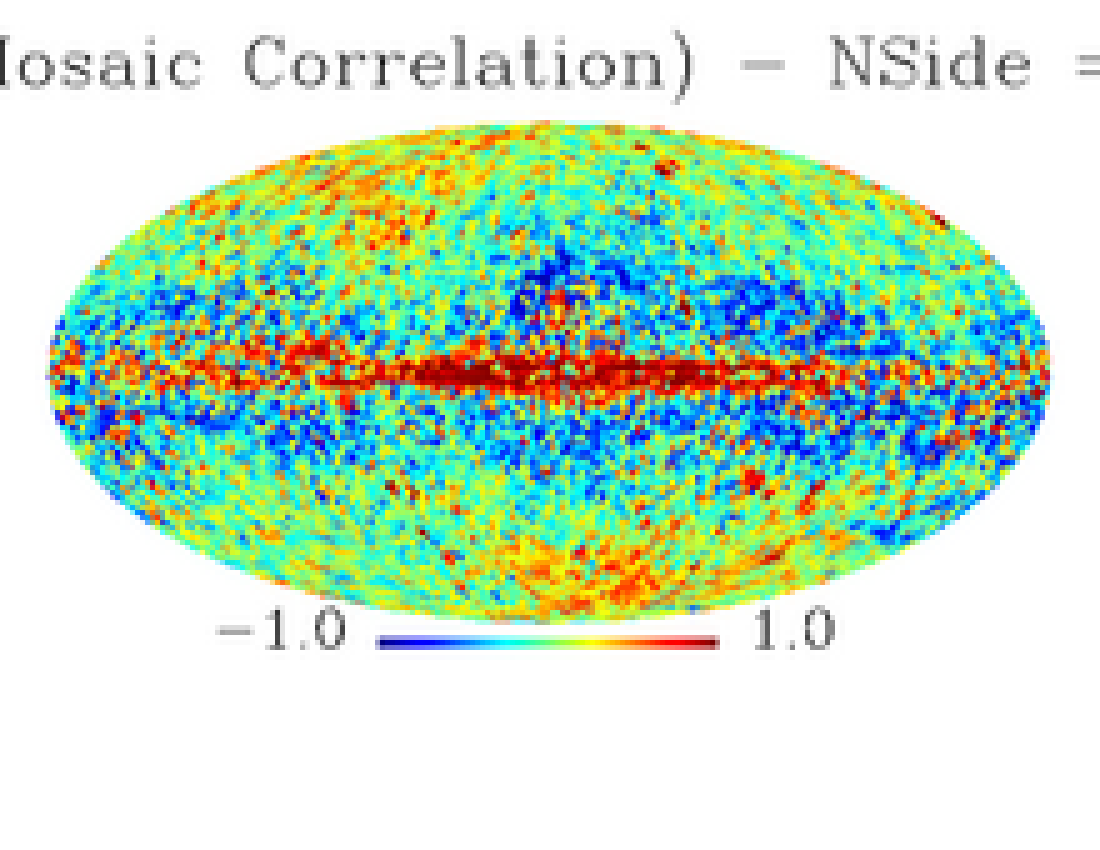}
\end{minipage}
\hfill
\begin{minipage}{0.24\textwidth}
\centering
\includegraphics[trim=0cm 1.8cm 0cm 1.2cm,clip=true,width=4cm]{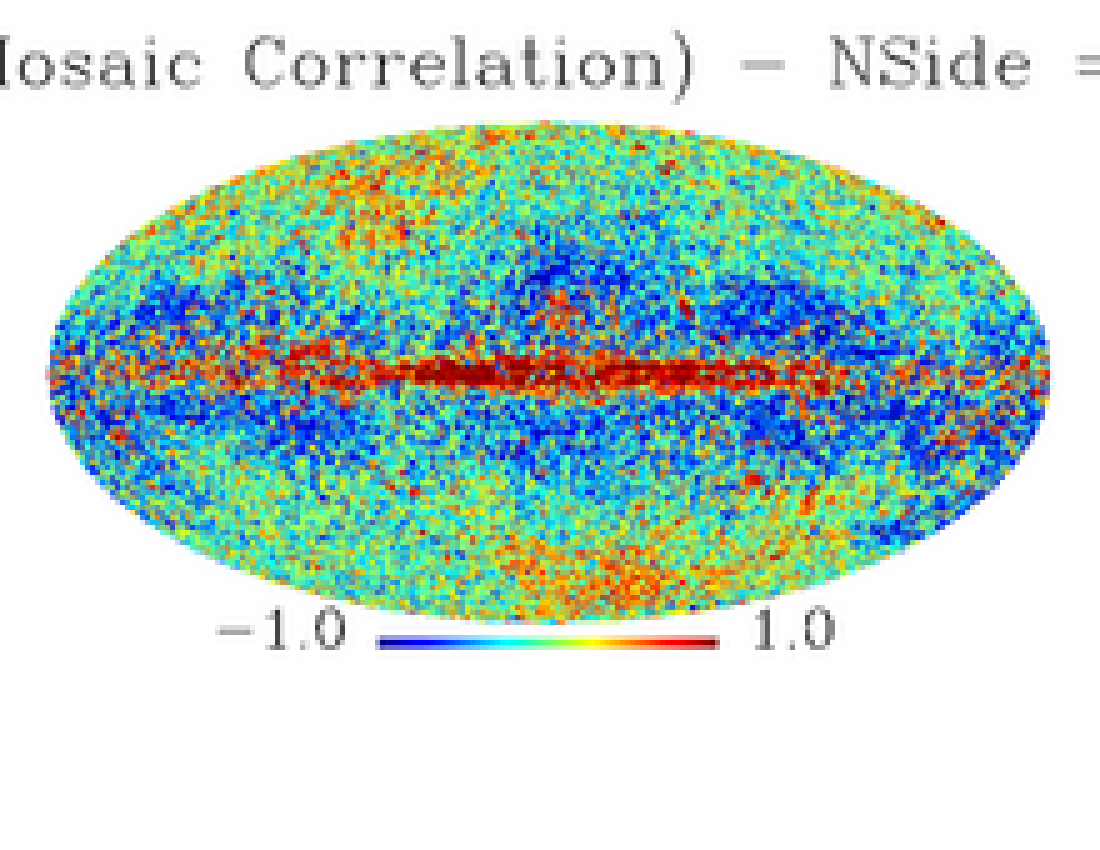}
\end{minipage}
\centering
\begin{minipage}{0.24\textwidth}
\centering
\includegraphics[trim=0cm 1.8cm 0cm 1.2cm,clip=true,width=4cm]{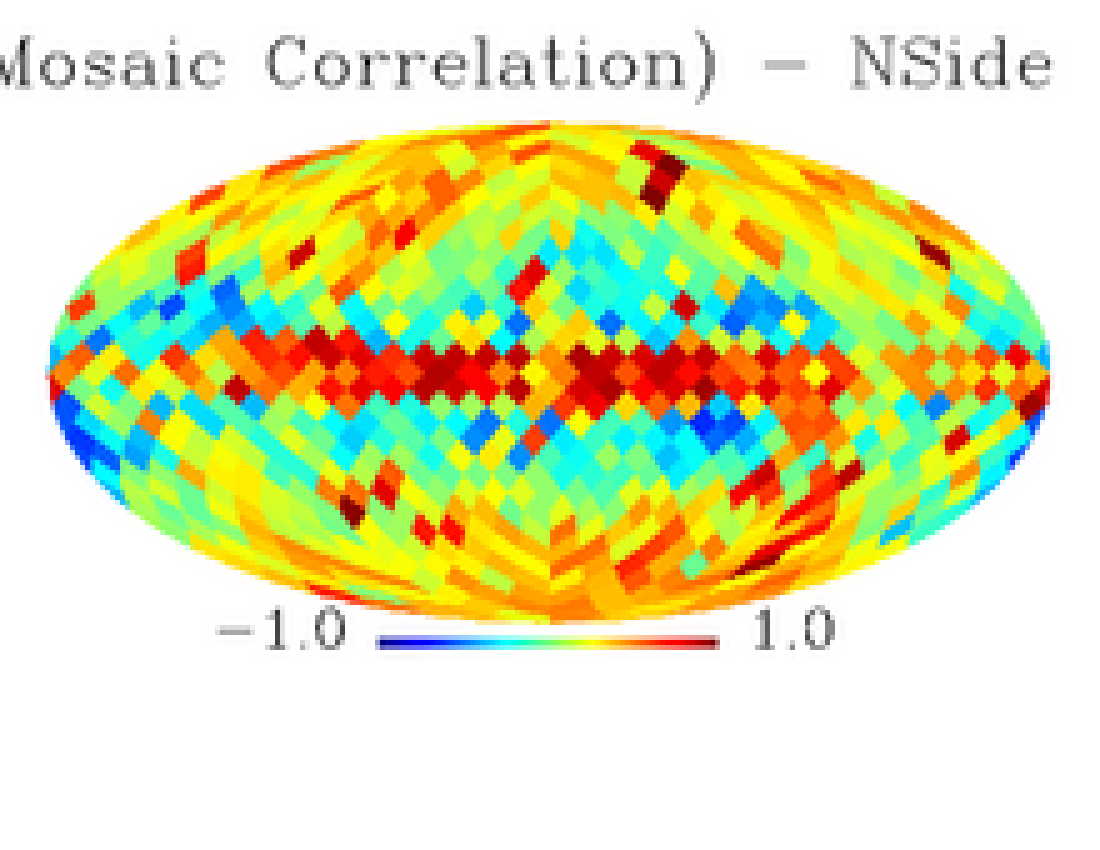}
\end{minipage}
\hfill
\begin{minipage}{0.24\textwidth}
\centering
\includegraphics[trim=0cm 1.8cm 0cm 1.2cm,clip=true,width=4cm]{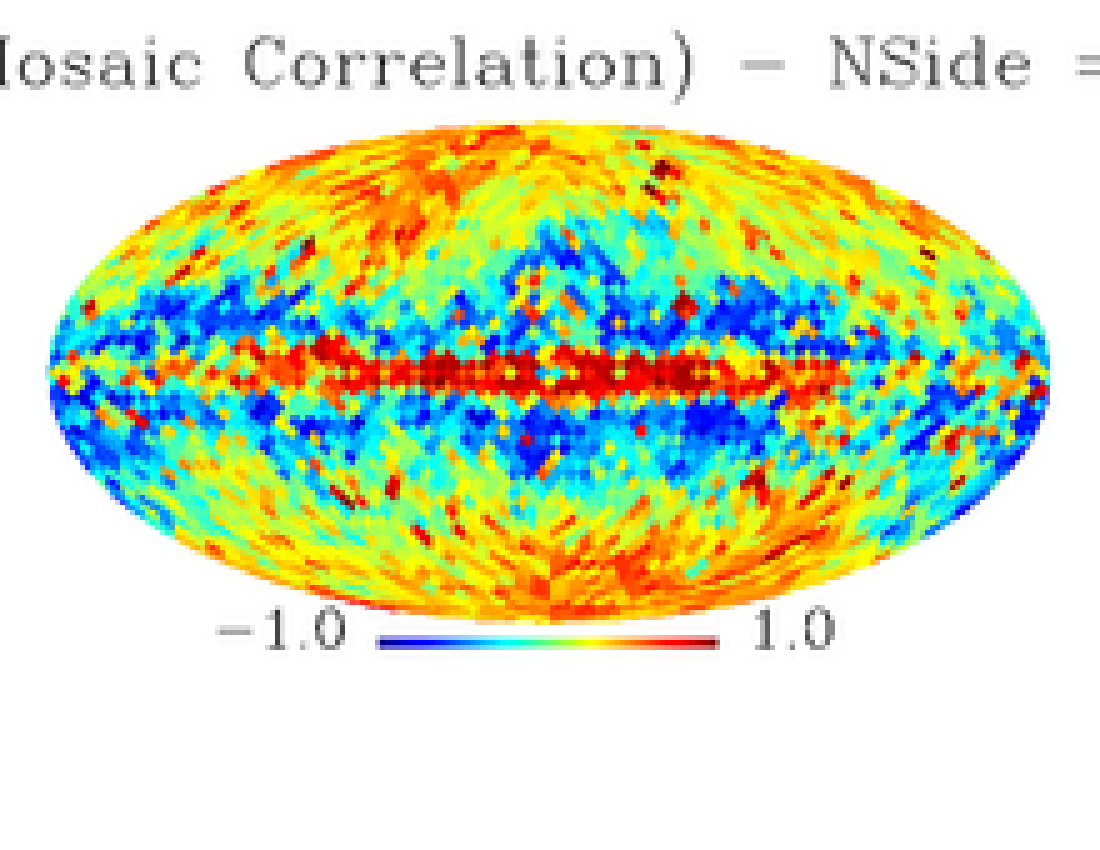}
\end{minipage}
\hfill
\centering
\begin{minipage}{0.24\textwidth}
\centering
\includegraphics[trim=0cm 1.8cm 0cm 1.2cm,clip=true,width=4cm]{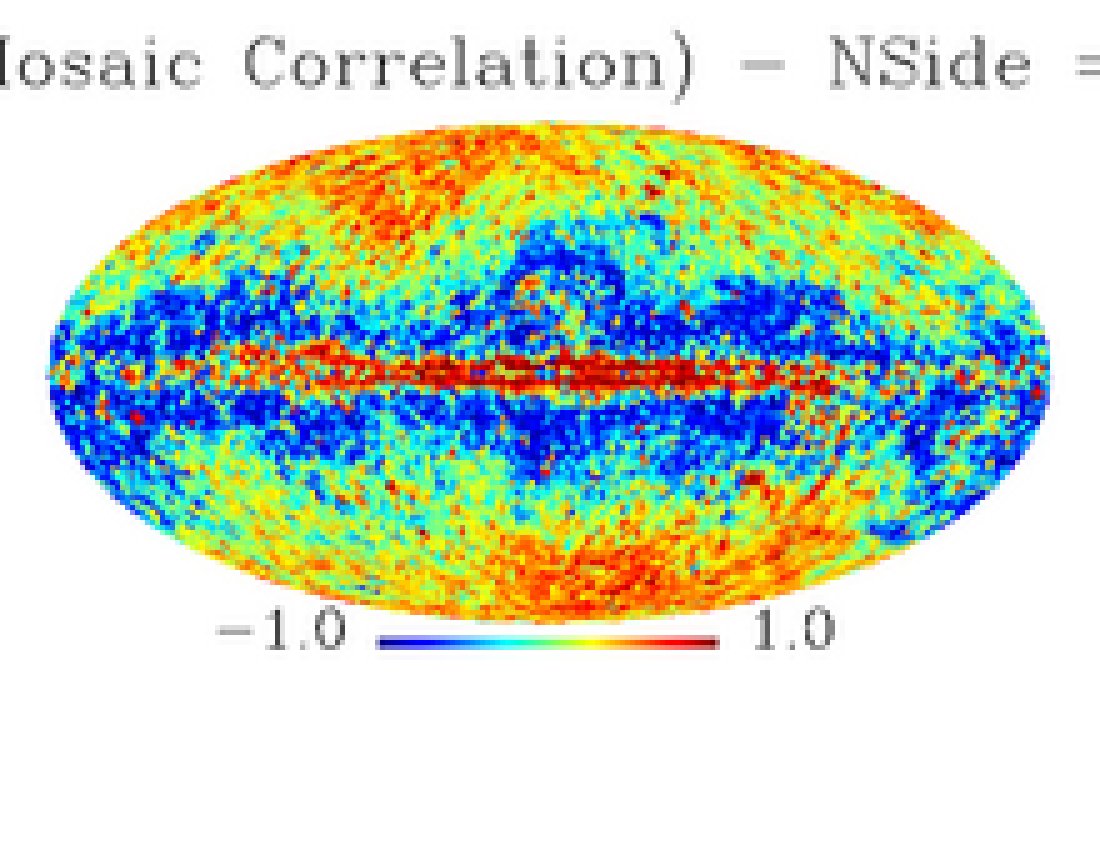}
\end{minipage}
\hfill
\begin{minipage}{0.24\textwidth}
\centering
\includegraphics[trim=0cm 1.8cm 0cm 1.2cm,clip=true,width=4cm]{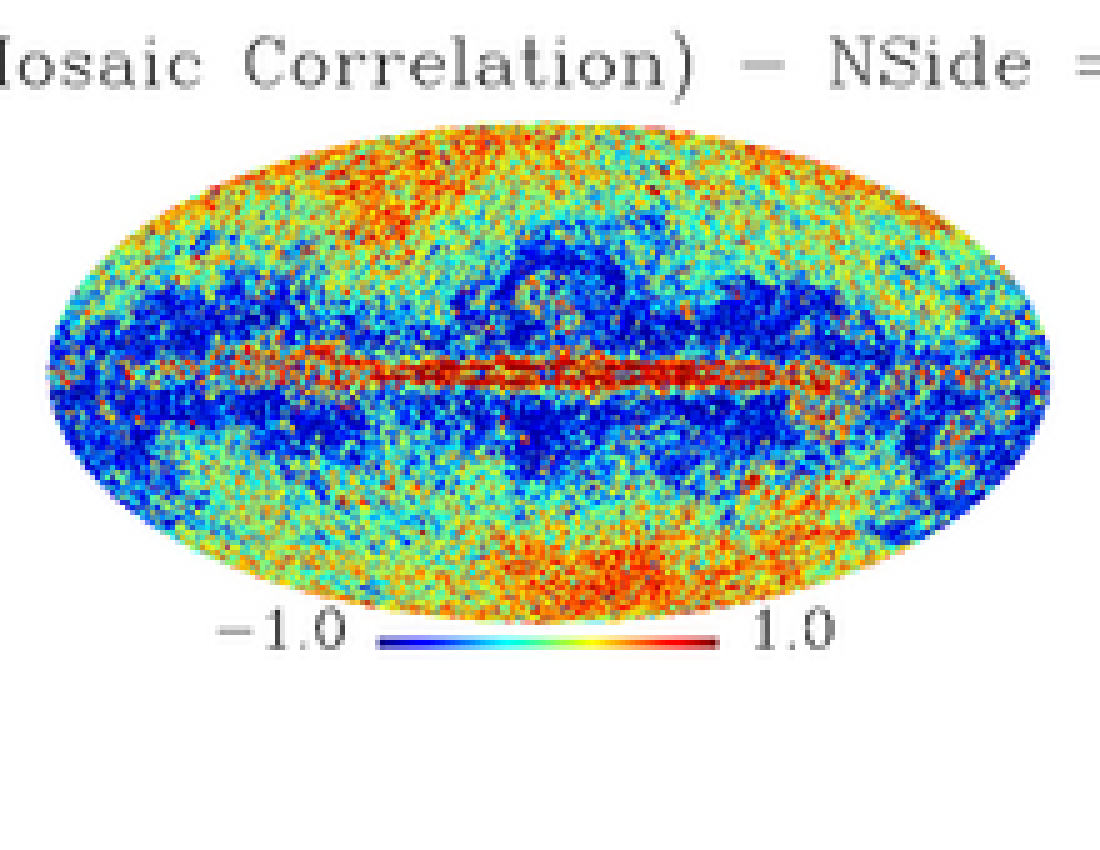}
\end{minipage}
\caption{Same as figure \ref{corrmaps1}, but here, weighted with weighting coefficients $w_i(n=2)$.}
\label{corrmaps2}
\end{figure}

\begin{figure}[H]
\centering
\begin{minipage}{0.24\textwidth}
\centering
\includegraphics[trim=0cm 1.8cm 0cm 1.2cm,clip=true,width=4cm]{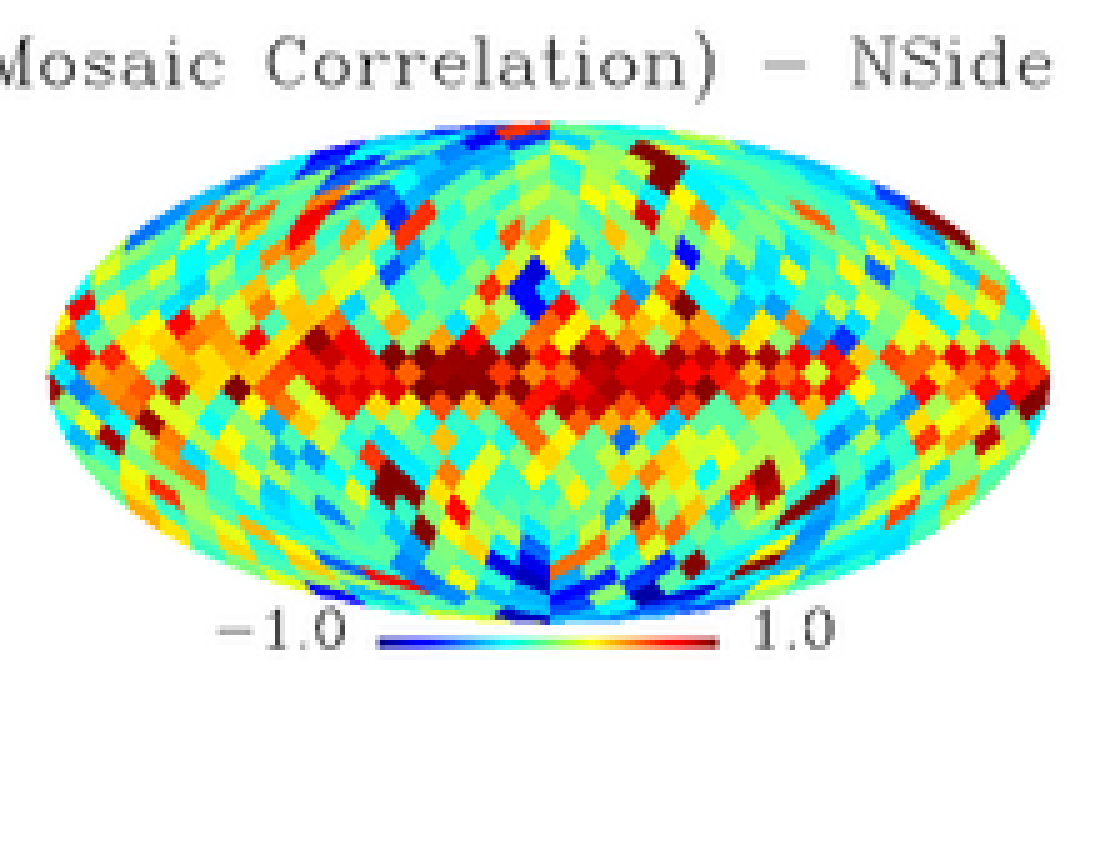}
\end{minipage}
\hfill
\begin{minipage}{0.24\textwidth}
\centering
\includegraphics[trim=0cm 1.8cm 0cm 1.2cm,clip=true,width=4cm]{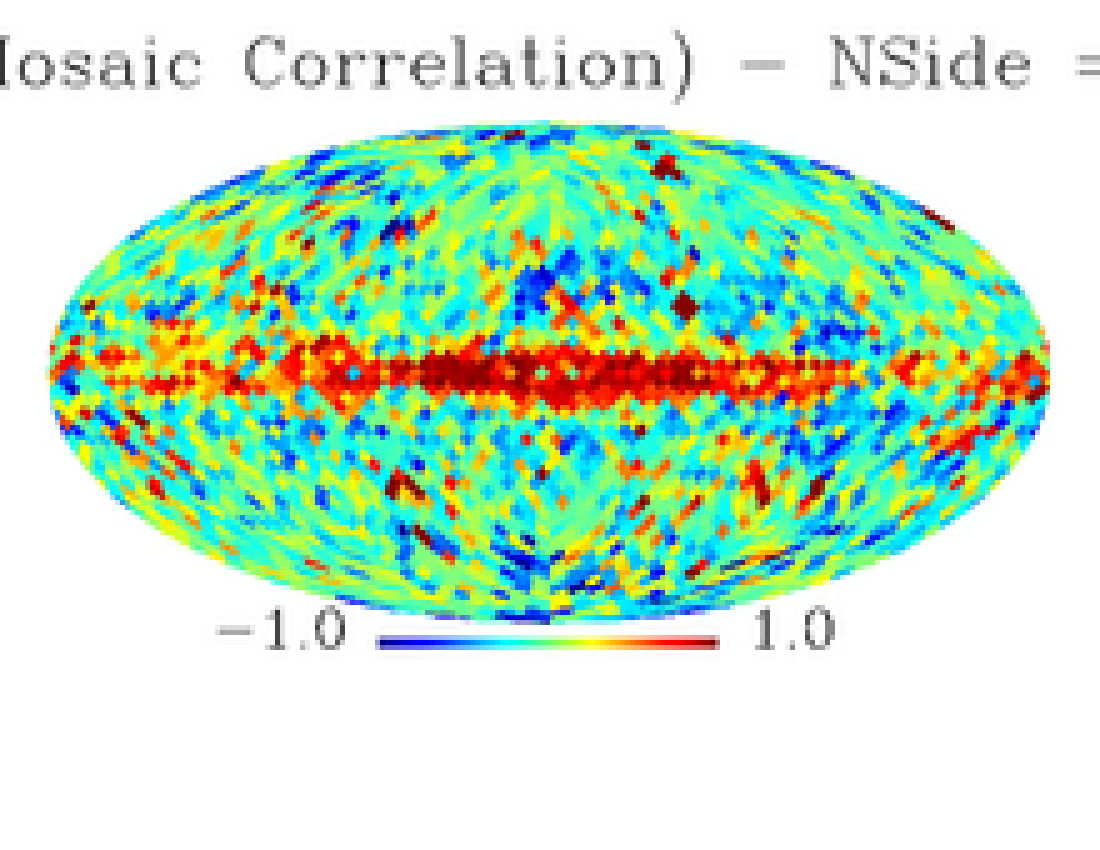}
\end{minipage}
\hfill
\centering
\begin{minipage}{0.24\textwidth}
\centering
\includegraphics[trim=0cm 1.8cm 0cm 1.2cm,clip=true,width=4cm]{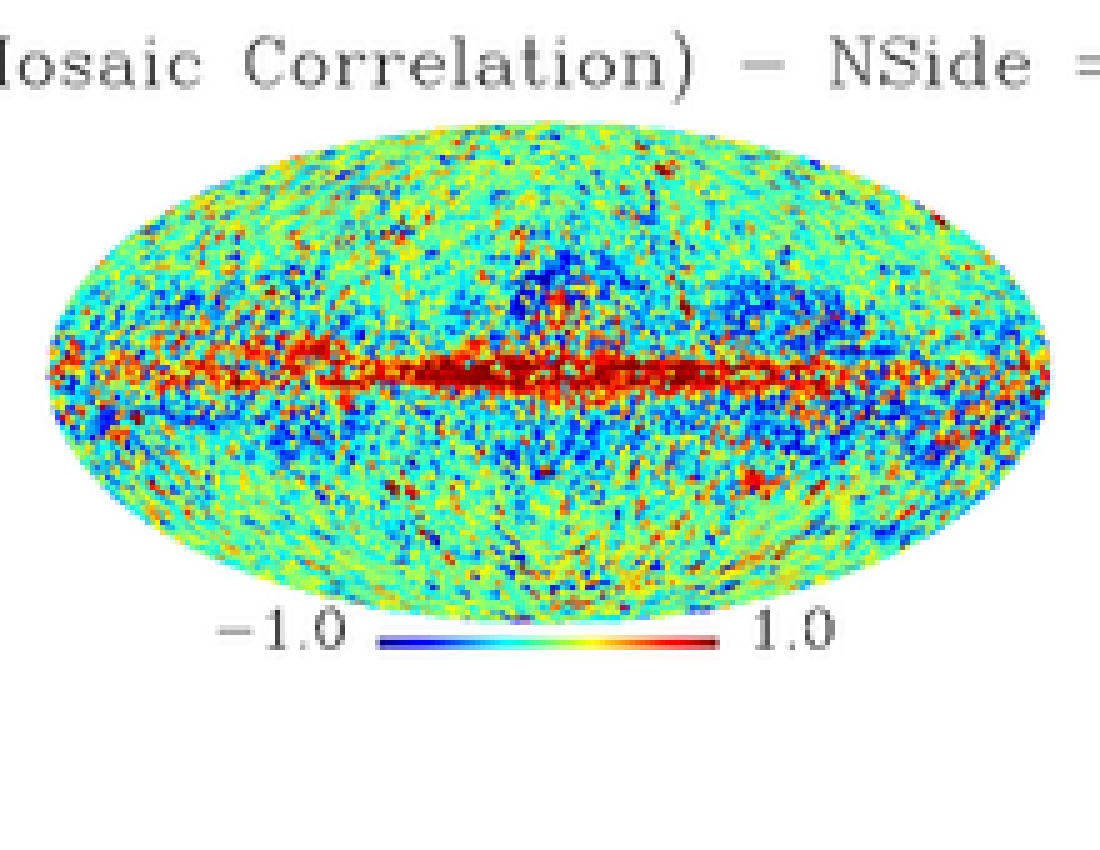}
\end{minipage}
\hfill
\begin{minipage}{0.24\textwidth}
\centering
\includegraphics[trim=0cm 1.8cm 0cm 1.2cm,clip=true,width=4cm]{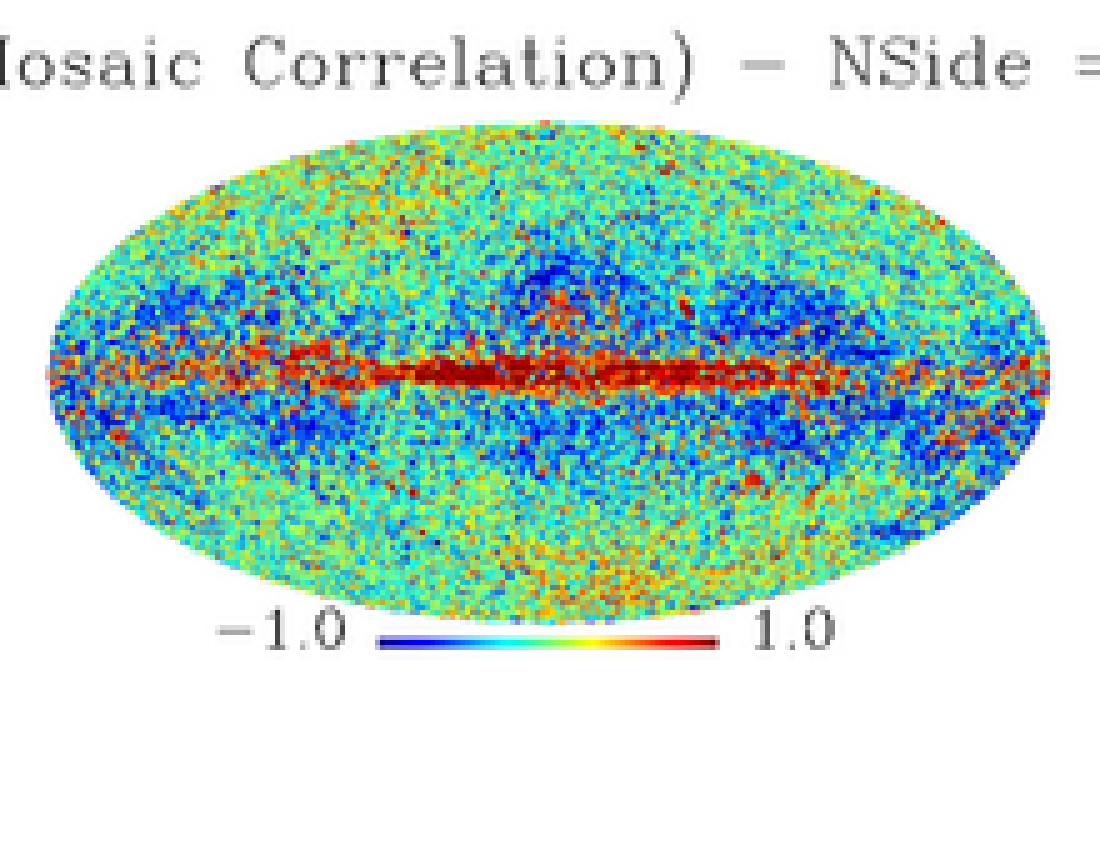}
\end{minipage}
\centering
\begin{minipage}{0.24\textwidth}
\centering
\includegraphics[trim=0cm 1.8cm 0cm 1.2cm,clip=true,width=4cm]{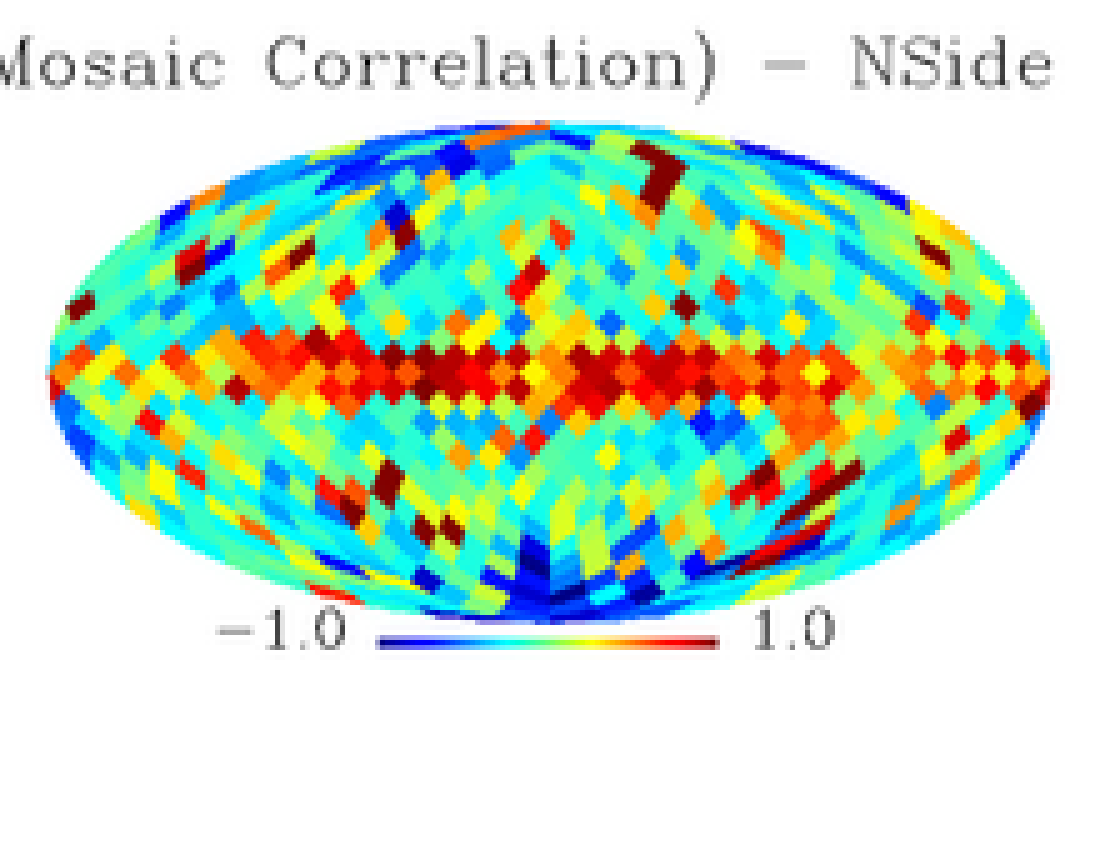}
\end{minipage}
\hfill
\begin{minipage}{0.24\textwidth}
\centering
\includegraphics[trim=0cm 1.8cm 0cm 1.2cm,clip=true,width=4cm]{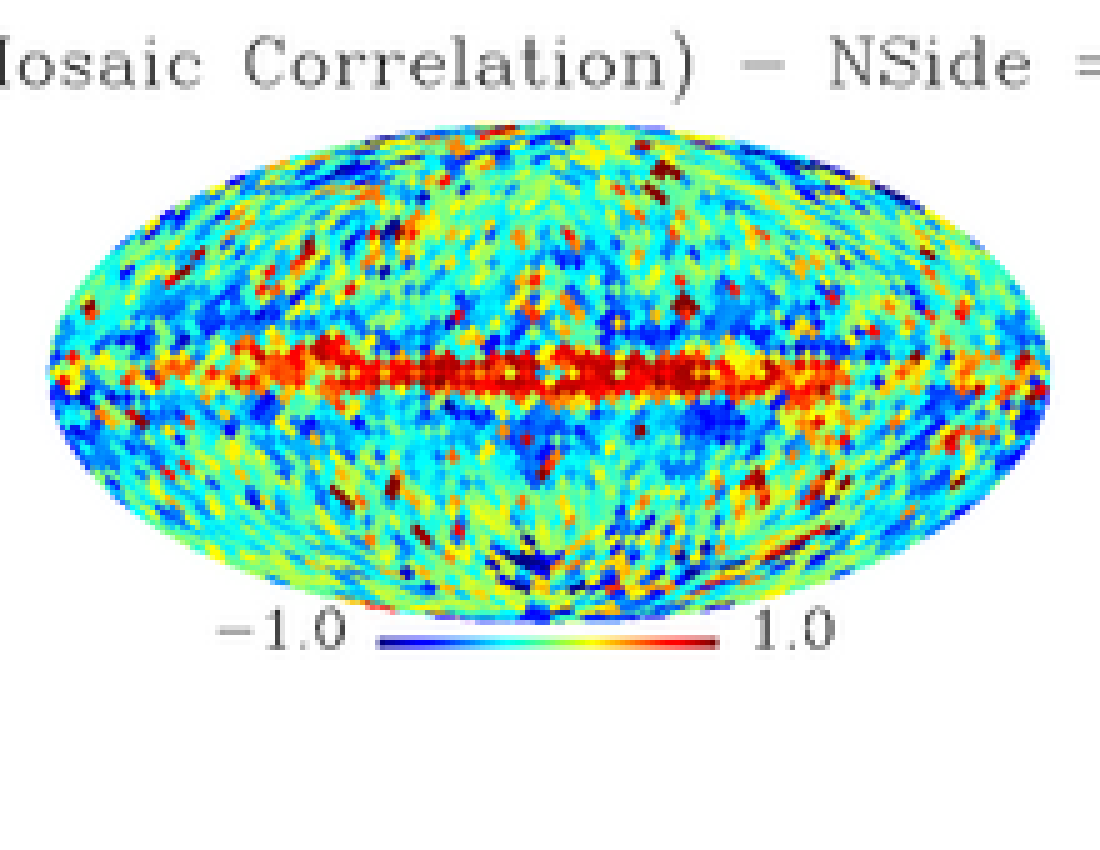}
\end{minipage}
\hfill
\centering
\begin{minipage}{0.24\textwidth}
\centering
\includegraphics[trim=0cm 1.8cm 0cm 1.2cm,clip=true,width=4cm]{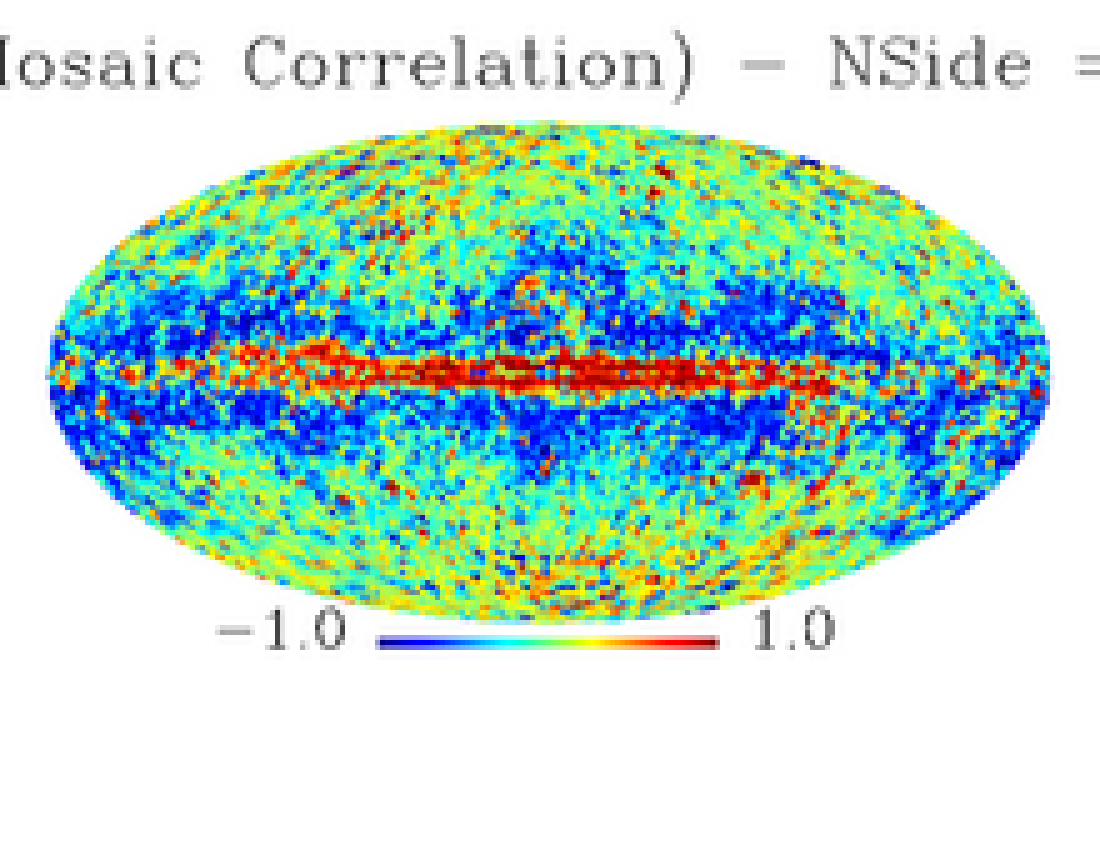}
\end{minipage}
\hfill
\begin{minipage}{0.24\textwidth}
\centering
\includegraphics[trim=0cm 1.8cm 0cm 1.2cm,clip=true,width=4cm]{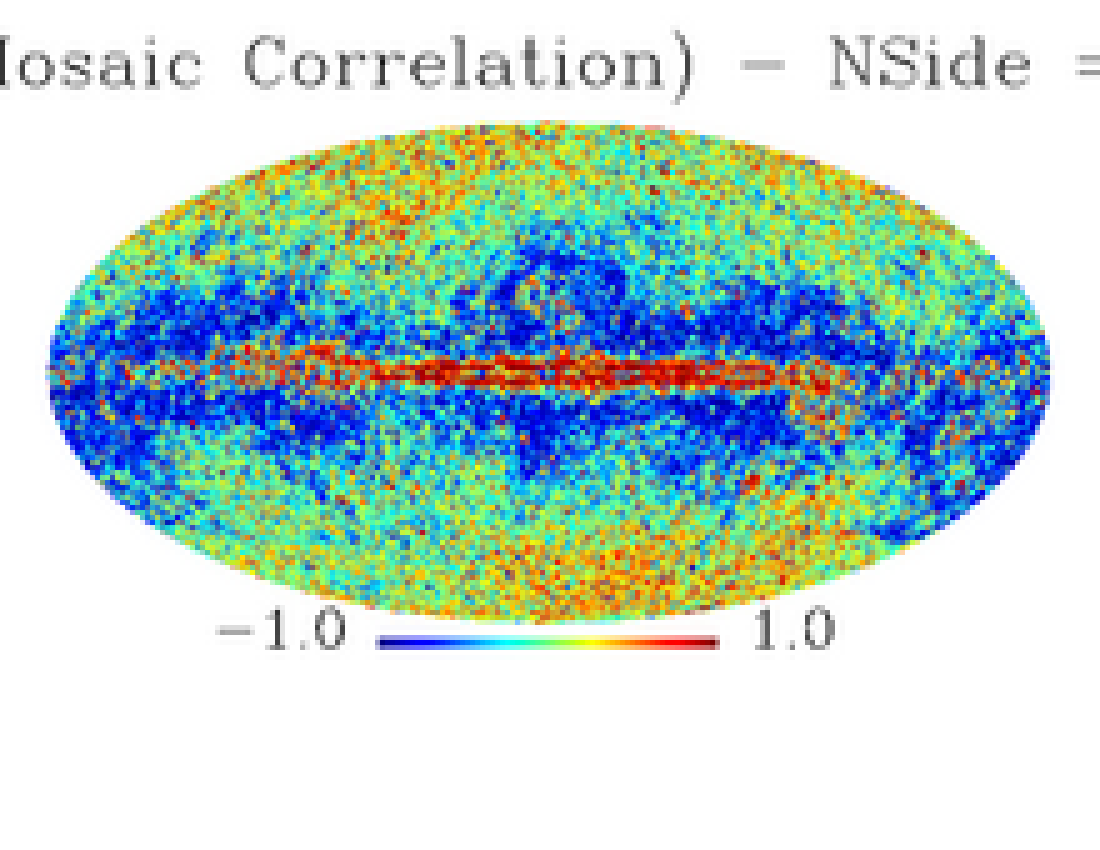}
\end{minipage}
\caption{Same as figure \ref{corrmaps1}, but here, weighted with weighting coefficients $w_i(n=10)$.}
\label{corrmaps10}
\end{figure}

\begin{figure}[H]
\begin{center}
\begin{minipage}{0.2\textwidth}
\centering
\includegraphics[trim=0cm 0cm 0cm 1cm,clip=true,width=4.1cm]{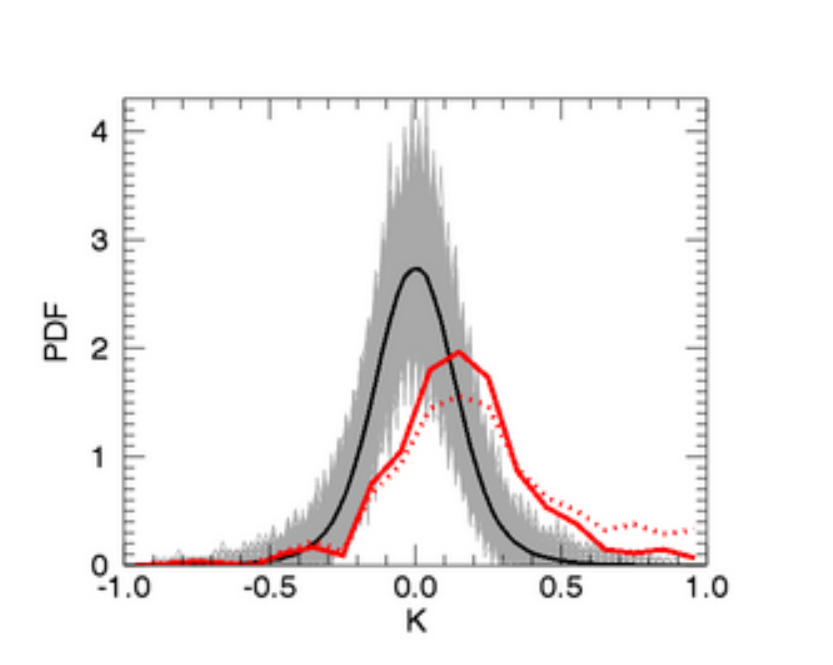}
\end{minipage}
\hspace{0.5cm}
\begin{minipage}{0.2\textwidth}
\centering
\includegraphics[trim=0cm 0cm 0cm 1cm,clip=true,width=4.1cm]{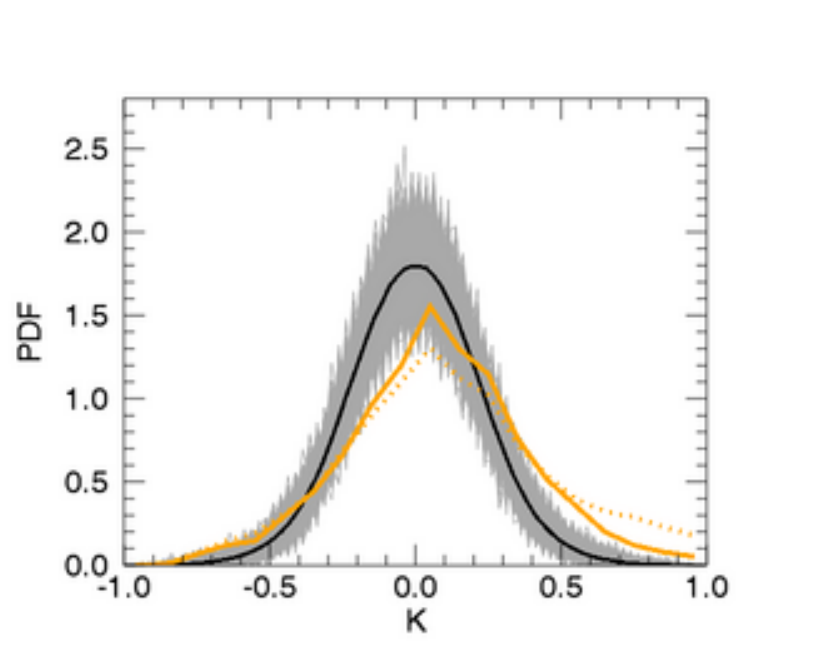}
\end{minipage}
\hspace{0.5cm}
\centering
\begin{minipage}{0.2\textwidth}
\centering
\includegraphics[trim=0cm 0cm 0cm 1cm,clip=true,width=4.1cm]{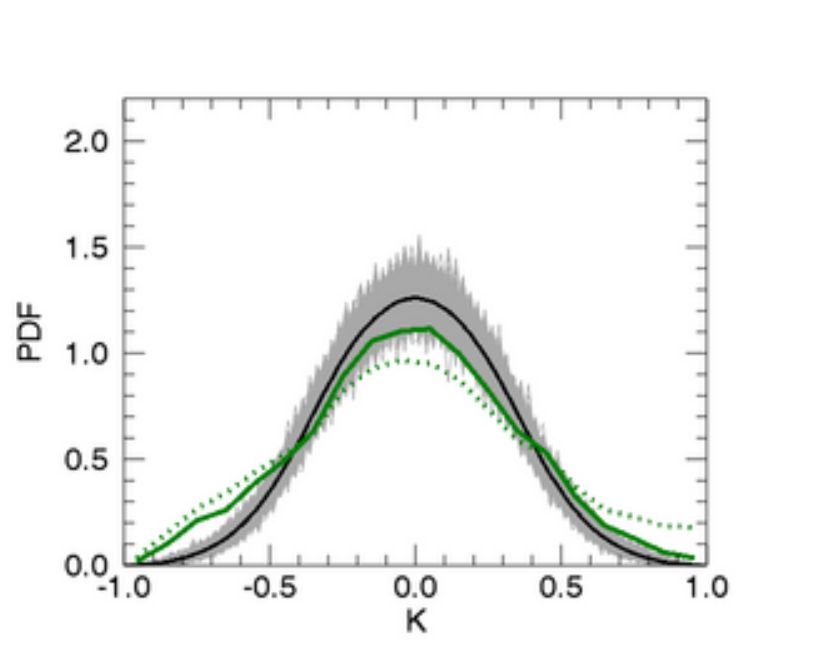}
\end{minipage}
\hspace{0.5cm}
\begin{minipage}{0.2\textwidth}
\centering
\includegraphics[trim=0cm 0cm 0cm 1cm,clip=true,width=4.1cm]{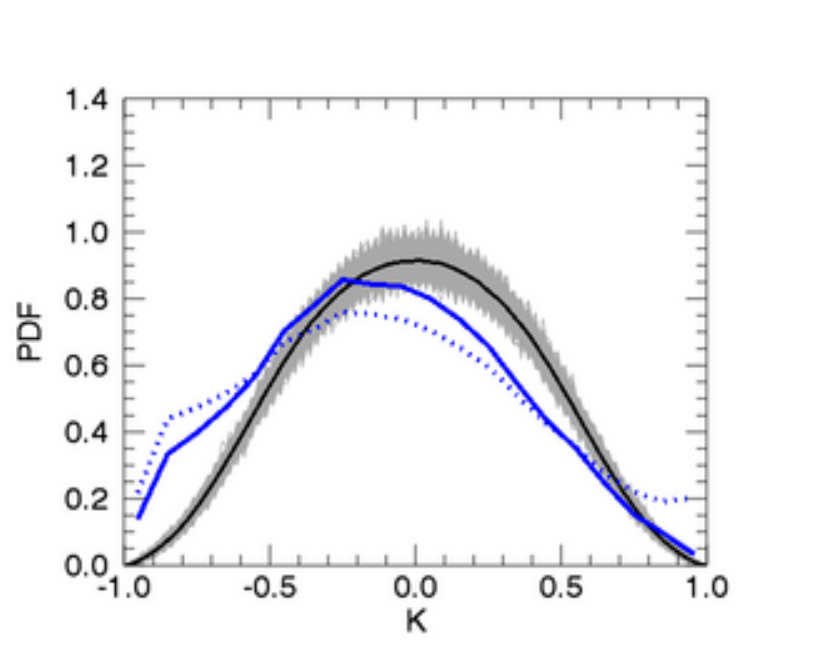}
\end{minipage}
\vspace{-0.8cm}
\begin{minipage}{0.2\textwidth}
\centering
\includegraphics[trim=0cm 0cm 0cm 1cm,clip=true,width=4.1cm]{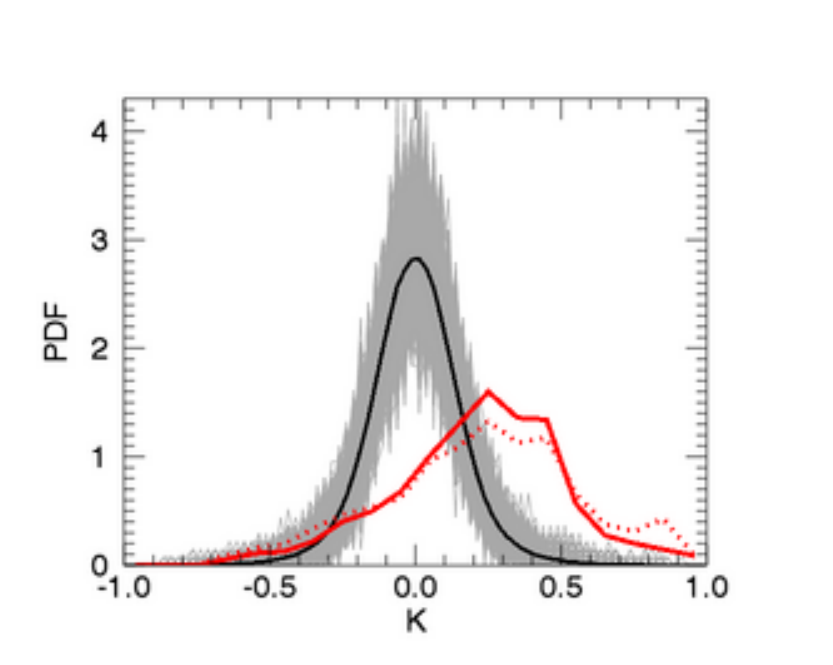}
\end{minipage}
\hspace{0.5cm}
\begin{minipage}{0.2\textwidth}
\centering
\includegraphics[trim=0cm 0cm 0cm 1cm,clip=true,width=4.1cm]{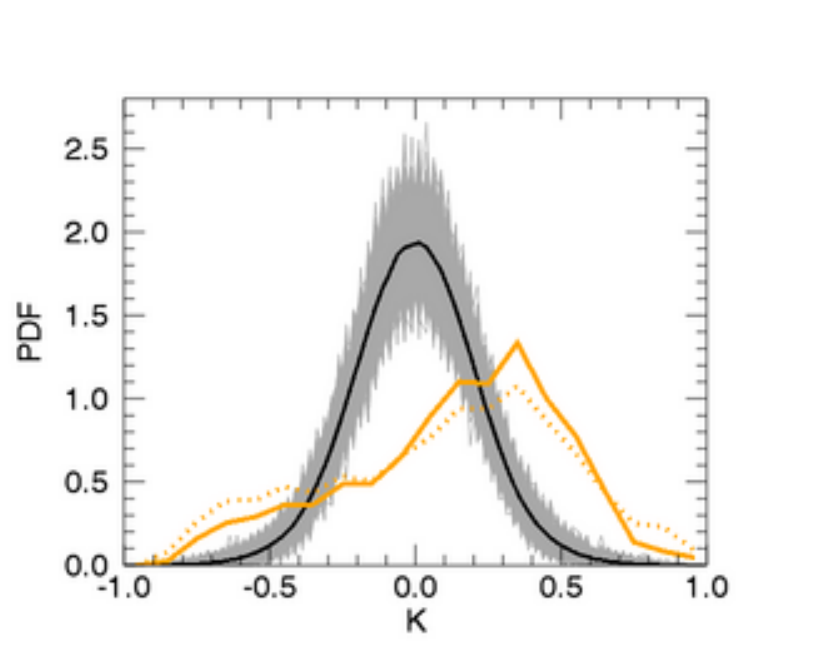}
\end{minipage}
\hspace{0.5cm}
\begin{minipage}{0.2\textwidth}
\centering
\includegraphics[trim=0cm 0cm 0cm 1cm,clip=true,width=4.1cm]{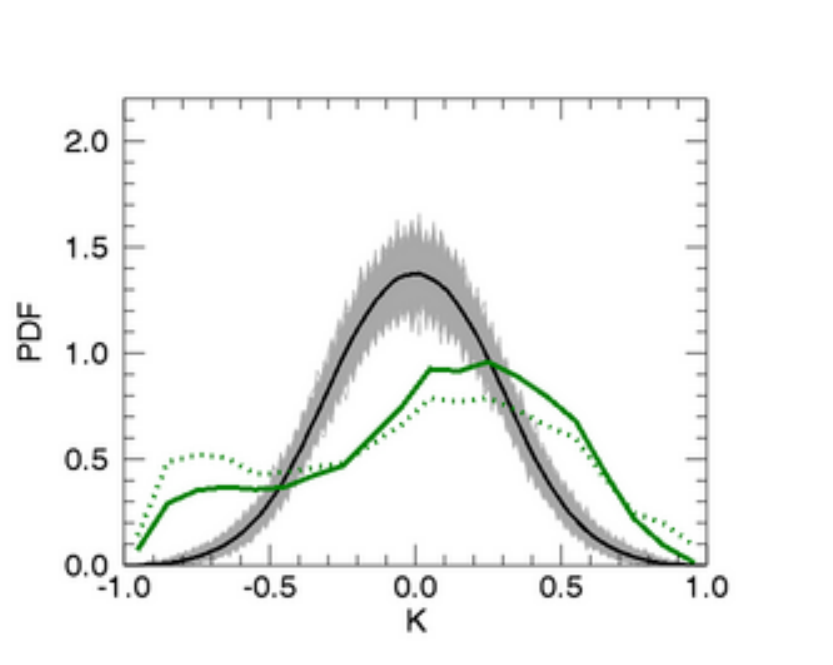}
\end{minipage}
\hspace{0.5cm}
\begin{minipage}{0.2\textwidth}
\centering
\includegraphics[trim=0cm 0cm 0cm 1cm,clip=true,width=4.1cm]{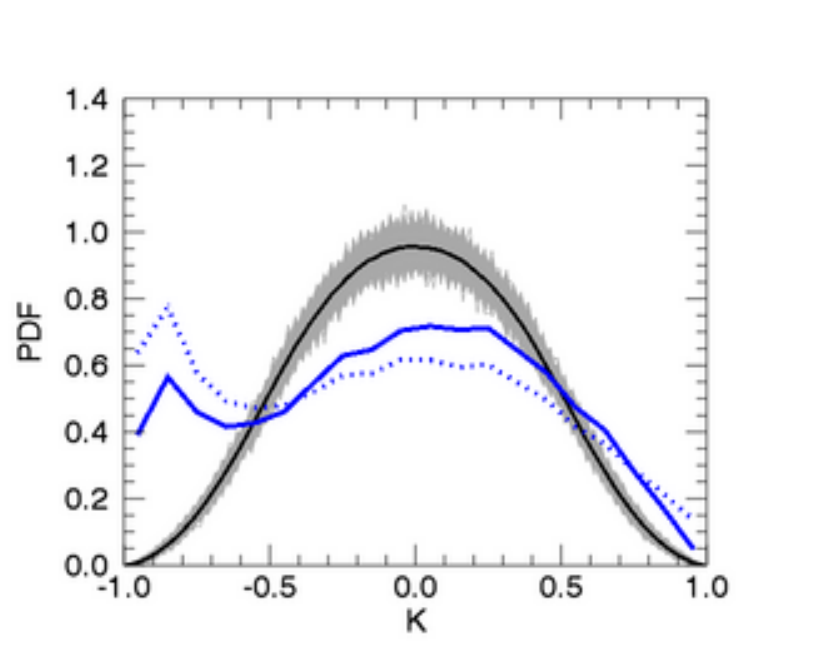}
\end{minipage}
\end{center}
\vspace{0.3cm}
\caption{PDFs of weighted correlation coefficients K (with $w_i(n=1)$) for different combinations
of foregrounds and each of the $\Omega$ sizes.
From top to bottom: Correlation between AME1 and Free-free emission, AME2 and Free-free emission. 
Colored lines give distributions of K's from the actual foregrounds, solid for masked sky, dotted for unmasked sky. The gray lines
show all distributions of Ks resulting from random simulations of AME1 or AME2 correlated with the free-free emission signal, and their averages (black line). Simulations are shown for the masked sky.
From left to right: $\Omega$ contains 1024, 256, 64 and 16 pixels.}
\label{dist1}
\end{figure}

\begin{figure}[H]
\begin{center}
\begin{minipage}{0.2\textwidth}
\centering
\includegraphics[trim=0cm 0cm 0cm 1cm,clip=true,width=4.1cm]{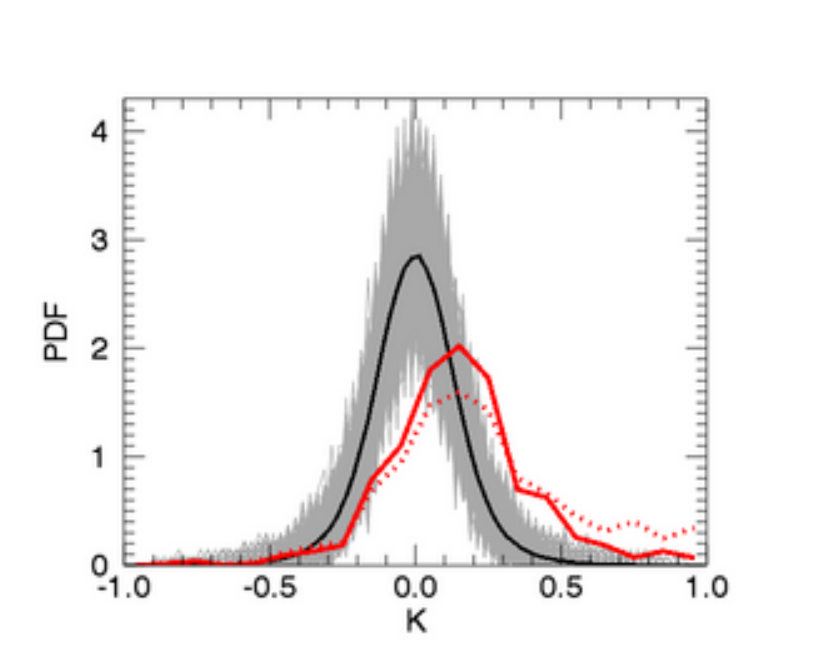}
\end{minipage}
\hspace{0.5cm}
\begin{minipage}{0.2\textwidth}
\centering
\includegraphics[trim=0cm 0cm 0cm 1cm,clip=true,width=4.1cm]{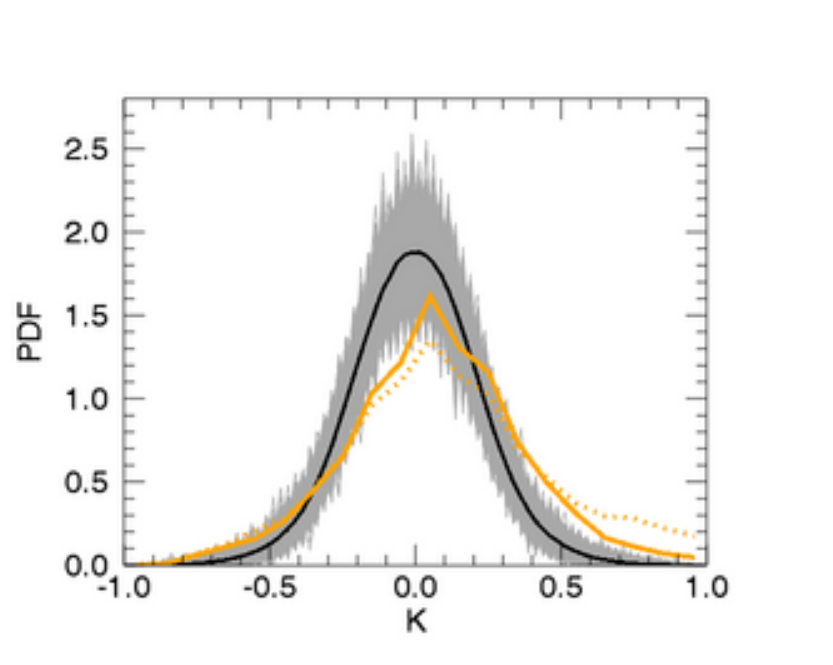}
\end{minipage}
\hspace{0.5cm}
\centering
\begin{minipage}{0.2\textwidth}
\centering
\includegraphics[trim=0cm 0cm 0cm 1cm,clip=true,width=4.1cm]{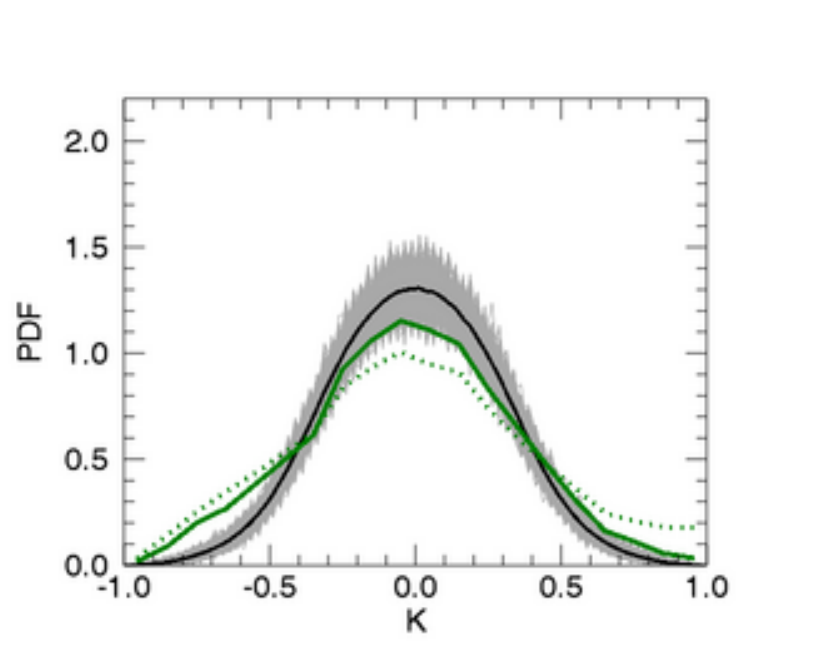}
\end{minipage}
\hspace{0.5cm}
\begin{minipage}{0.2\textwidth}
\centering
\includegraphics[trim=0cm 0cm 0cm 1cm,clip=true,width=4.1cm]{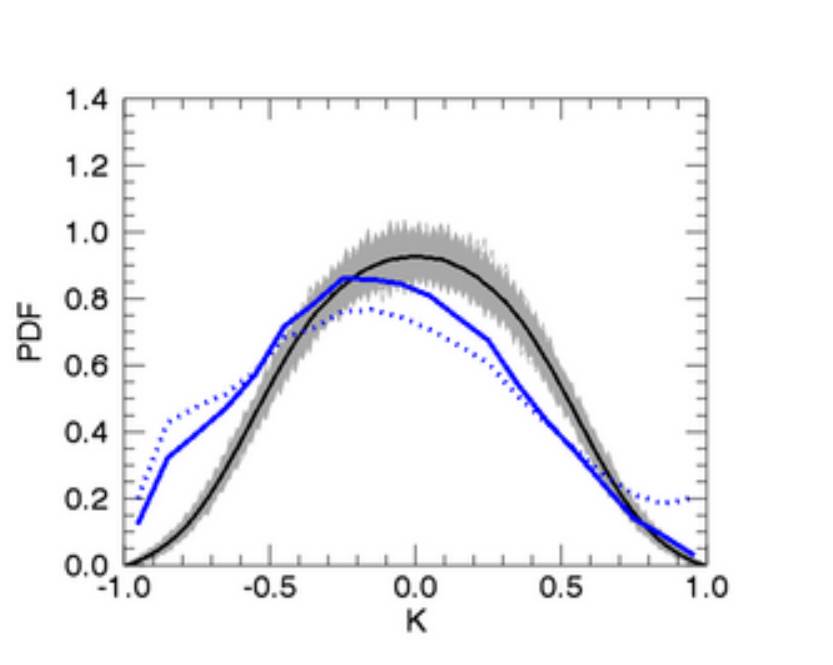}
\end{minipage}
\vspace{-0.5cm}
\begin{minipage}{0.2\textwidth}
\centering
\includegraphics[trim=0cm 0cm 0cm 1cm,clip=true,width=4.1cm]{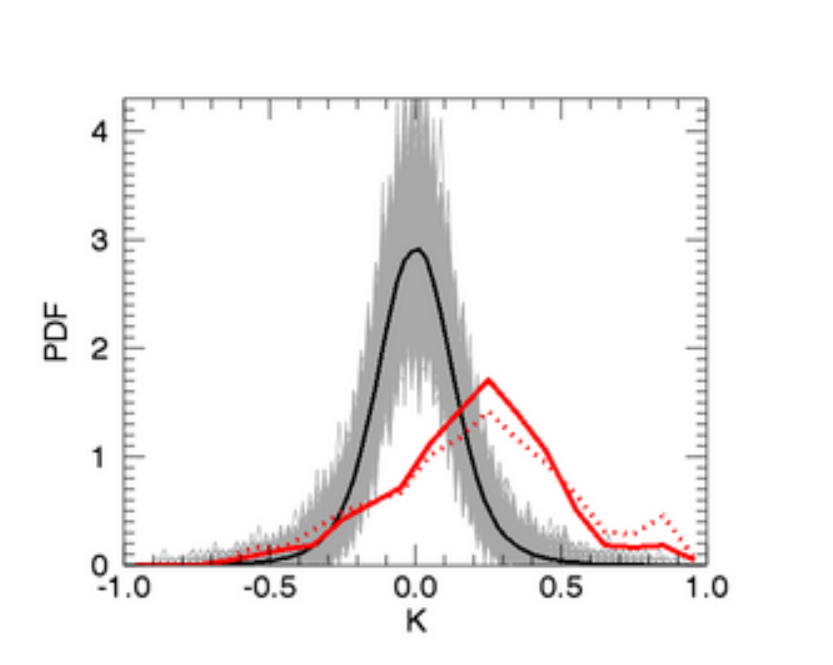}
\end{minipage}
\hspace{0.5cm}
\begin{minipage}{0.2\textwidth}
\centering
\includegraphics[trim=0cm 0cm 0cm 1cm,clip=true,width=4.1cm]{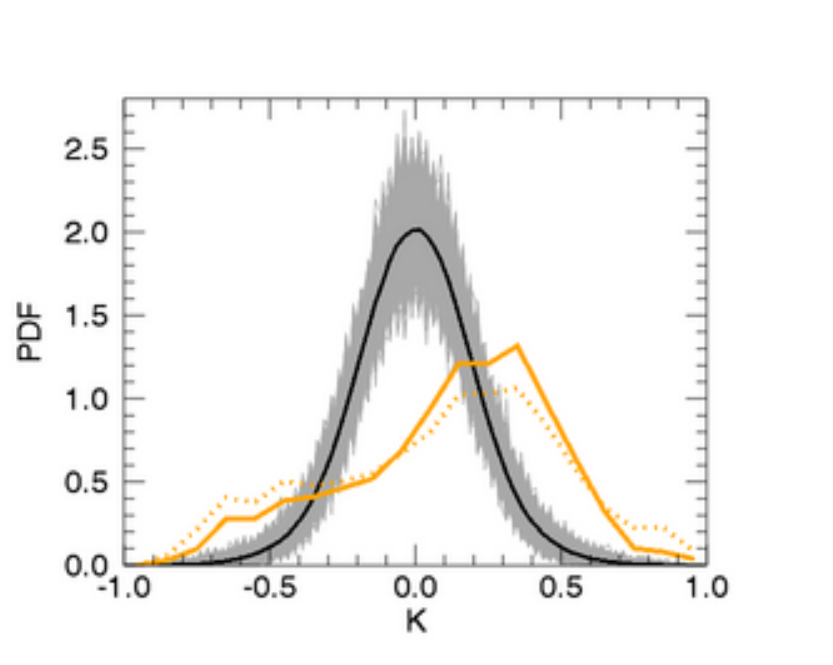}
\end{minipage}
\hspace{0.5cm}
\begin{minipage}{0.2\textwidth}
\centering
\includegraphics[trim=0cm 0cm 0cm 1cm,clip=true,width=4.1cm]{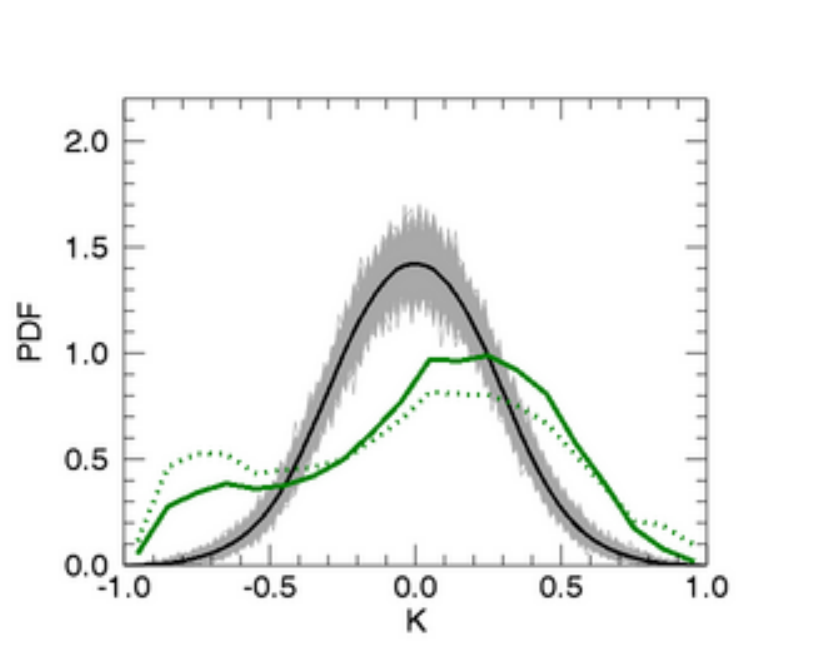}
\end{minipage}
\hspace{0.5cm}
\begin{minipage}{0.2\textwidth}
\centering
\includegraphics[trim=0cm 0cm 0cm 1cm,clip=true,width=4.1cm]{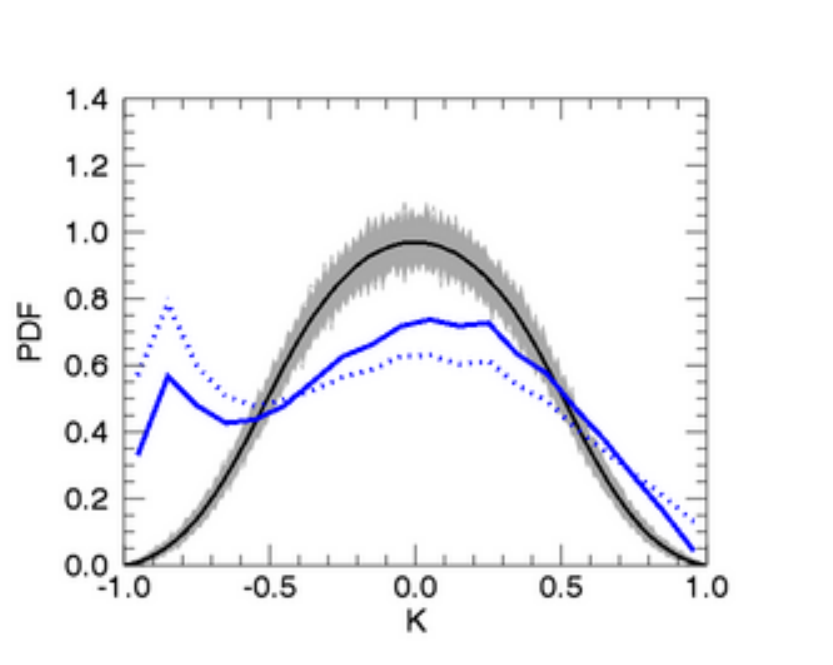}
\end{minipage}
\end{center}
\caption{Same as figure \ref{dist1}, but here, weighted with weighting coefficients $w_i(n=2)$}
\label{dist2}
\end{figure}

\begin{figure}[H]
\begin{center}
\begin{minipage}{0.2\textwidth}
\centering
\includegraphics[trim=0cm 0cm 0cm 1cm,clip=true,width=4.1cm]{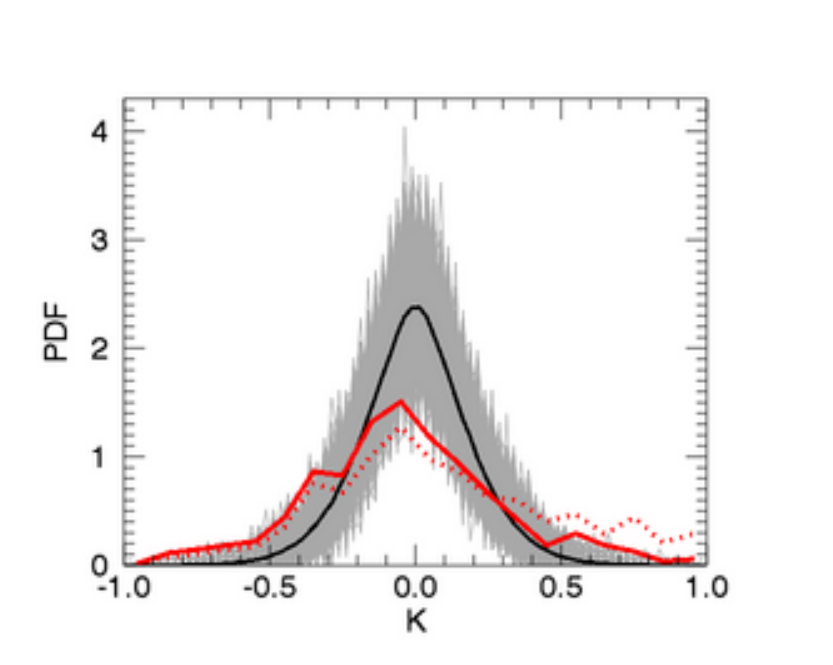}
\end{minipage}
\hspace{0.5cm}
\begin{minipage}{0.2\textwidth}
\centering
\includegraphics[trim=0cm 0cm 0cm 1cm,clip=true,width=4.1cm]{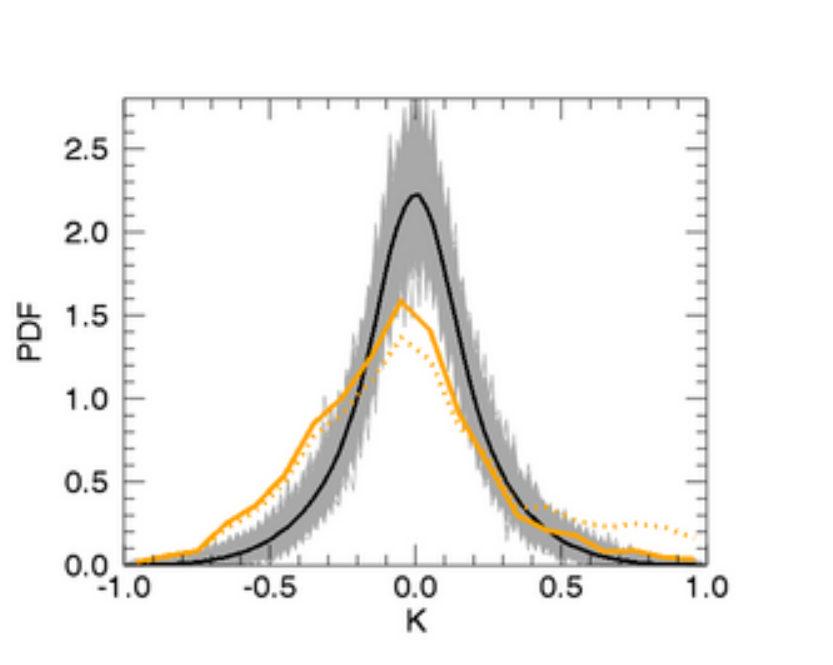}
\end{minipage}
\hspace{0.5cm}
\centering
\begin{minipage}{0.2\textwidth}
\centering
\includegraphics[trim=0cm 0cm 0cm 1cm,clip=true,width=4.1cm]{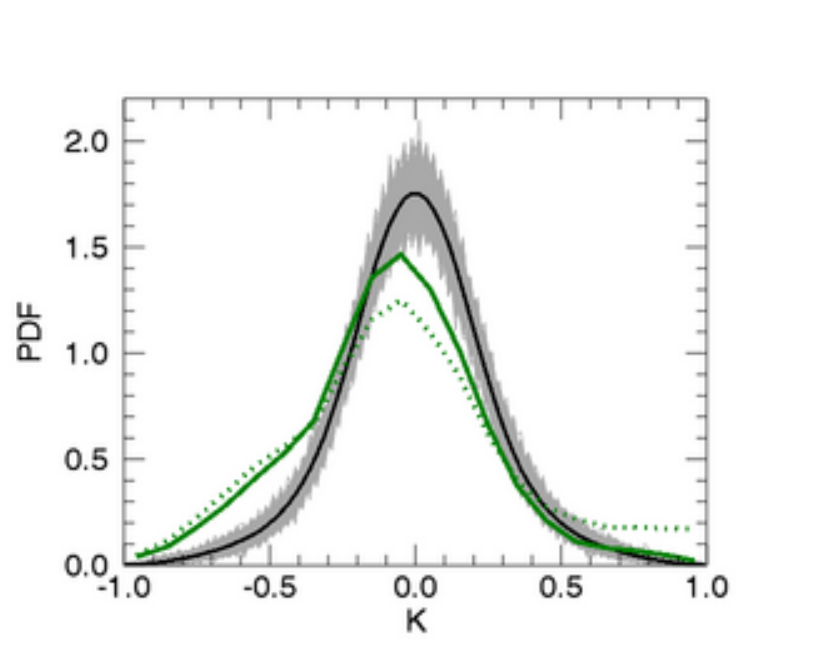}
\end{minipage}
\hspace{0.5cm}
\begin{minipage}{0.2\textwidth}
\centering
\includegraphics[trim=0cm 0cm 0cm 1cm,clip=true,width=4.1cm]{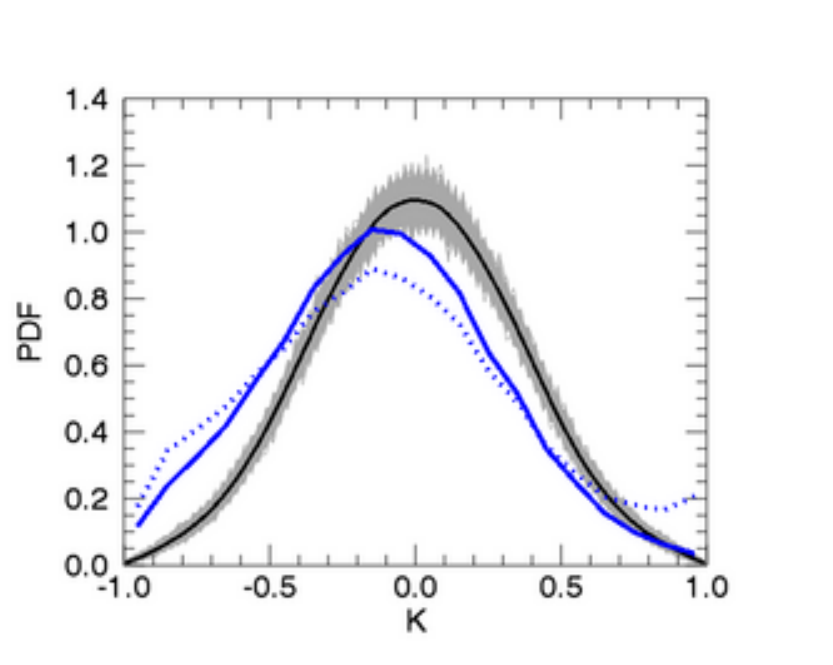}
\end{minipage}
\vspace{-0.5cm}
\begin{minipage}{0.2\textwidth}
\centering
\includegraphics[trim=0cm 0cm 0cm 1cm,clip=true,width=4.1cm]{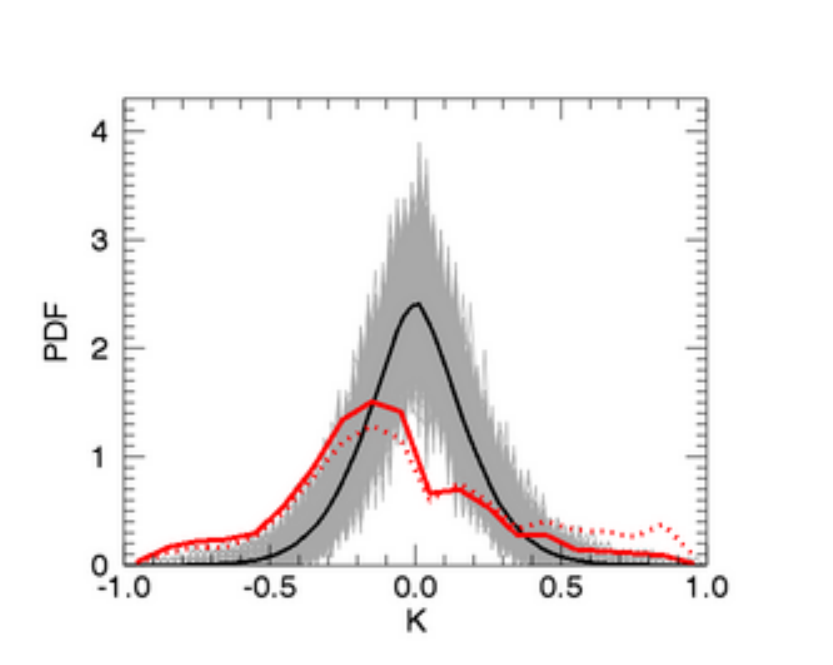}
\end{minipage}
\hspace{0.5cm}
\begin{minipage}{0.2\textwidth}
\centering
\includegraphics[trim=0cm 0cm 0cm 1cm,clip=true,width=4.1cm]{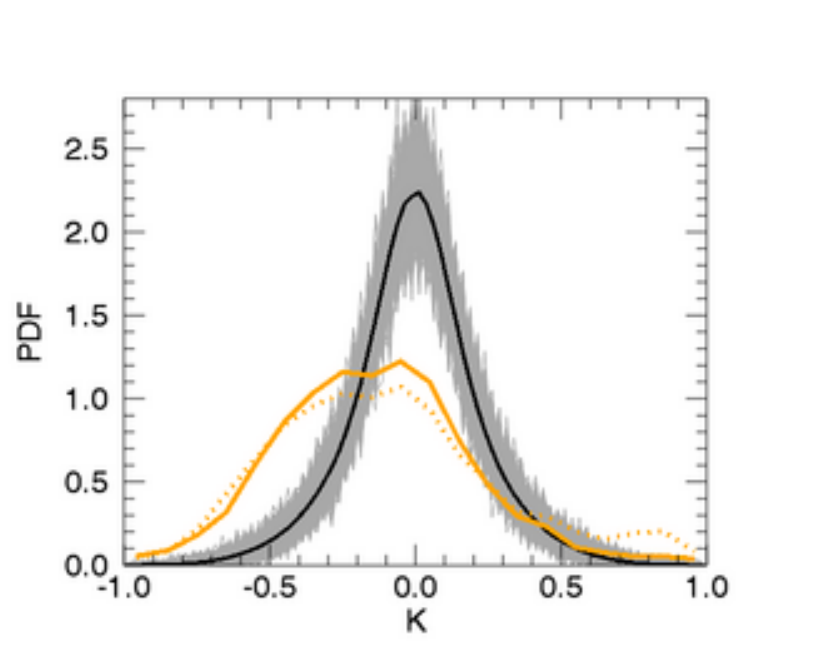}
\end{minipage}
\hspace{0.5cm}
\begin{minipage}{0.2\textwidth}
\centering
\includegraphics[trim=0cm 0cm 0cm 1cm,clip=true,width=4.1cm]{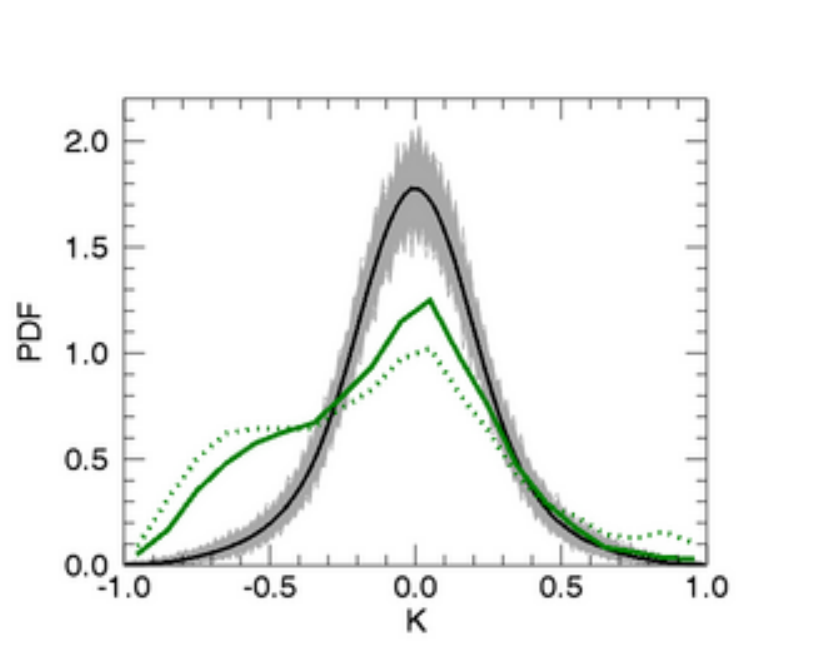}
\end{minipage}
\hspace{0.5cm}
\begin{minipage}{0.2\textwidth}
\centering
\includegraphics[trim=0cm 0cm 0cm 1cm,clip=true,width=4.1cm]{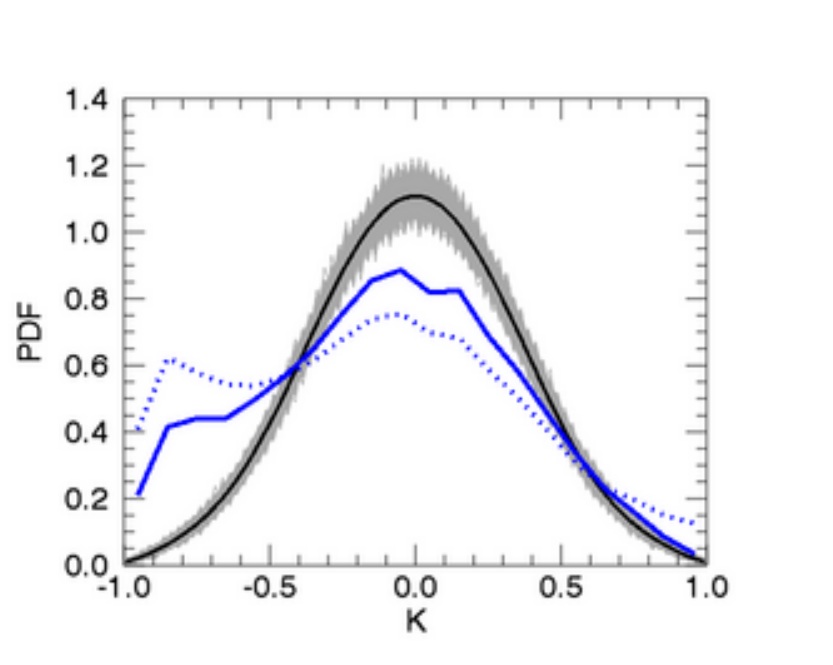}
\end{minipage}
\end{center}
\caption{Same as figure \ref{dist1}, but here, weighted with weighting coefficients $w_i(n=10)$}
\label{dist10}
\end{figure}

Now, we include the errors $\sigma_i(p)$ and therefore use the mean value maps as $S_i(p)$ instead of those of maximum likelihood. We saw in figure \ref{weights} that the weights corresponding to the free-free map will have by far the largest impact on the correlations, compared to the other foregrounds' weights. Indeed, we see that after taking the weights under consideration only the correlations of AME with free-free emission deviate significantly from the unweighted ones. Therefore, we here only show the results of those correlations, while we move the results regarding synchrotron radiation and thermal dust emission to the appendix (figures \ref{Acorrmaps1} - \ref{Adist10}). The correlation maps for $n=1,2$ and $10$ can be seen in figures \ref{corrmaps1}, \ref{corrmaps2} and \ref{corrmaps10}, while the corresponding PDFs are shown in figures \ref{dist1}, \ref{dist2} and \ref{dist10}, respectively.

\vspace{-0.5cm}

\vspace{-0.5cm}

\subsection{Comparison of weighted and unweighted results}
\label{sec:expect}

\begin{figure}
\centering
\begin{minipage}{0.46\textwidth}
\hspace{-1cm}
\includegraphics[trim=0cm 0cm 0.5cm 1cm,clip=true,width=7.5cm]{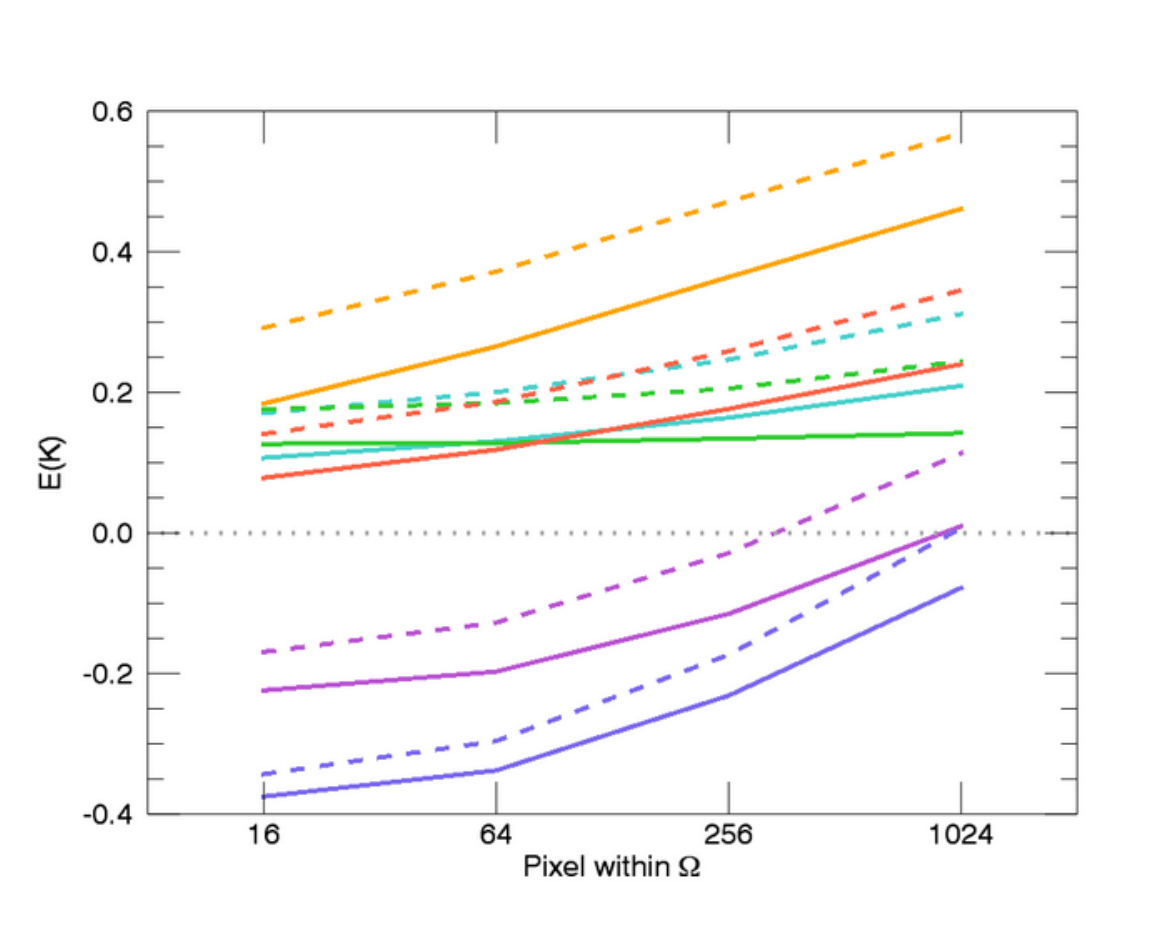}
\end{minipage}
\begin{minipage}{0.46\textwidth}
\includegraphics[trim=0cm 0cm 0.5cm 1cm,clip=true,width=7.5cm]{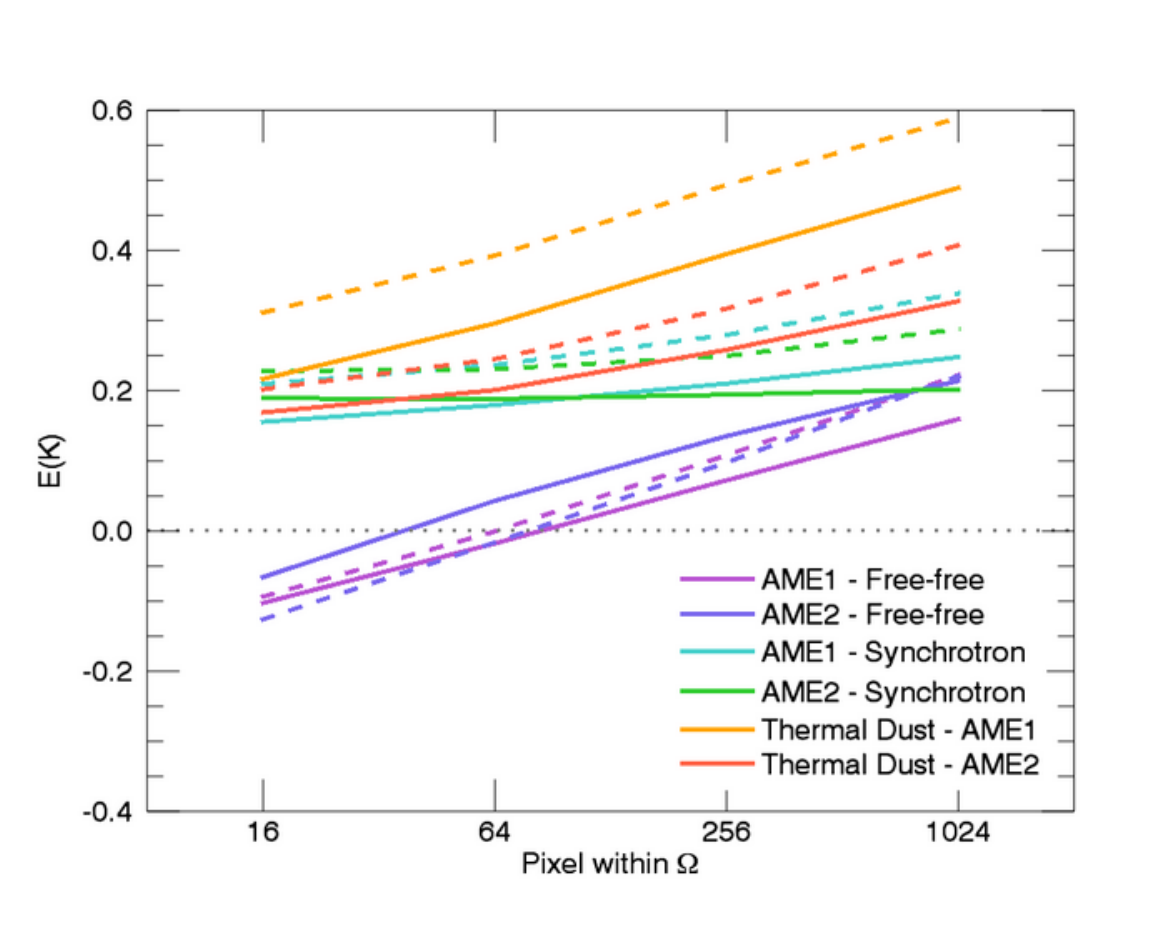}
\end{minipage}
\begin{minipage}{0.46\textwidth}
\hspace{-1cm}
\includegraphics[trim=0cm 0cm 0.5cm 1cm,clip=true,width=7.5cm]{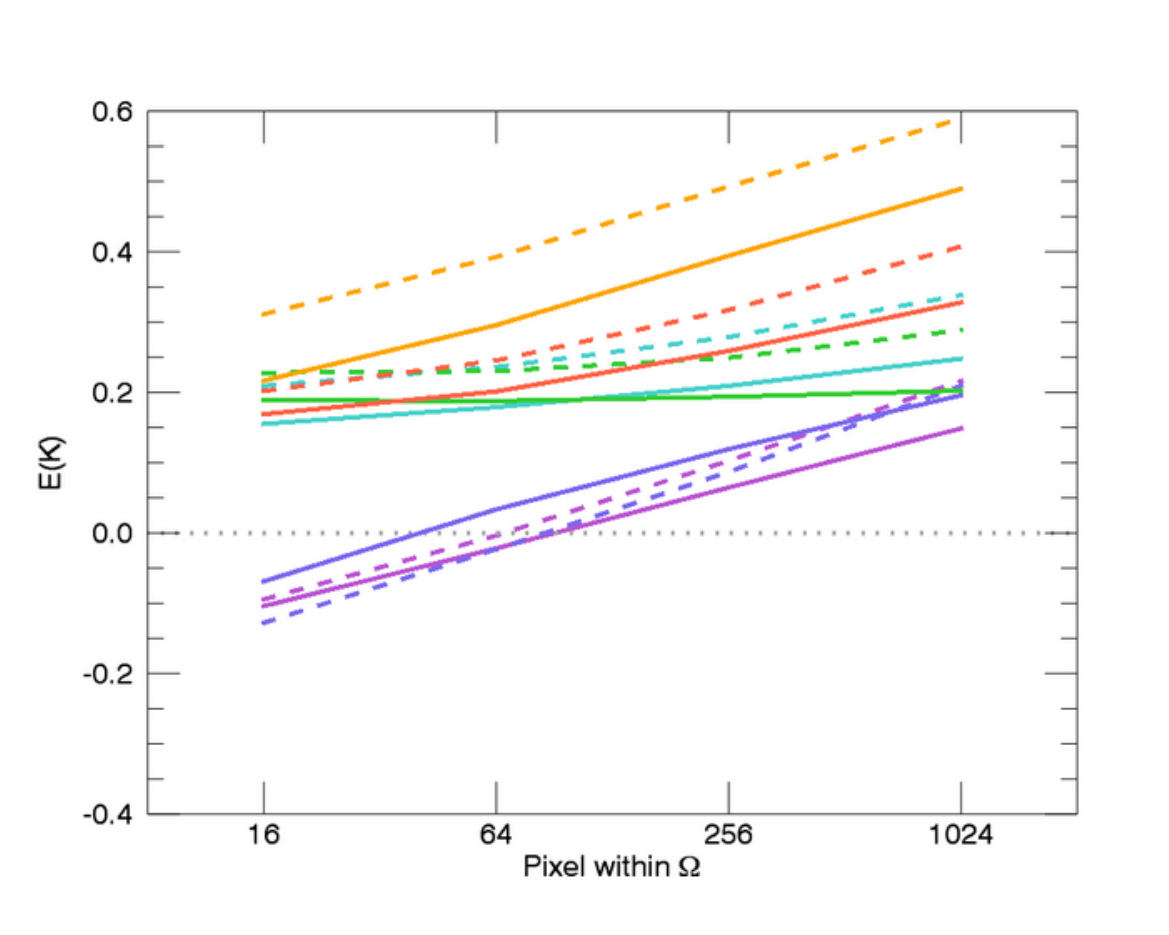}
\end{minipage}
\begin{minipage}{0.46\textwidth}
\includegraphics[trim=0cm 0cm 0.5cm 1cm,clip=true,width=7.5cm]{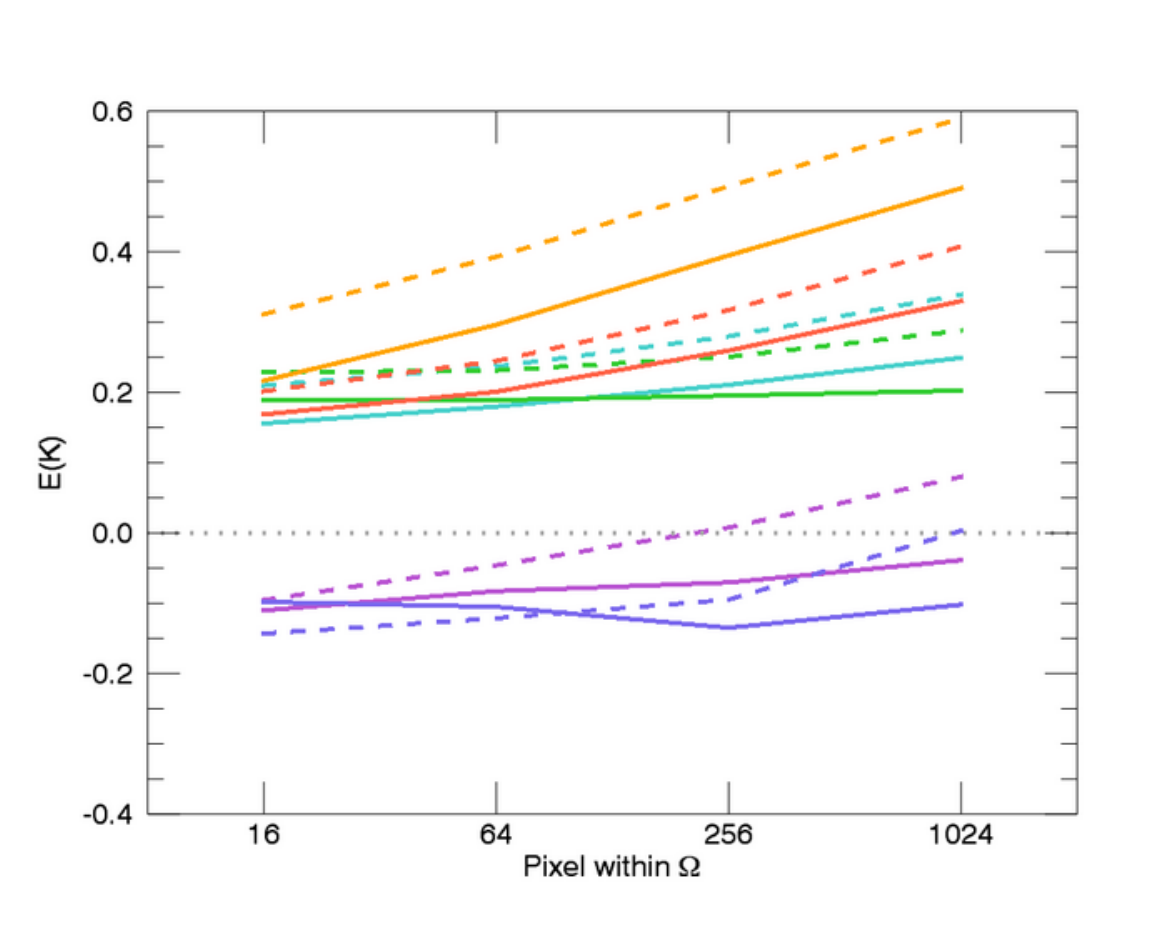}
\end{minipage}
\caption{The mean value of the cross-correlation coefficients $ E(K)$ for all the combinations of foregrounds. Top left: unweighted correlations, top right: weighted correlations with $w_i(n=1)$, bottom left: weighted correlations with $w_i(n=2)$ and bottom right: weighted correlations with $w_i(n=10)$. The legend for all plots can be found in the upper right panel. Dashed lines correspond to values from the entire sky, whereas solid lines correspond to the masked sky.} 
\label{mean}
\end{figure}

It is apparent that the simulations as well as the actual cross-correlations change their distribution function according to the size of the mother pixels $\Omega$. In order to portray this fact and also to compare all correlations, in figure \ref{mean} we simply show the mean value of the coefficients $ E(K)=\sum_K K\cdot P(K)$, where $P(K)$ is the  PDF for a given resolution. $E(K)$ acts as the average similarity of the AME components in respect to the other foregrounds from the smallest angular scale of about $\theta_p\simeq 0.9^{\circ}$ up to $\theta_p\approx 7.4^{\circ}$. The corresponding range 
of multipoles for those scales can be estimated as $l\cdot\theta_p\le 1$, which gives $l\le 126$ for $N_{\Omega}=16$ and $l\le 16$ for $N_{\Omega}=1024$.

As one could already see in the PDFs (figures \ref{distno},\ref{dist1}
-\ref{dist10},\ref{Adist1}
-\ref{Adist1}) and now more clearly from figure \ref{mean},
the correlations of foreground components can be separated into correlations, more pronounced at small (as AME and thermal dust) and at large angular scales (as AME and free-free).
The most distinct positive correlations ($E(K)\approx10\,-\,50\%$) are detected for AME1 and AME2 with thermal dust at all scales, increasing for larger scales. Hardly influenced by scale, the correlations between AME and synchrotron emission lie steady around $10\,-20\,\%$. As already mentioned, for both, AME with thermal dust and AME with synchrotron, weighting does not change their behavior noticeably.

The unweighted correlation between AME and free-free emission shows an opposite trend to that of the thermal dust correlation. Here $E(K)$ ranges between -20 and -40$\%$ for small scales and goes to zero for larger ones. When applying weights to the data, the strong negative correlation is pushed to higher values. For large scales the values even grow to be positive up to $+\,20\,\%$. 

\section{Conclusion}

We have computed weighted mosaic correlations between recently released foreground templates of the Planck Collaboration. This method allows for the determination of correlations within defined regions, filling out the entire sky. Since the method only takes place in pixel space, it can be applied to inhomogeneous, anisotropic signals on the sky, as is the case here. When altering the size of the pixels, one obtains correlations corresponding to different scales. We then simply chose the expectation value of the distribution of correlation coefficients for each probed scale as a rough estimator of the manifestation of correlation and thereby obtained a scale dependency of the correlation.
Furthermore, we included the errors on the signals, provided by the Planck data product, by introducing appropriate weights. These could be tuned to vary the emphasis put on data points with signal-to-error ratios below or above 1.

A strong positive correlation was detected for AME and thermal dust, for both, the unweighted and weighted case. Also the correlation of both AME components with synchrotron radiation give a pronounced positive correlation. As mentioned before, the positive correlation with thermal dust is in accordance with the earliest studies of AME. Due to the common relation of both - the dust particles causing AME and electrons causing synchrotron radiation - to the galactic magnetic fields, the correlation with synchrotron emission was also to be expected.

Both AME components show a strong anti-correlation with free-free emission at small angular scales for the unweighted case, especially at the smallest angular scales. We noticed an even stronger anti-correlation for AME2 with free-free emission than for AME1. After including the weights in the calculation, the anti-correlation decreases noticeably on small scales, while we even see positive values on larger scales. Negative correlation is restored again on all scales for the choice of the most rigorous weights at $n=10$.
The weighted correlations differ substantially from the unweighted ones when we regard the free-free template. Due to its very small signal-to-error ratios on medium to high galactic latitudes, the weights \mbox{$w_i(n=1,2)$} seem to induce positive correlations among AME and free-free emission, formerly anti-correlated. Values of those pixels with a low signal-to-error ratio are pushed down towards zero by the weights. We therefore obtain positive correlations with the AME skies because they as well have low values in exactly those regions. The choice of $w_i(n=10)$ basically forces all pixels with signal-to-error ratios less than 1 completely to zero. It appears that those pixels causing the positive correlation for the choice of $n=1,2$ are now contributing to the negative correlation. Our conclusion must be that the positive correlations at high galactic latitudes should not be taken seriously. Moreover, we should acknowledge that the negative correlations closer to the galactic plane are robust as they survive even the most aggressive weights.

We interpret these results by claiming that the nano-particles causing AME are pushed out of the hot zones within our galaxy. The electrons, responsible for free-free emission, remain in these zones, perhaps being held back by abundant ions.

Understanding the composition and the dynamics of the ISM is essential in separating galactic foregrounds and in the long run is vital for a correct insight in the early Universe, including the detecting of a cosmological gravitational wave background. Assuming that the AME skies provided by the Planck Collaboration describe the true anomalous emission sufficiently well, we believe that our result should be taken into account in future work on models of AME and the ISM.


\acknowledgments

We are very appreciative of helpful comments by K. Górski and A. Frejsel and thank especially P. Naselsky for his support and useful discussions in this work.\\
This work is based on observations obtained with Planck\footnote{http://www.esa.int/Planck}, an ESA science mission with instruments and contributions directly funded by ESA Member States, NASA, and Canada. The development of Planck has been supported by: ESA; CNES and CNRS / INSU-IN2P3-INP (France); ASI, CNR, and INAF (Italy); NASA and DoE (USA); STFC and UKSA (UK); CSIC, MICINN and JA (Spain); Tekes, AoF and CSC (Finland); DLR and MPG (Germany); CSA (Canada); DTU Space (Denmark); SER/SSO (Switzerland); RCN (Norway); SFI (Ireland); FCT/MCTES (Portugal); and PRACE (EU). A description of the Planck Collaboration and a list of its members, including the technical or scientific activities in which they have been involved, can be found at the Planck web page.\footnote{http://www.cosmos.esa.int/web/planck/planck-collaboration}\\
We acknowledge use of the \texttt{HEALPix}\footnote{http://healpix.sourceforge.net} package~\cite{Healpix}. This work is supported by Danmarks Grundforskningsfond which allowed the establishment of the Danish Discovery Center, the Villum Foundation in support of the Deep Space project, the National Natural Science Foundation of China (Grant No. 11033003), the National Natural Science Foundation for Young Scientists of China (Grant No. 11203024) and the Youth Innovation Promotion Association.

\newpage

\appendix

\section{Additional Figures}

\begin{figure}[H]
\centering
\begin{minipage}{0.24\textwidth}
\centering
\includegraphics[trim=0cm 1.5cm 0cm 1.2cm,clip=true,width=4cm]{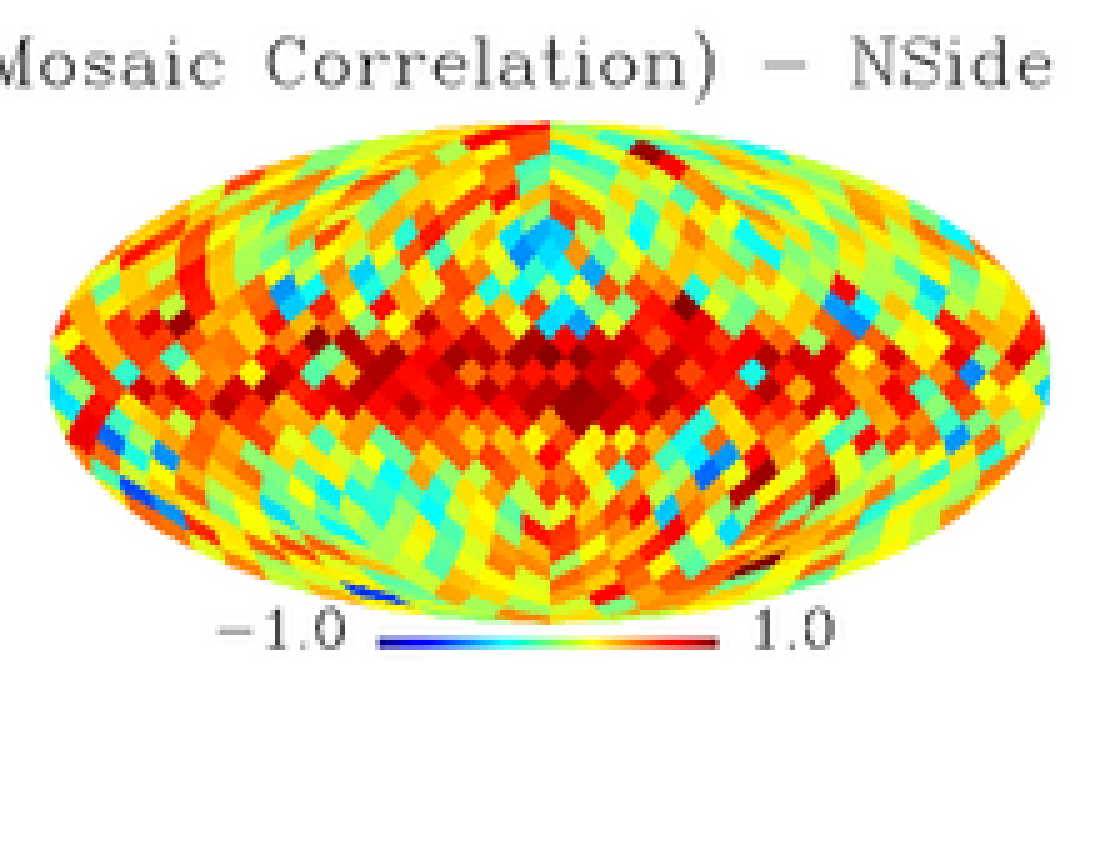}
\end{minipage}
\hfill
\begin{minipage}{0.24\textwidth}
\centering
\includegraphics[trim=0cm 1.5cm 0cm 1.2cm,clip=true,width=4cm]{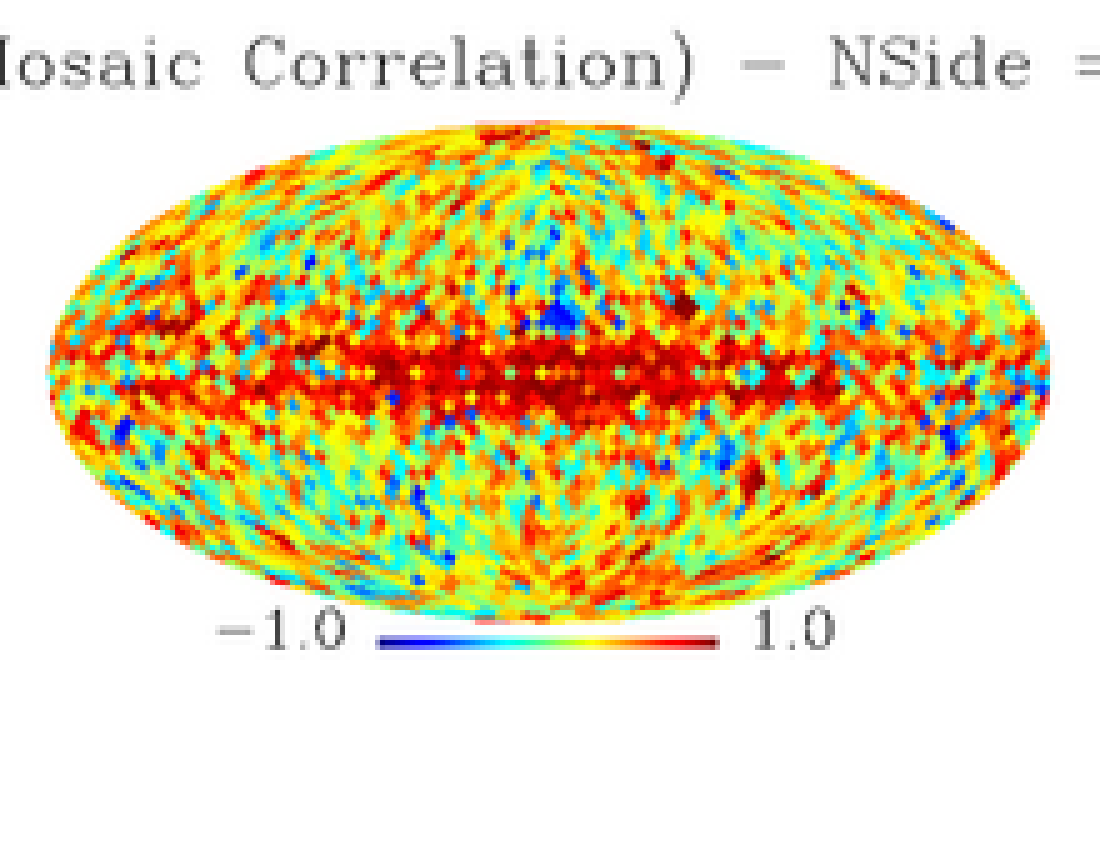}
\end{minipage}
\hfill
\centering
\begin{minipage}{0.24\textwidth}
\centering
\includegraphics[trim=0cm 1.5cm 0cm 1.2cm,clip=true,width=4cm]{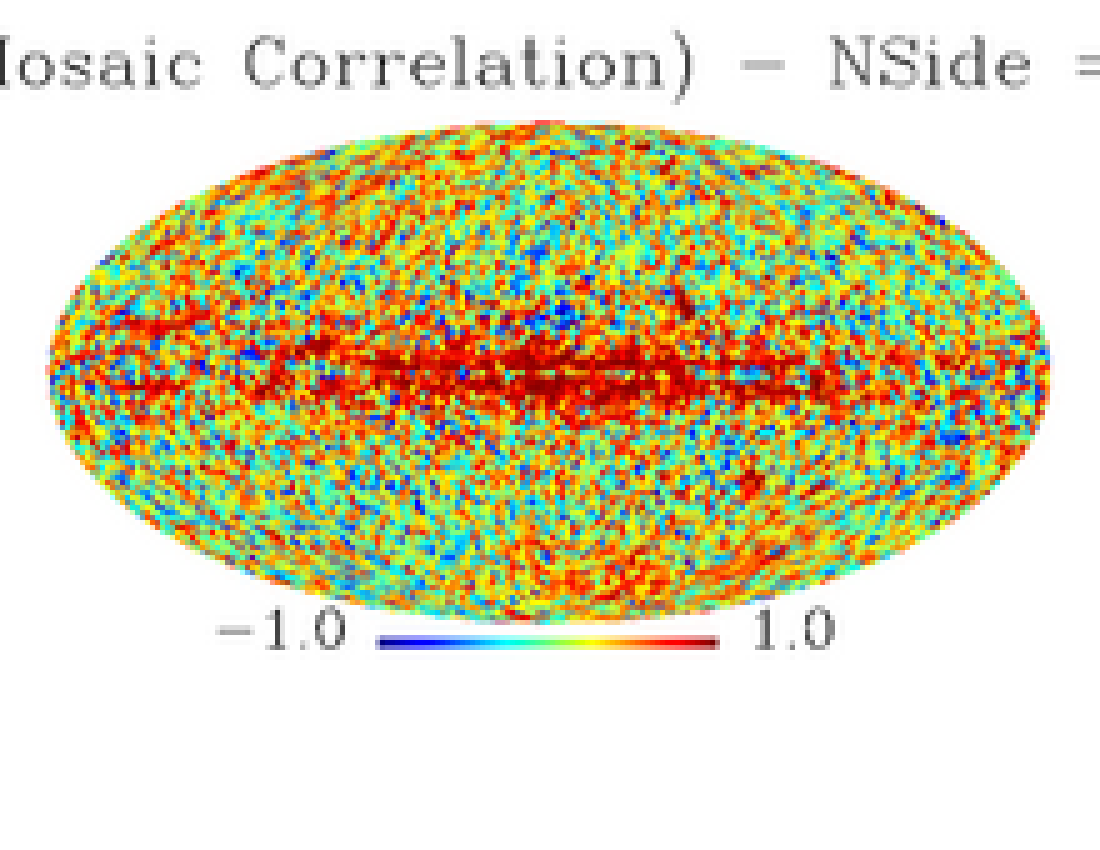}
\end{minipage}
\hfill
\begin{minipage}{0.24\textwidth}
\centering
\includegraphics[trim=0cm 1.5cm 0cm 1.2cm,clip=true,width=4cm]{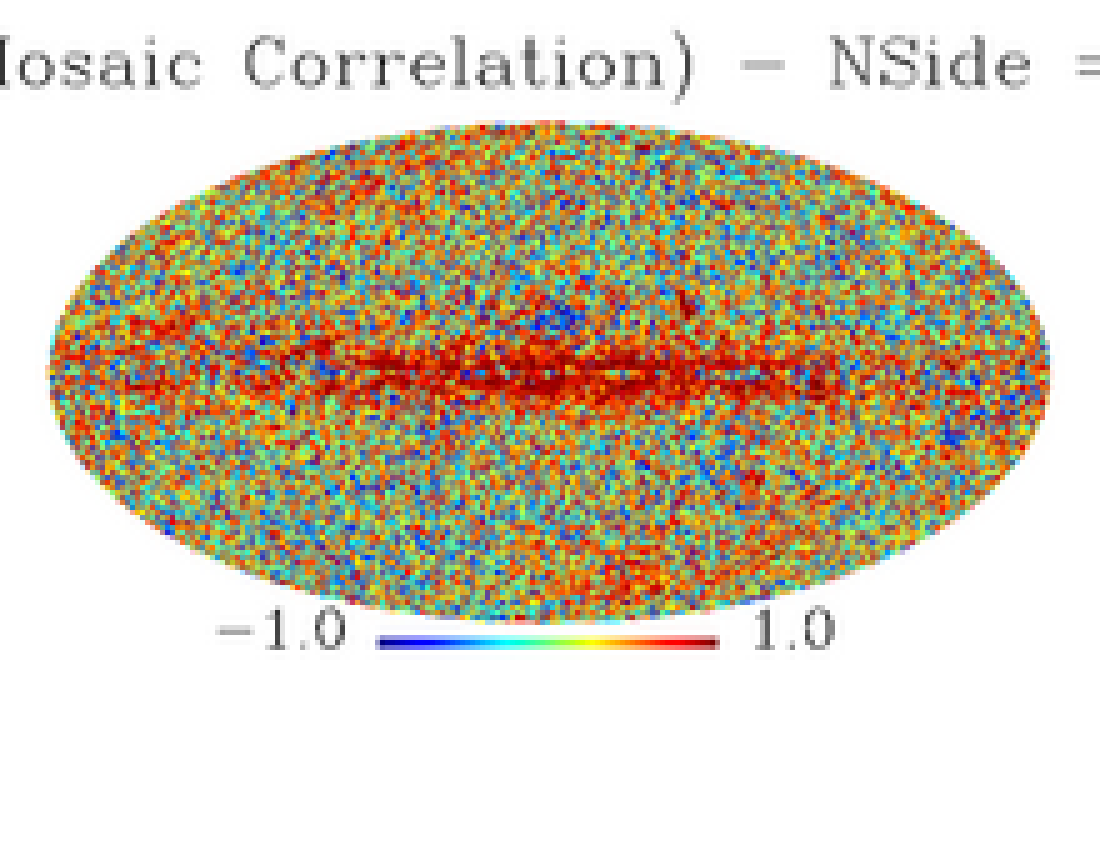}
\end{minipage}
\centering
\begin{minipage}{0.24\textwidth}
\centering
\includegraphics[trim=0cm 1.5cm 0cm 1.2cm,clip=true,width=4cm]{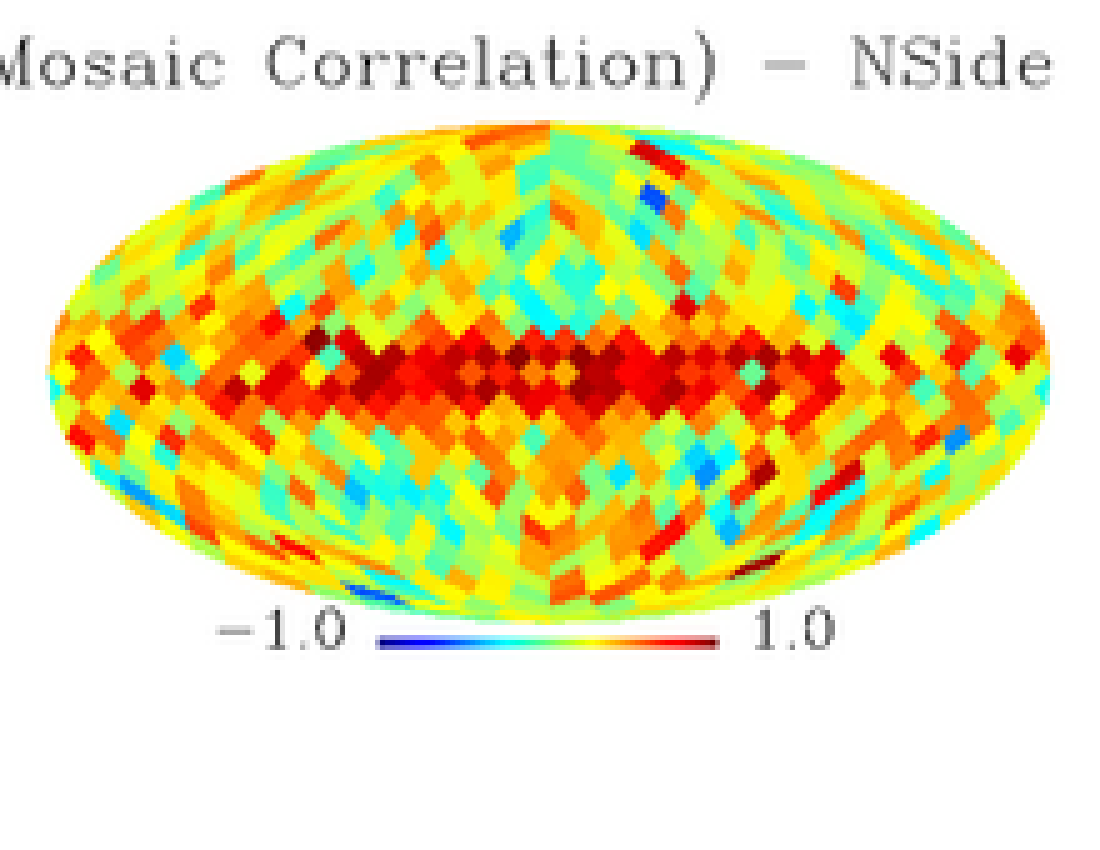}
\end{minipage}
\hfill
\begin{minipage}{0.24\textwidth}
\centering
\includegraphics[trim=0cm 1.5cm 0cm 1.2cm,clip=true,width=4cm]{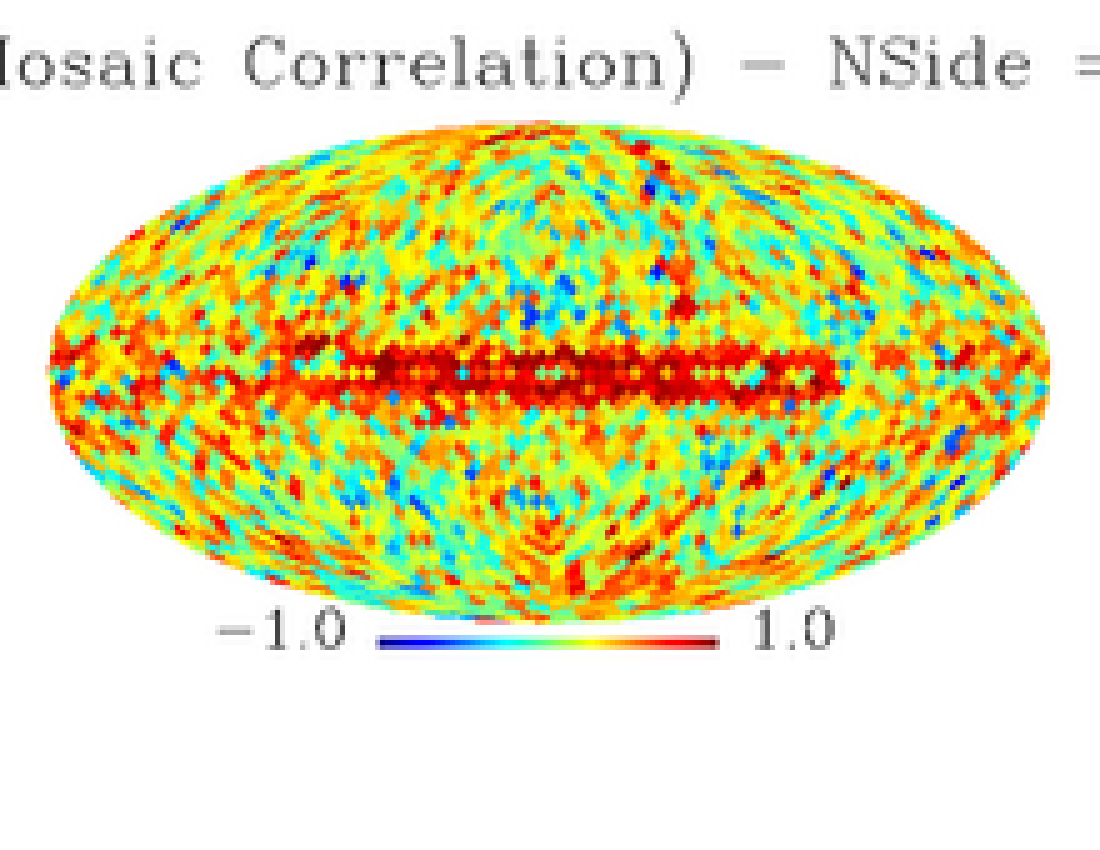}
\end{minipage}
\hfill
\centering
\begin{minipage}{0.24\textwidth}
\centering
\includegraphics[trim=0cm 1.5cm 0cm 1.2cm,clip=true,width=4cm]{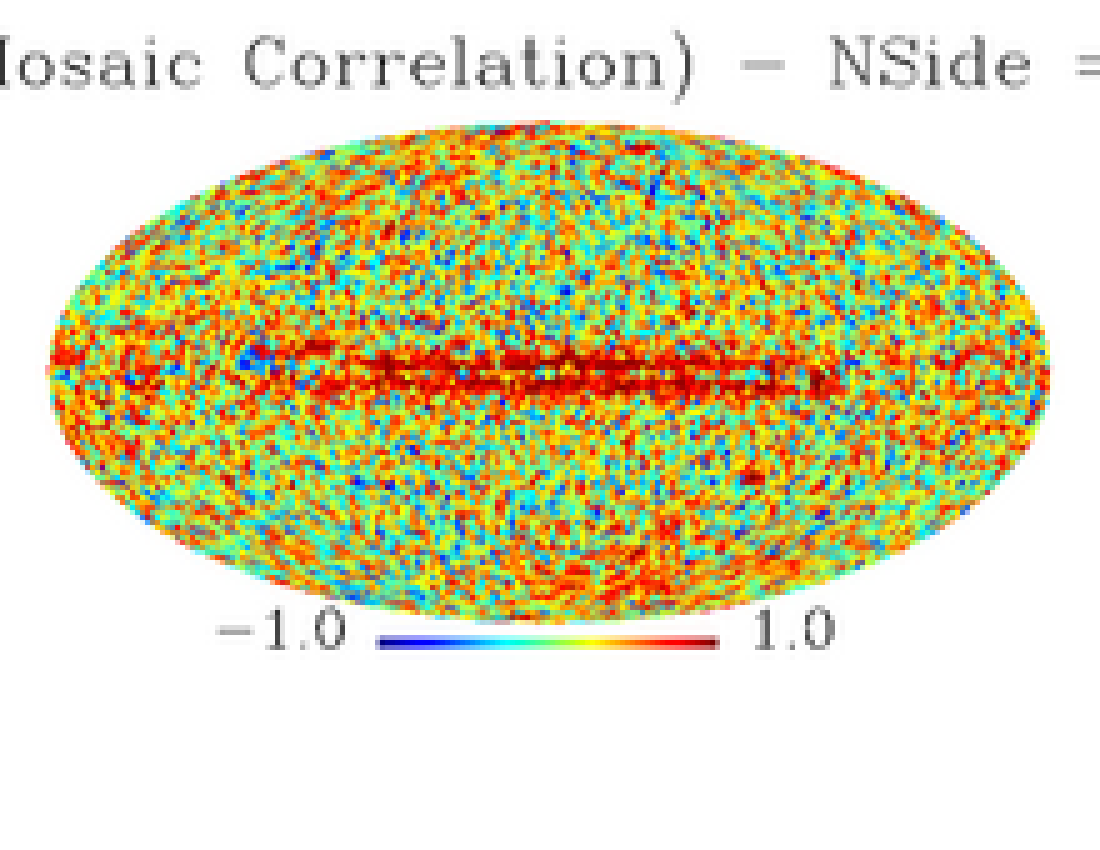}
\end{minipage}
\hfill
\begin{minipage}{0.24\textwidth}
\centering
\includegraphics[trim=0cm 1.5cm 0cm 1.2cm,clip=true,width=4cm]{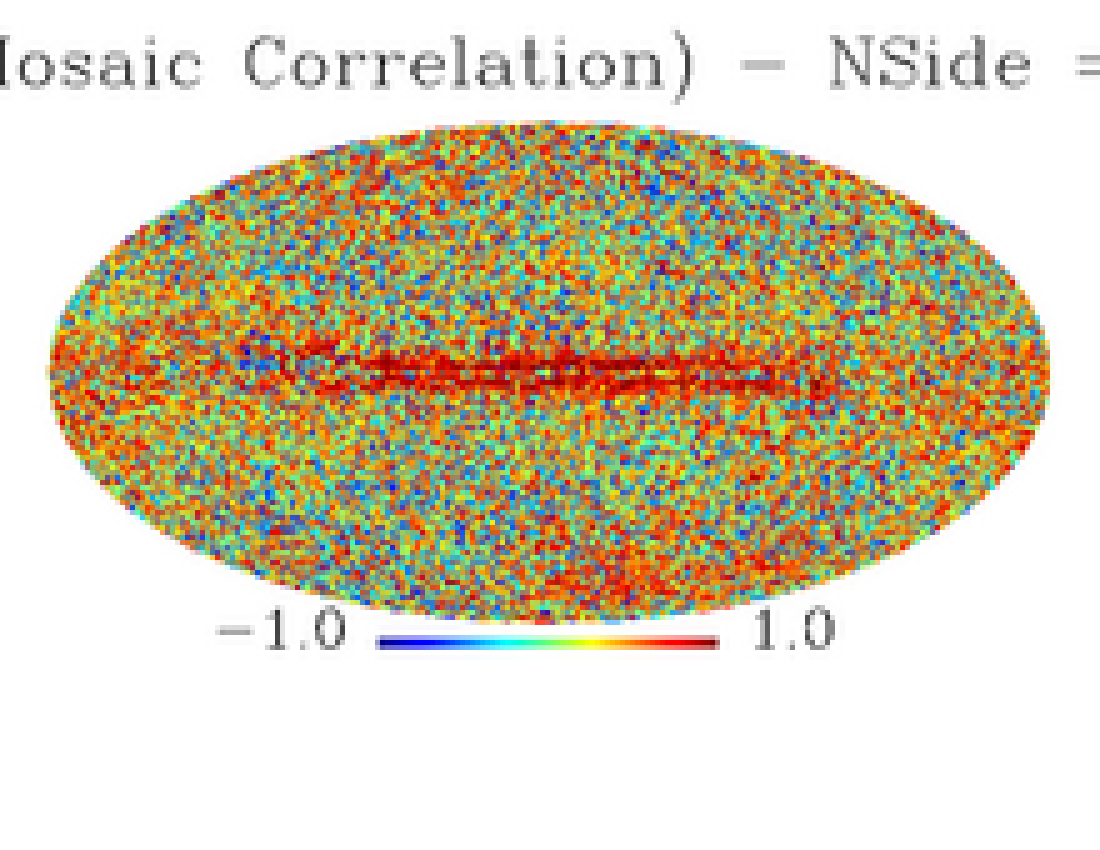}
\end{minipage}
\centering
\begin{minipage}{0.24\textwidth}
\centering
\includegraphics[trim=0cm 1.5cm 0cm 1.2cm,clip=true,width=4cm]{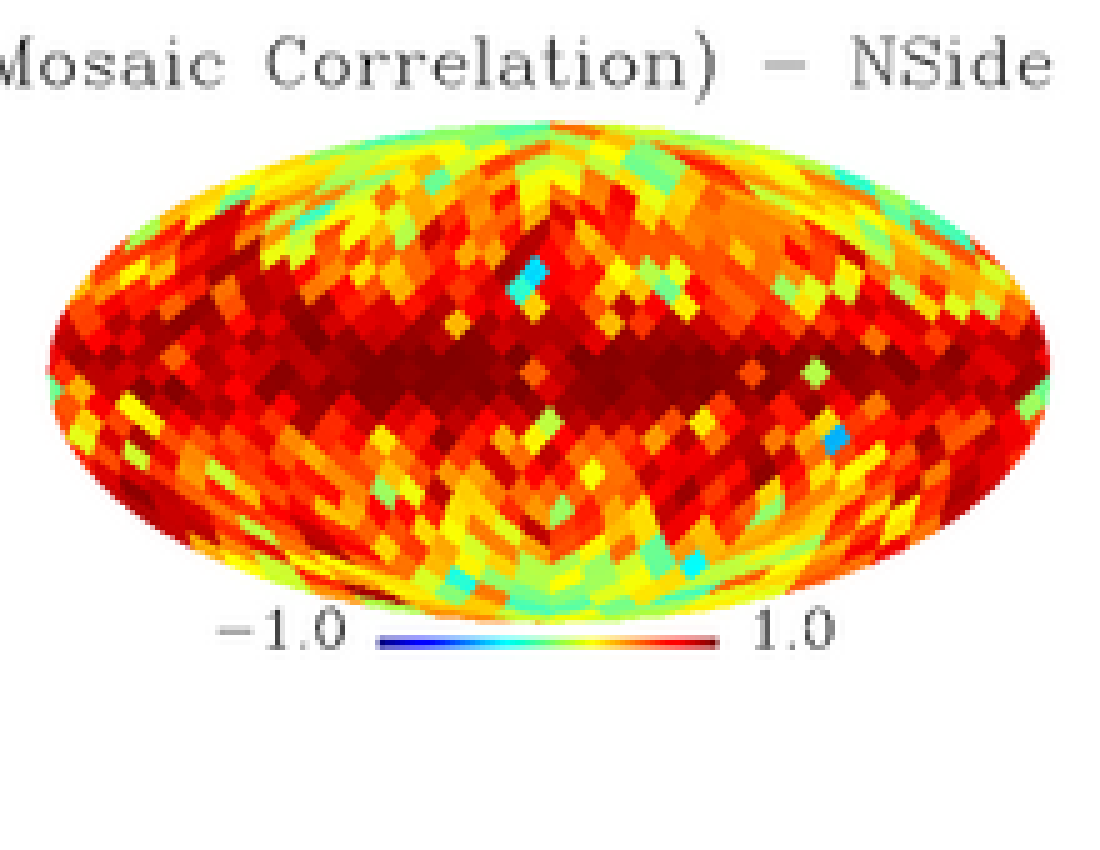}
\end{minipage}
\hfill
\begin{minipage}{0.24\textwidth}
\centering
\includegraphics[trim=0cm 1.5cm 0cm 1.2cm,clip=true,width=4cm]{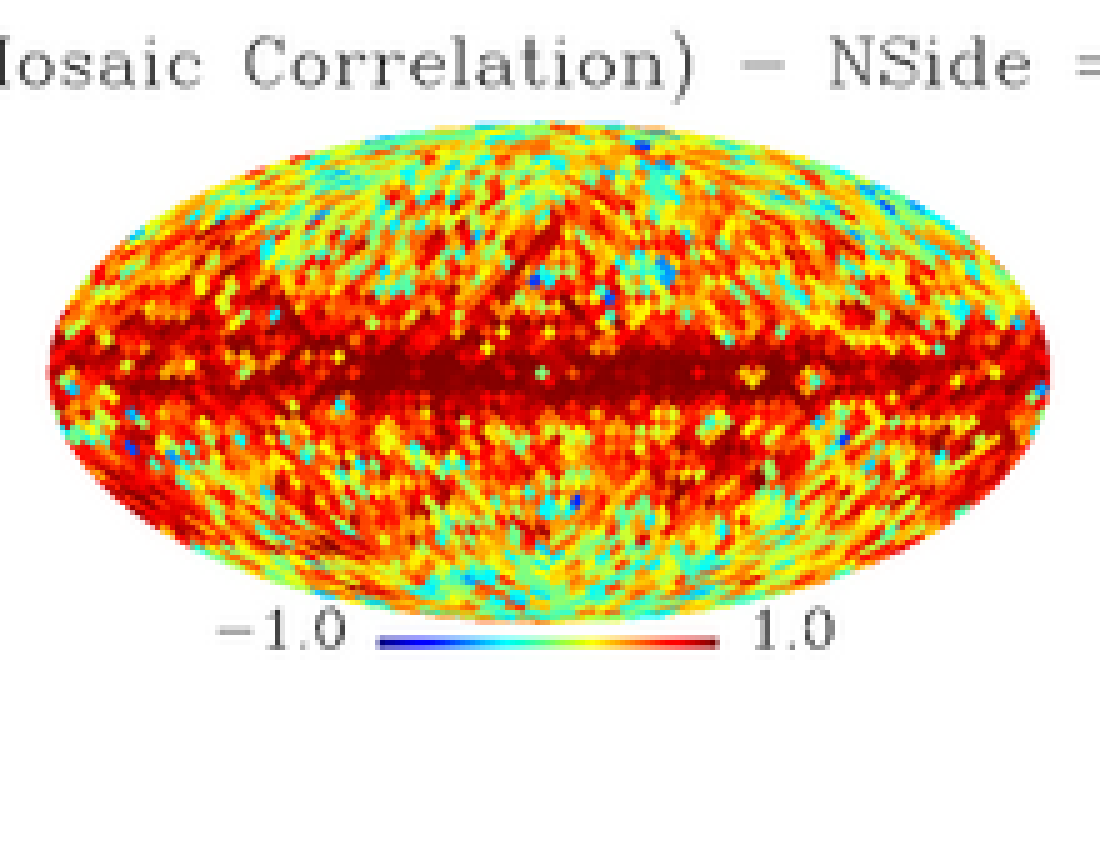}
\end{minipage}
\hfill
\centering
\begin{minipage}{0.24\textwidth}
\centering
\includegraphics[trim=0cm 1.5cm 0cm 1.2cm,clip=true,width=4cm]{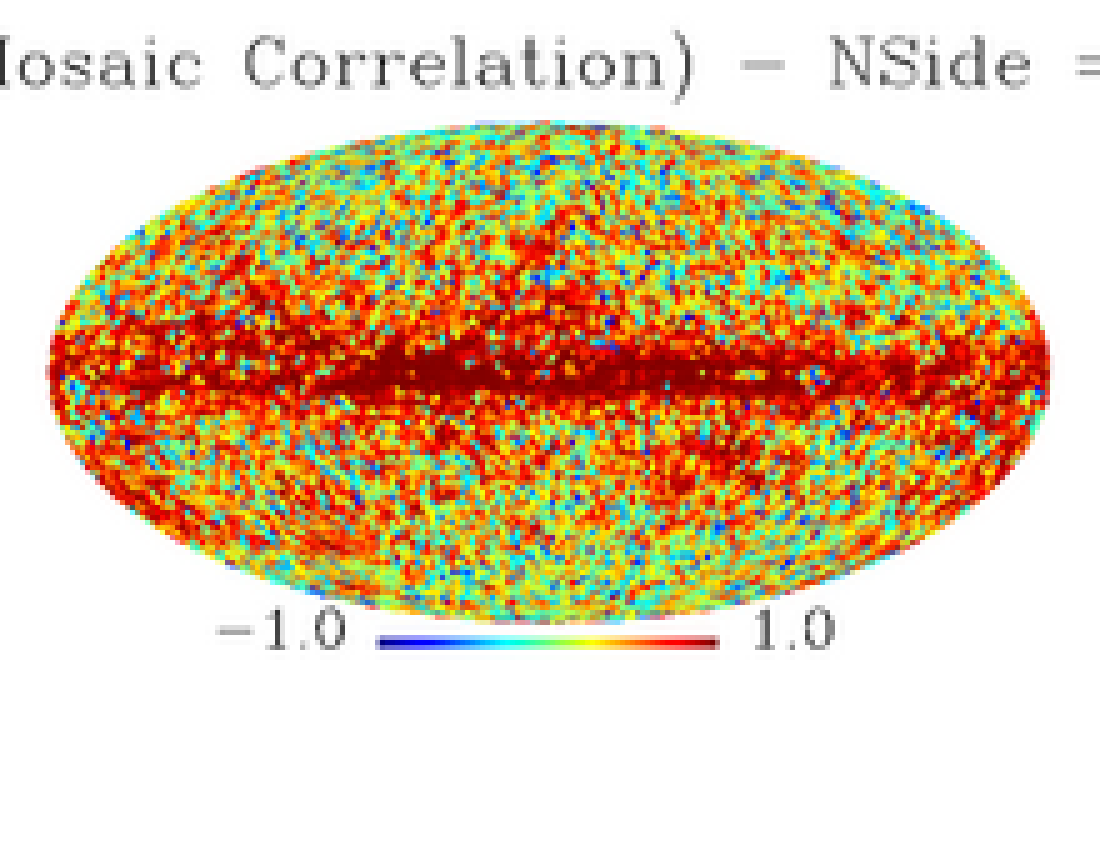}
\end{minipage}
\hfill
\begin{minipage}{0.24\textwidth}
\centering
\includegraphics[trim=0cm 1.5cm 0cm 1.2cm,clip=true,width=4cm]{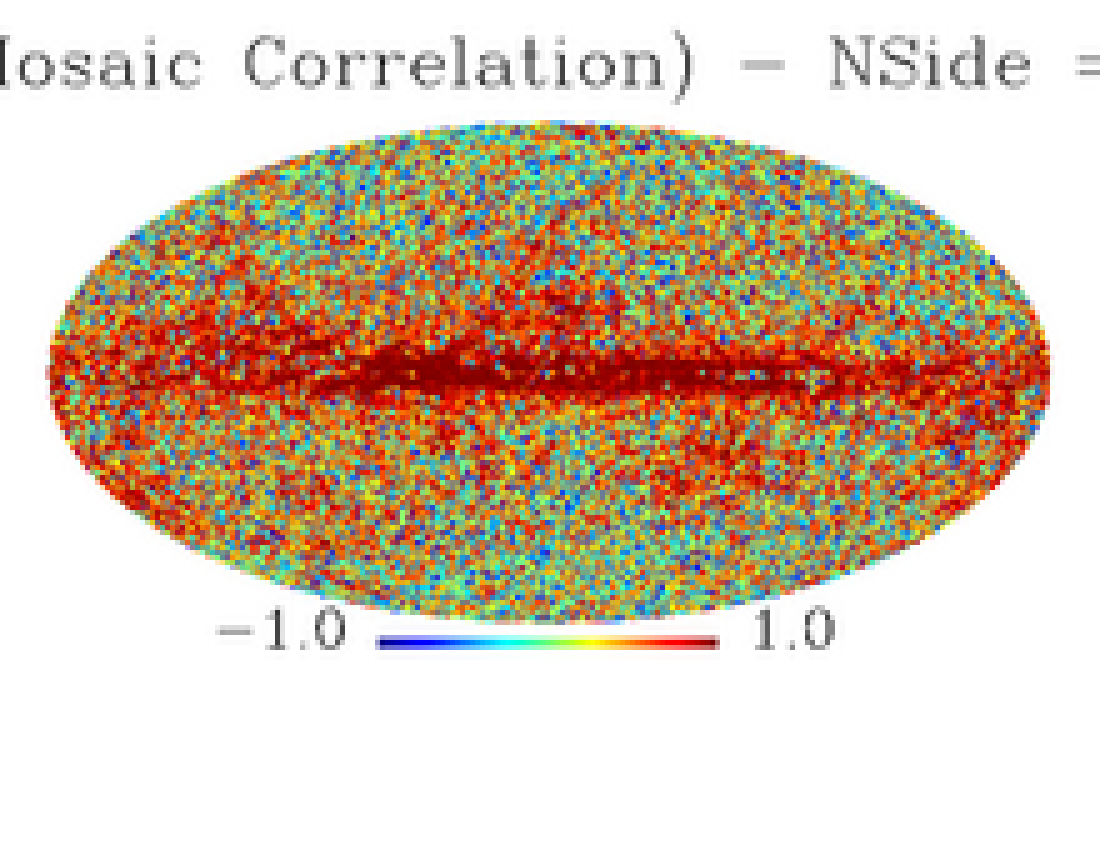}
\end{minipage}
\centering
\begin{minipage}{0.24\textwidth}
\centering
\includegraphics[trim=0cm 1.5cm 0cm 1.2cm,clip=true,width=4cm]{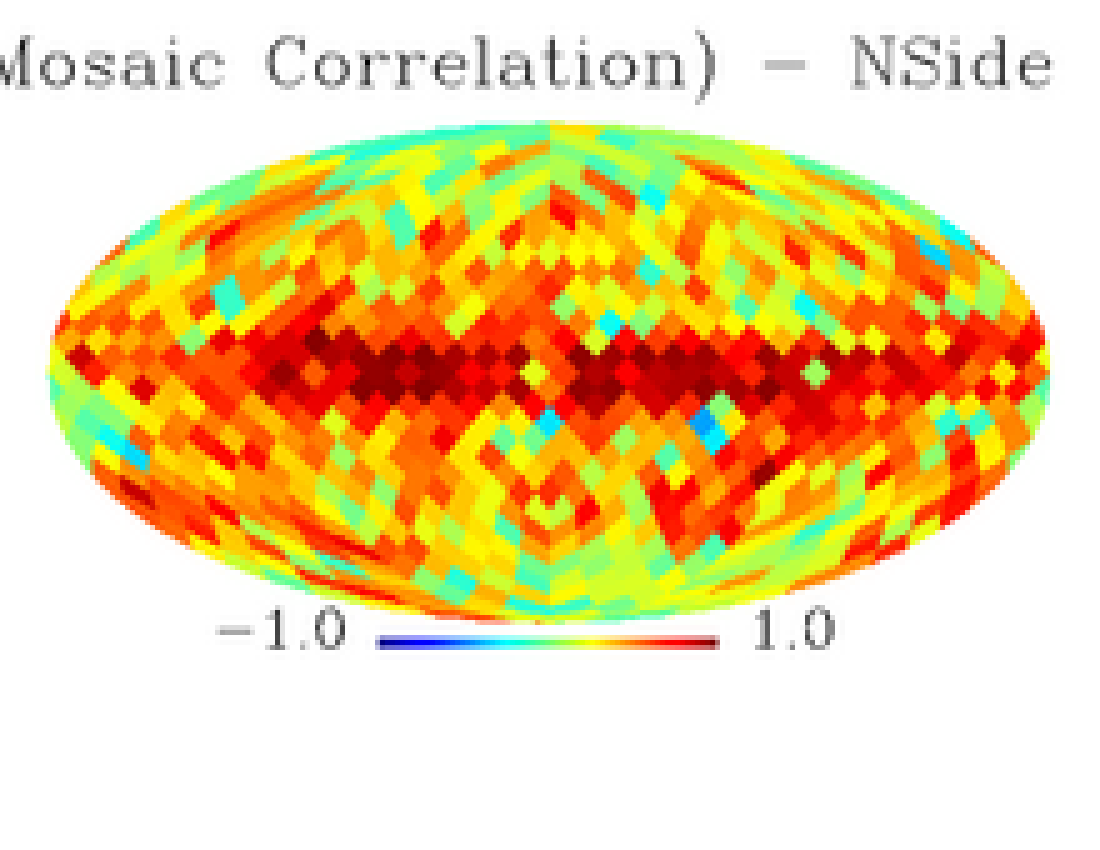}
\end{minipage}
\hfill
\begin{minipage}{0.24\textwidth}
\centering
\includegraphics[trim=0cm 1.5cm 0cm 1.2cm,clip=true,width=4cm]{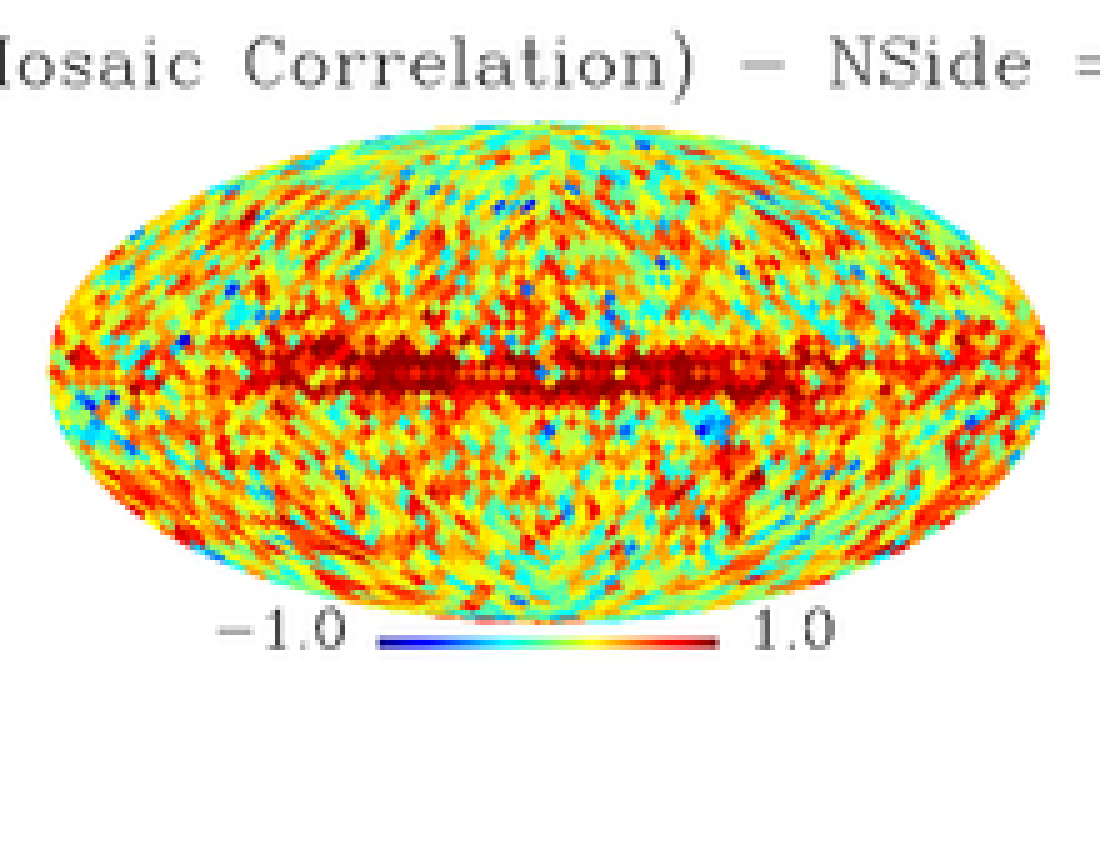}
\end{minipage}
\hfill
\centering
\begin{minipage}{0.24\textwidth}
\centering
\includegraphics[trim=0cm 1.5cm 0cm 1.2cm,clip=true,width=4cm]{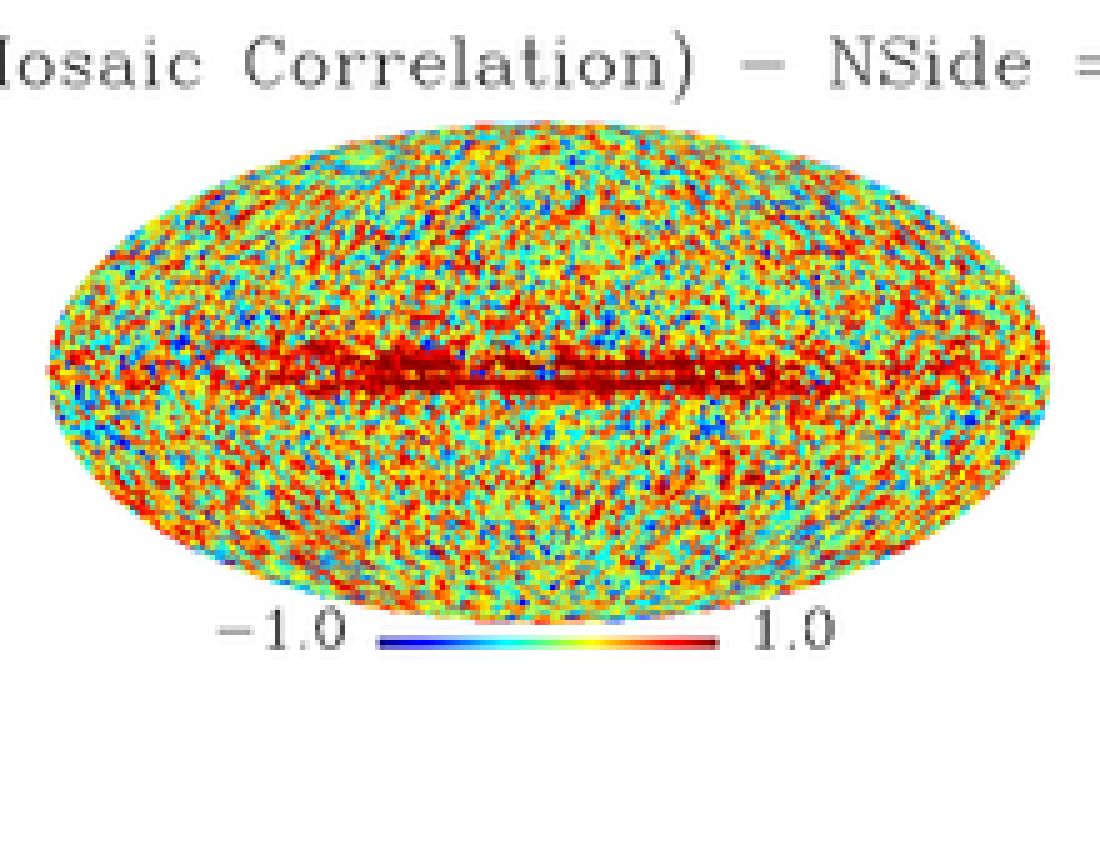}
\end{minipage}
\hfill
\begin{minipage}{0.24\textwidth}
\centering
\includegraphics[trim=0cm 1.5cm 0cm 1.2cm,clip=true,width=4cm]{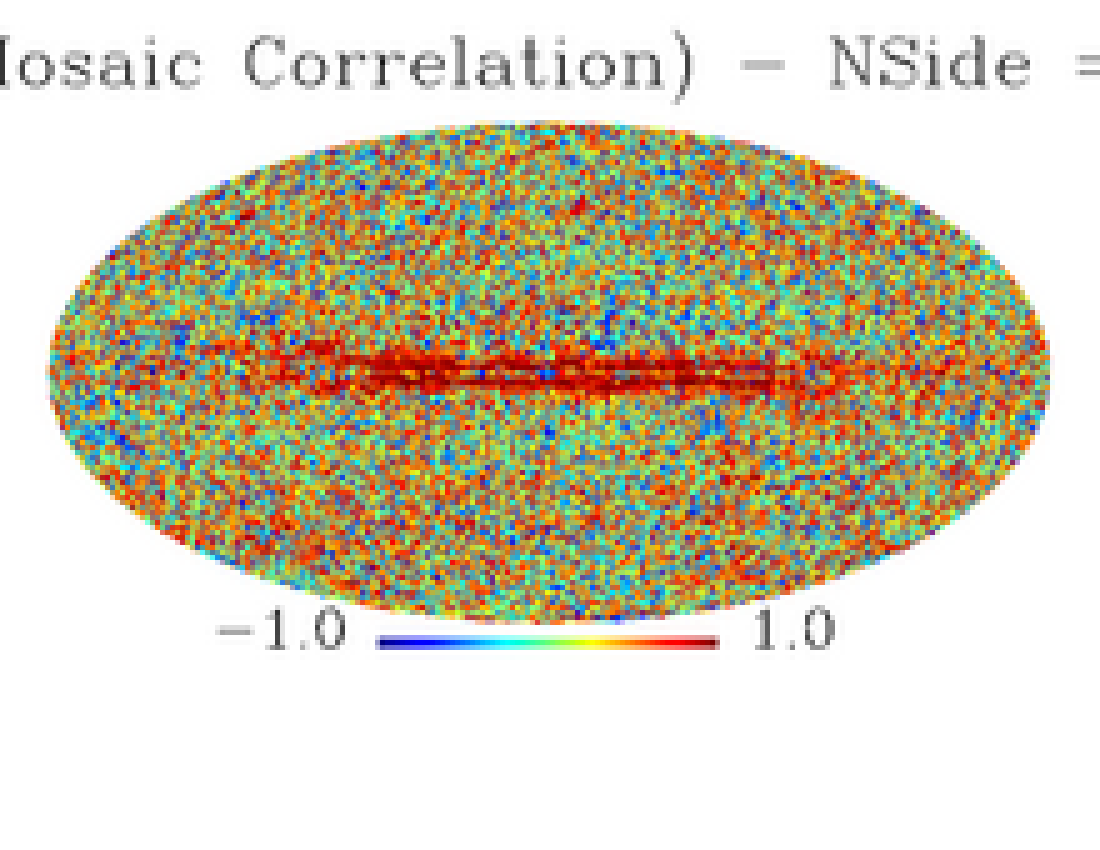}
\end{minipage}
\caption{From top to bottom: Weighted-Correlation maps between AME1 and synchrotron radiation, AME2 and synchrotron radiation, AME1 and thermal dust emission, AME2 and thermal dust emission. From left to right: $\Omega$ contains 1024, 256, 64 and 16 pixel. Here, we weighted with $w_i(n=1)$.}
\label{Acorrmaps1}
\end{figure}

\begin{figure}[H]
\centering
\begin{minipage}{0.24\textwidth}
\centering
\includegraphics[trim=0cm 1.5cm 0cm 1.2cm,clip=true,width=4cm]{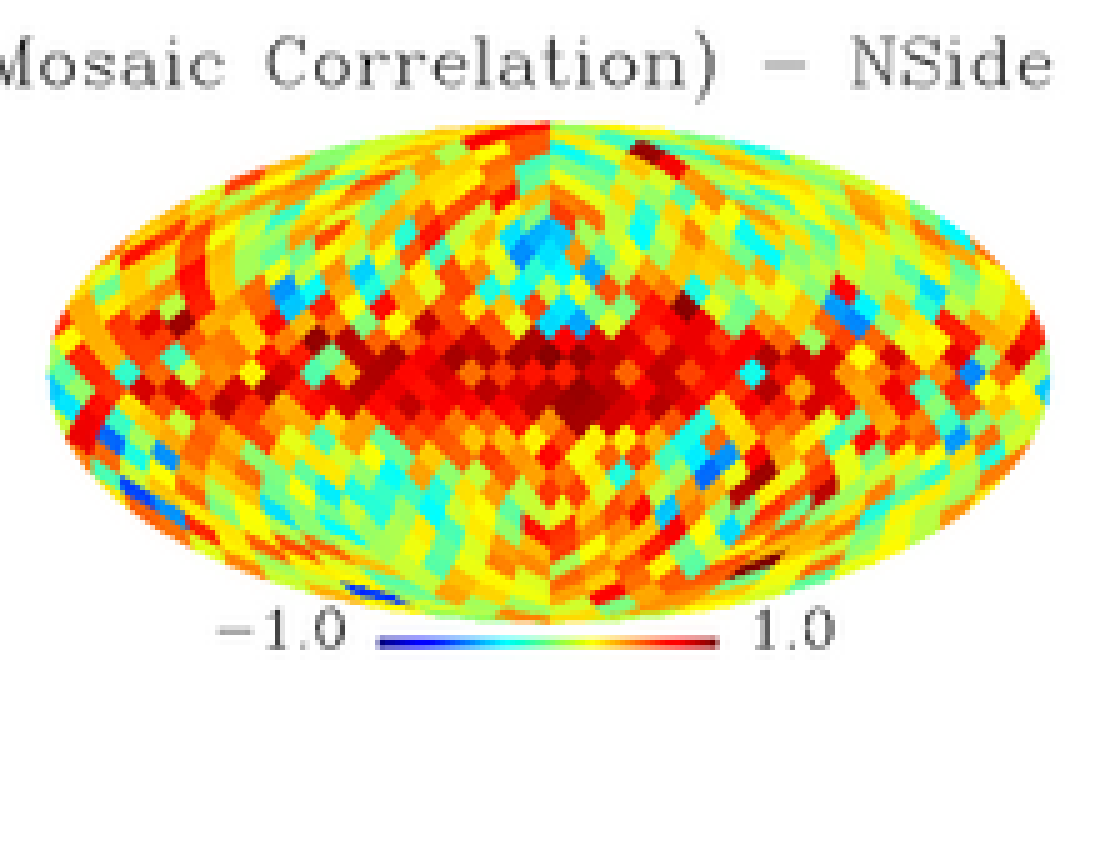}
\end{minipage}
\hfill
\begin{minipage}{0.24\textwidth}
\centering
\includegraphics[trim=0cm 1.5cm 0cm 1.2cm,clip=true,width=4cm]{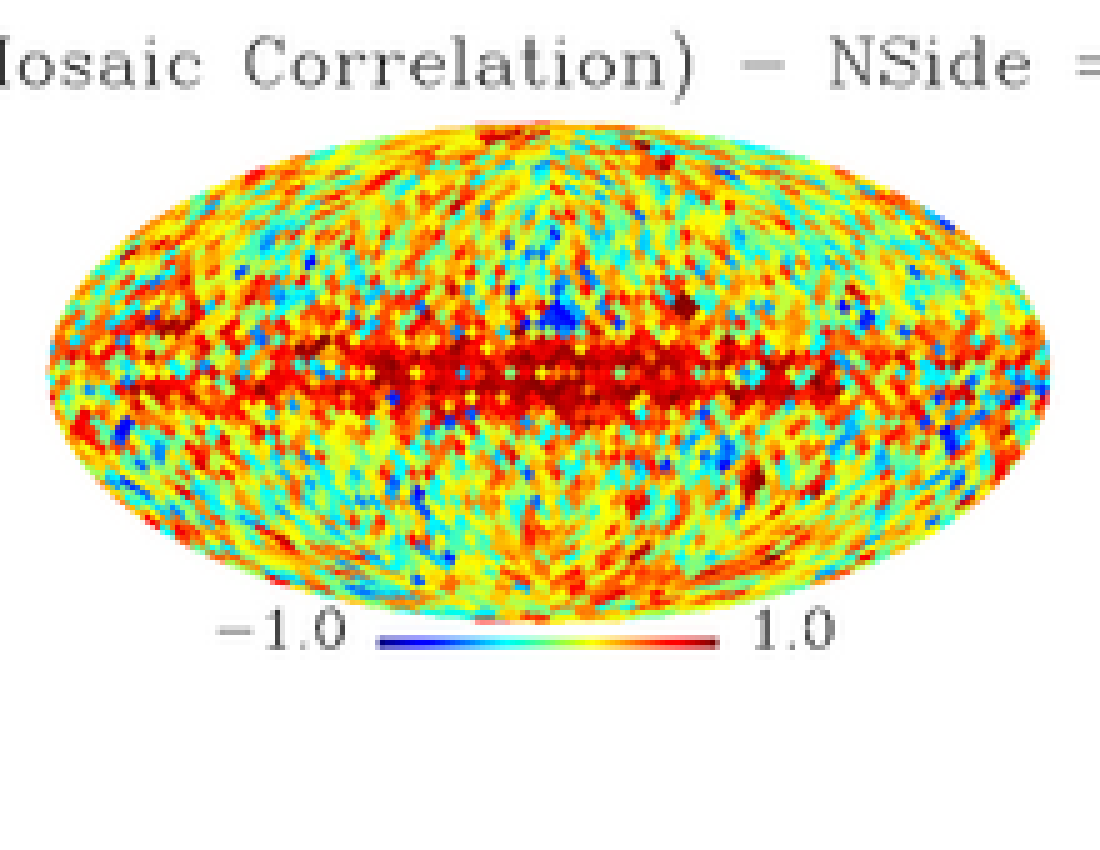}
\end{minipage}
\hfill
\centering
\begin{minipage}{0.24\textwidth}
\centering
\includegraphics[trim=0cm 1.5cm 0cm 1.2cm,clip=true,width=4cm]{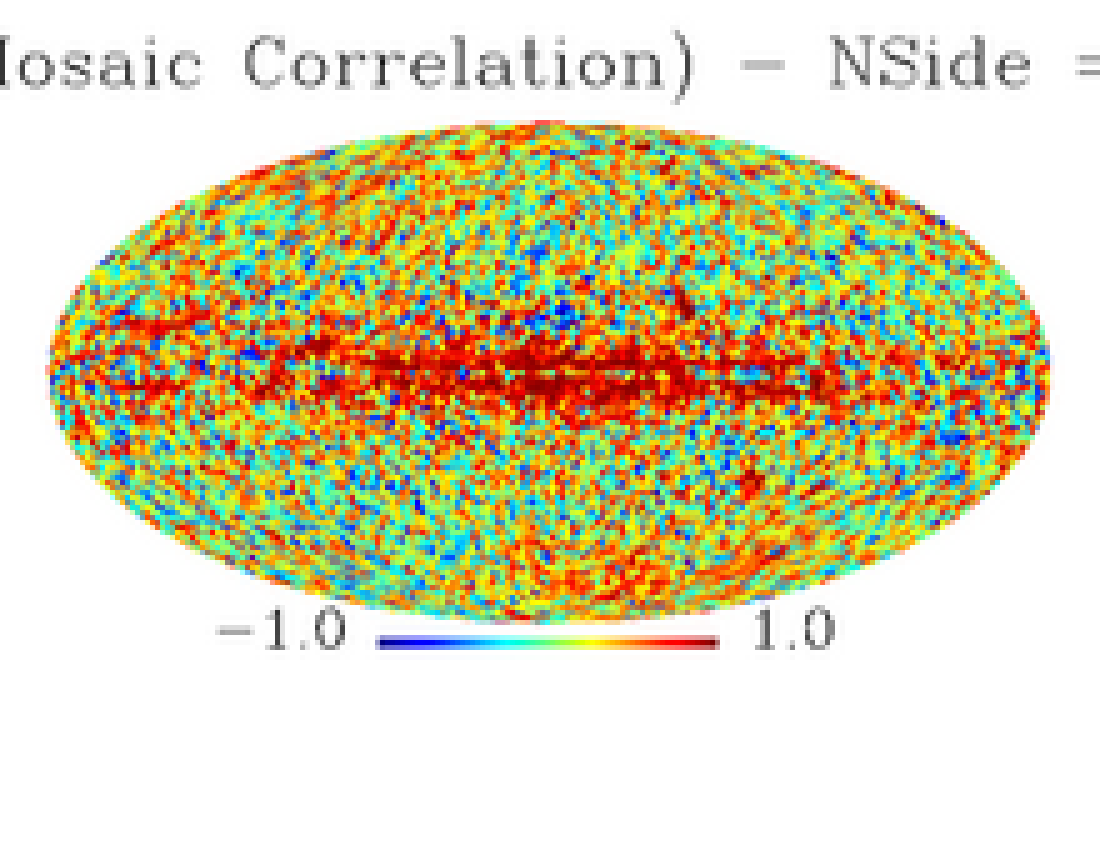}
\end{minipage}
\hfill
\begin{minipage}{0.24\textwidth}
\centering
\includegraphics[trim=0cm 1.5cm 0cm 1.2cm,clip=true,width=4cm]{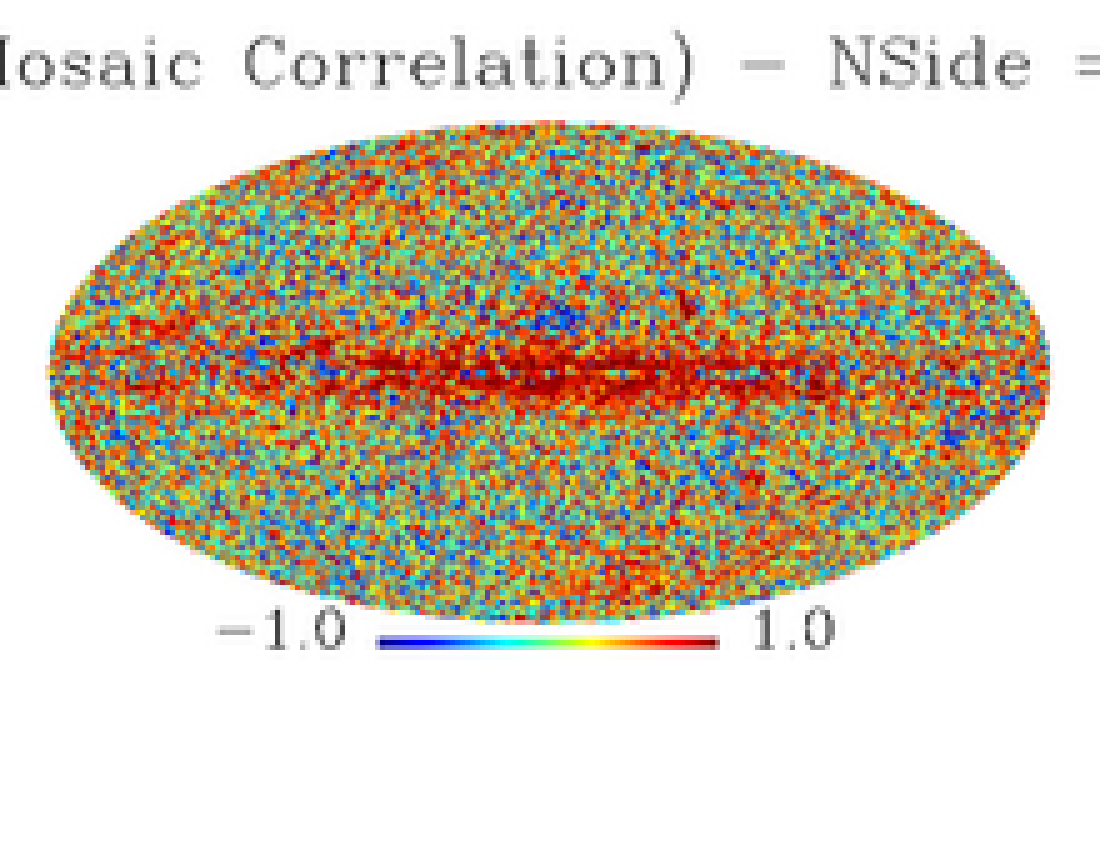}
\end{minipage}
\centering
\begin{minipage}{0.24\textwidth}
\centering
\includegraphics[trim=0cm 1.5cm 0cm 1.2cm,clip=true,width=4cm]{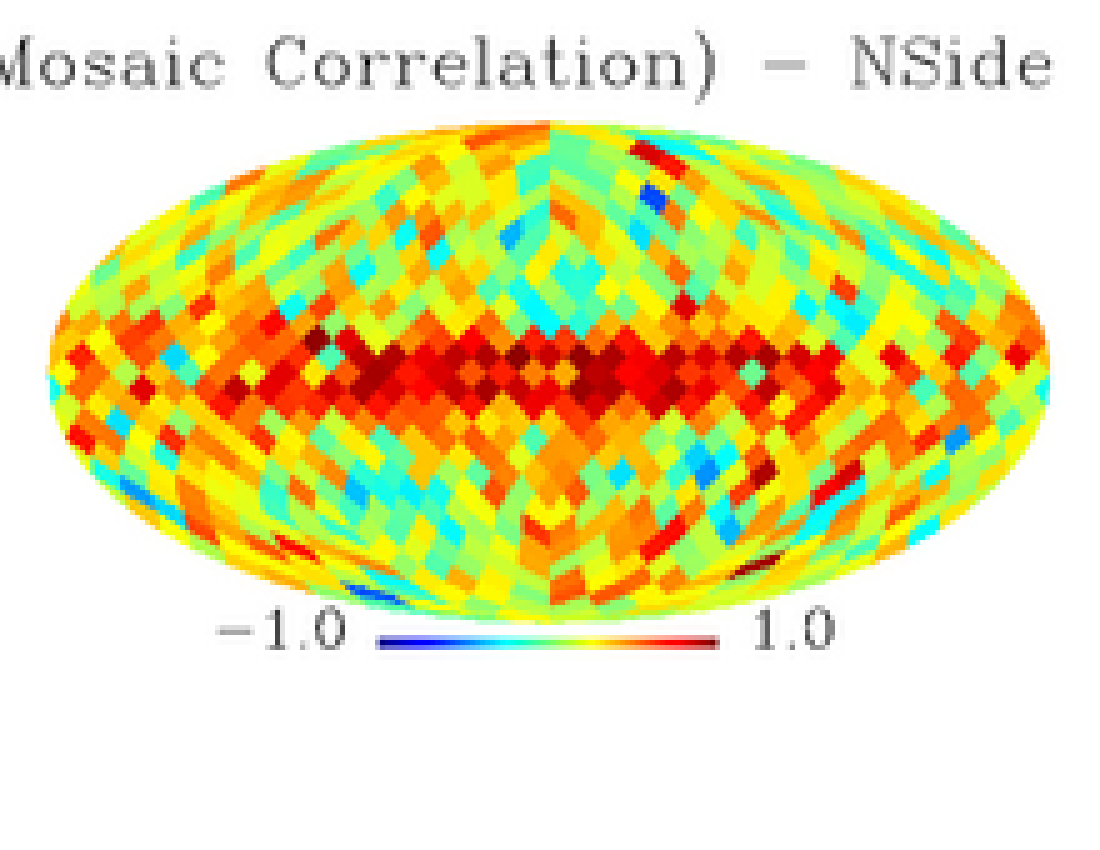}
\end{minipage}
\hfill
\begin{minipage}{0.24\textwidth}
\centering
\includegraphics[trim=0cm 1.5cm 0cm 1.2cm,clip=true,width=4cm]{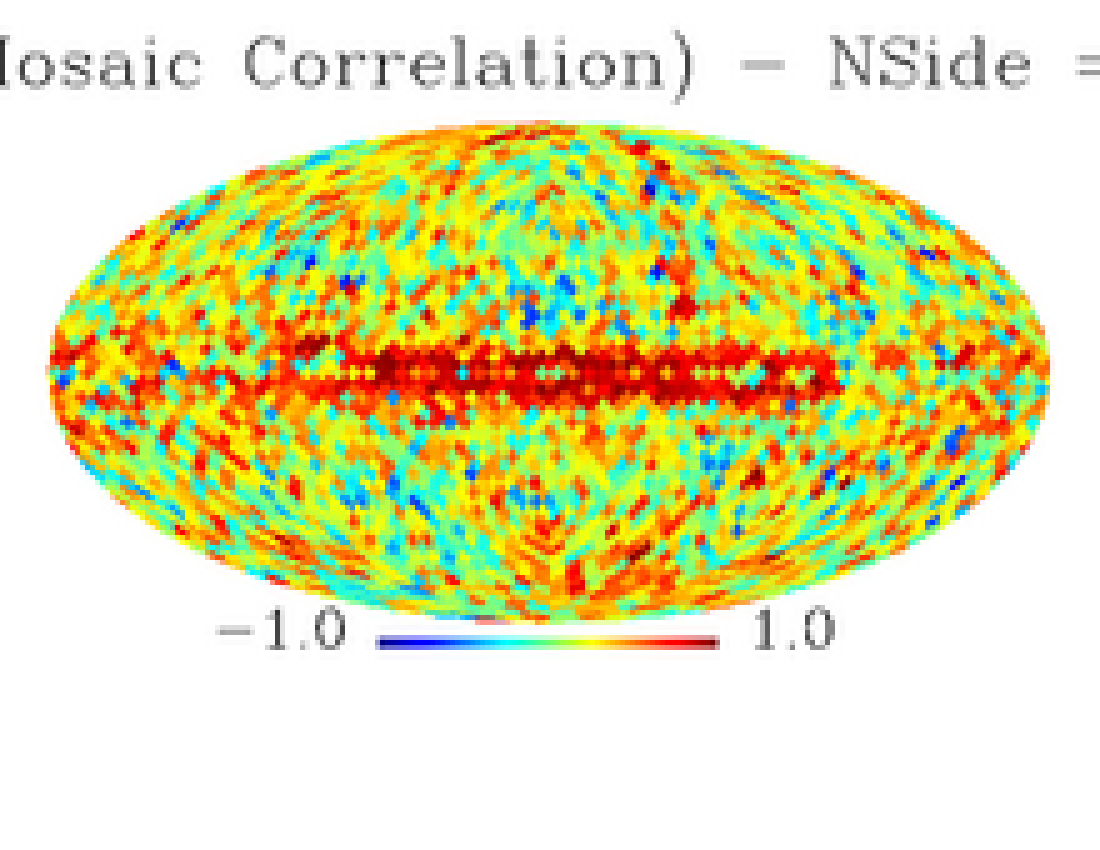}
\end{minipage}
\hfill
\centering
\begin{minipage}{0.24\textwidth}
\centering
\includegraphics[trim=0cm 1.5cm 0cm 1.2cm,clip=true,width=4cm]{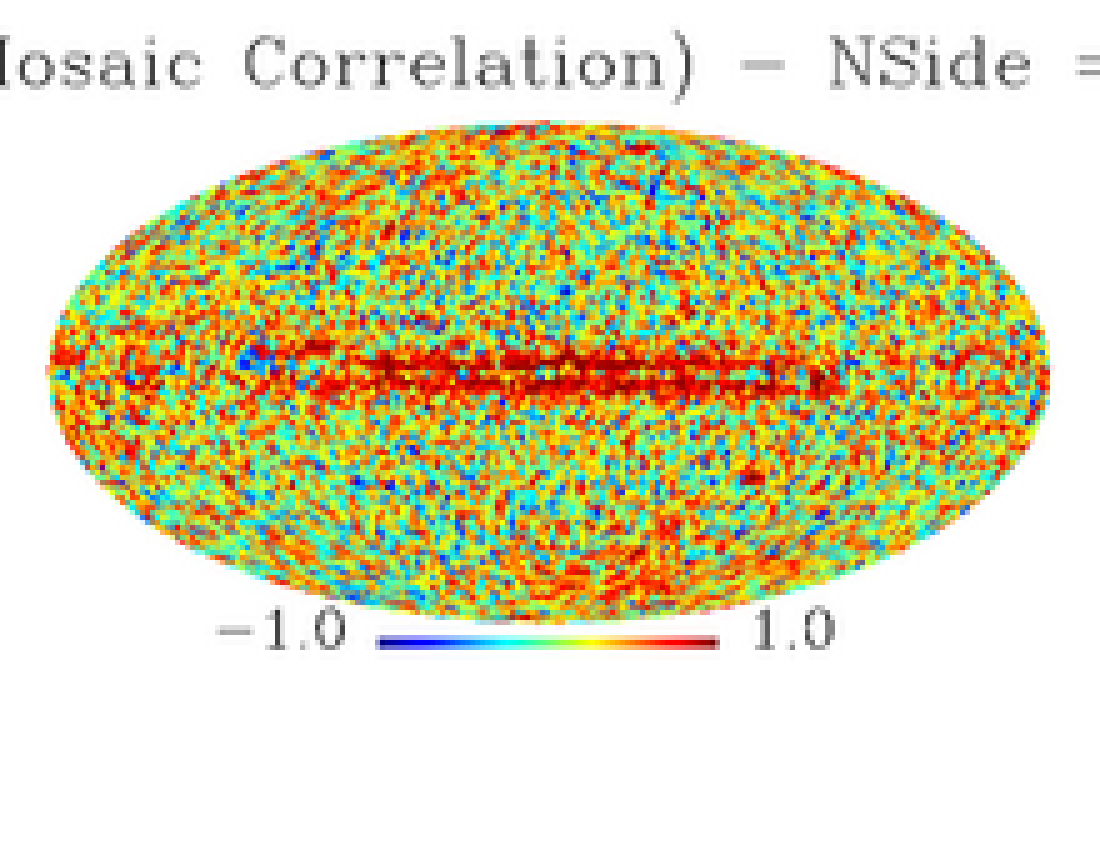}
\end{minipage}
\hfill
\begin{minipage}{0.24\textwidth}
\centering
\includegraphics[trim=0cm 1.5cm 0cm 1.2cm,clip=true,width=4cm]{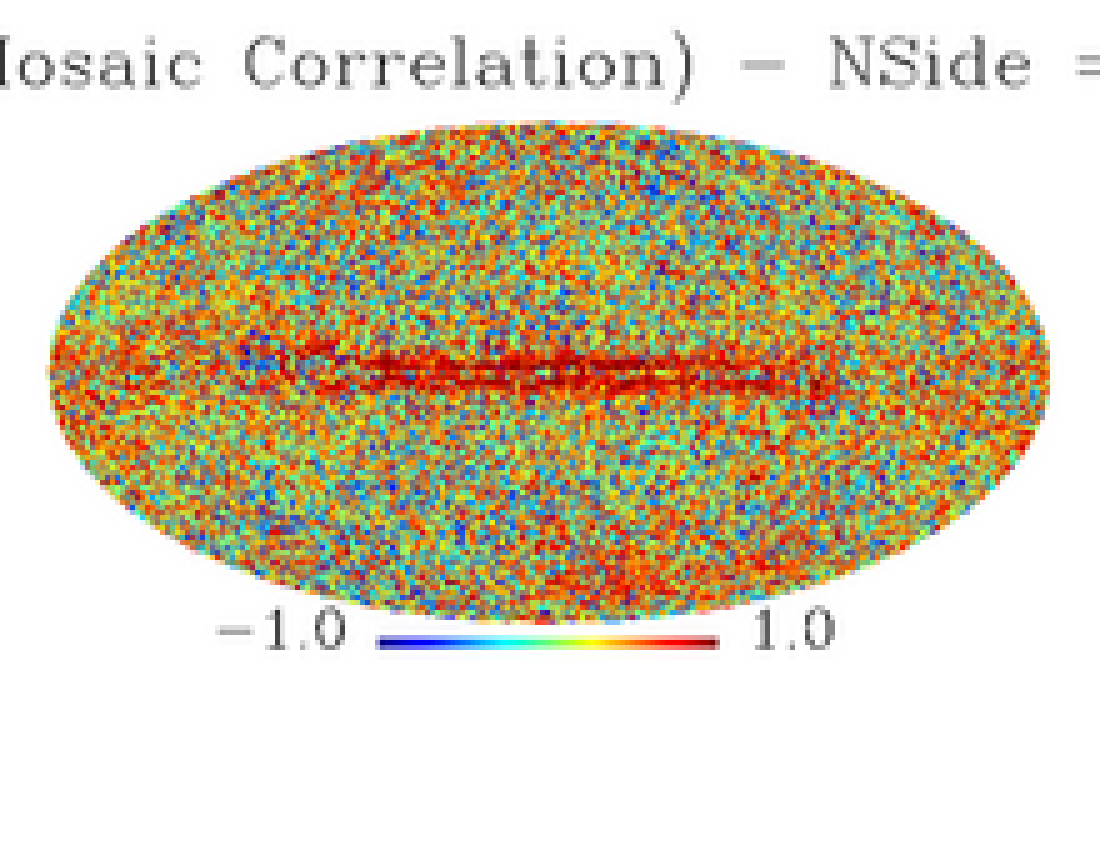}
\end{minipage}
\centering
\begin{minipage}{0.24\textwidth}
\centering
\includegraphics[trim=0cm 1.5cm 0cm 1.2cm,clip=true,width=4cm]{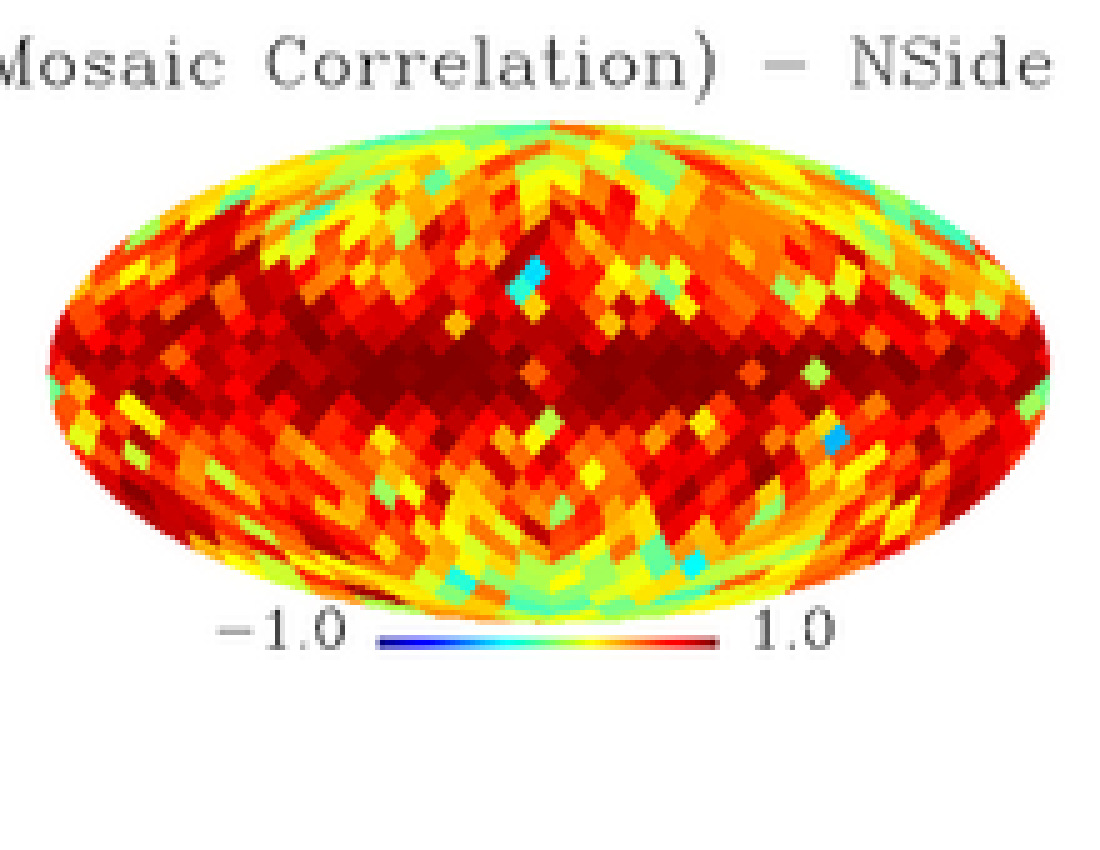}
\end{minipage}
\hfill
\begin{minipage}{0.24\textwidth}
\centering
\includegraphics[trim=0cm 1.5cm 0cm 1.2cm,clip=true,width=4cm]{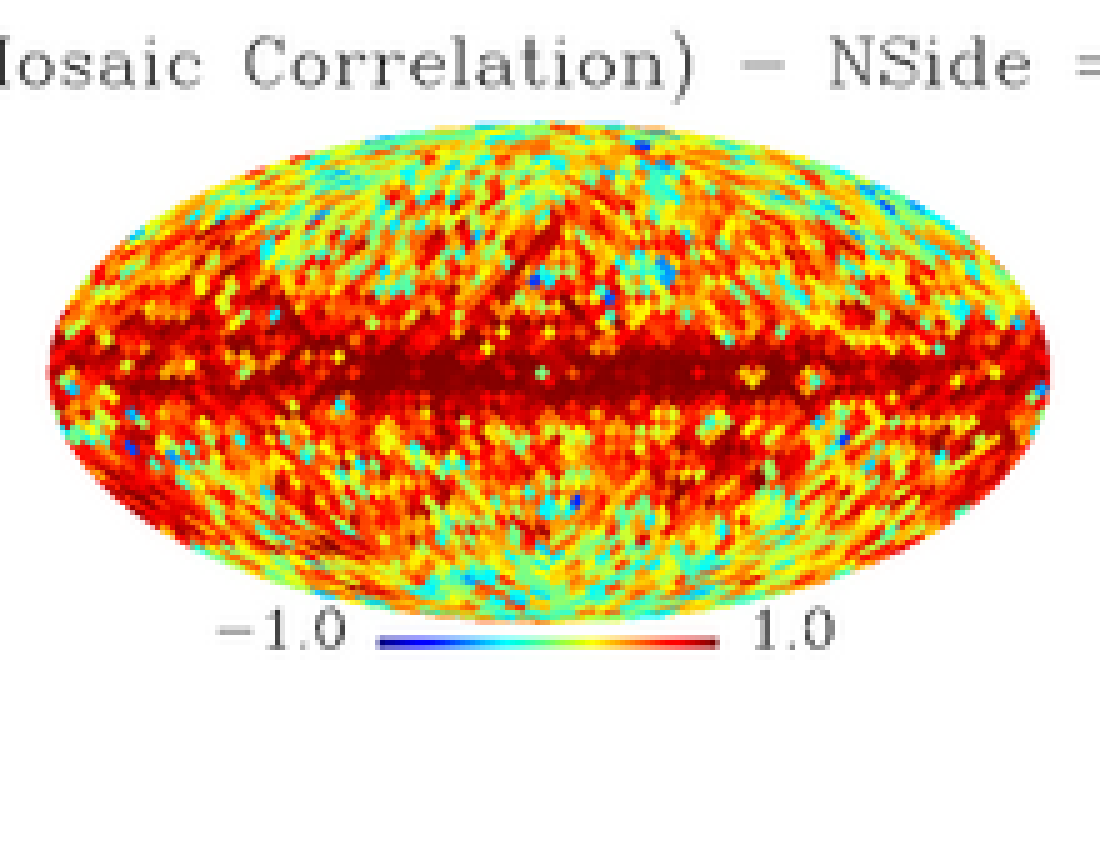}
\end{minipage}
\hfill
\centering
\begin{minipage}{0.24\textwidth}
\centering
\includegraphics[trim=0cm 1.5cm 0cm 1.2cm,clip=true,width=4cm]{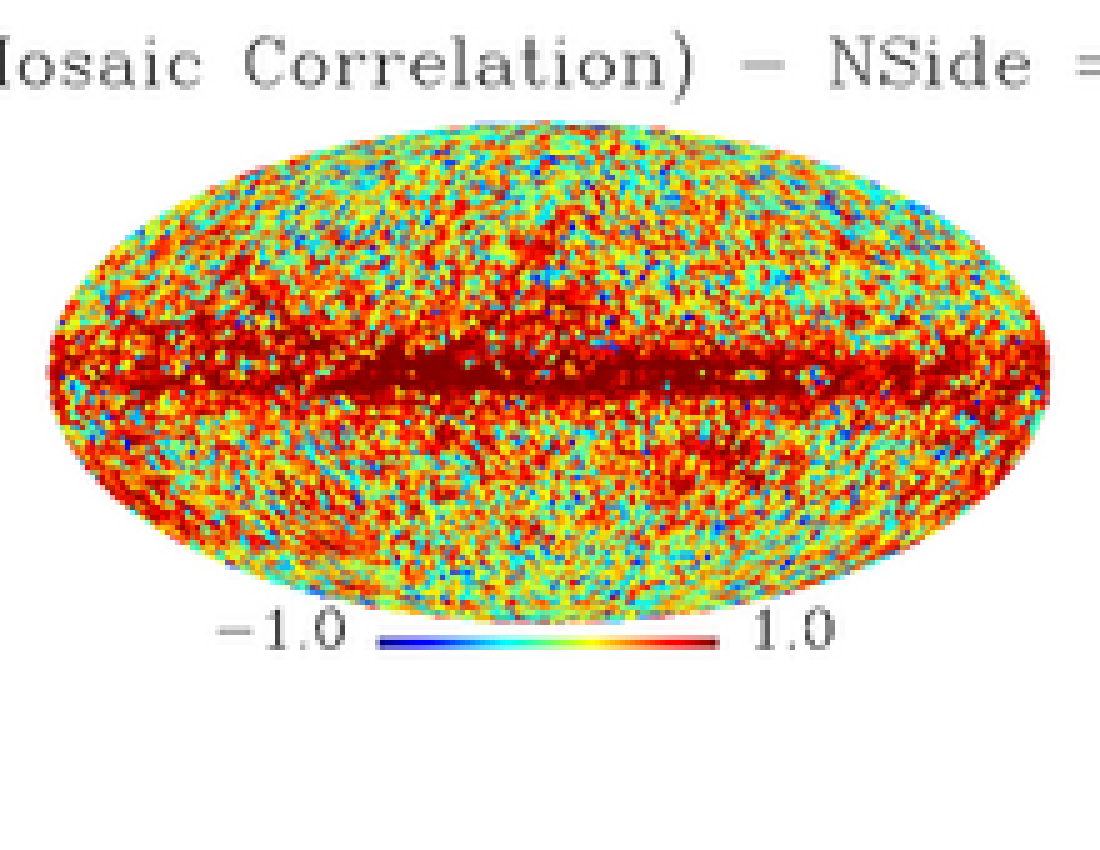}
\end{minipage}
\hfill
\begin{minipage}{0.24\textwidth}
\centering
\includegraphics[trim=0cm 1.5cm 0cm 1.2cm,clip=true,width=4cm]{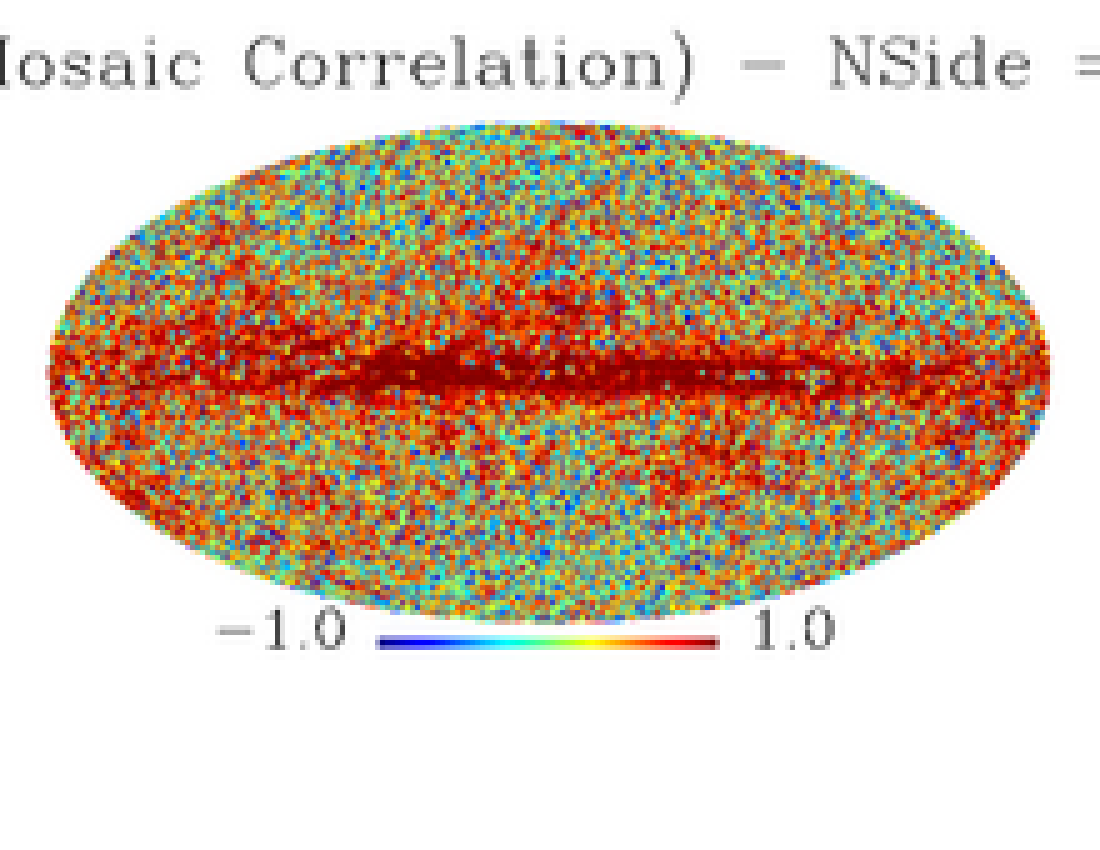}
\end{minipage}
\centering
\begin{minipage}{0.24\textwidth}
\centering
\includegraphics[trim=0cm 1.5cm 0cm 1.2cm,clip=true,width=4cm]{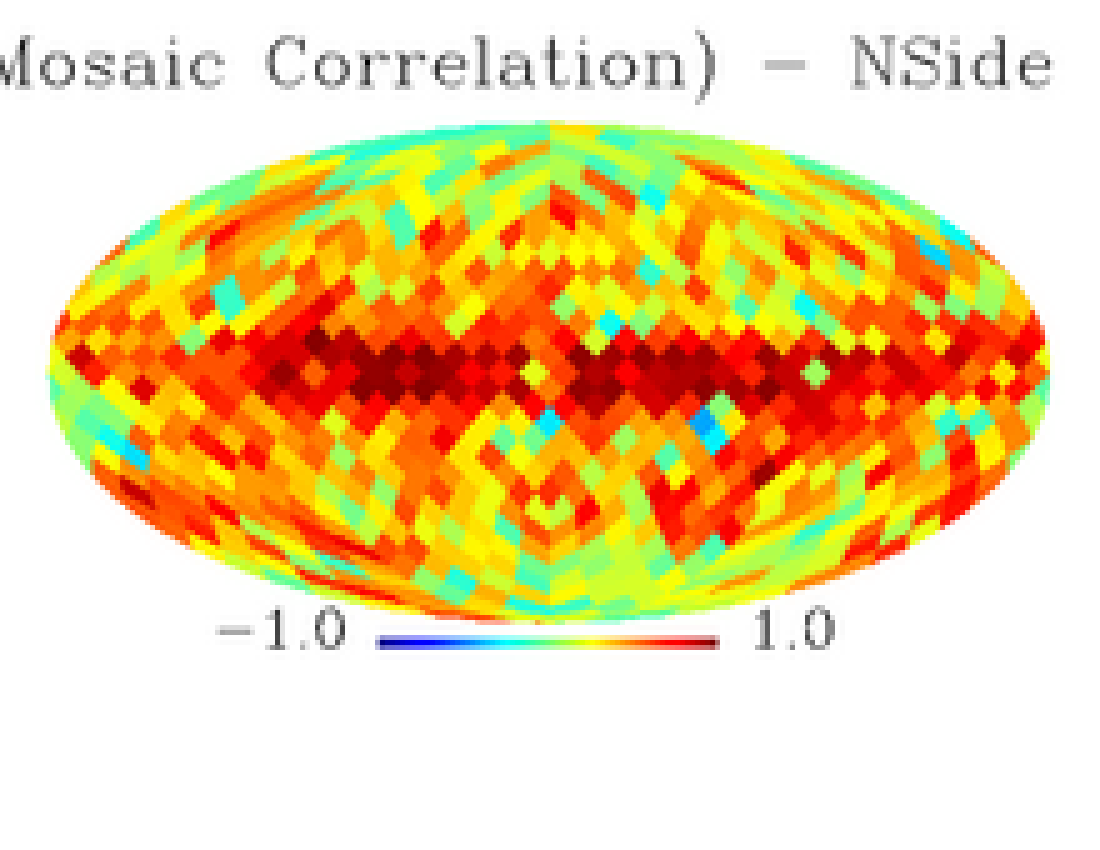}
\end{minipage}
\hfill
\begin{minipage}{0.24\textwidth}
\centering
\includegraphics[trim=0cm 1.5cm 0cm 1.2cm,clip=true,width=4cm]{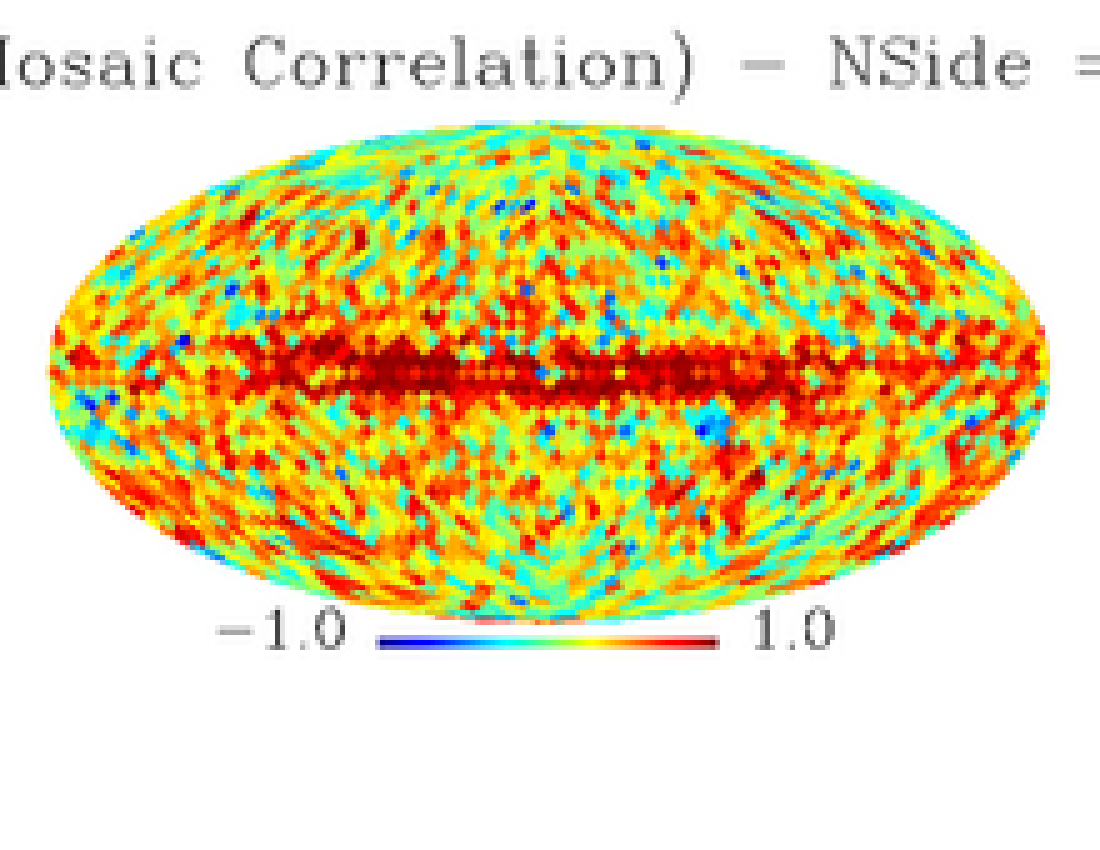}
\end{minipage}
\hfill
\centering
\begin{minipage}{0.24\textwidth}
\centering
\includegraphics[trim=0cm 1.5cm 0cm 1.2cm,clip=true,width=4cm]{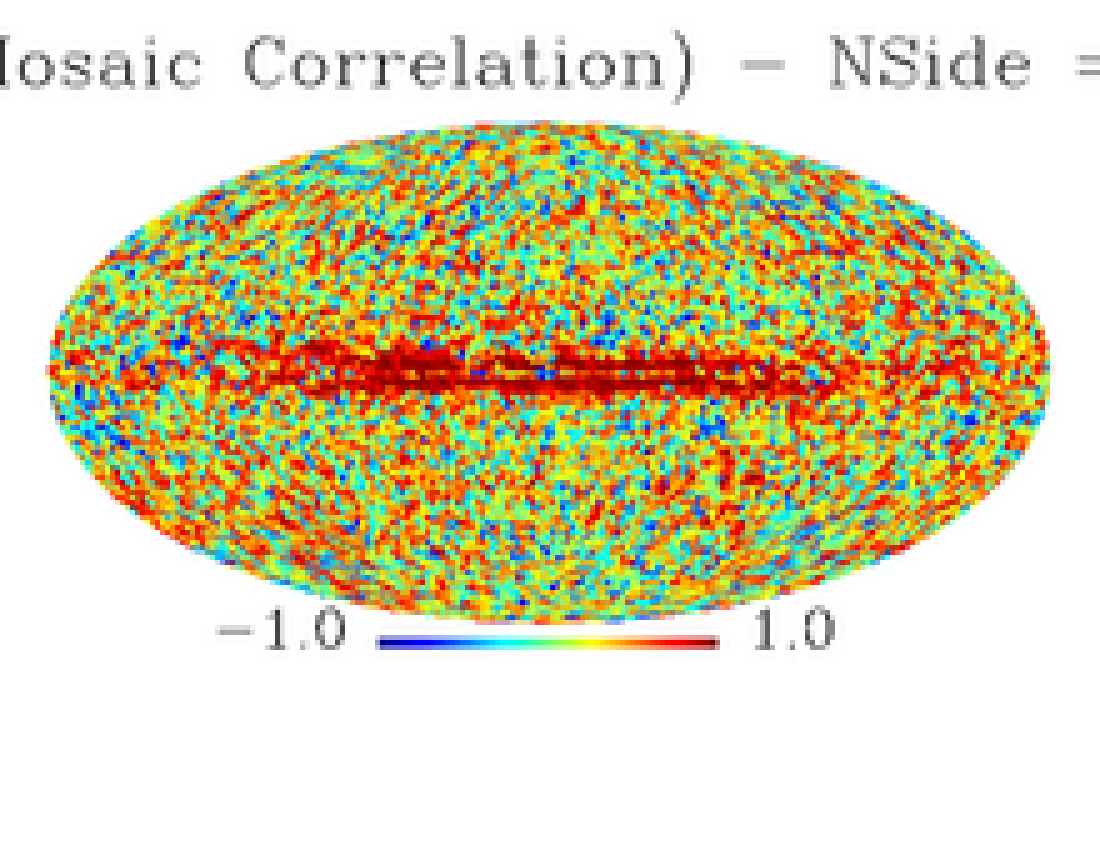}
\end{minipage}
\hfill
\begin{minipage}{0.24\textwidth}
\centering
\includegraphics[trim=0cm 1.5cm 0cm 1.2cm,clip=true,width=4cm]{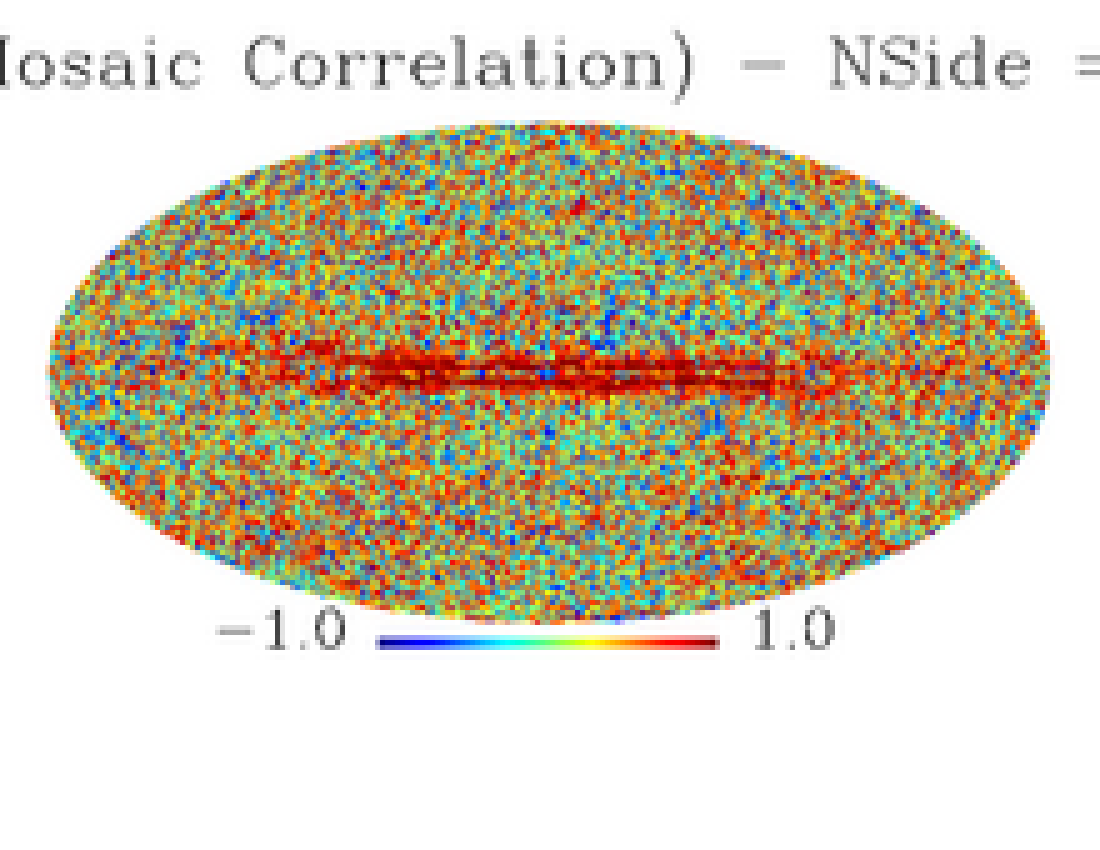}
\end{minipage}
\caption{Same as figure \ref{Acorrmaps1}, but here, weighted with weighting coefficients $w_i(n=2)$.}
\label{Acorrmaps2}
\end{figure}

\begin{figure}[H]
\centering
\begin{minipage}{0.24\textwidth}
\centering
\includegraphics[trim=0cm 1.5cm 0cm 1.2cm,clip=true,width=4cm]{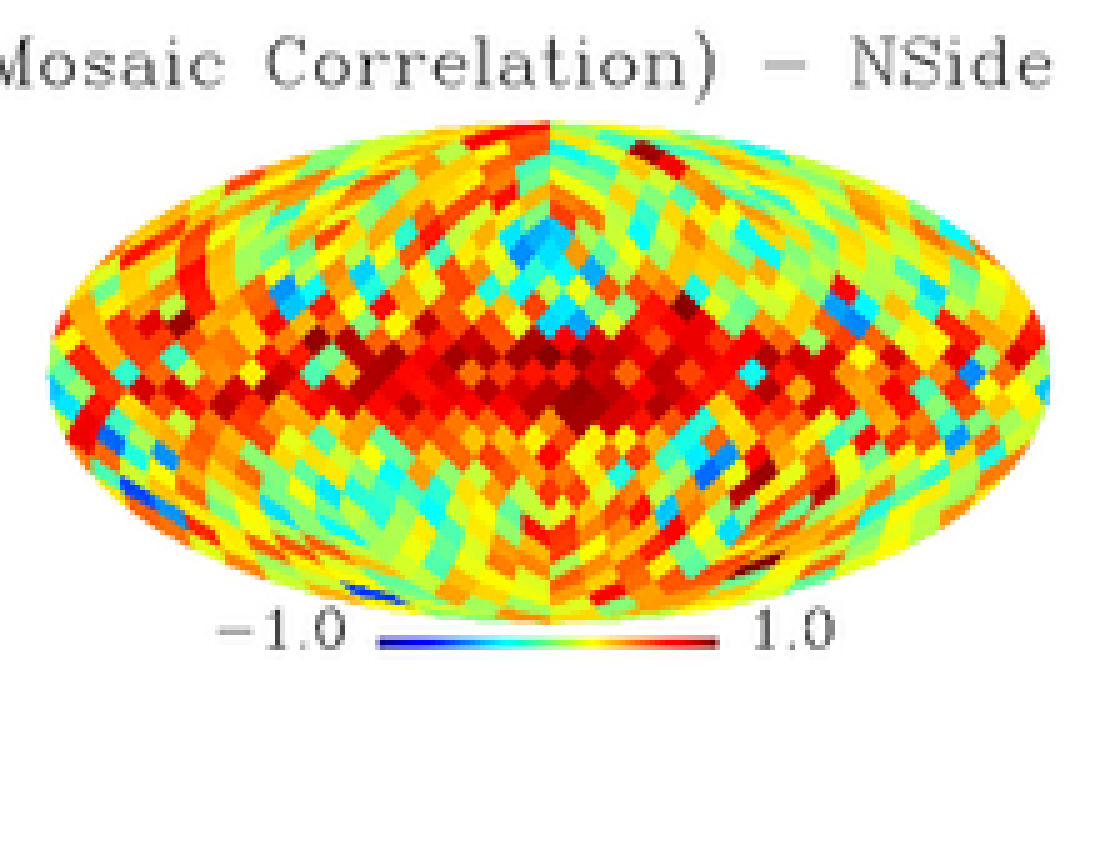}
\end{minipage}
\hfill
\begin{minipage}{0.24\textwidth}
\centering
\includegraphics[trim=0cm 1.5cm 0cm 1.2cm,clip=true,width=4cm]{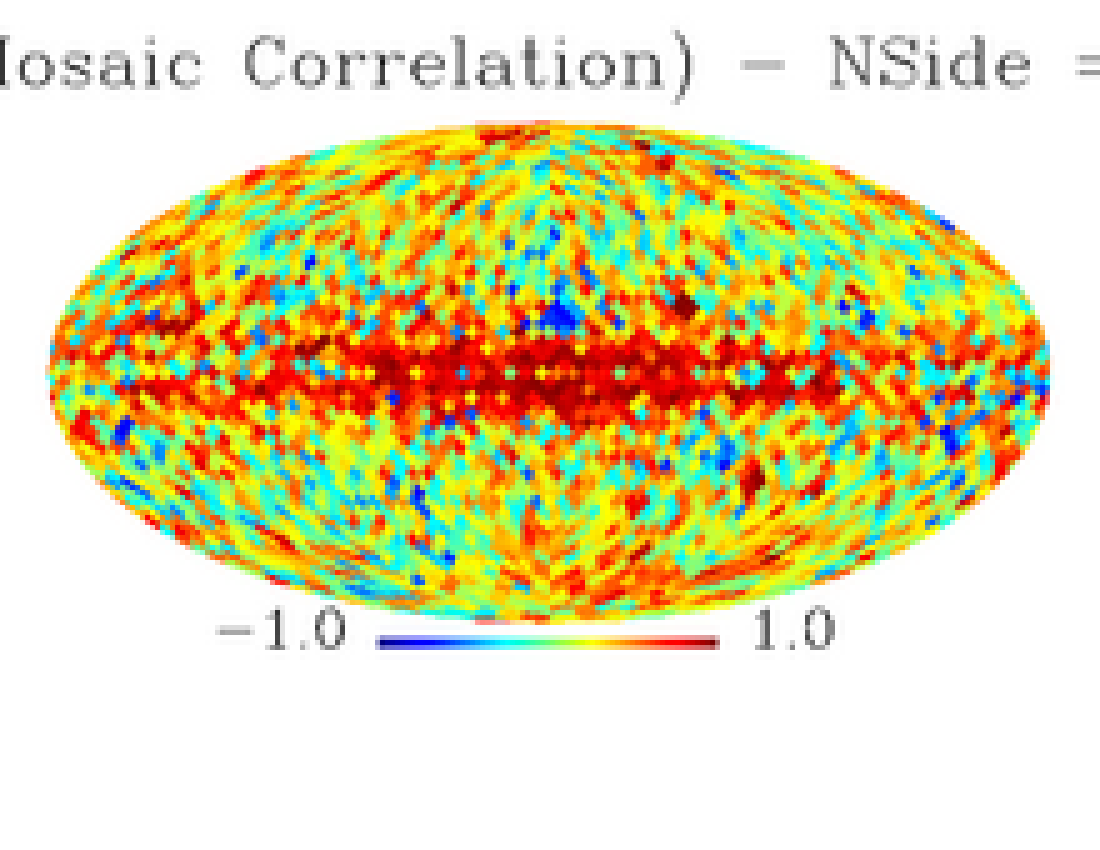}
\end{minipage}
\hfill
\centering
\begin{minipage}{0.24\textwidth}
\centering
\includegraphics[trim=0cm 1.5cm 0cm 1.2cm,clip=true,width=4cm]{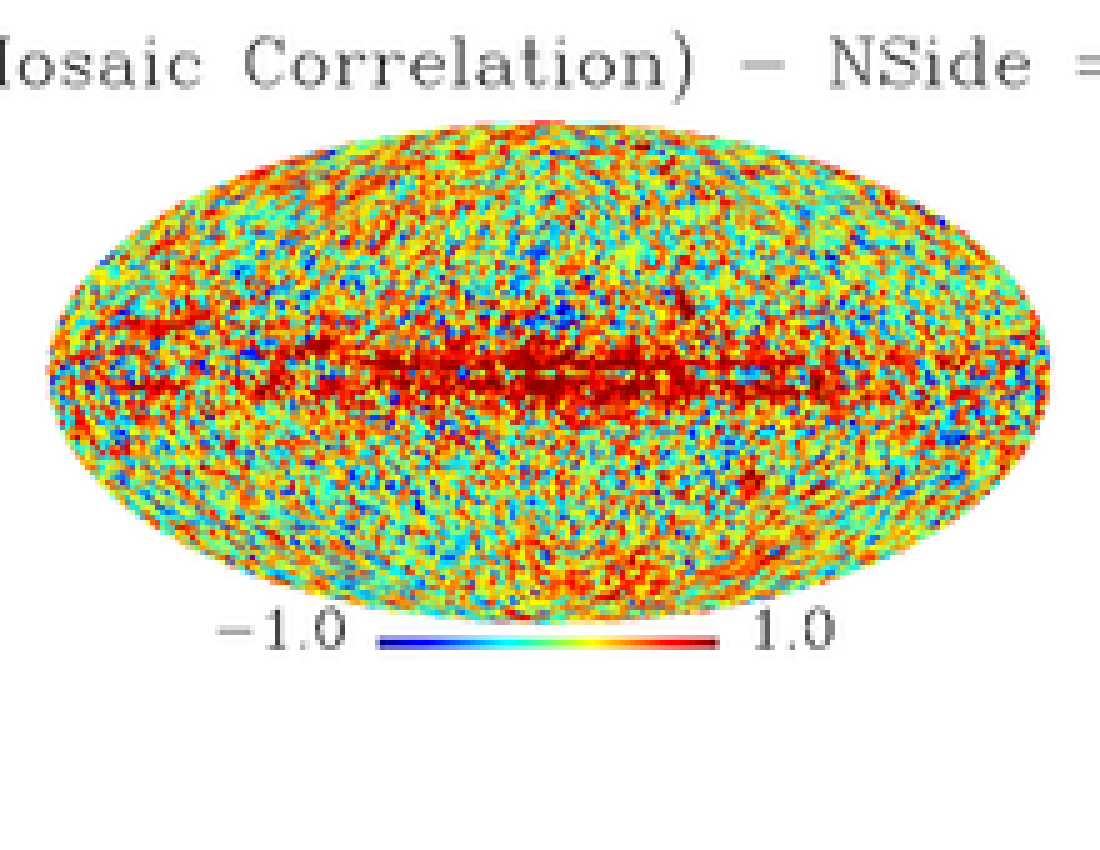}
\end{minipage}
\hfill
\begin{minipage}{0.24\textwidth}
\centering
\includegraphics[trim=0cm 1.5cm 0cm 1.2cm,clip=true,width=4cm]{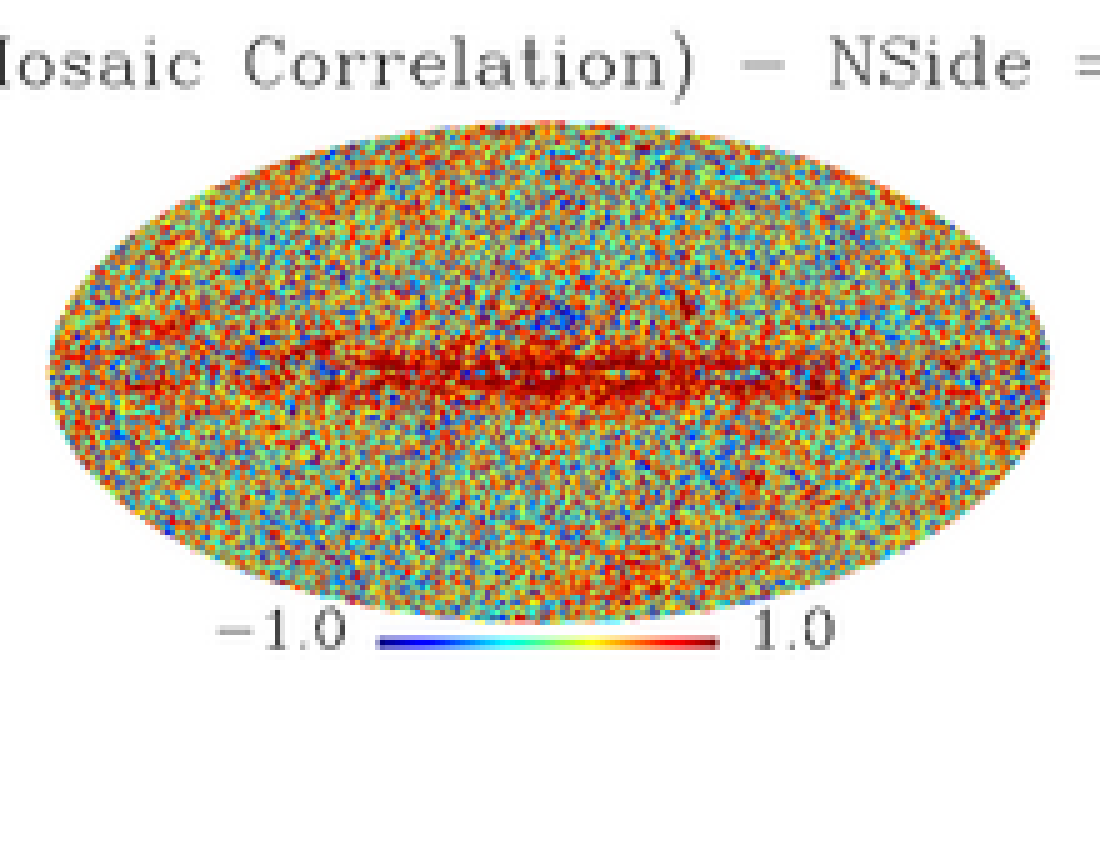}
\end{minipage}
\centering
\begin{minipage}{0.24\textwidth}
\centering
\includegraphics[trim=0cm 1.5cm 0cm 1.2cm,clip=true,width=4cm]{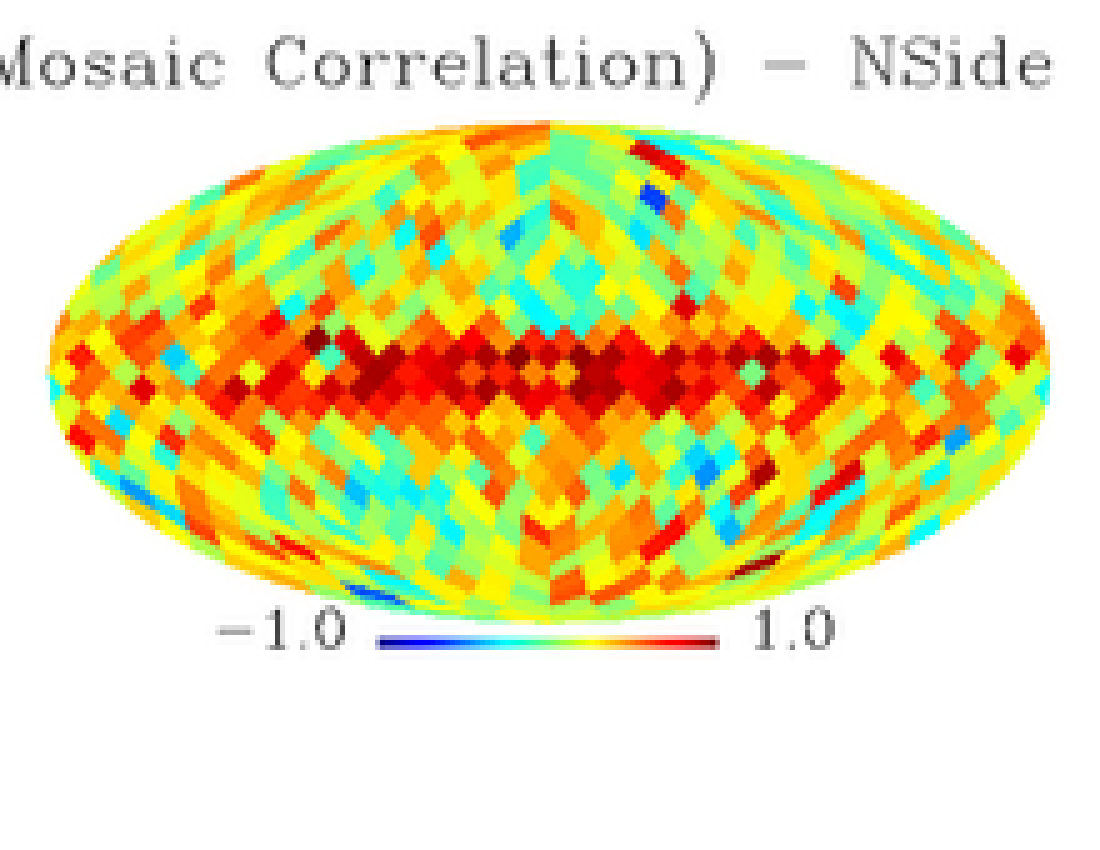}
\end{minipage}
\hfill
\begin{minipage}{0.24\textwidth}
\centering
\includegraphics[trim=0cm 1.5cm 0cm 1.2cm,clip=true,width=4cm]{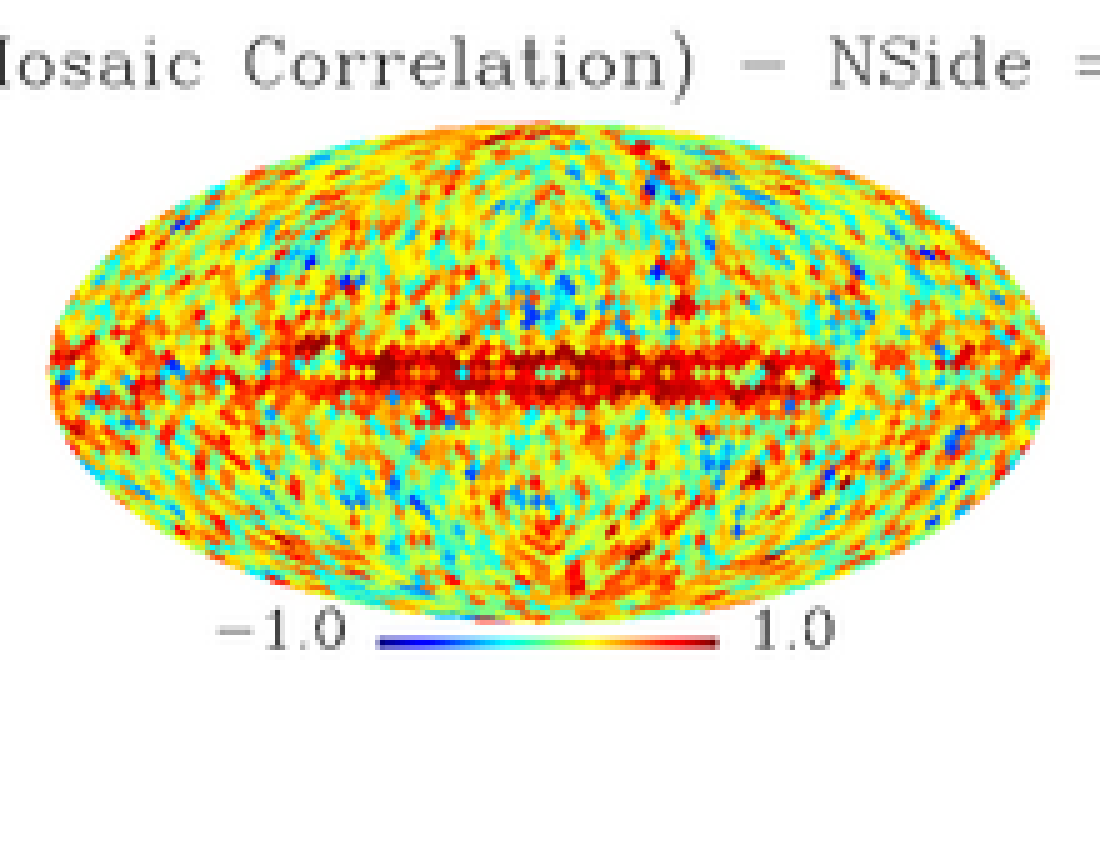}
\end{minipage}
\hfill
\centering
\begin{minipage}{0.24\textwidth}
\centering
\includegraphics[trim=0cm 1.5cm 0cm 1.2cm,clip=true,width=4cm]{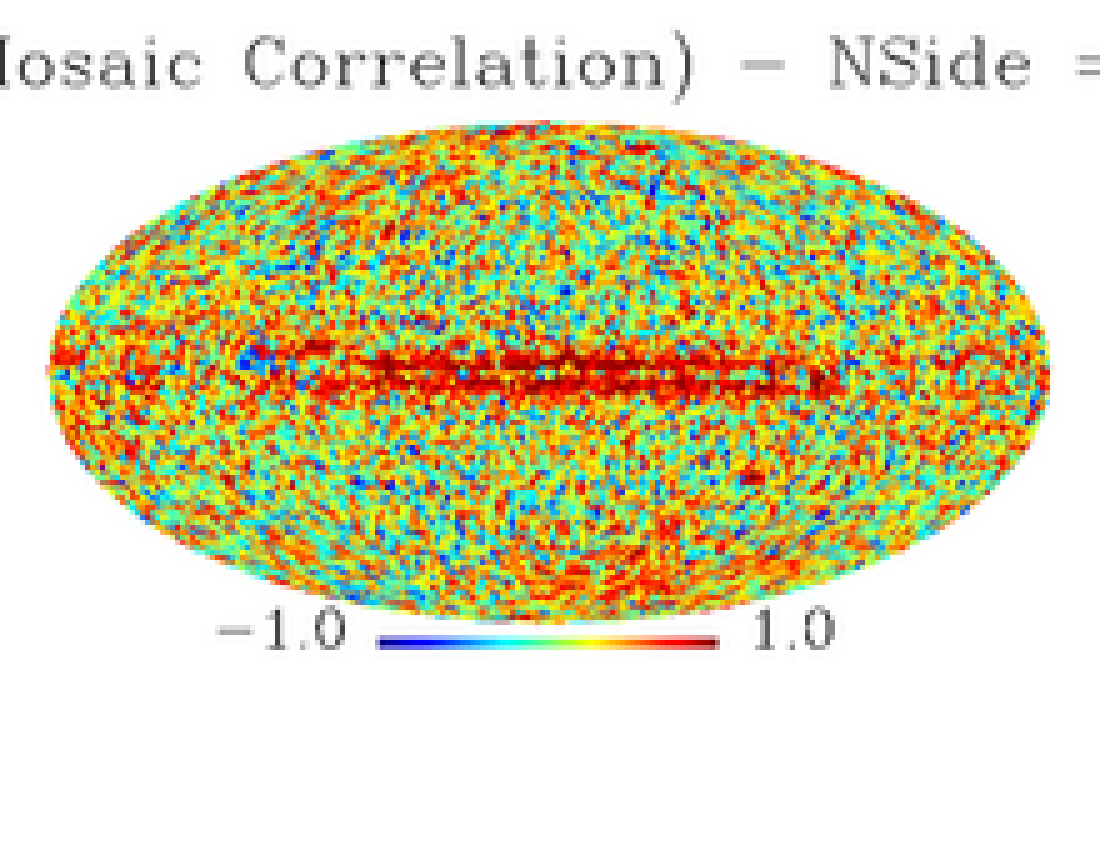}
\end{minipage}
\hfill
\begin{minipage}{0.24\textwidth}
\centering
\includegraphics[trim=0cm 1.5cm 0cm 1.2cm,clip=true,width=4cm]{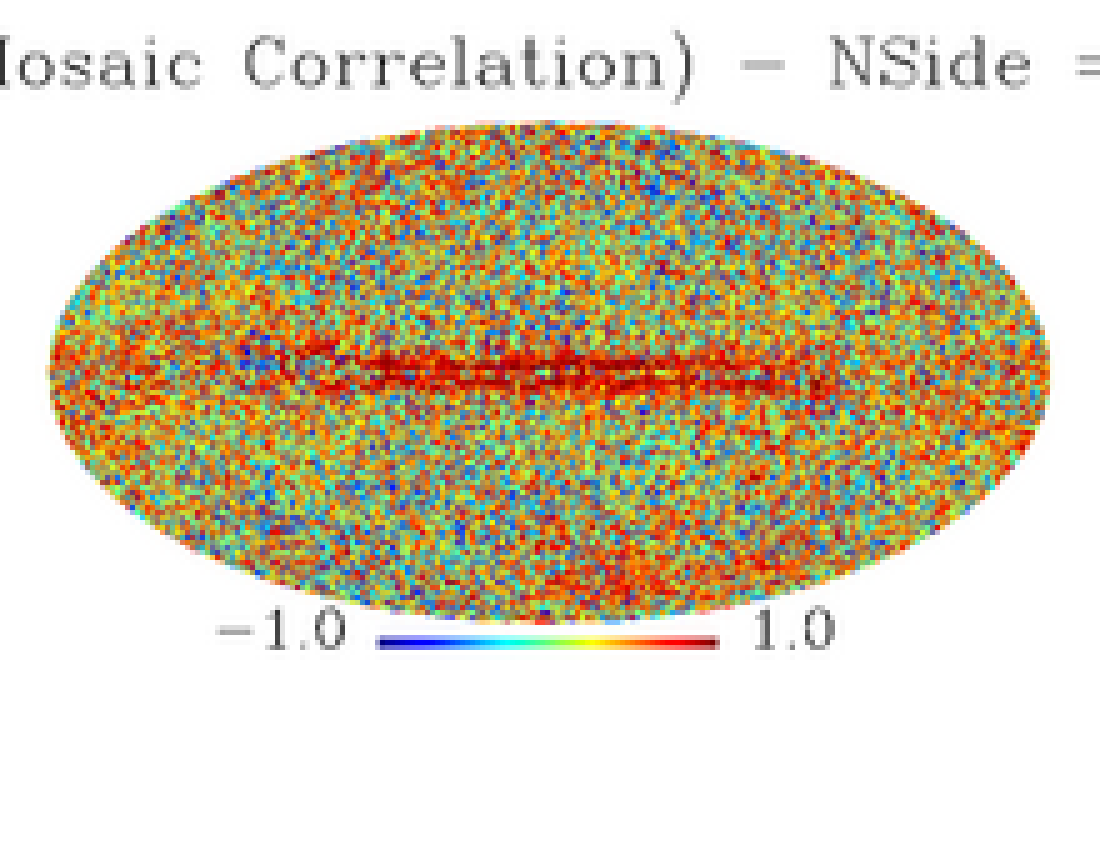}
\end{minipage}
\centering
\begin{minipage}{0.24\textwidth}
\centering
\includegraphics[trim=0cm 1.5cm 0cm 1.2cm,clip=true,width=4cm]{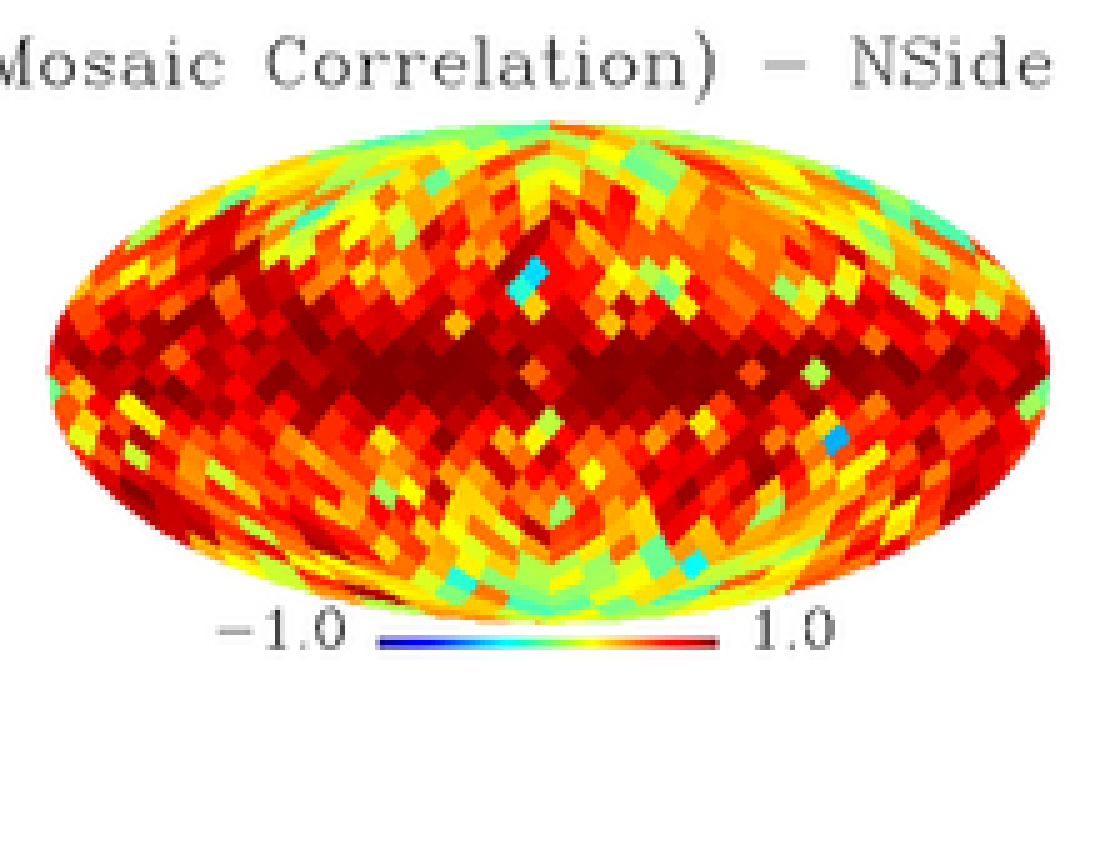}
\end{minipage}
\hfill
\begin{minipage}{0.24\textwidth}
\centering
\includegraphics[trim=0cm 1.5cm 0cm 1.2cm,clip=true,width=4cm]{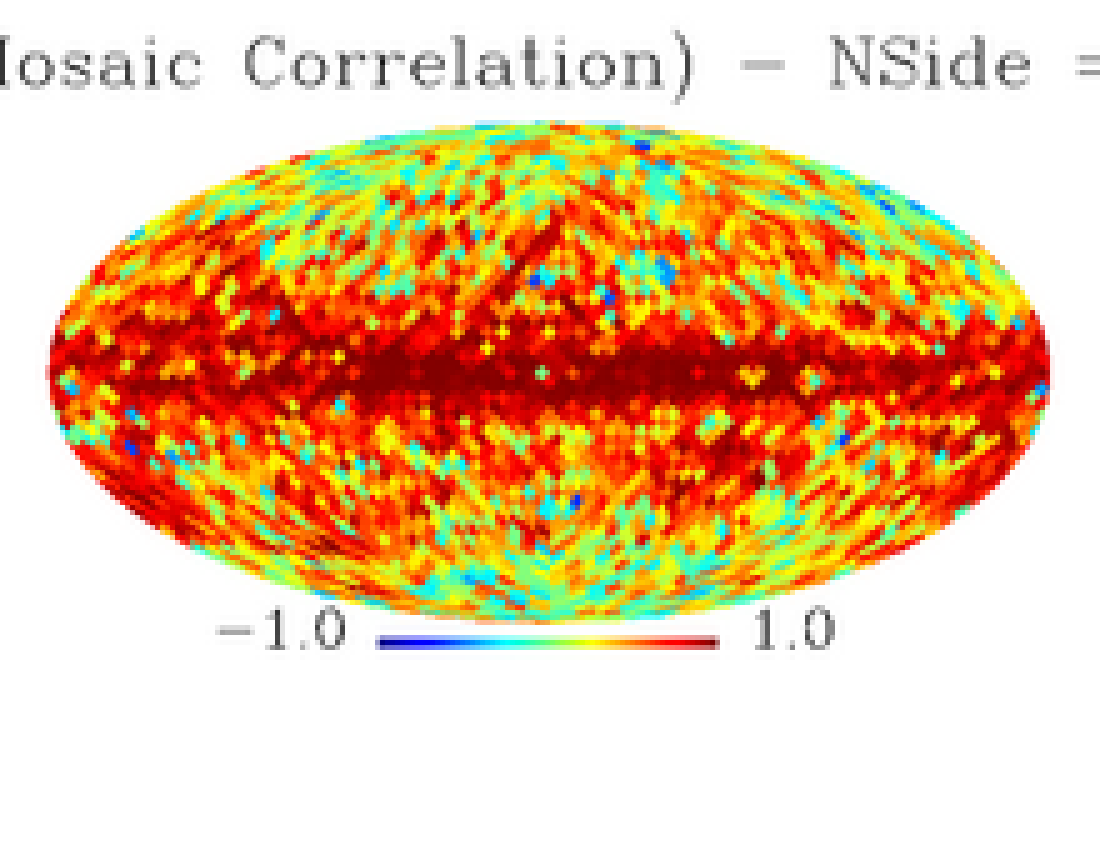}
\end{minipage}
\hfill
\centering
\begin{minipage}{0.24\textwidth}
\centering
\includegraphics[trim=0cm 1.5cm 0cm 1.2cm,clip=true,width=4cm]{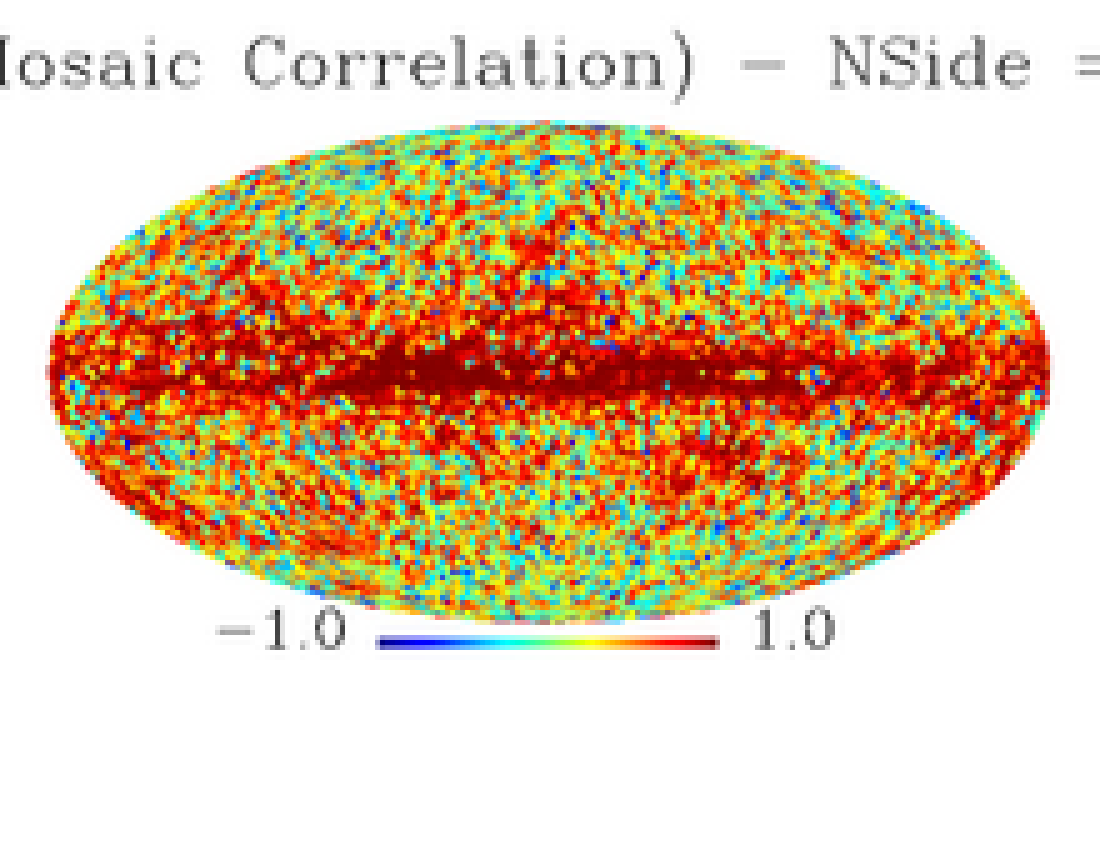}
\end{minipage}
\hfill
\begin{minipage}{0.24\textwidth}
\centering
\includegraphics[trim=0cm 1.5cm 0cm 1.2cm,clip=true,width=4cm]{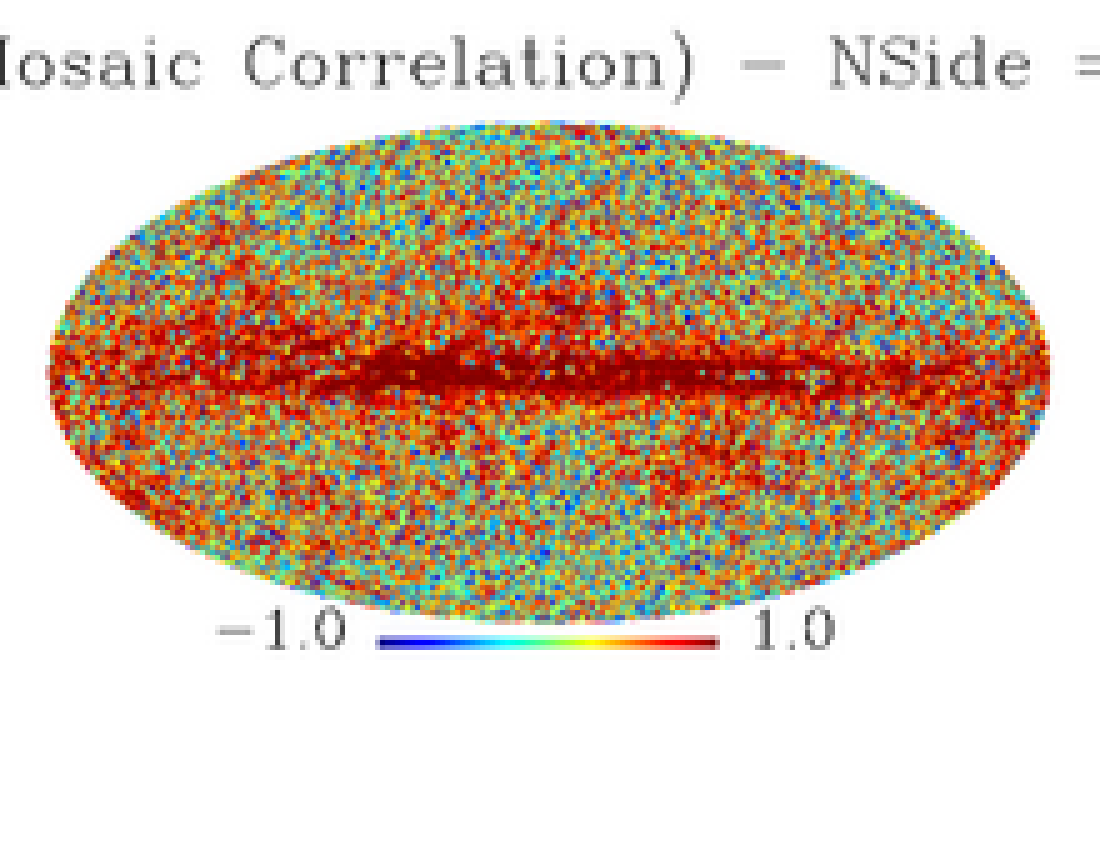}
\end{minipage}
\centering
\begin{minipage}{0.24\textwidth}
\centering
\includegraphics[trim=0cm 1.5cm 0cm 1.2cm,clip=true,width=4cm]{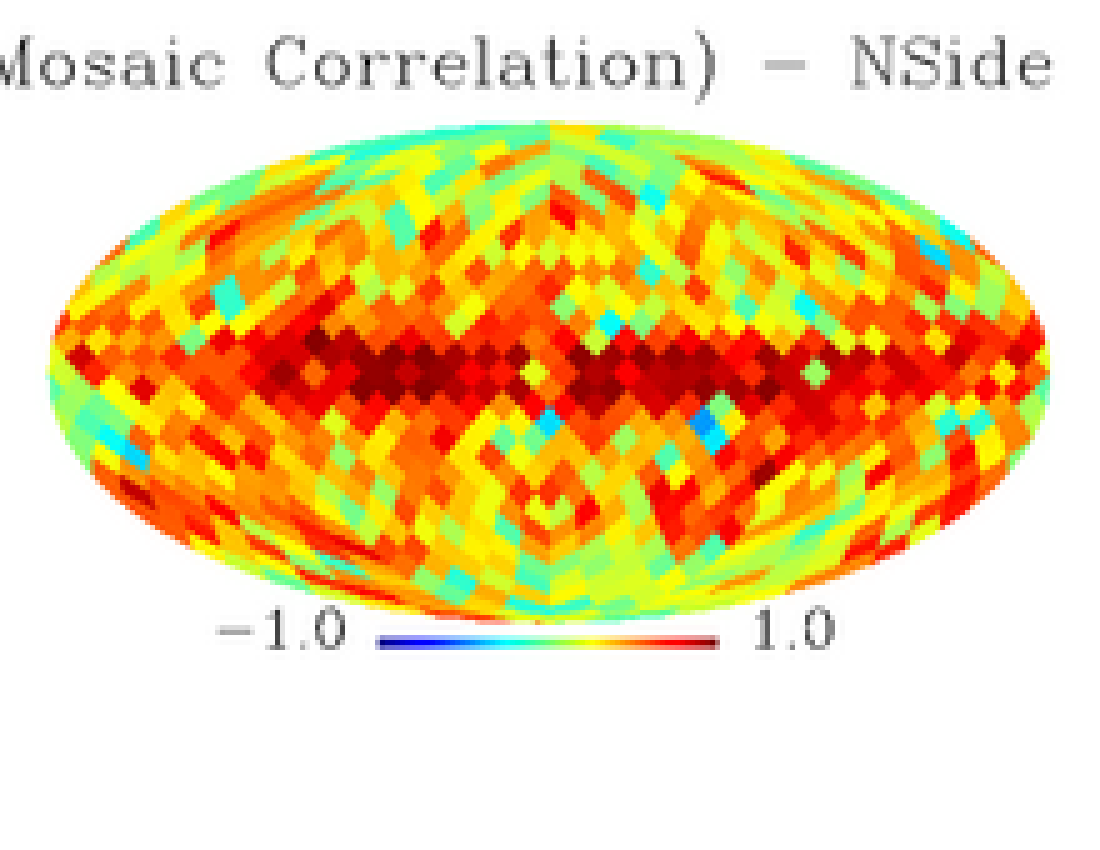}
\end{minipage}
\hfill
\begin{minipage}{0.24\textwidth}
\centering
\includegraphics[trim=0cm 1.5cm 0cm 1.2cm,clip=true,width=4cm]{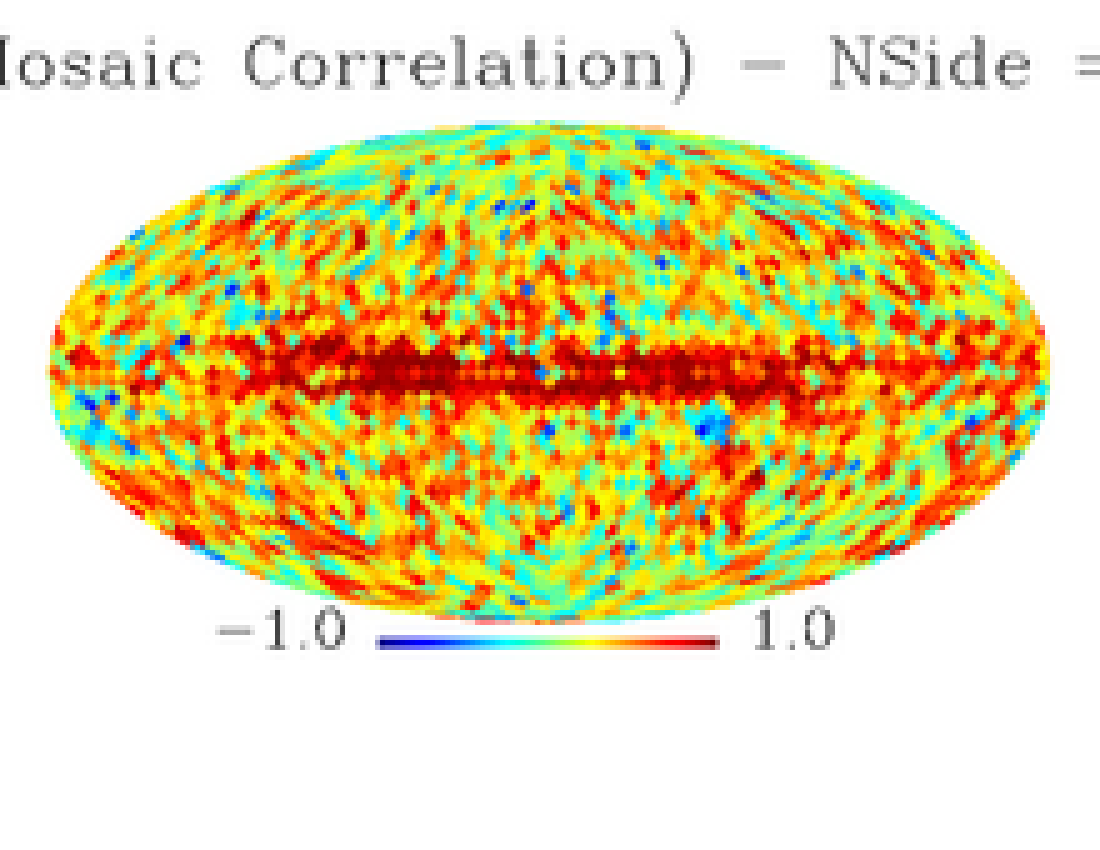}
\end{minipage}
\hfill
\centering
\begin{minipage}{0.24\textwidth}
\centering
\includegraphics[trim=0cm 1.5cm 0cm 1.2cm,clip=true,width=4cm]{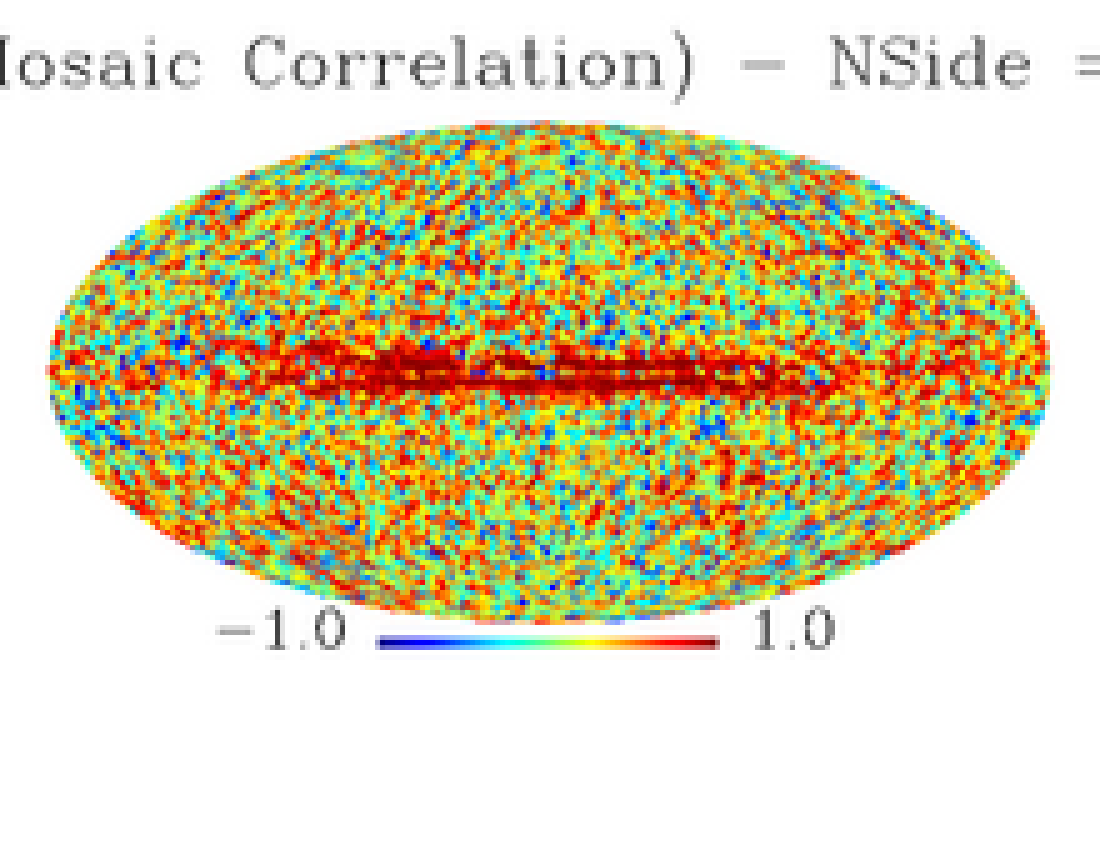}
\end{minipage}
\hfill
\begin{minipage}{0.24\textwidth}
\centering
\includegraphics[trim=0cm 1.5cm 0cm 1.2cm,clip=true,width=4cm]{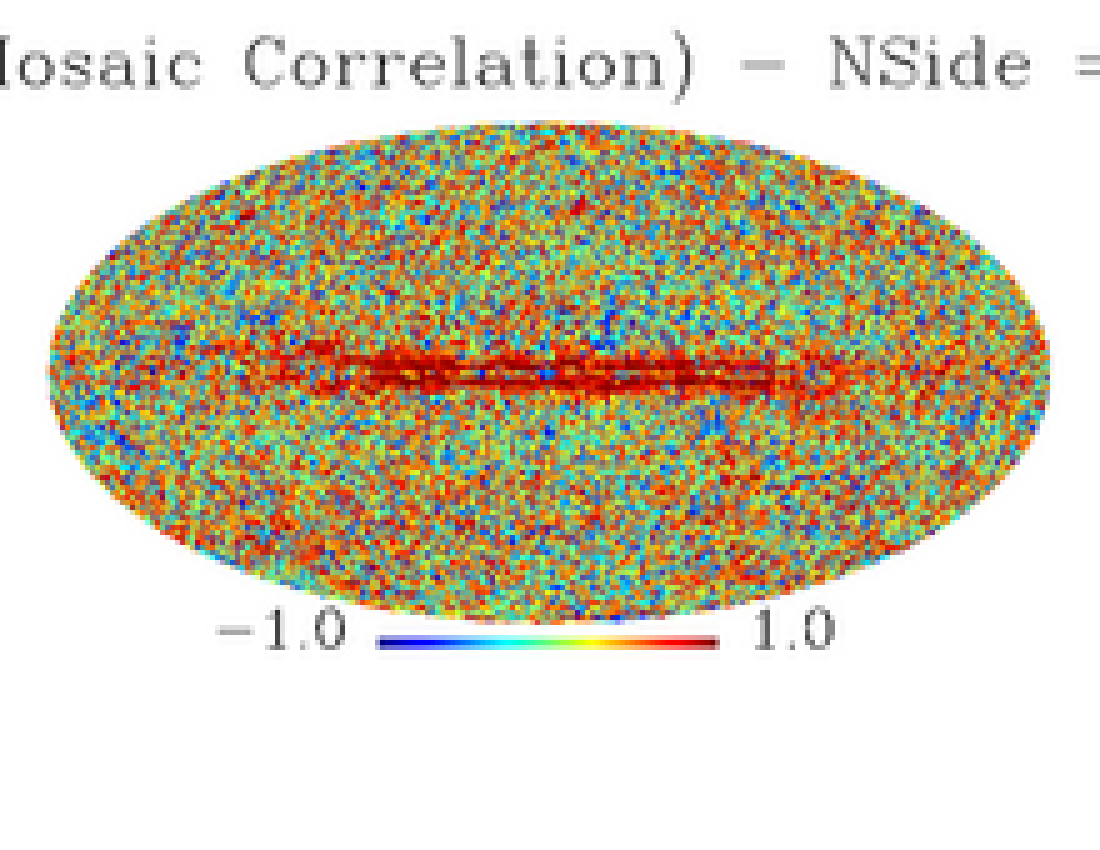}
\end{minipage}
\caption{Same as figure \ref{Acorrmaps1}, but here, weighted with weighting coefficients $w_i(n=10)$.}
\label{Acorrmaps10}
\end{figure}

\begin{figure}[H]
\begin{center}
\begin{minipage}{0.2\textwidth}
\centering
\includegraphics[trim=0cm 0cm 0cm 0cm,clip=true,width=4.1cm]{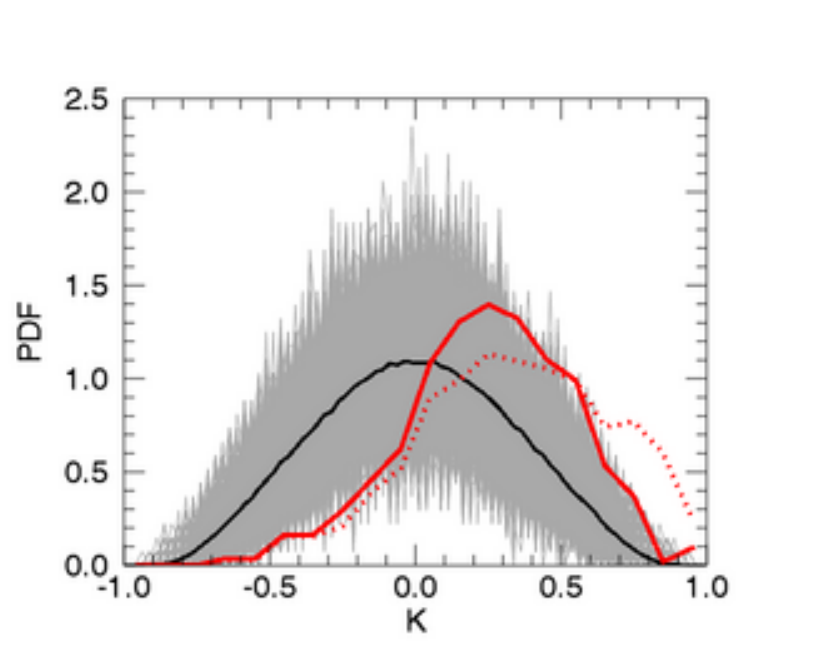}
\end{minipage}
\hspace{0.5cm}
\begin{minipage}{0.2\textwidth}
\centering
\includegraphics[trim=0cm 0cm 0cm 0cm,clip=true,width=4.1cm]{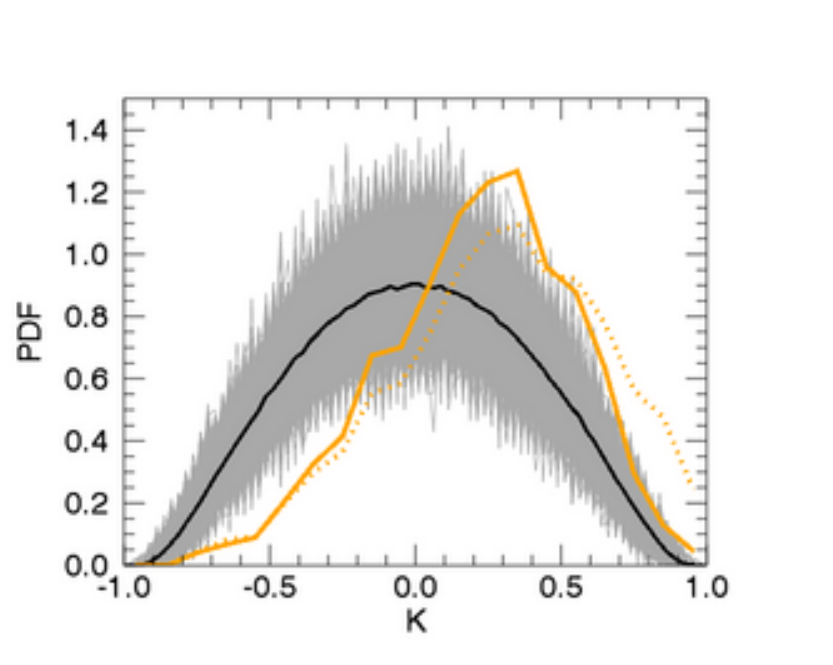}
\end{minipage}
\hspace{0.5cm}
\begin{minipage}{0.2\textwidth}
\centering
\includegraphics[trim=0cm 0cm 0cm 0cm,clip=true,width=4.1cm]{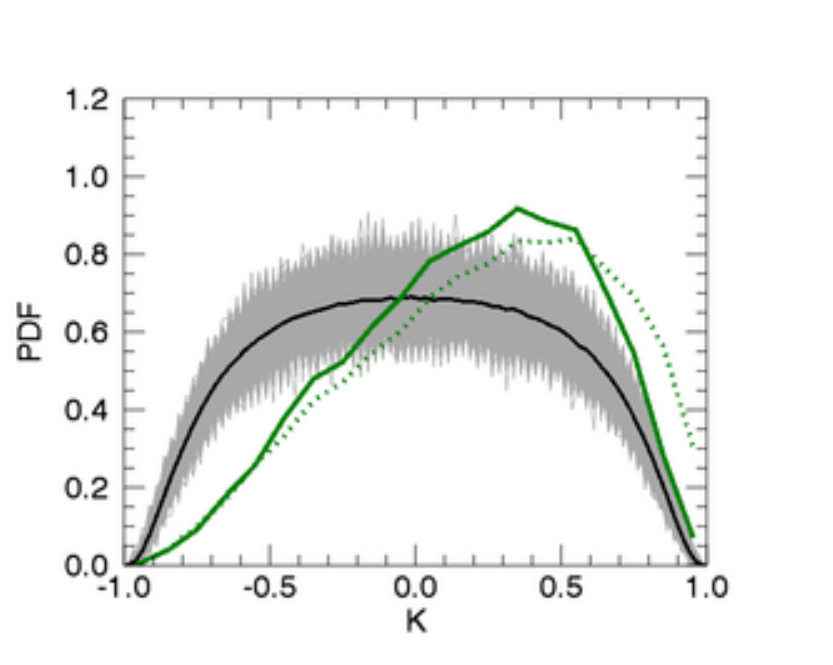}
\end{minipage}
\hspace{0.5cm}
\begin{minipage}{0.2\textwidth}
\centering
\includegraphics[trim=0cm 0cm 0cm 0cm,clip=true,width=4.1cm]{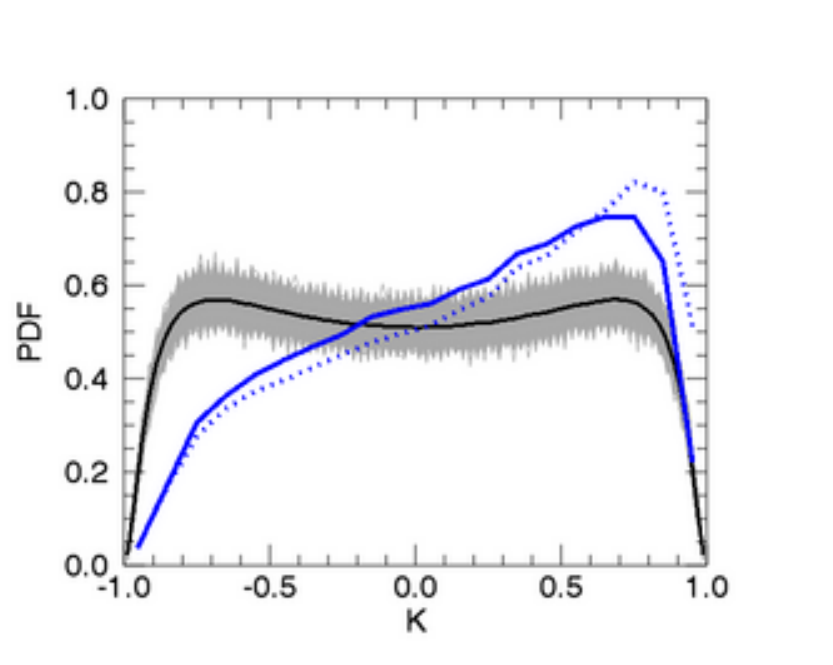}
\end{minipage}
\vspace{-0.2cm}
\begin{minipage}{0.2\textwidth}
\centering
\includegraphics[trim=0cm 0cm 0cm 0cm,clip=true,width=4.1cm]{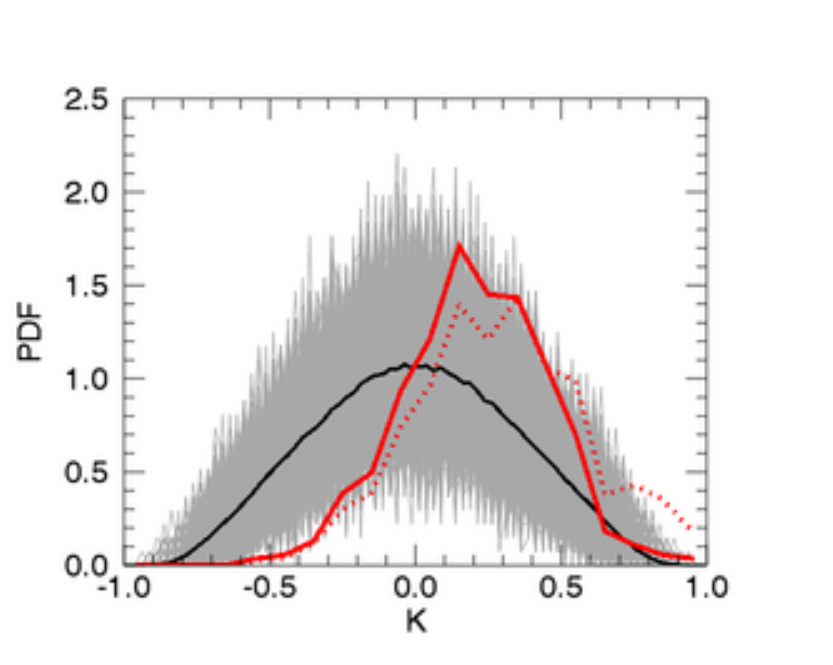}
\end{minipage}
\hspace{0.5cm}
\begin{minipage}{0.2\textwidth}
\centering
\includegraphics[trim=0cm 0cm 0cm 0cm,clip=true,width=4.1cm]{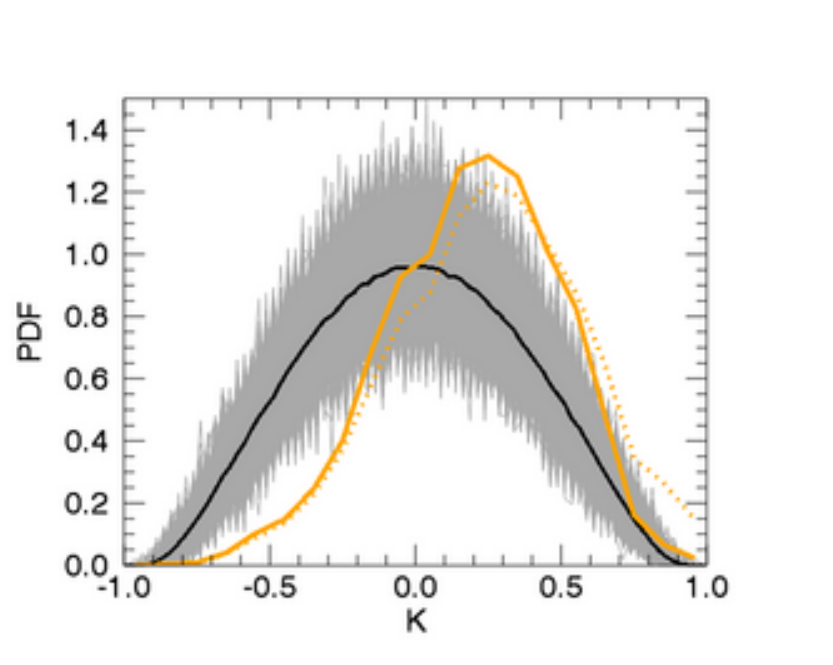}
\end{minipage}
\hspace{0.5cm}
\begin{minipage}{0.2\textwidth}
\centering
\includegraphics[trim=0cm 0cm 0cm 0cm,clip=true,width=4.1cm]{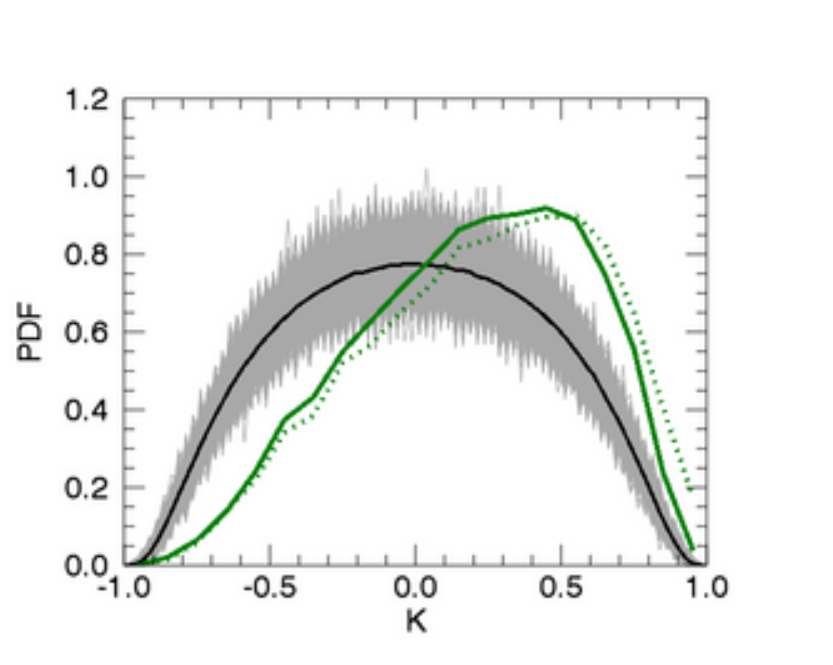}
\end{minipage}
\hspace{0.5cm}
\begin{minipage}{0.2\textwidth}
\centering
\includegraphics[trim=0cm 0cm 0cm 0cm,clip=true,width=4.1cm]{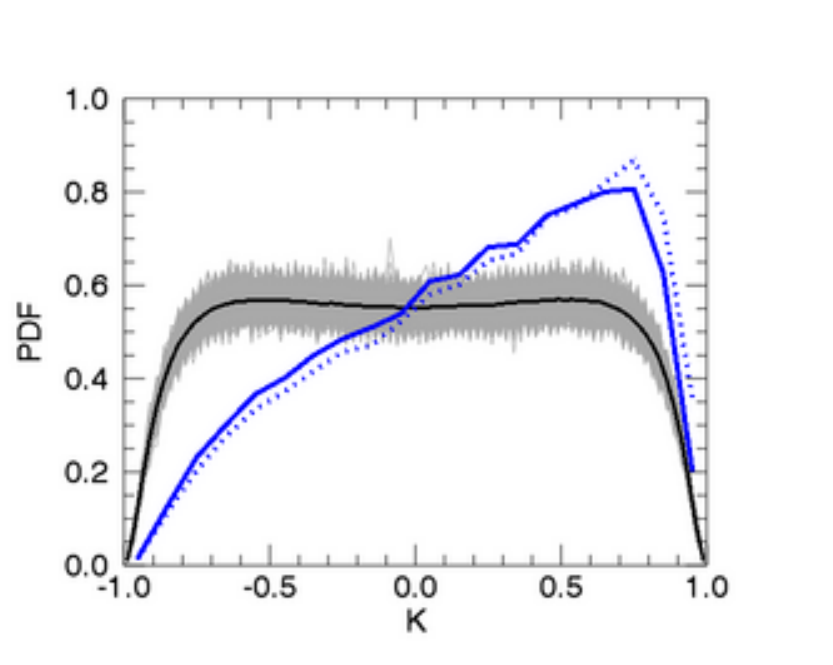}
\end{minipage}
\vspace{-0.2cm}
\begin{minipage}{0.2\textwidth}
\centering
\includegraphics[trim=0cm 0cm 0cm 0cm,clip=true,width=4.1cm]{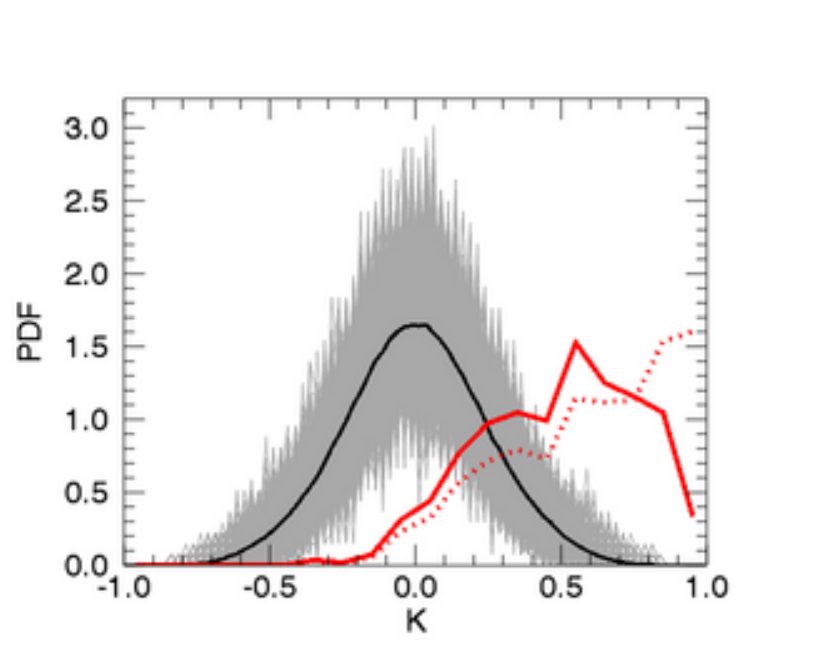}
\end{minipage}
\hspace{0.5cm}
\begin{minipage}{0.2\textwidth}
\centering
\includegraphics[trim=0cm 0cm 0cm 0cm,clip=true,width=4.1cm]{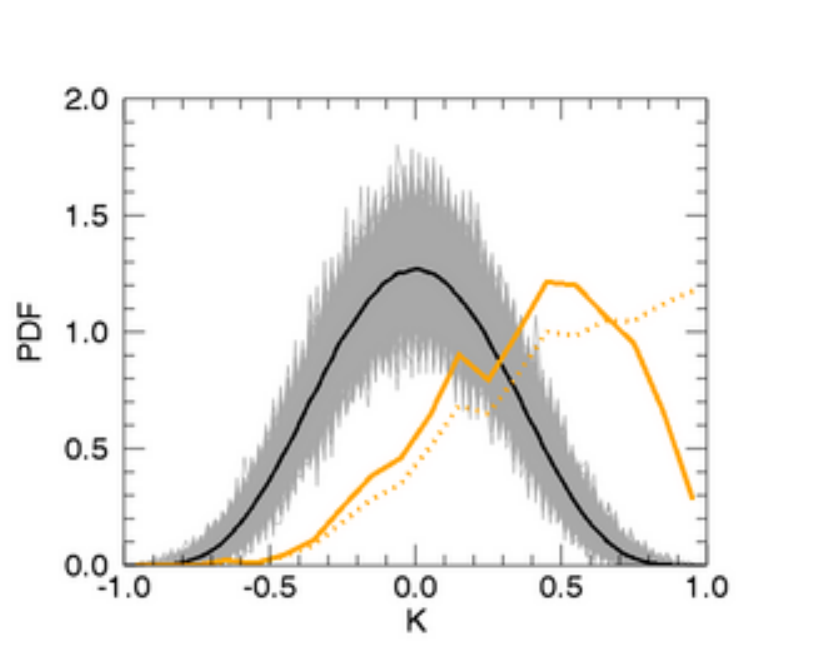}
\end{minipage}
\hspace{0.5cm}
\begin{minipage}{0.2\textwidth}
\centering
\includegraphics[trim=0cm 0cm 0cm 0cm,clip=true,width=4.1cm]{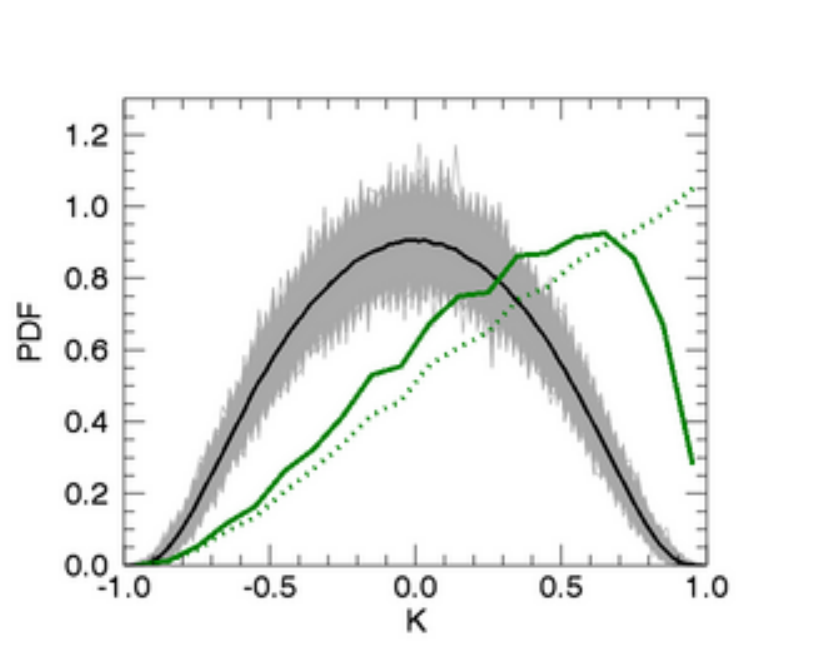}
\end{minipage}
\hspace{0.5cm}
\begin{minipage}{0.2\textwidth}
\centering
\includegraphics[trim=0cm 0cm 0cm 0cm,clip=true,width=4.1cm]{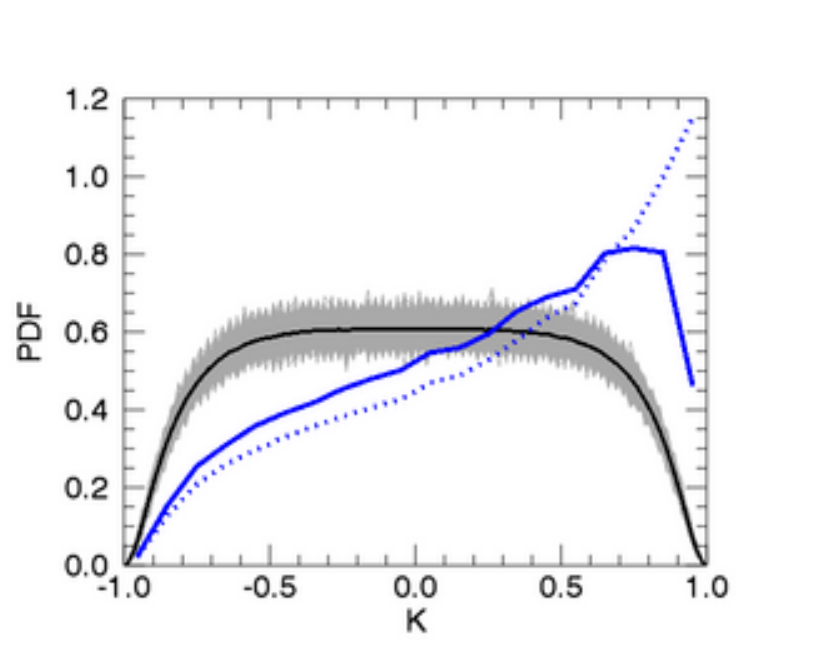}
\end{minipage}
\vspace{-0.2cm}
\begin{minipage}{0.2\textwidth}
\centering
\includegraphics[trim=0cm 0cm 0cm 0cm,clip=true,width=4.1cm]{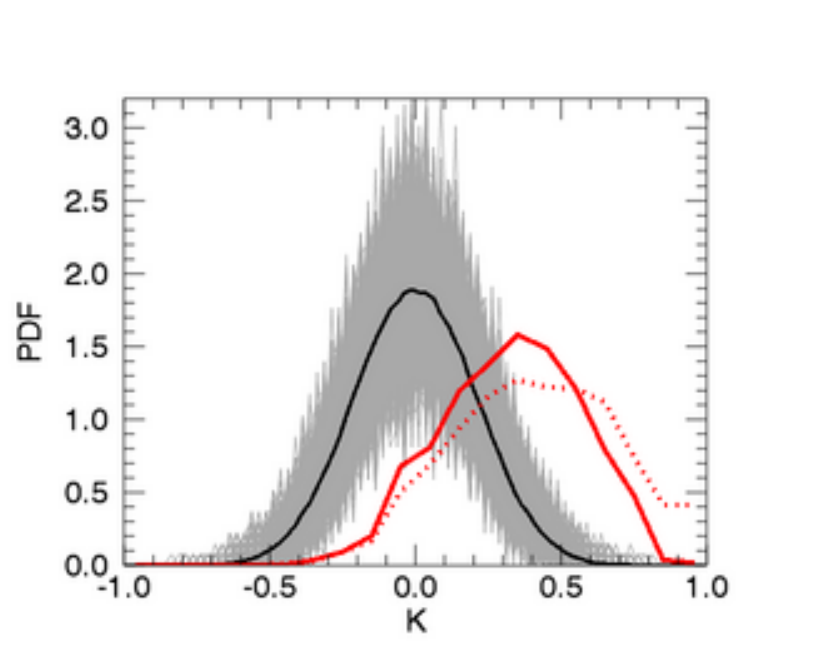}
\end{minipage}
\hspace{0.5cm}
\begin{minipage}{0.2\textwidth}
\centering
\includegraphics[trim=0cm 0cm 0cm 0cm,clip=true,width=4.1cm]{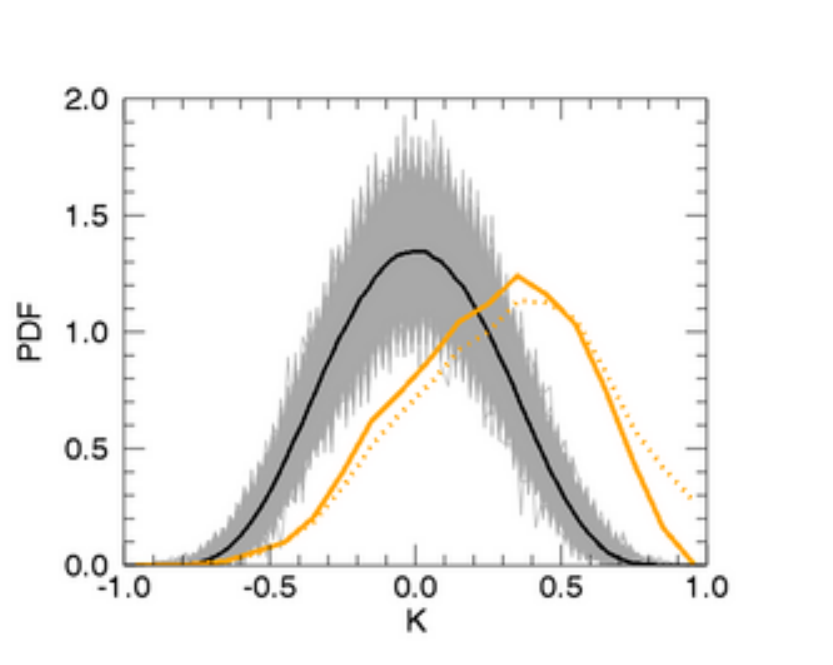}
\end{minipage}
\hspace{0.5cm}
\begin{minipage}{0.2\textwidth}
\centering
\includegraphics[trim=0cm 0cm 0cm 0cm,clip=true,width=4.1cm]{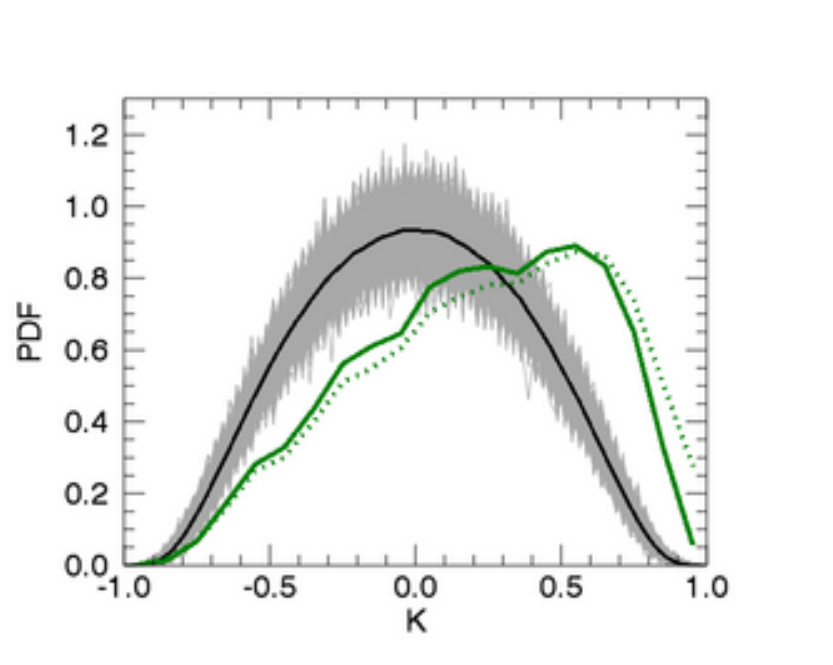}
\end{minipage}
\hspace{0.5cm}
\begin{minipage}{0.2\textwidth}
\centering
\includegraphics[trim=0cm 0cm 0cm 0cm,clip=true,width=4.1cm]{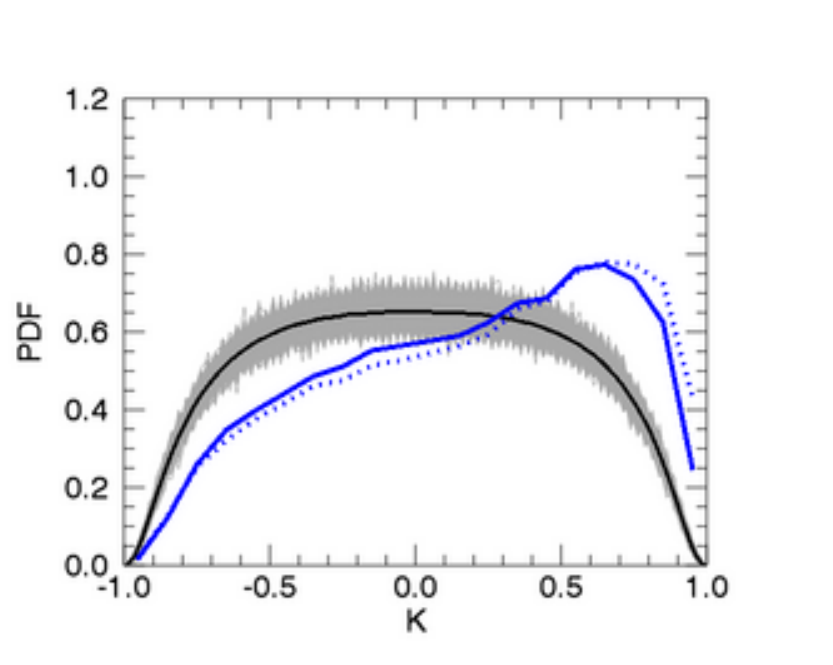}
\end{minipage}
\end{center}
\caption{PDFs of weighted correlation coefficients K (with $w_i(n=1)$) for different combinations
of foregrounds and each of the $\Omega$ sizes.
From top to bottom: AME1 and Synchrotron radiation, AME2 and Synchrotron radiation, AME1 and Thermal Dust emission, AME2 and Thermal Dust emission. 
Colored lines give distributions of K's from the actual foregrounds, solid for masked sky, dotted for unmasked sky. The gray lines
show all distributions of K's resulting from random simulations of AME1 or AME2 correlated with the respective foregrounds. 
Black lines are averages over all 1000 random realizations. Simulations are shown for the masked sky.
From left to right: $\Omega$ contains 1024, 256, 64 and 16 pixels.}
\label{Adist1}
\end{figure}

\begin{figure}[H]
\begin{center}
\begin{minipage}{0.2\textwidth}
\centering
\includegraphics[trim=0cm 0cm 0cm 0cm,clip=true,width=4.1cm]{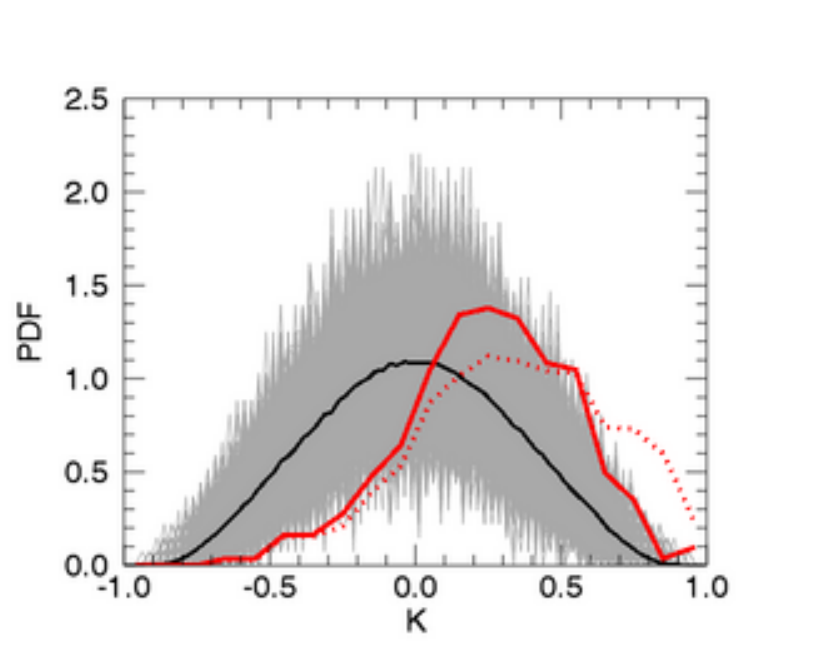}
\end{minipage}
\hspace{0.5cm}
\begin{minipage}{0.2\textwidth}
\centering
\includegraphics[trim=0cm 0cm 0cm 0cm,clip=true,width=4.1cm]{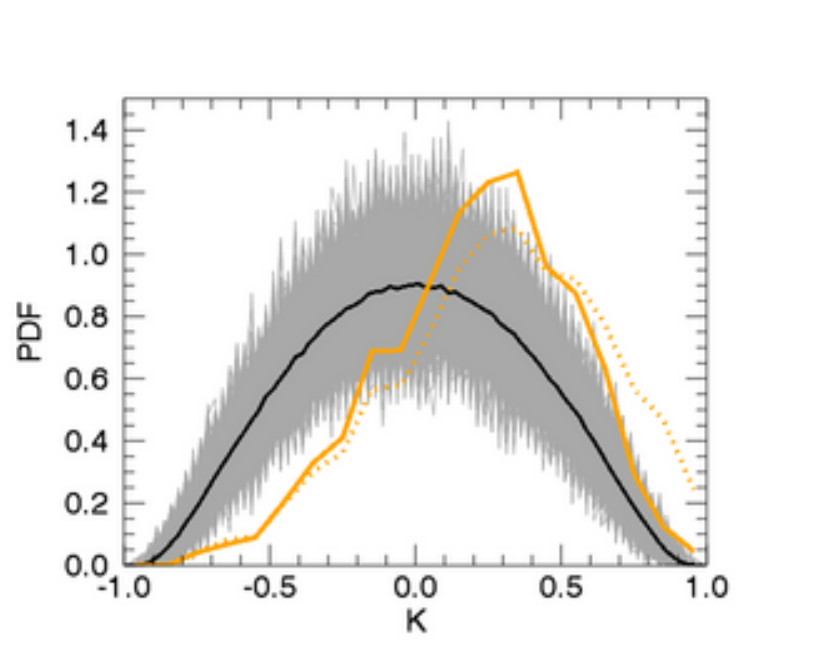}
\end{minipage}
\hspace{0.5cm}
\begin{minipage}{0.2\textwidth}
\centering
\includegraphics[trim=0cm 0cm 0cm 0cm,clip=true,width=4.1cm]{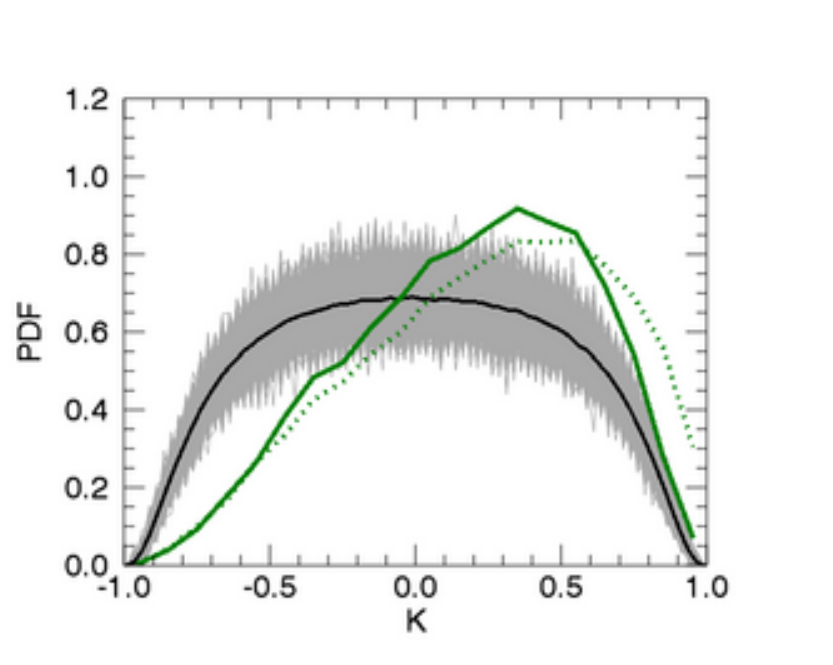}
\end{minipage}
\hspace{0.5cm}
\begin{minipage}{0.2\textwidth}
\centering
\includegraphics[trim=0cm 0cm 0cm 0cm,clip=true,width=4.1cm]{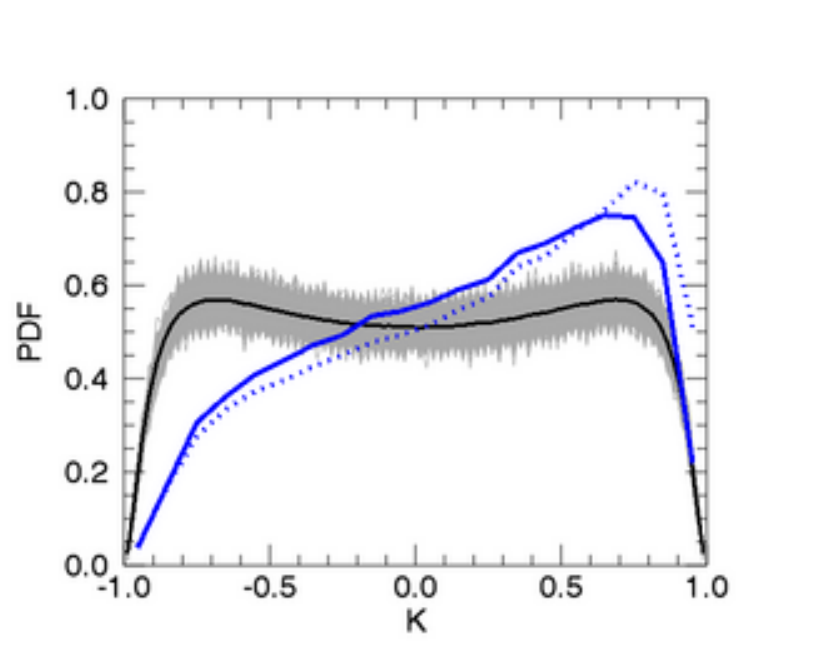}
\end{minipage}
\vspace{-0.2cm}
\begin{minipage}{0.2\textwidth}
\centering
\includegraphics[trim=0cm 0cm 0cm 0cm,clip=true,width=4.1cm]{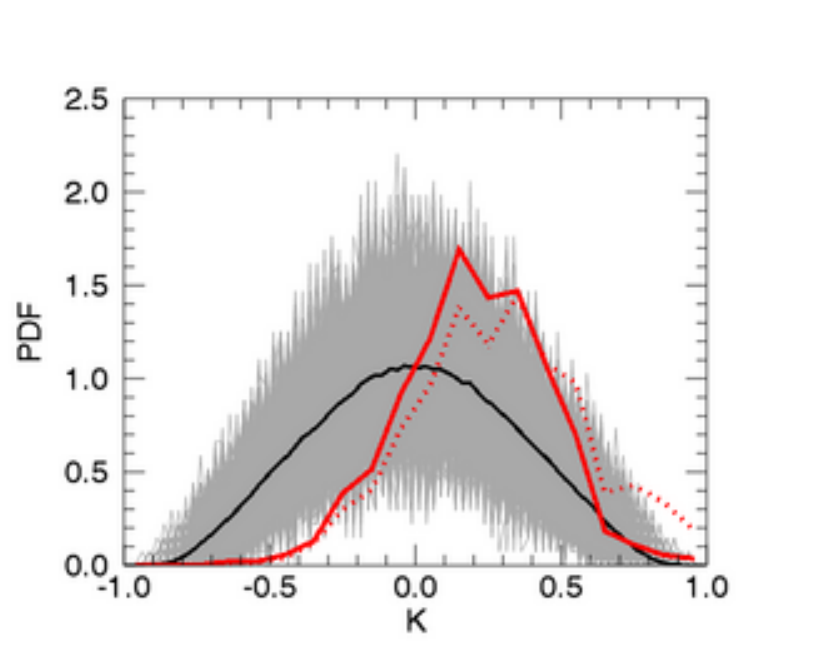}
\end{minipage}
\hspace{0.5cm}
\begin{minipage}{0.2\textwidth}
\centering
\includegraphics[trim=0cm 0cm 0cm 0cm,clip=true,width=4.1cm]{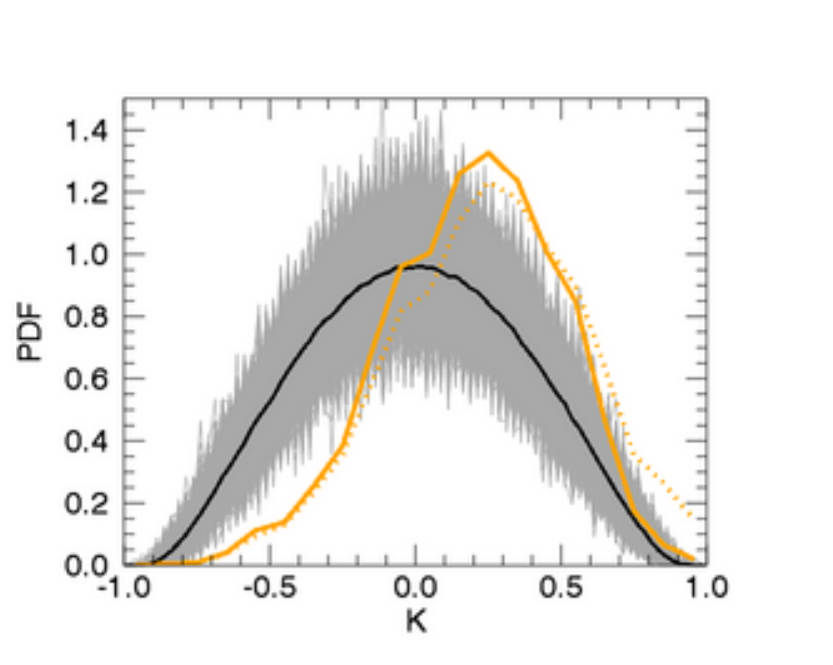}
\end{minipage}
\hspace{0.5cm}
\begin{minipage}{0.2\textwidth}
\centering
\includegraphics[trim=0cm 0cm 0cm 0cm,clip=true,width=4.1cm]{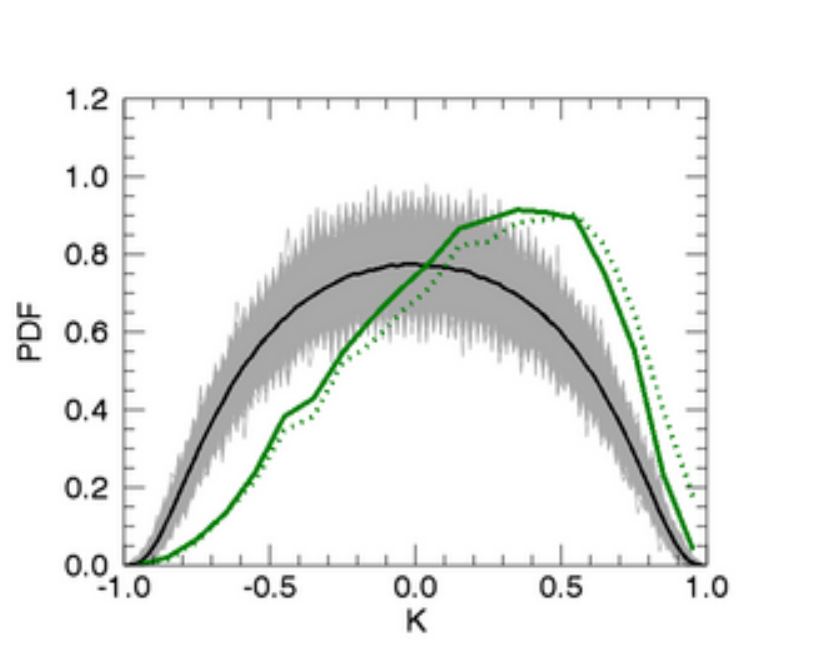}
\end{minipage}
\hspace{0.5cm}
\begin{minipage}{0.2\textwidth}
\centering
\includegraphics[trim=0cm 0cm 0cm 0cm,clip=true,width=4.1cm]{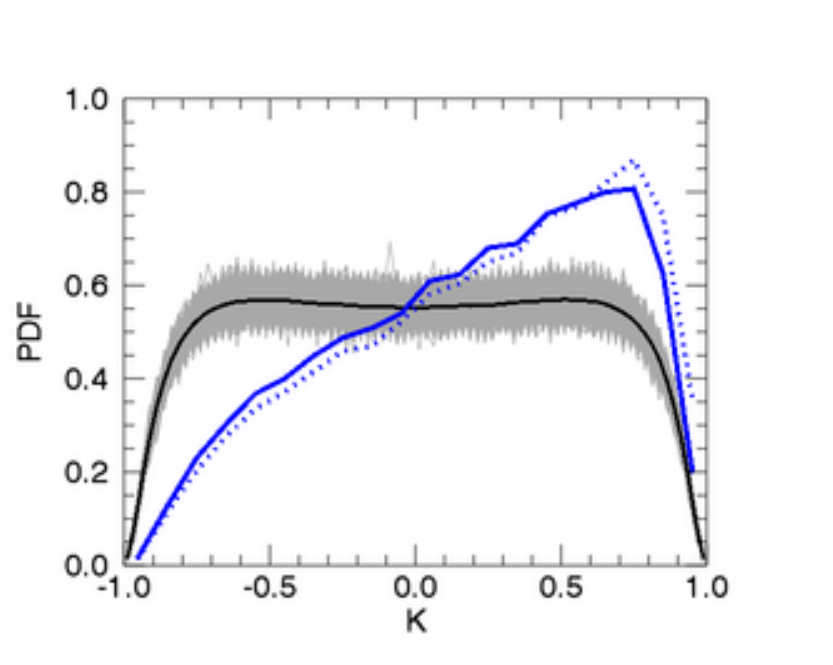}
\end{minipage}
\vspace{-0.2cm}
\begin{minipage}{0.2\textwidth}
\centering
\includegraphics[trim=0cm 0cm 0cm 0cm,clip=true,width=4.1cm]{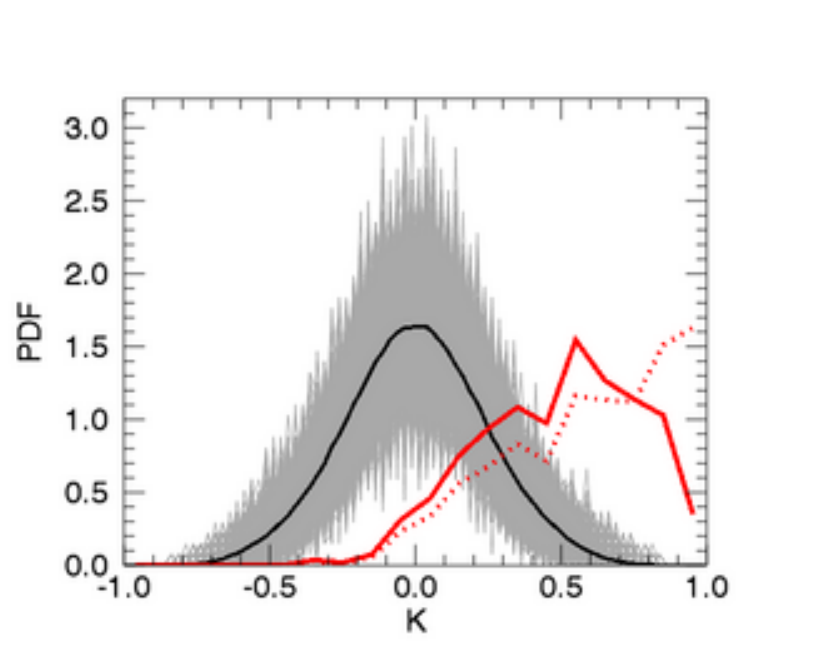}
\end{minipage}
\hspace{0.5cm}
\begin{minipage}{0.2\textwidth}
\centering
\includegraphics[trim=0cm 0cm 0cm 0cm,clip=true,width=4.1cm]{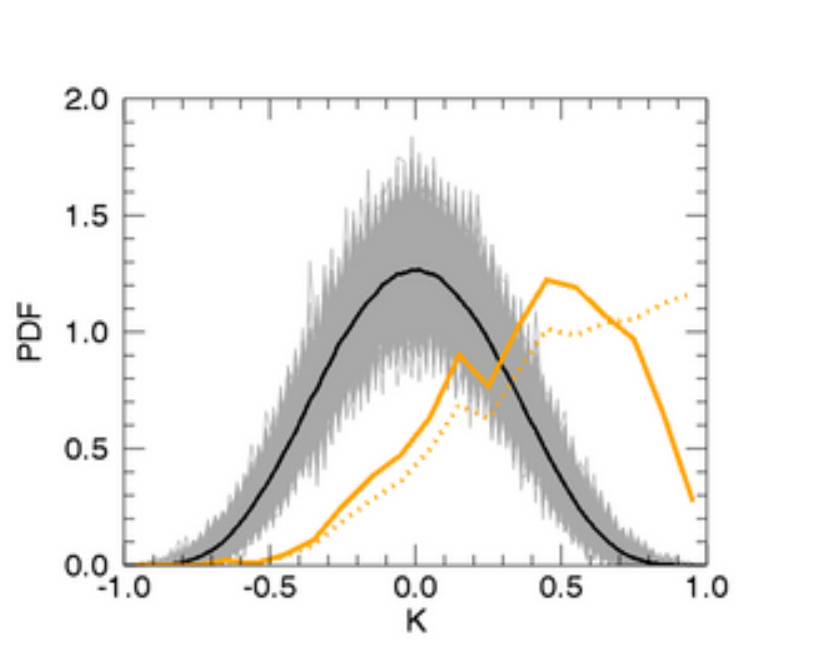}
\end{minipage}
\hspace{0.5cm}
\begin{minipage}{0.2\textwidth}
\centering
\includegraphics[trim=0cm 0cm 0cm 0cm,clip=true,width=4.1cm]{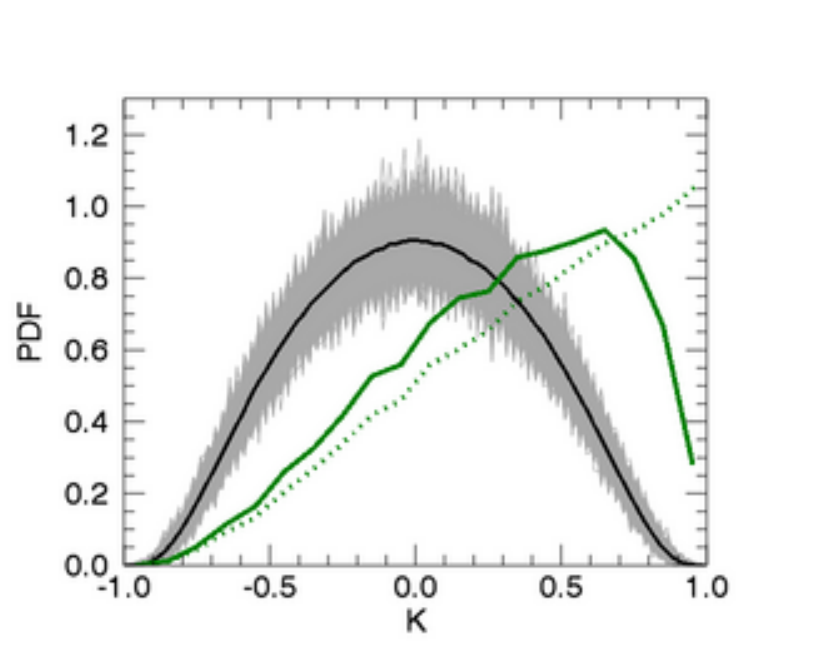}
\end{minipage}
\hspace{0.5cm}
\begin{minipage}{0.2\textwidth}
\centering
\includegraphics[trim=0cm 0cm 0cm 0cm,clip=true,width=4.1cm]{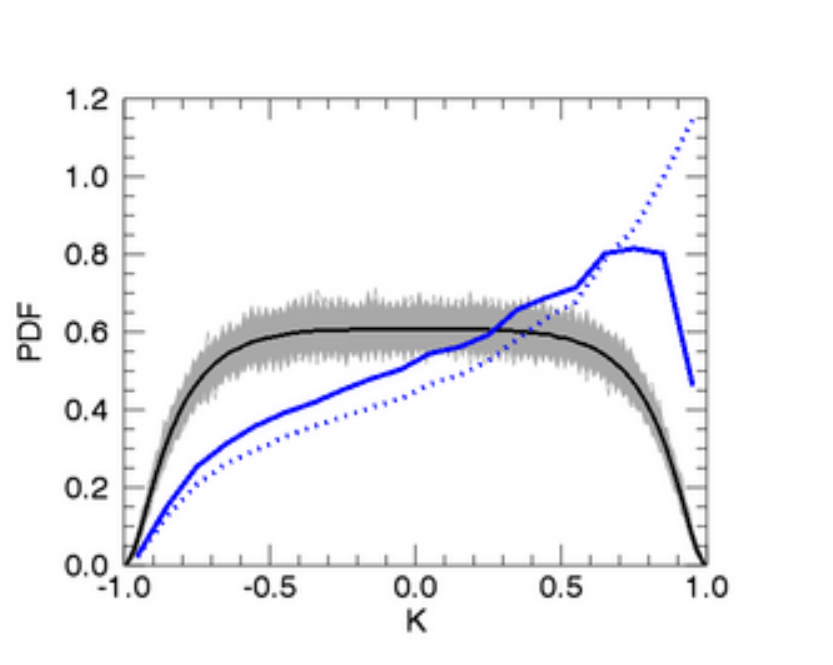}
\end{minipage}
\vspace{-0.2cm}
\begin{minipage}{0.2\textwidth}
\centering
\includegraphics[trim=0cm 0cm 0cm 0cm,clip=true,width=4.1cm]{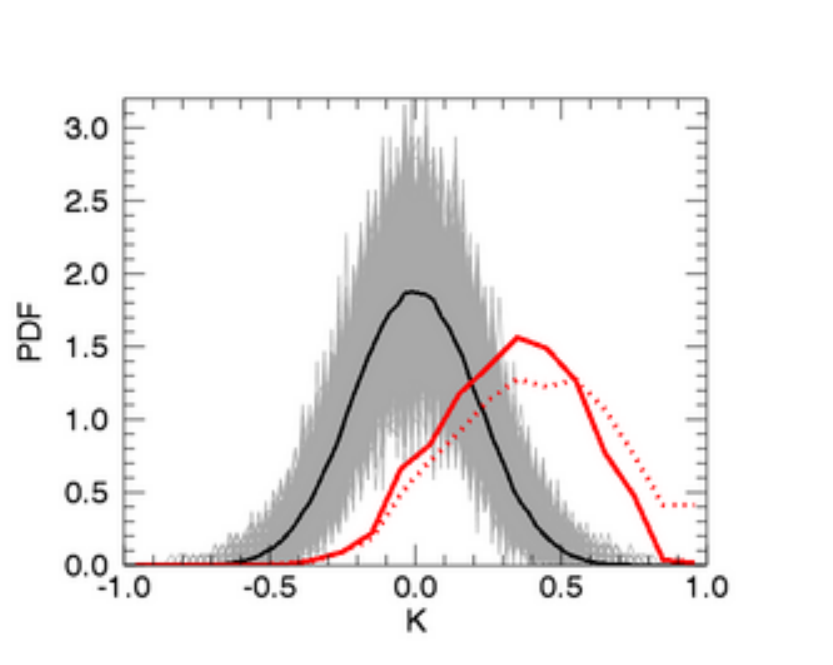}
\end{minipage}
\hspace{0.5cm}
\begin{minipage}{0.2\textwidth}
\centering
\includegraphics[trim=0cm 0cm 0cm 0cm,clip=true,width=4.1cm]{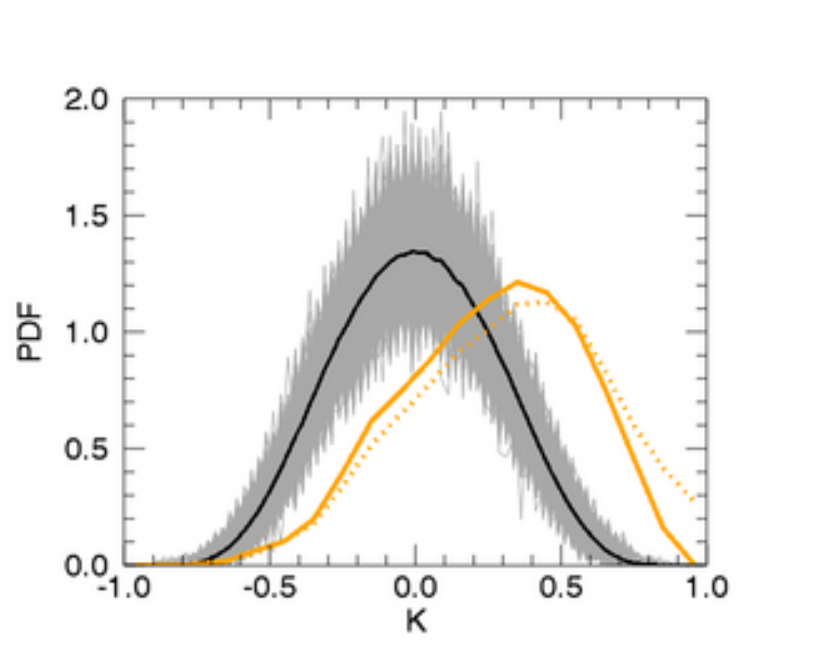}
\end{minipage}
\hspace{0.5cm}
\begin{minipage}{0.2\textwidth}
\centering
\includegraphics[trim=0cm 0cm 0cm 0cm,clip=true,width=4.1cm]{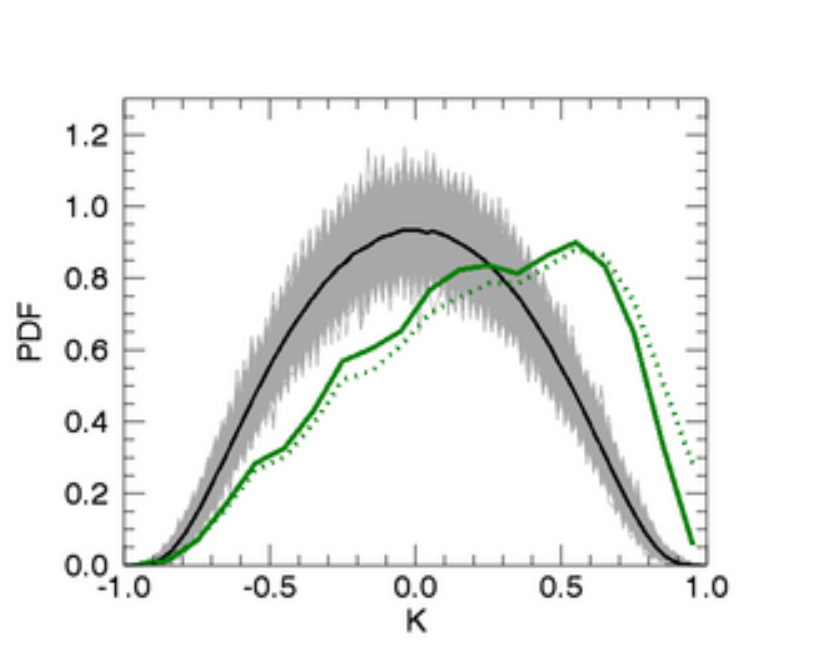}
\end{minipage}
\hspace{0.5cm}
\begin{minipage}{0.2\textwidth}
\centering
\includegraphics[trim=0cm 0cm 0cm 0cm,clip=true,width=4.1cm]{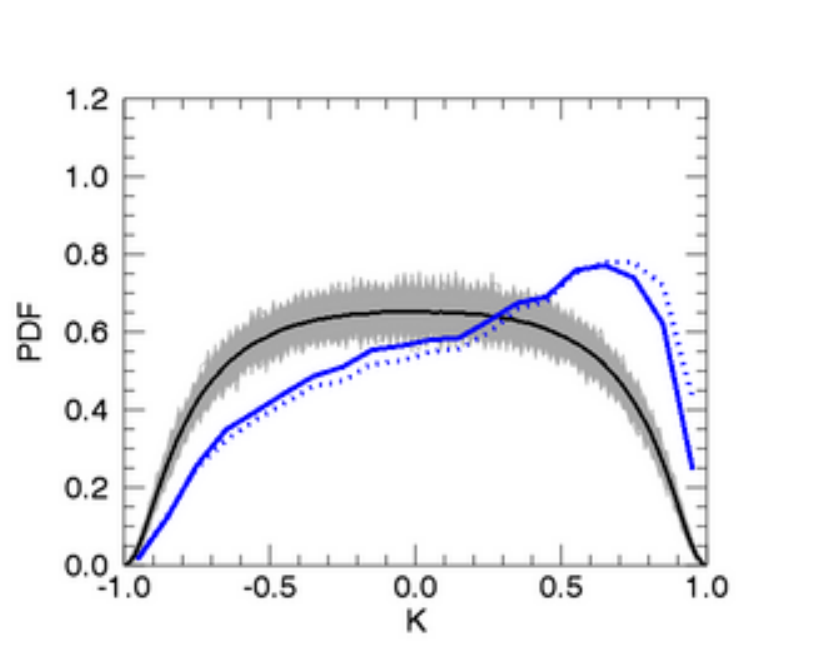}
\end{minipage}
\end{center}
\caption{Same as figure \ref{Adist1}, but here, weighted with weighting coefficients $w_i(n=2)$}
\label{Adist2}
\end{figure}

\begin{figure}[H]
\begin{center}
\begin{minipage}{0.2\textwidth}
\centering
\includegraphics[trim=0cm 0cm 0cm 0cm,clip=true,width=4.1cm]{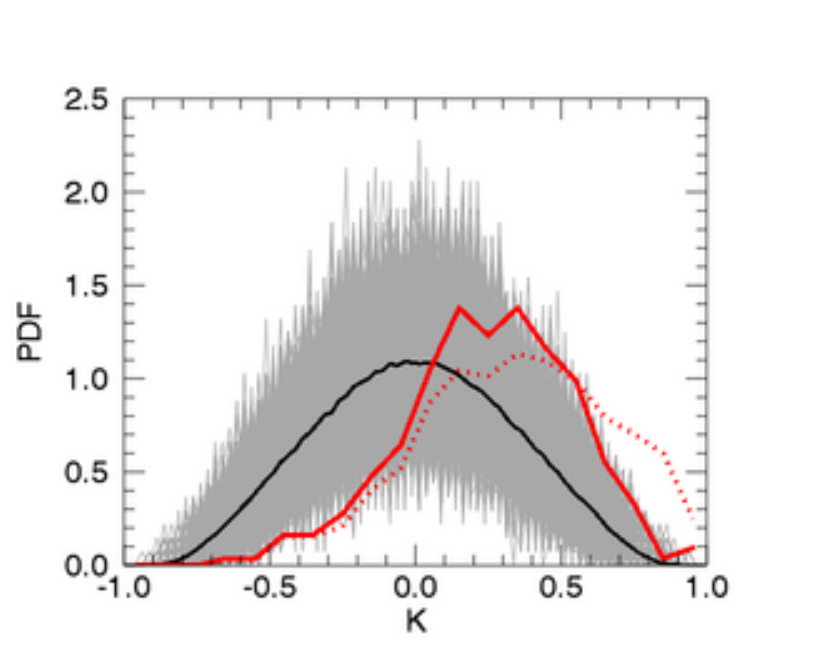}
\end{minipage}
\hspace{0.5cm}
\begin{minipage}{0.2\textwidth}
\centering
\includegraphics[trim=0cm 0cm 0cm 0cm,clip=true,width=4.1cm]{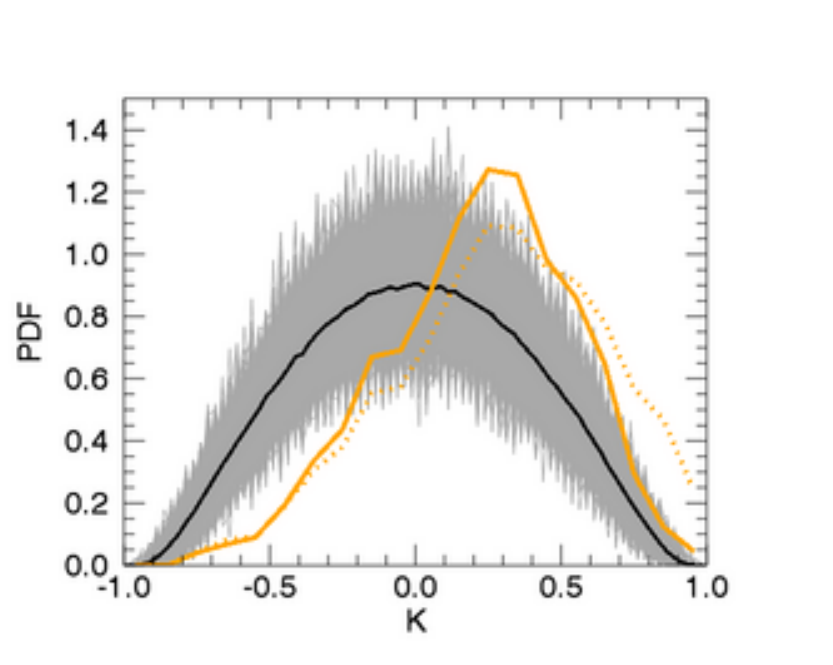}
\end{minipage}
\hspace{0.5cm}
\begin{minipage}{0.2\textwidth}
\centering
\includegraphics[trim=0cm 0cm 0cm 0cm,clip=true,width=4.1cm]{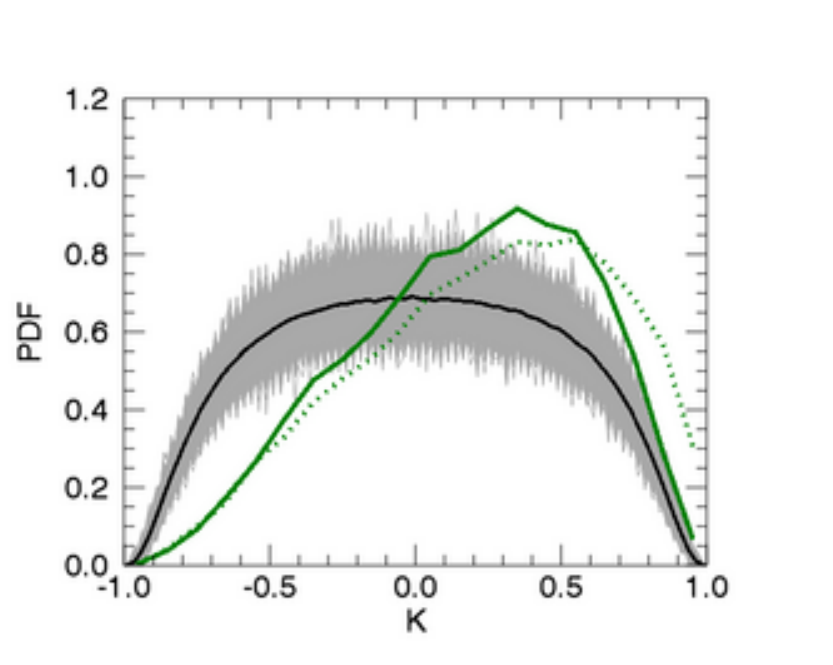}
\end{minipage}
\hspace{0.5cm}
\begin{minipage}{0.2\textwidth}
\centering
\includegraphics[trim=0cm 0cm 0cm 0cm,clip=true,width=4.1cm]{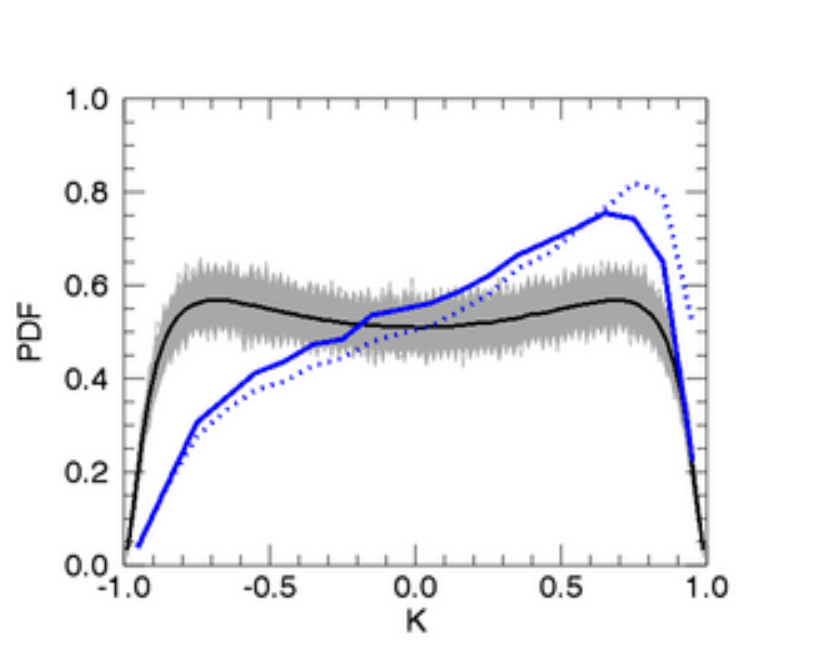}
\end{minipage}
\vspace{-0.2cm}
\begin{minipage}{0.2\textwidth}
\centering
\includegraphics[trim=0cm 0cm 0cm 0cm,clip=true,width=4.1cm]{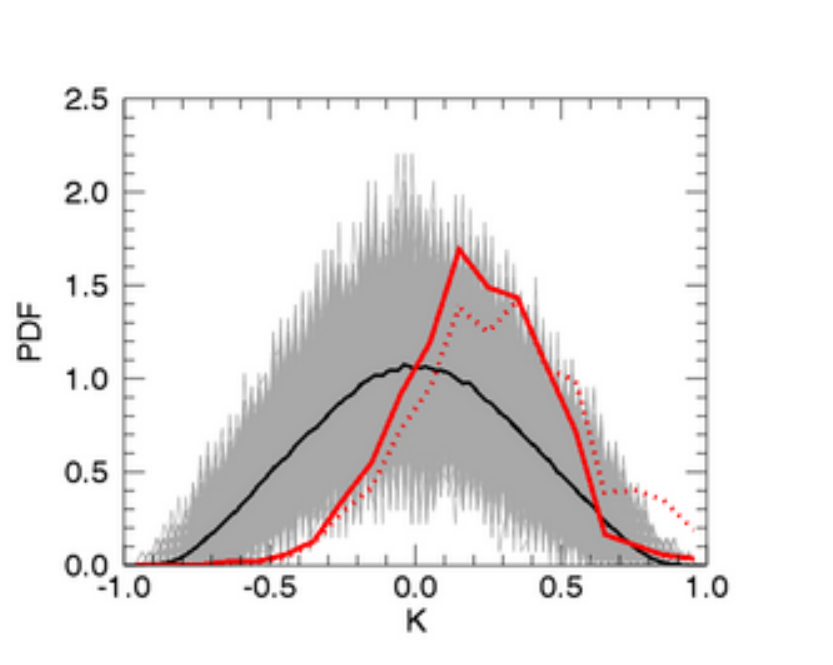}
\end{minipage}
\hspace{0.5cm}
\begin{minipage}{0.2\textwidth}
\centering
\includegraphics[trim=0cm 0cm 0cm 0cm,clip=true,width=4.1cm]{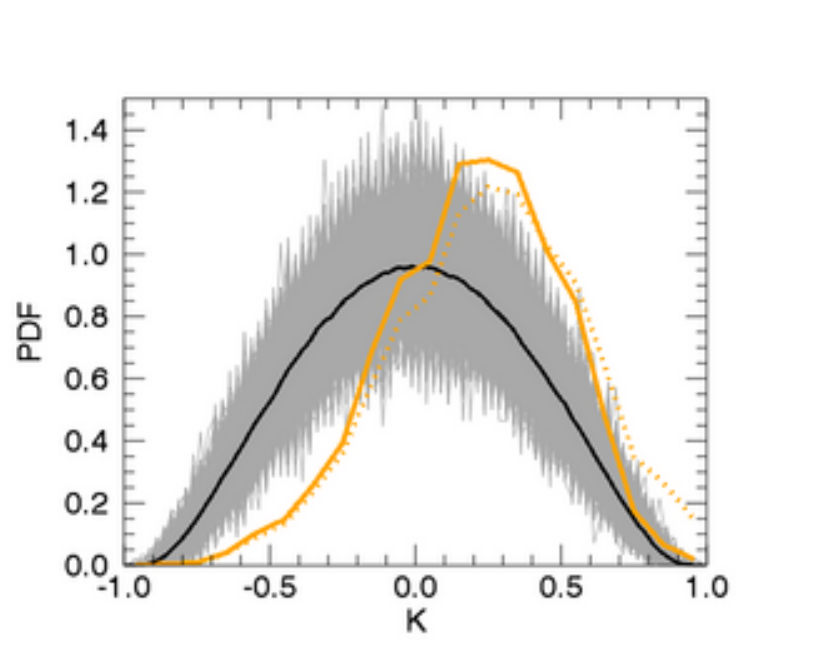}
\end{minipage}
\hspace{0.5cm}
\begin{minipage}{0.2\textwidth}
\centering
\includegraphics[trim=0cm 0cm 0cm 0cm,clip=true,width=4.1cm]{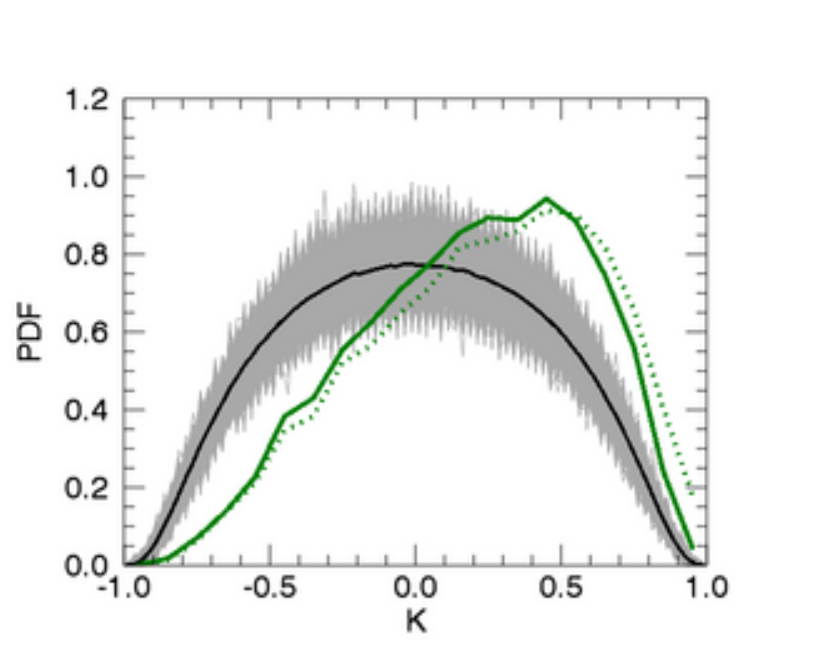}
\end{minipage}
\hspace{0.5cm}
\begin{minipage}{0.2\textwidth}
\centering
\includegraphics[trim=0cm 0cm 0cm 0cm,clip=true,width=4.1cm]{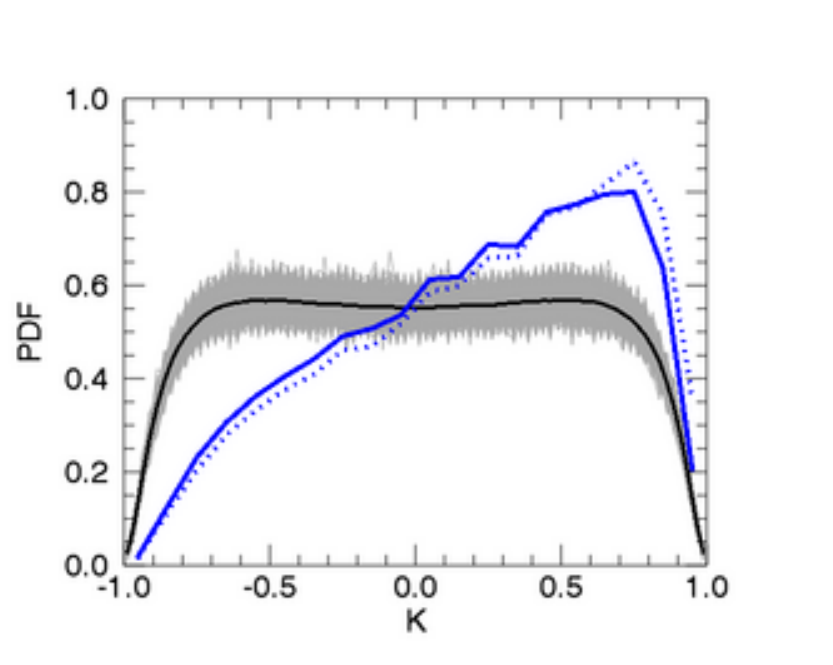}
\end{minipage}
\vspace{-0.2cm}
\begin{minipage}{0.2\textwidth}
\centering
\includegraphics[trim=0cm 0cm 0cm 0cm,clip=true,width=4.1cm]{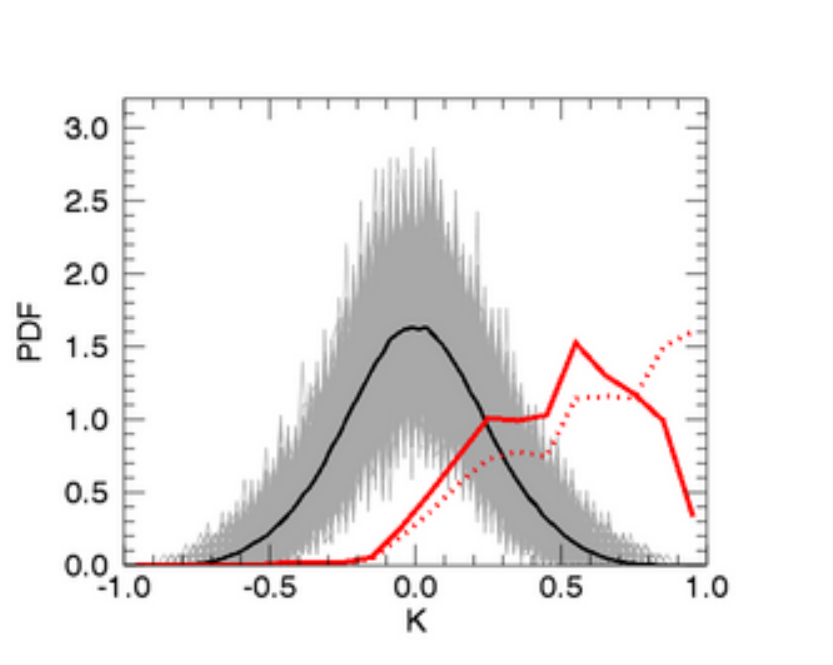}
\end{minipage}
\hspace{0.5cm}
\begin{minipage}{0.2\textwidth}
\centering
\includegraphics[trim=0cm 0cm 0cm 0cm,clip=true,width=4.1cm]{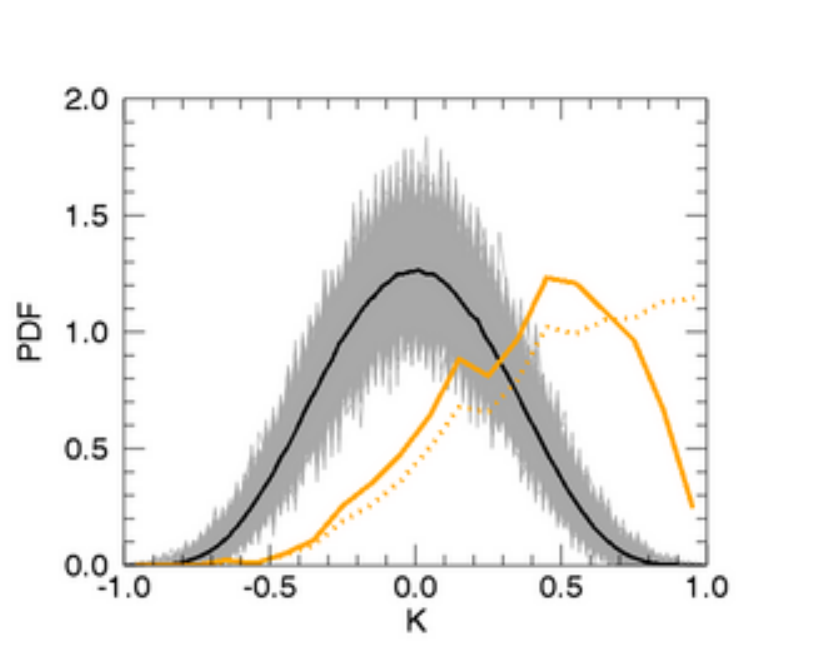}
\end{minipage}
\hspace{0.5cm}
\begin{minipage}{0.2\textwidth}
\centering
\includegraphics[trim=0cm 0cm 0cm 0cm,clip=true,width=4.1cm]{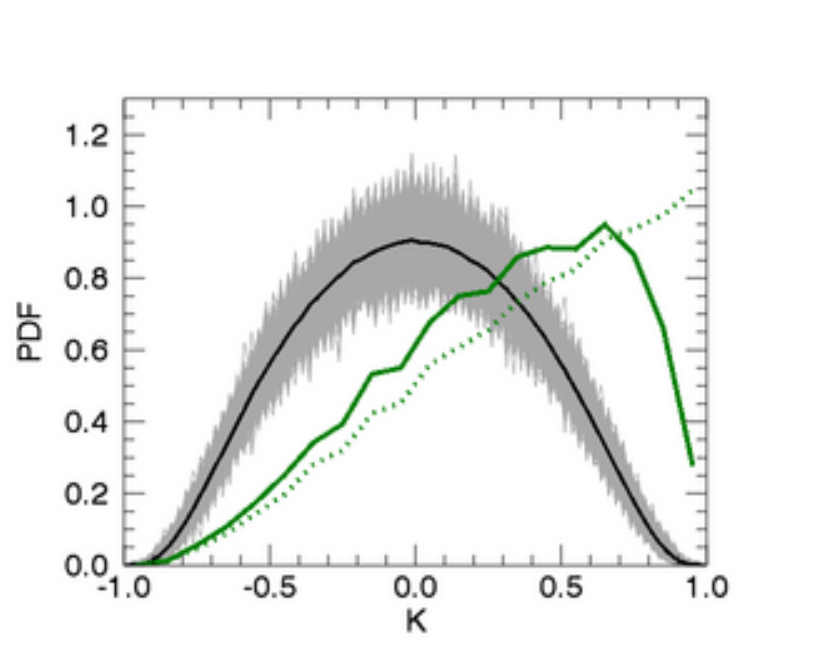}
\end{minipage}
\hspace{0.5cm}
\begin{minipage}{0.2\textwidth}
\centering
\includegraphics[trim=0cm 0cm 0cm 0cm,clip=true,width=4.1cm]{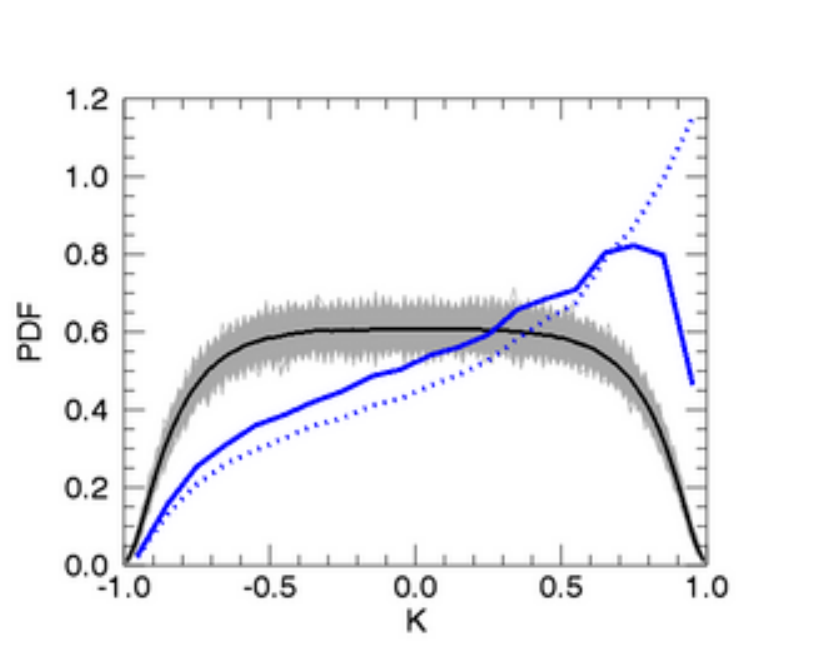}
\end{minipage}
\vspace{-0.2cm}
\begin{minipage}{0.2\textwidth}
\centering
\includegraphics[trim=0cm 0cm 0cm 0cm,clip=true,width=4.1cm]{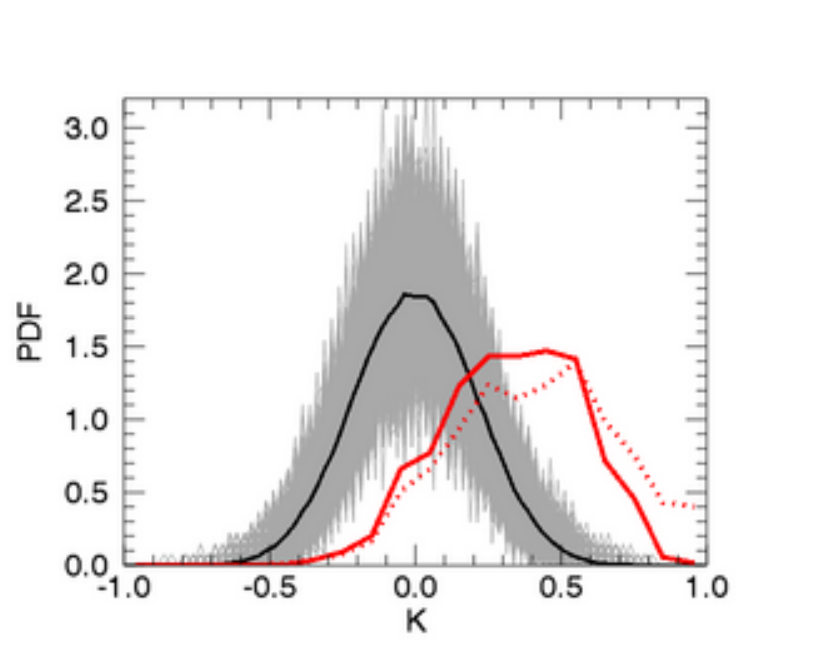}
\end{minipage}
\hspace{0.5cm}
\begin{minipage}{0.2\textwidth}
\centering
\includegraphics[trim=0cm 0cm 0cm 0cm,clip=true,width=4.1cm]{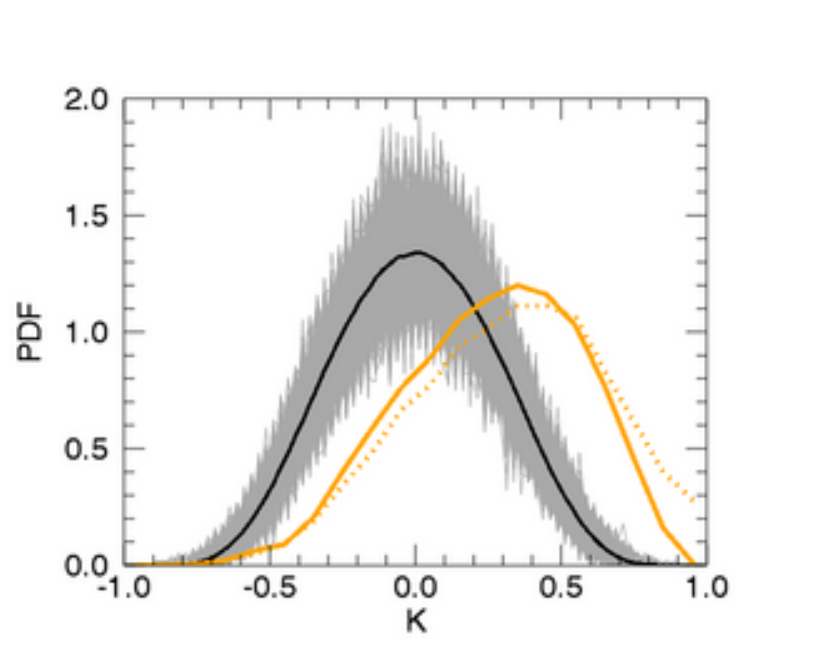}
\end{minipage}
\hspace{0.5cm}
\begin{minipage}{0.2\textwidth}
\centering
\includegraphics[trim=0cm 0cm 0cm 0cm,clip=true,width=4.1cm]{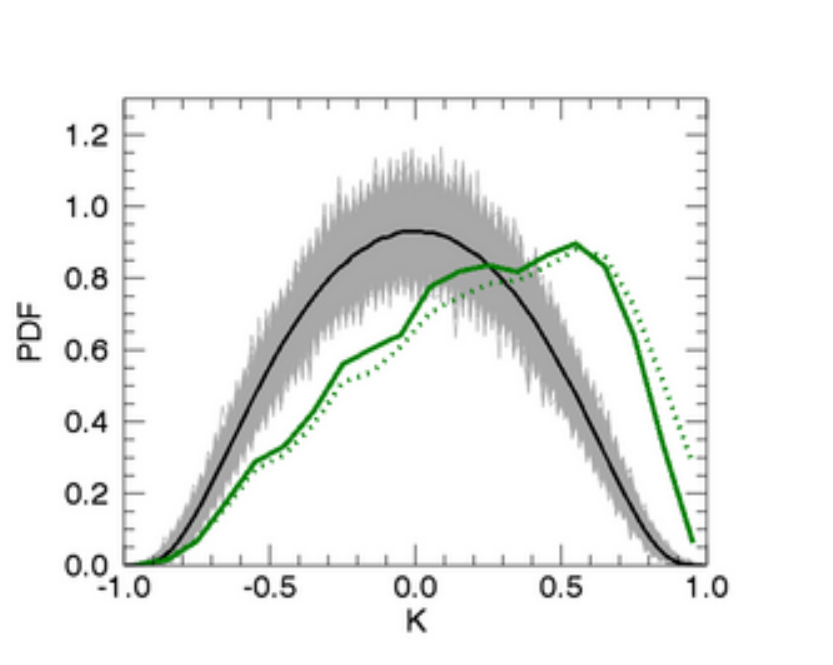}
\end{minipage}
\hspace{0.5cm}
\begin{minipage}{0.2\textwidth}
\centering
\includegraphics[trim=0cm 0cm 0cm 0cm,clip=true,width=4.1cm]{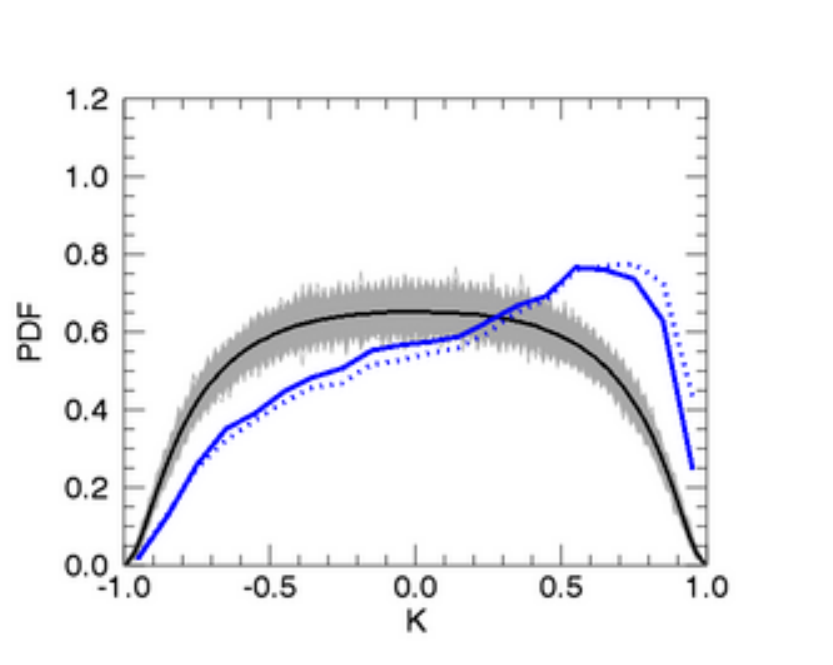}
\end{minipage}
\end{center}
\caption{Same as figure \ref{Adist1}, but here, weighted with weighting coefficients $w_i(n=10)$}
\label{Adist10}
\end{figure}

\end{document}